%% file: Tim_paper.tex
\documentclass[a4paper,11pt]{article}
\pdfoutput=1 

\usepackage{jcappub} 

\usepackage{graphicx}
\usepackage[dvipsnames]{xcolor}
\usepackage{soul}
\usepackage{amstext}
\usepackage{bm}				
\usepackage{booktabs}		
\usepackage{cleveref}		
\usepackage[section]{placeins}	
\usepackage{pgfplots}		
\usepackage{tabularx}
\usepackage{multirow}
\usepackage{arydshln}
\newcolumntype{H}{>{\centering\arraybackslash}X}

\newcommand{\dd}{\mathrm{d}} 

\newcommand{\be}{\begin{equation}}
\newcommand{\ee}{\end{equation}}
\newcommand{\ba}{\begin{eqnarray}}
\newcommand{\ea}{\end{eqnarray}}

\title{Cosmology in the era of Euclid and the Square Kilometre Array}

\author[a]{Tim Sprenger}
\author[a]{, Maria Archidiacono} 
\author[a]{, Thejs Brinckmann}
\author[b,c,a]{, S\'ebastien Clesse}
\author[a]{and Julien Lesgourgues} 

\affiliation[a]{Institute for Theoretical Particle Physics and Cosmology (TTK), RWTH Aachen University, D-52056 Aachen, Germany}
\affiliation[b]{Centre for Cosmology, Particle Physics and Phenomenology (CP3), Institut de Recherche en Mathematique et Physique (IRMP), Louvain University, 2 Chemin du Cyclotron, 1348 Louvain-la-Neuve, Belgium}
\affiliation[c]{Namur Center of Complex Systems (naXys), Department of Mathematics, University of Namur, Rempart de la Vierge 8, 5000 Namur, Belgium}

\emailAdd{tsprenger@physik.rwth-aachen.de} 
\emailAdd{archidiacono@physik.rwth-aachen.de} 
\emailAdd{brinckmann@physik.rwth-aachen.de}
\emailAdd{sebastien.clesse@uclouvain.be, sebastien.clesse@unamur.be}
\emailAdd{Julien.Lesgourgues@physik.rwth-aachen.de} 

\date{\today}

\abstract{
Theoretical uncertainties on non-linear scales are among the main obstacles to exploit the sensitivity of forthcoming galaxy and hydrogen surveys like Euclid or the Square Kilometre Array (SKA).
Here, we devise a new method to model the theoretical error that goes beyond the usual cut-off on small scales.
The advantage of this more efficient implementation of the non-linear uncertainties is tested through
a Markov--Chain--Monte--Carlo (MCMC) forecast of the sensitivity of Euclid and SKA to the parameters of the standard $\Lambda$CDM model, including massive neutrinos with total mass $M_\nu$, and to 3 extended scenarios, including 1) additional relativistic degrees of freedom ($\Lambda$CDM + $M_\nu$ + $N_\mathrm{eff}$), 2) a deviation from the cosmological constant ($\Lambda$CDM + $M_\nu$ + $w_0$), and 3) a time-varying dark energy equation of state parameter ($\Lambda$CDM + $M_\nu$ + $\left(w_0,w_a \right)$).
We compare the sensitivity of 14 different combinations of cosmological probes and experimental configurations.
For Euclid combined with Planck, assuming a plain cosmological constant, our method gives robust predictions for a high sensitivity to the primordial spectral index $n_{\rm s}$ ($\sigma(n_s)=0.00085$), the Hubble constant $H_0$ ($\sigma(H_0)=0.141 \, {\rm km/s/Mpc}$), the total neutrino mass $M_\nu$ ($\sigma(M_\nu)=0.020 \, {\rm eV}$). Assuming dynamical dark energy we get $\sigma(M_\nu)=0.030 \, {\rm eV}$ for the mass and $(\sigma(w_0), \sigma(w_a)) = (0.0214, 0.071)$ for the equation of state parameters.
The predicted sensitivity to $M_\nu$ is mostly stable against the extensions of the cosmological model considered here.
Interestingly, a significant improvement of the constraints on the extended model parameters is also obtained when combining Euclid with a low redshift HI intensity mapping survey by SKA1, demonstrating the importance of the synergy of Euclid and SKA.
}

\begin{document}

\hfill{\small TTK-18-04, CP3-18-06}

\maketitle

\input{Intro}

\input{GC}
\input{CS}
\input{IM}
\input{theoerr}
\input{datasets}
\input{Mnu}
\input{ext}

\input{Conclusions}

\section*{Acknowledgement}
We would like to thank Andrej Obuljen, Emanuele Castorina, Francisco Villaescusa-Navarro and Matteo Viel for pointing us at the importance of defining the neutrino galaxy bias with respect to the $P_{cb}(k,z)$ spectrum.
Simulations were performed with computing resources granted by RWTH Aachen University under project \texttt{thes0287} and with computing resources granted by JARA-HPC from RWTH Aachen University under \texttt{jara0184}.
The work of SC is supported by the Belgian Fund for Research FRS-FNRS, under a \textit{Charg\'e de Recherche} grant, and that of TB by the Deutsche ForschungsGemeinschaft (DFG).

\bibliography{references,refs_montepython,refs_core,refs_mnu_tau,refs_euclid_ska,refs_bias,refs_intro,refs_extra_v3}

\bibliographystyle{JHEP}

\appendix
\input{AppGC}

\input{AppCS}
\input{AppIM} 

\end{document}

%% file: Intro.tex
\section{Introduction} 

During the last three decades, the increasingly accurate observations of the Cosmic Microwave Background (CMB) anisotropies have promoted Cosmology to a precision science.
  Combined with other probes such as Type-1a supernovae, the statistical distribution of Large Scale Structures (LSS), the weak gravitational lensing and the Lyman-$\alpha$ forest, the Planck satellite has measured the standard cosmological parameters with an accuracy down to the percent level~\cite{Ade:2015xua}.   Until now, however, the nature of Dark Energy and Dark Matter accounting for about 95\% of the density of the Universe still remains a deep mystery.   In the next decade, galaxy surveys like Euclid~\cite{Laureijs:2011gra,Amendola:2012ys} and the Square Kilometre Array (SKA)~\cite{Maartens:2015mra,Santos:2015gra} will take over and will probe the growth of LSS with an unprecedented precision, up to redshifts $z\sim 3$.    Compared to the CMB that is a snapshot of the early Universe, galaxy surveys will achieve a tomography of the Universe over its last twelve billion years.   In addition, the SKA will also achieve a precise map of neutral hydrogen through 21-cm intensity mapping, a tracer of the LSS distribution, back to the reionization era and even the cosmic dawn~\cite{Pritchard:2015fia}, up to redshift $z\sim 20$.    Euclid and the SKA will detect billions of galaxies, over a large fraction of the sky, and will set unprecedented constraints on the various cosmological scenarios, through galaxy clustering and weak gravitational lensing.   Three major challenges of Euclid and SKA are to reveal the properties of Dark Energy and Dark Matter, and to measure the cosmological neutrino mass.   
  
Euclid and SKA will reach the required sensitivity and angular resolution to probe the non-linear growth of structures on small scales.  Their figure of merit will therefore strongly depend on our understanding of the various, often complex physical processes at play on these scales.  This includes general relativistic corrections to nonlinear structure formation (see e.g.~\cite{Tansella:2017rpi} and references therein), the accuracy of Newtonian N-body simulations~\cite{Schneider:2015yka} and fitting methods~\cite{Takahashi:2012em,Mead:2015yca} or emulators~\cite{Casarini:2016ysv}, the galaxy non-linear bias~\cite{Jennings:2015lea}, the baryonic feedback~\cite{vanDaalen:2011xb,Schneider:2015wta,Rabold:2017qdp}, the intrinsic alignment of galaxies~\cite{Hilbert:2016ylf}, etc...  Usually, when doing forecasts, these considerations lead to the introduction of a maximal wavenumber $k_{\rm max}$, below which one trusts the theoretical prediction of the matter power spectrum, and above which the physical uncertainties are expected to exceed the experimental noise.   For instance, $k_{\rm max} = 0.2 h \rm{Mpc}^{-1}$ ($h$ defining the Hubble expansion rate today $H_0 = h \times 100 \rm{km/s/Mpc}$) was often used in Euclid forecasts~\cite{Laureijs:2011gra,Amendola:2012ys}.   Introducing such a cut-off scale means that all the information obtained by the experiment on smaller scales is simply unexploited.     However, this information is crucial: Indeed,  the non-linear growth of structure can be significantly altered in theories of modified gravity~\cite{Brax:2013mua,Brax:2012nk,Winther:2015wla}, interacting/decaying dark matter, massive neutrinos~\cite{Wright:2017dkw}. Moreover, a larger lever arm allows to get better constraints on the initial shape of the power spectrum of density fluctuations from inflation~\cite{Basse:2014qqa}.   

Therefore, new methods need to be developed and tested in order to take into account the non-linear theoretical uncertainties, while optimizing the amount of information relevant for cosmology.  For instance, it has been proposed to introduce either an error on the power spectrum, totally uncorrelated between wavelength modes, or on the contrary, a correlated error increasing at small scales~\cite{Audren:2012vy}.  Significant differences in forecasts were found between these two extremes.
Both cases are actually unrealistic. Correlated errors assume such a good understanding of all sources of theoretical errors that they can be reduced to one (or a few) templates with known dependence on $(k,z)$ but unknown normalization. On the other hand, uncorrelated errors allow for erratic deviations between the predicted and fitted theoretical spectra, that no inaccuracy in the physical modeling of the observables could justify, and tend to fit random noise features in the observational data. Moreover, when introducing uncorrelated errors, it is very difficult to find a well-motivated prescription for assigning a standard error deviation to each independent degree of freedom, as we shall discuss later in Section~\ref{sec:GC_IM_error}. Reference~\cite{Baldauf:2016sjb} advocates a third approach based on the notion of correlation length of the error in $(k,z)$ space, which has clear physical motivation, but is computationally expensive and difficult to implement in an efficient parameter extraction pipeline.

In this paper, we introduce a new numerical method to take into account realistically the theoretical uncertainties on the non-linear spectra. Two cases, based on current knowledge and expected improvements, will be presented:  first, a \textit{conservative} case, based on the present theoretical non-linear uncertainties combined with a conservative redshift dependent cut-off scale, and second, a \textit{realistic} case in which we consider some expected and realistic refinement in the modelisation of nonlinear effects, e.g. simply due to the increasing numerical resources by the time the real data will be available in the 2020's.   These two cases allow us to present what we think are the most realistic range for the future constraints on the cosmological parameters of the standard cosmological model, including the sum of neutrino masses, as well as on the parameters of common extended cosmological scenarios.   For the first time, we use such a method to derive realistic forecasts for both Euclid and the SKA, and for the combination of them, using three probes: galaxy clustering power spectrum, cosmic shear angular power spectrum, and 21-cm intensity mapping subsequent to reionization.  For this purpose, we have used the Bayesian Markov--Chains--Monte--Carlo (MCMC) technique, rather than the Fisher matrix formalism that might be subject to numerical instabilities particularly in non-standard cosmological scenarios.

The paper is organized as follows:  In Sections 2, 3 and 4, we introduce respectively the calculations of the galaxy clustering power spectrum, the weak lensing angular power spectrum, and the 21cm intensity mapping, as well as the related experimental uncertainties due to Euclid and SKA specifications.   In Section 5, the method used to model the nonlinear theoretical uncertainties is described.    After summarizing the considered datasets in Section 6, forecasts for the baseline $\Lambda$CDM + $M_\nu$ model are presented in Section 7, as well as for a series of extended models in Section 8.    We summarize our results and discuss some perspectives in the conclusion.  Finally, a description of the construction of the different likelihoods is included in the Appendix.

%% file: GC.tex
\section{Galaxy clustering}
\label{sec:GC}

\subsection{Galaxy power spectrum}

The spatial distribution of galaxies represents a biased tracer of the underlying dark matter distribution. Therefore, various effects have to be taken into account when converting the 
matter power spectrum\footnote{As explained later in sec.~\ref{sec:baseline}, when the model features massive neutrinos, we don't plug in here the total matter power spectrum $P_m(k,z)$, but only the power spectrum of baryons and CDM $P_{cb}(k,z)$, since the galaxy power spectrum is more a tracer of the latter quantity, see e.g.~\cite{Castorina:2013wga,Castorina:2015bma,Vagnozzi:2018pwo}.} $P_m$
into the
observed galaxy power spectrum $P_g$:
\begin{equation}
\label{P_g-def}
P_g(k,\mu,z) =  f_{\text{AP}}(z) \times f_{\text{res}}(k,\mu,z) \times f_{\text{RSD}}(\hat{k},\hat{\mu},z) \times b^2(z) \times P_m(\hat{k},z) \ .
\end{equation}
Before proceeding to explain the different effects and the associated functions $f_i$ contributing to this formula,
let us notice that we have employed a flat-sky approximation~\cite{Lemos:2017arq,Asgari:2016txw} that allows for an unambiguous definition of the angle between the Fourier modes $\bm{k}$ and the line of sight distance vector $\bm{r}$. The observer's fixed point of view breaks the isotropy of the matter power spectrum, but symmetry in perpendicular directions to the line of sight is preserved. Hence, the following coordinates are sufficient to describe all the effects:
\begin{equation}
k = \vert\bm{k}\vert \ , \ \mu = \frac{\bm{k}\cdot\bm{r}}{k r} \ .
\end{equation}
The parallel part of a mode is given by $k_{\shortparallel} = \mu k$ and the perpendicular one by $k_{\perp} = k\sqrt{1-\mu^2}$.

Since we can observe only the redshift and the position in the sky, in order to get a distribution in three-dimensional space, we need to make assumptions on the underlying cosmology\footnote{This is actually one of the reasons for which alternative methods to express the 2-point statistics of galaxy distributions are being discussed in the literature, see e.g.\,\cite{Tansella:2017rpi} and references therein.}. However, physical quantities calculated within this fiducial cosmology may differ from the corresponding values in the true/real cosmology (hereafter denoted by $\hat{}$, e.g.\,$\hat{H}$). The Fourier modes of real space can be related to those of the fiducial space via
\begin{equation}
\label{k^s}
\hat{k}^2 = \left[\left(\frac{\hat{H}}{H}\right)^2\mu^2 + \left(\frac{D_A}{\hat{D}_A}\right)^2\left(1-\mu^2\right)\right] k^2
\end{equation}
and
\begin{equation}
\label{mu^s}
\hat{\mu}^2 = \left(\frac{\hat{H}}{H}\right)^2 \mu^2 \cdot \left[\left(\frac{\hat{H}}{H}\right)^2 \mu^2 + \left(\frac{D_A}{\hat{D}_A}\right)^2\left(1-\mu^2\right)\right]^{-1} \ ,
\end{equation}
where $H$ and $D_A$ are, respectively, the Hubble parameter and the angular diameter distance as functions of redshift $z$. The change in the power spectrum when extracted from the same data but assuming different cosmologies, the so-called Alcock-Paczinsky effect, gives rise to the first term in Eq.~\ref{P_g-def}:
\begin{equation}
f_{\text{AP}}(z) = \frac{D_A^2 \hat{H}}{\hat{D}_A^2 H} \ .
\end{equation}
The second term in Eq.~\ref{P_g-def} is due to the limited resolution of instruments suppressing the apparent perturbations on small scales. Assuming Gaussian errors $\sigma_{\shortparallel}(z)$ and $\sigma_{\perp}(z)$ on coordinates parallel and perpendicular to the line of sight at redshift $z$, the suppression factor turns out to be exponential:
\begin{equation}
f_{\text{res}}(k,\mu,z) = \exp\left(-k^2\left[\mu^2\cdot\left(\sigma_{\shortparallel}^2(z)-\sigma_{\perp}^2(z)\right)+\sigma_{\perp}^2(z)\right]\right) \ .
\end{equation}
Since Fourier modes scale inversely w.r.t. spatial distances under a change of cosmology, the above factor is independent of the assumed cosmology.
 
The cosmological redshift, that is used to obtain the spatial coordinates, is not the only source of redshift. The classical Doppler effect induces an apparent anisotropy in the redshift-space power spectrum. On large scales within the linear regime, this effect is described by the Kaiser formula~\cite{Kaiser:1987qv}. On top of this large scale infall, additional random peculiar velocities of the galaxies further distort the redshift information on smaller scales, leading to features in redshift-space called fingers of God
\cite{Jackson:2008yv,1978IAUS...79...31T}. Following Ref.~\cite{Bull:2014rha}, we describe this additional suppression with an exponential factor. To sum up, the redshift effects encoded in the third term of Eq.~\ref{P_g-def}
are given by
\begin{equation}
f_{\text{RSD}}(\hat{k},\hat{\mu},z) = \left( 1+\beta(\hat{k},z) \, {\hat{\mu}}^2 \right)^2 e^{-{\hat{k}}^2{\hat{\mu}}^2\sigma_{\text{NL}}^2} \ ,
\end{equation}
where the first term in parentheses corresponds to the Kaiser formula and the exponential accounts for the fingers of God. In particular, $\sigma_{\text{NL}}$ has a fiducial value of 7\,Mpc and we allow it to vary between 4-10\,Mpc in our forecasts,
while $\beta$ is the (possibly scale-dependent) growth rate $f(\hat{k},z)$ corrected by the galaxy bias $b(z)$:
\begin{equation}
\label{eq:beta}
\beta(\hat{k},z) \equiv \frac{f(\hat{k},z)}{b(z)} \equiv \frac{1}{b(z)}\cdot\frac{\dd \ln\left(\sqrt{P_m(\hat{k},z)}\right)}{\dd \ln a} = -\frac{1+z}{2b(z)}\cdot\frac{\dd \ln P_m(\hat{k},z)}{\dd z} \ .
\end{equation}
The bias is a function of redshift which relates density perturbations in the galaxy field to dark matter density perturbations. We will assume the linear approximation $\delta_g = b(z) \times \delta_m$ where the bias is scale independent\footnote{We will take into account the consequences of non-linear bias later, either in our choice of a cut-off $k_{\rm max}$ or through our ansatz for the theoretical error function: this will be discussed in section~\ref{sec:GC_IM_error}.}. Approximate formulas for the bias are obtained by populating cosmological simulations with galaxies which will then be measured. In this case, $\delta_g$ and $\delta_m$ can be identified separately~\cite{Raccanelli:2017kht,LoVerde:2014pxa,Castorina:2013wga}.

We divide the surveys into bins of width $\Delta z = 0.1$ with mean redshift $\bar{z}$. Correlation functions are defined inside the bin's data and are approximated to probe the power spectrum at a fixed redshift $\bar{z}$. The volume of one redshift bin can be computed via
\begin{equation}
V_r(\bar{z}) = 4\pi f_{\text{sky}}\cdot\int_{\Delta r(\bar{z})}r^2 { \dd}r = \frac{4\pi}{3} f_{\text{sky}} \cdot\left[r^3\left(\bar{z}+\frac{\Delta z}{2}\right)-r^3\left(\bar{z}-\frac{\Delta z}{2}\right)\right] \ ,
\label{eq:Vr}
\end{equation}
where $f_{\text{sky}}$ is the fraction of the sky covered by the survey.
The distribution of galaxies is discrete, rather than continuous like the density field $\delta_g$.
Thus we have to take into account the experimental shot noise in each redshift bin:
\begin{equation}
P_{N}(\bar{z}) = \frac{1}{\bar{n}(\bar{z})} = \frac{V_r(\bar{z})}{N(\bar{z})} \ ,
\end{equation}
where $N(\bar{z})$ is the number of galaxies in the bin, $V_r(\bar{z})$ the volume of the bin and $\bar{n}(\bar{z})$ the galaxy number density. 
Taking this shot noise into account, the quantity actually measured by the experiment in each bin is
\begin{equation}
\label{eq:pobs}
P_\mathrm{obs}(k,\mu,\bar{z}) = P_g(k,\mu,\bar{z}) + P_{N}(\bar{z})~.
\end{equation}

\subsection{Euclid specifications}
\label{sec:GC:spec}

\begin{table}
\begin{center}
\caption{Euclid specifications.}
\label{EuclidSpec}
\begin{tabular}{c|ccccc} 
 \toprule 
parameter & $z_{\text{min}}$ & $z_{\text{max}}$ & $f_{\text{sky}}$ & $\sigma_{z}$ &$\sigma_{\theta}$ [$''$]\\
\midrule 
Euclid & 0.7 & 2.0 & 0.3636 & $0.001(1+z)$ & 0 \\
\bottomrule 
\end{tabular}
\end{center}
\end{table}
The redshift range accessible to Euclid is roughly $0.7 - 2.0$. Hence, mean redshifts of $\bar{z} = 0.75, \, 0.85$, ..., 1.95 are used. The error on spectroscopic redshift measurements is assumed to be $\sigma_z = 0.001(1+z)$, as in Refs.~\cite{Audren:2012vy,Amendola:2012ys}. The effect of angular resolution is neglected. Thus, $\sigma_{\perp}$ is set to 0. The specifications for Euclid are summarized in Table~\ref{EuclidSpec}. The redshift error can be propagated to the error on radial distance:
\begin{equation}
\sigma_{\shortparallel} = \frac{c}{H} \sigma_z \ .
\end{equation}
The galaxy number count distribution $\frac{\dd N(z)/\dd z}{1\text{deg}^2}$ has been taken from Table 3 of Ref.~\cite{Pozzetti:2016cch} (model 1) assuming a limiting flux of $3 \times 10^{-16}$\,erg\,s$^{-1}$\,cm$^{-2}$. Ref.~\cite{Pozzetti:2016cch} updates the results of Ref.~\cite{Geach:2009tm} which where used for the forecasts of Ref.~\cite{Audren:2012vy} and in the Euclid Definition Study Report (Ref.~\cite{Laureijs:2011gra}).

We use a sky fraction of $f_{\text{sky}}=0.3636$. The total number of detected galaxies in a given redshift bin can be inferred from the given values:
\begin{equation}
N(\bar{z}) = 41253 f_{\text{sky}} \text{deg}^2 \cdot\int_{\bar{z}-\frac{\Delta z}{2}}^{\bar{z}+\frac{\Delta z}{2}}\frac{\dd N(z)/\dd z}{1\text{deg}^2}\dd z \ .
\end{equation}
As done in Ref.~\cite{Audren:2012vy}, the bias factor corresponding to galaxies detected by Euclid is assumed to be close to the simple relation:
\begin{equation}
b(z) = \sqrt{1+z} \ .
\end{equation}
In order to account for inaccuracies in this relation, we have introduced two nuisance parameters with mean value $1$:
\begin{equation}
b(z) = \beta_0^{\text{Euclid}}(1+z)^{0.5\beta_1^{\text{Euclid}}} \ .
\end{equation}
A 5\%-precision (2$\sigma$) is taken as a prior on the $\beta$-factors.

\subsection{SKA specifications}

Currently, SKA1-MID Band 2 is the most promising option for a galaxy survey with SKA1. We use specifications according to the baseline design defined in Ref.~\cite{SKA_BaselineV2} (see Table~\ref{SKAspec}). We assume a survey area $S_{\text{area}} = f_{\text{sky}} \times 41253~\text{deg}^2$ in agreement with the optimization procedure described in Ref.~\cite{Yahya:2014yva}, while the frequency range of SKA2 is the same of Ref.~\cite{Bull:2015lja}.

The target signal of SKA is the HI line or 21cm line of cold neutral hydrogen with a rest frequency of $\nu_{0} = 1420\,\text{MHz}$. The frequency range translates through the redshifting of $\nu_{0}$ into a redshift range which has been rounded to fit redshift bins of width $\Delta z = 0.1$. The frequency $\nu$ and its error translate into redshifts:
\begin{align}
z &= \frac{\nu_{0}}{\nu}-1 \ , \\
\label{sigzsignu}
\sigma_z &= (1+z)^2 \frac{\sigma_{\nu}}{\nu_{0}} \ .
\end{align}
These relations are independent of cosmology. Therefore it is valid to treat $z$ as a direct observable as was done in the case of Euclid.\\
The number counts of detected galaxies and their bias w.r.t. the underlying dark matter distribution have to be extracted from simulations. This was done in Ref.~\cite{Yahya:2014yva} using the following fitting formula:
\begin{align}
\frac{\dd N(z)/\dd z}{1\text{deg}^2} &= 10^{c_1}z^{c_2}\exp(-c_3 z) \ , \\
b_{\text{HI}}(z) &= c_4 \exp(c_5 z) \ .
\end{align}
Ref.~\cite{Bull:2015lja} adapted the results to the current baseline design, obtaining the parameters listed in Table~\ref{c_fitpar}. Ref.~\cite{Yahya:2014yva} used a frequency resolution of 10\,kHz. Band 2 divided into 64,000 channels (see~\cite{SKA_BaselineV2}) yields a bandwidth of $\delta \nu = 12.7$\,kHz per channel, which verifies this number. The same approach yields $\delta \nu = 12.8$\,kHz for SKA2. By equating $\delta \nu$ to the full width at half maximum (FWHM), the approximation of a Gaussian error $\sigma_{\nu} = \delta \nu / \sqrt{8\ln2}$ can be made. This determines the error on the redshift measurement as described in Eq.~\ref{sigzsignu}.  Note that we are equating here the frequency sensitivity to the sensitivity of the mean frequency of a galaxy's signal. However, the fingers of God described by $\sigma_{NL}$, whose effect is indistinguishable from the redshift resolution, dominate the suppression of the power spectrum. Hence, this approximation is good enough.

The inaccuracy of the theoretical bias formula can be accommodated for with similar nuisance parameters as in the case of Euclid:
\begin{equation}
b(z) = c_4 \beta_0^{\text{SKA1/2}} \exp(c_5 \beta_1^{\text{SKA1/2}} z)~,
\end{equation}
where $\beta_0^{\text{SKA1/2}}$, $\beta_1^{\text{SKA1/2}}$ are assigned gaussian priors with mean value $1$ and standard deviation 0.025.
We also include the effect of angular resolution as a Gaussian error:
\begin{align}
\sigma_{\perp} &= (1+z) D_A \sigma_{\theta} \ , \\
\sigma_{\theta} &= \frac{1}{\sqrt{8\ln2}} \frac{\lambda_0}{B} (1+z) \ .
\end{align}
The FWHM of an interferometer is approximately given by the wavelength divided by the maximum baseline $B$. In the case of SKA, the wavelength is the redshifted rest wavelength $\lambda_{0} = 21.11$\,cm. The maximum baseline is $B\approx 150$\,km for SKA1 and $B\approx 3000$\,km for SKA2. However, given the large sky fraction, the survey is not expected to exploit the maximum resolution the array is capable of. The simulated number counts are valid for a 10,000 hour survey. Taking a conservative approach, we use a maximum baseline of 5\,km for both SKA1 and SKA2, corresponding to the diameter of the inner core of the array with a high density of dishes. Even with this approach, the effect of angular resolution remains insignificant for a galaxy survey.

\begin{table}
\centering
\caption{Fitting parameters. \cite{Bull:2015lja}}
\label{c_fitpar}
\begin{tabular}{c|ccccc}
\toprule
parameter & $c_1$ & $c_2$ & $c_3$ & $c_4$ & $c_5$ \\
\hline
SKA1 band 2 (5$\sigma$) & 5.450 & 1.310 & 14.394 & 0.616 & 1.017 \\
SKA2 (10$\sigma$) & 6.319 & 1.736 & 5.424 & 0.554 & 0.783 \\
\bottomrule
\end{tabular}
\end{table}

\begin{table}
\centering
\caption{SKA specifications. \cite{SKA_BaselineV2,Bull:2015lja}}
\label{SKAspec}
\begin{tabular}{c|ccccccc}
\toprule
parameter & $\nu_{\text{min}}$\,[MHz] & $\nu_{\text{max}}$\,[MHz] & $z_{\text{min}}$ & $z_{\text{max}}$ & $S_{\text{area}}$\,[deg$^2$] & $\delta \nu$\,[kHz] & $B$ [km]\\
\hline
SKA1 band 2 & 950 & 1760 & 0.00 & 0.50 & 5,000 & 12.7 & 150 (5) \\
SKA2 & 470 & 1290 & 0.10 & 2.00 & 30,000 & 12.8 & 3000 (5) \\
\bottomrule
\end{tabular}
\end{table}

%% file: CS.tex
\section{Cosmic shear}
\label{sec:CS}
\subsection{Angular power spectrum}
A cosmic shear survey maps the alignments of galaxies induced by weak gravitational lensing caused by large scale structures along the line of sight.
The cosmological information is extracted from auto- and cross-correlations of alignment maps at different redshifts. For SKA, a cosmic shear survey will be possible thanks to the detection of the continuum emission of galaxies.

The projected shear power spectrum of the redshift bins $i$ and $j$ at multipoles $\ell$ can be inferred from the three-dimensional matter power spectrum via
\begin{equation}
\label{Cl_Pk}
C_\ell^{ij} = \frac{9}{16}\Omega_m^2H_0^4 \int_0^{\infty}\frac{\dd r}{r^2} g_i(r) g_j(r) P\left(k=\frac{\ell}{r},z(r)\right) \ .
\end{equation}
The functions $g_i(r)$ depend on the radial distribution of galaxies in the redshift bin $i$, i.e. on the convolution of the distribution of detected galaxies with the corresponding redshift errors:
\begin{align}
g_i(r) &= 2 r (1+z(r)) \int_r^\infty \dd r' \frac{\eta_i(r') (r'-r)}{r'} \ , \\
\eta_i(r) &= H(r) n_i(z(r)) \ , \\
n_i(z) &= \frac{D_i(z)}{\int_0^\infty D_i(z') \dd z'} \ , \\
D_i(z) &= \int_{z_i^{\rm min}}^{z_i^{\rm max}}  {\cal P}(z,z') \, \frac{\dd n_{\text{gal}}}{\dd z}(z') \, \dd z' \ .
\end{align}
Due to the intrinsic alignment of galaxies, there is also a noise contribution $N_{\ell}$. The noise spectrum added to the theoretical $C_\ell^{ij}$ is
\begin{equation}
N_{\ell}^{ij} = \delta_{ij}\sigma_{\text{shear}}^2 n_i^{-1} \ ,
\end{equation}
where $\sigma_{\text{shear}}$ is the root mean square of the galaxy intrinsic ellipticity and is set to $0.3$,
and $n_i$ is the number of galaxies per steradian in the $i$'th redshift bin. We divide the redshift range into ten redshift bins with equal number of galaxies. Therefore, for every redshift bin we have:
\begin{equation}
n_i = \frac{n_{\text{gal}}}{10} \times 3600\left(\frac{180}{\pi}\right)^2 \ .
\end{equation}
\\

\subsection{Euclid and SKA specifications}

\begin{table}
\caption{Sky coverage and cosmic shear specifications for Euclid (see ~\cite{Audren:2012vy}) and for SKA (see~\cite{Harrison:2016stv}). Here $n_{\text{gal}}$ is in units of arcmin$^{-2}$.}
\label{LensingNumbers}
\footnotesize
\centering
\begin{tabular}{c|ccccccccccc}
\toprule
Experiment & $f_{\text{sky}}$ & $n_{\text{gal}}$ & $z_m$ & $\alpha$ & $\beta$ & $\gamma$ & $f_{\text{spec-z}}$ & $z_{\text{spec-max}}$ & $\sigma_{\text{photo-z}}$ & $z_{\text{photo-max}}$ & $\sigma_{\text{no-z}}$\\
\hline
SKA1 & 0.1212 & 2.7 & 1.1 & $\sqrt{2}$ & 2 & 1.25 & 0.15 & 0.6 & 0.05 & 2.0 & 0.3 \\
SKA2 & 0.7272 & 10 & 1.3 & $\sqrt{2}$ & 2 & 1.25 & 0.5 & 2.0 & 0.03 & 2.0 & 0.3 \\
Euclid & 0.3636 & 30 & 0.9 & $\sqrt{2}$ & 2 & 1.5 & 0.0 & 0.0 & 0.05 & 4.0 & 0.3 \\
\bottomrule
\end{tabular}
\end{table}

The number density of sources and the corresponding redshift errors for Euclid and SKA are taken, respectively, from Ref.~\cite{Audren:2012vy} and from Ref.~\cite{Harrison:2016stv}\footnote{Ref.~\cite{Harrison:2016stv} provides values for Euclid which differ only in the value of $\sigma_{\text{photo-z}}$}. Note that the real redshift uncertainties are the numbers from \cref{LensingNumbers} of \cite{Harrison:2016stv} multiplied by $(1+z)$. The meaning and motivation for these numbers is described in section 3.1 of Ref.~\cite{Harrison:2016stv} and in the end of section 4.2 of Ref.~\cite{Bonaldi:2016lbd}. 

The unnormalized redshift number density distribution is given by:
\begin{equation}
\frac{\dd n_{\text{gal}}}{\dd z} = z^{\beta} \exp\left[-\left(\frac{z}{\alpha z_m}\right)^{\gamma}\right] \ .
\end{equation}
The redshift uncertainty is parameterized as follows:
\begin{equation}
{\cal P}(z,z') = \begin{cases}
\dfrac{1-f_{\text{spec-z}}}{\sqrt{2\pi}\sigma_{\text{photo-z}}(1+z)}\exp\left[-\dfrac{(z-z')^2}{2\sigma^2_{\text{photo-z}}(1+z)^2}\right] + f_{\text{spec-z}}\delta(z-z'),&z\leq z_{\text{spec-max}}\\[15pt]
\dfrac{1}{\sqrt{2\pi}\sigma_{\text{photo-z}}(1+z)}\exp\left[-\dfrac{(z-z')^2}{2\sigma^2_{\text{photo-z}}(1+z)^2}\right],&z\leq z_{\text{photo-max}} \\[15pt]
\dfrac{1}{\sqrt{2\pi}\sigma_{\text{no-z}}(1+z)}\exp\left[-\dfrac{(z-z')^2}{2\sigma^2_{\text{no-z}}(1+z)^2}\right],&z\geq z_{\text{photo-max}} ,
\end{cases}
\end{equation}
where the measured redshift is denoted as $z'$, while the true one is denoted as $z$. Some of these Gaussians span a big redshift range, but they are anyway multiplied by the distribution $\frac{\dd n_{\text{gal}}}{\dd z}$ that is almost zero outside the range of interest.

The number counts of Euclid are negligible above $z=3.5$, so here the error function becomes as easy as
\begin{equation}
{\cal P}(z,z') = \frac{1}{\sqrt{2\pi}\sigma_{\text{photo-z}}(1+z)}\exp\left[-\frac{(z-z')^2}{2\sigma^2_{\text{photo-z}}(1+z)^2}\right] \ .
\end{equation}
The sky coverage $f_{\text{sky}}$ is the same as in the case of galaxy clustering (\cref{LensingNumbers}).\\

%% file: IM.tex
\section{21cm intensity mapping}
\label{sec:IM}
\subsection{21cm power spectrum}
\label{subsection:21cm power spectrum}
 
The goal of 21cm intensity mapping experiments is to measure the differential brightness temperature $\Delta T_{\rm b}$ defined as the difference between the observed brightness temperature $T_{\rm b}$ and the one expected for CMB photons only,  $T_{\gamma}$.  
The signal coming from 21cm hyperfine transitions of neutral hydrogen (HI) atoms is emitted at the frequency $\nu_{0} = 1420.4057\,\text{MHz}$. So the frequency measured today can be directly related to the redshift, 
\begin{equation}
\Delta T_{\rm b} \equiv \frac{T_{\rm b}(z)-T_{\gamma}(z)}{1+z} \ .
\end{equation}

Here, we focus only on the low redshift signal coming from the neutral hydrogen inside galaxies.  A detailed description of cosmology with the 21cm-signal at high redshifts, from the reionization and even from the cosmic dawn and the dark ages, and its interest to probe modified gravity, can be found in Refs.~\cite{Furlanetto:2006jb,Mao:2008ug,Lewis:2007kz,Clesse:2012th,Brax:2012cr}.  At low frequencies the mean differential brightness temperature is given by
\begin{equation}
\label{DelTb189}
\overline{\Delta T_b} \simeq 189\left[\frac{H_0(1+z)^2}{H(z)}\right]\Omega_{\text{HI}(z)} \, h \, \text{mK} \ ,
\end{equation}
where $H_0$ is the Hubble constant $H_0 = h \times 100$\,km/(s\,Mpc) and $\Omega_{\text{HI}}(z) = \rho_{\text{HI}}(z)/\rho_c$ is the mass density of neutral hydrogen divided by the critical density of the present-day universe. This relation is derived in detail in Appendix~\ref{Appendix:21cm_likelihood}.
Deviations from this value are proportional to the density perturbations in neutral hydrogen, which can be related to the dark matter density perturbations:
\begin{equation}
\Delta T_b - \overline{\Delta T_b} = \overline{\Delta T_b} \delta_{\text{HI}} = \overline{\Delta T_b} b_{\text{HI}} \delta_m \ .
\end{equation}
Thus, the power spectrum of these deviations is proportional to the matter power spectrum:
\begin{equation}
\label{eq:P21_1}
P_{21} = b_{21}^2 P_m
\end{equation}
with $b_{21} \equiv \overline{\Delta T_b} b_{\text{HI}}$. 
The redshift dependence of $\Omega_{\text{HI}}(z)$ and $b_{\text{HI}}(z)$ are modelled following Ref.~\cite{Villaescusa-Navarro:2016kbz}:
\begin{align}
\Omega_{\text{HI}}(z) &= \Omega_{\text{HI,0}}(1+z)^{\alpha_{\text{HI}}} \\
b_{\text{HI}}(z) &= 0.904+0.135(1+z)^{1.696} \ ,
\end{align}
with $\Omega_{\text{HI,0}}$ and $\alpha_{\text{HI}}$ set to the fiducial values $4\times10^{-4}$ and $0.6$, respectively, and allowed to vary in the forecast.
As in the case of galaxy clustering, we use nuisance parameters to describe the future accuracy of bias modeling:
\begin{equation}
b_{\text{HI}}(z) = \beta_0^{\text{IM}}\left[0.904+0.135(1+z)^{1.696\beta_1^{\text{IM}}}\right] \ ,
\end{equation}
with mean value zero and a prior corresponding to a rms of $0.025$.

In addition we have to consider observational effects analogous to those of the galaxy power spectrum of Eq.~\ref{P_g-def}:
\begin{equation}
P_{21}(k,\mu,z) =  f_{\text{AP}}(z) \times f_{\text{res}}(k,\mu,z) \times f_{\text{RSD}}(\hat{k},\hat{\mu},z) \times b_{21}^2(z) \times P_m(\hat{k},z) \ .
\label{eq:P21}
\end{equation}
The prefactors are the same as in the case of galaxy surveys because the signal dominantly consists of radiation originating from galaxies. Thus, it is affected by redshifting according to the movement of the galaxies. A power spectrum reconstructed from this map of intensities therefore suffers from redshift-space distortions, limited resolution and the Alcock-Paczinsky effect in the same way as one reconstructed from a map of galaxy positions. 

The observed power spectrum also includes a noise spectrum:
\begin{equation}
P_{21,\mathrm{obs}}(k,\mu,z) = P_{21}(k,\mu,z)  + P_N(z)
\label{eq:P21_PN}
\end{equation}
where $P_N(z)$ will be described in the next subsection. By identifying galaxies, all unwanted contributions to the power spectrum except for the shot noise could be removed. Instead, the correlations of unprocessed intensity include correlations in the foregrounds and in random noise in the sky or in the experimental setup itself.

Hence, the biggest disadvantage of 21cm intensity mapping surveys in comparison to 21cm galaxy surveys is the high contamination of the signal with telescope noise and foreground signals.  If the latter are sufficiently smooth in frequency, they are removable, but it is likely that residual foreground contamination will prevent to fully exploit the lowest radial modes, thus potentially degrading the constraints on cosmological parameters. 
This effect can be studied by introducing a radial mode cut-off $k_{\parallel {\rm min} }$, but it is difficult to determine quantitatively the value of the cut-off and the amplitude of the degradation. According to~\cite{Mao:2008ug}, it could reach up to a factor three for intensity mapping at reionization.   Most of the forecasts released so far did not include such a cut-off, not even the recent update of the SKA red book~\cite{Bacon:2018dui} presenting the latest performance forecasts.   In order to ease the comparison with other works, and to avoid an additional complexity by adding additional nuisance parameters, we did not include a modeling of this effect.   This will be done separately in a future work that will also evaluate the performance degradation in different cases.

\subsection{SKA specifications}

We start by describing the noise power spectrum $P_N$ in Eq.~\ref{eq:P21_PN}.
\begin{table}[h!]
\centering
\caption{IM specifications. \cite{SKA_BaselineV2,Olivari:2017bfv}}
\label{SKAspecIM}
\begin{tabular}{c|cccccccc}
\toprule
parameter & $\nu_{\text{min}}$\,[MHz] & $\nu_{\text{max}}$\,[MHz] & $z_{\text{min}}$ & $z_{\text{max}}$ & $\delta \nu$\,[kHz] & $T_{\text{inst}}$\,[K] \\
\hline
SKA1 band 1 & $\sim$400 (350) & $\sim$1000 (1050) & 0.45 & 2.65 & 10.9 & 23 \\
SKA1 band 2 & $\sim$1000 (950) & 1421 (1760) & 0.05 & 0.45 & 12.7 & 15.5 \\
\bottomrule
\end{tabular}
\end{table}
We consider here a survey executed in \textit{single dish} mode; this enhances the speed of the survey, but dismisses the advantages of radio interferometry. In this case the noise power is given by
\begin{equation}
P_N(z) = T^2_{\text{sys}} \frac{4\pi f_{\text{sky}}r^2(z)\,(1+z)^2}{2H(z)t_{\text{tot}}\nu_0 N_{\text{dish}}} \ ,
\label{eq:PN}
\end{equation}
where $T_{\text{sys}}$ is the system temperature, $t_{\text{tot}}$ is the total observation time and $N_{\text{dish}}$ is the number of dishes. The noise power originates from random uncorrelated fluctuations in the intensity of single pixels. Their amplitude is given by $T_{\text{sys}}$. Since this noise is independent of the signal it can be described as an additional perturbation field whose power spectrum $P_N$ is added to the power spectrum of the signal.  We adopt $t_{\text{tot}} = 10000$\,h and $N_{\text{dish}} = 200$, i.e. the same noise power as in Ref.~\cite{Villaescusa-Navarro:2016kbz}, where one can also find  a derivation of Eq.~\ref{eq:PN}.

Since there is no need to resolve a single galaxy, SKA1 has access to signals from higher redshift. Thus, band 1 can also be used for intensity mapping. Following Ref.~\cite{Olivari:2017bfv},
the system temperature is defined as the sum of the instrument's temperature $T_{\text{inst}}$ and the sky temperature
\begin{equation}
T_{\text{sky}} = 20\,\text{K}\,\left(\frac{408\,\text{MHz}}{\nu}\right)^{2.75} \ .
\end{equation}
The Gaussian suppressions of the power spectrum are quantified using the relation between the full width at half maximum to the rms, given by $\text{FWHM} = \sqrt{8\ln2} \ \sigma$. In the case of frequency, the channel width due to band separation into 64,000 channels is used as FWHM:
\begin{align}
\sigma_{\theta} &= \frac{1}{\sqrt{8\ln2}} \frac{\lambda_0}{D} (1+z) \ , \\
\sigma_{\nu} &= \frac{\delta \nu}{\sqrt{8\ln2}} \ .
\end{align}
In single dish mode the angular resolution is determined by the diameter $D = 15$~m of a single dish. As a result, $\sigma_{\theta}$ is as big as $0.34\,^{\circ}(1+z)$.

As we already mentioned, foregrounds are expected to be much larger than the 21cm signal itself.  Nevertheless, foregrounds are expected to be sufficiently spectrally smooth to be removed.  Here, correlations between remnants of the foregrounds or artefacts of their removal are expected to be negligible. Yet, the effect of foreground removal is taken into account in two ways. First, the observed part of the sky is decreased to the regions with the lowest foreground intensity. Following Ref.~\cite{Villaescusa-Navarro:2016kbz} the probed fraction of the sky is reduced to $f_{\text{sky}} = 0.58$.
Second, as discussed in Ref.~\cite{Alonso:2014dhk}, foreground removal does not work close to the edges of the frequency band (i.e.\,$\lesssim 50$~MHz). Therefore, we reduce the redshift range to exclude information from the edges.

The specifications used for intensity mapping forecasts are listed in Table~\ref{SKAspecIM}.

%% file: theoerr.tex
\section{Non-linear theoretical uncertainty}

Euclid and SKA will survey a large sky volume and detect a huge number of galaxies. This will dramatically decrease the size of sampling variance and shot noise compared to current surveys. Therefore, on small scales, theoretical errors will be the leading source of uncertainty and the limiting factor for parameter extraction, at least for analyses based on three-dimensional power spectra, 
such as galaxy surveys and intensity mapping.
Here we describe our strategy for modelling the theoretical error. Since it is easier to deal with the theoretical error of the bi-dimensional angular power spectrum, we start by discussing the weak lensing case.

\subsection{Cosmic shear error modelling}
\label{sec:CS_error}
The simplest way to model the theoretical uncertainty is to introduce a cutoff. This means neglecting all theoretical uncertainties up to a wavenumber $k_{\text{NL}}$, while dismissing all information above that wavenumber. This scheme is a good approximation when the result does not depend strongly on the region where the uncertainty raises from almost zero to infinity. 
Since non-linear effects increase with time, the cutoff scale should then decrease with redshift. Following Ref.~\cite{Smith:2002dz}, the redshift dependence of non-linear effects can be parametrised as
\begin{equation}
k_{\text{NL}}(z) = k_{\text{NL}}(0)\cdot(1+z)^{2/(2+n_s)} \ .
\label{eq:k_NL_z}
\end{equation}
The quantity of interest for weak lensing surveys is the shear power spectrum $C_\ell$, which is given by Eq.~\ref{Cl_Pk} as a weighted integral of $P(k)$ convoluted with a window function spanning a large range in $k$. Thus, there is no simple equivalent $\ell_{\text{NL}}$ of $k_{\text{NL}}$. Our approach consists in identifying values of $\ell$ above which most information comes from wavenumbers $k>k_{\text{NL}}$. First, we find the value $r$ corresponding to the maximum of the product of the window functions for a pair ($i$, $j$) of redshift bins:
\begin{equation}
r^{ij}_{\text{peak}} = \frac{\displaystyle \int_0^{\infty}\frac{\dd r \cdot r}{r^2} g_i(r) g_j(r)}{\displaystyle \int_0^{\infty}\dfrac{\dd r}{r^2} g_i(r) g_j(r)} \ .
\end{equation}
This value mostly depends on the lower redshift bin of $i$ and $j$, so an average over higher bins can be performed to get $\bar{r}^{i}_{\text{peak}}\equiv (\sum_{j>i} r^{ij}_{\rm peak})/(N-i)$ where $N$ is the number of bins. This can be related to a maximum $\ell$: 
\begin{equation}
\label{eq:ell_max}
\ell_{\text{max}}^i = k_{\text{NL}}(z) \cdot \bar{r}^{i}_{\text{peak}} \ .
\end{equation}
All $C_\ell^{ij}$ at $\ell$ larger than $\ell_{\text{max}}^i$ or $\ell_{\text{max}}^j$ are discarded. The resulting $C_\ell$'s are still quadratic.

One way to better understand the likelihood consists in splitting it into contributions from each $\ell$. The likelihood for cosmic shear can be expanded as:
\begin{equation}
-2 \ln {\cal L} = \sum_{\ell} \Delta \chi_{\ell}^2(\Delta P) \ ,
\end{equation}
where $\Delta P$ is the difference between the fiducial and sampled power spectrum. To understand the weight of each multipole, we can plot $\chi_{\ell}^2(\Delta P)$ versus $\ell$ while assuming that the fiducial and sampled power spectra differ by the same relative factor $\Delta P = \Delta P(\bm{k},\bar{z}) = P-\hat{P} = 0.001 P$ for every redshift and wavenumber. The resulting contributions $\Delta \chi_{\ell}^2$ solely depend on the characteristics of the likelihood. 
\begin{figure}
	\centering
	\input{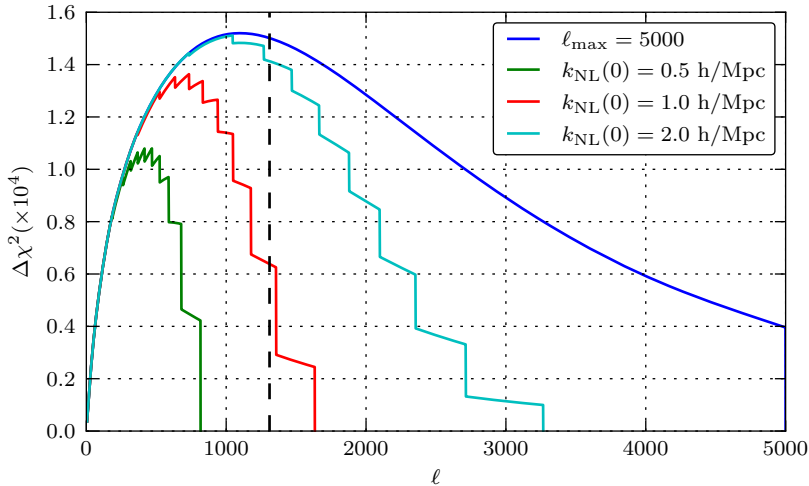}

	\caption{Euclid cosmic shear combined with Planck (see section~\ref{sec:datasets} for details): sensitivity to a 0.1\%-variation of $P(k)$ for different cutoff wavenumbers (always scaled with redshift). The flat $\ell_{max}=5000$ cut-off (blue) shows the amount of information available in absence of a cut-off. The second (green) and third (red) cases are more conservative than a sharp cut-off at $\ell=1310$ would be. For comparison, the dashed line marks $\ell=1310$, corresponding to the $\ell_{\rm max}$ used by the KiDS collaboration in Ref.~\cite{Kohlinger:2017sxk} as a reasonable cut-off producing stable results. The last case (cyan) is a little more constraining than this sharp cut-off, intended to reflect improvements in non-linear modeling in the analysis of future data. For our analysis we will use $k_{\text{NL}}(0)=0.5 h/\text{Mpc}$ (conservative) and $k_{\text{NL}}(0)=2.0 h/\text{Mpc}$ (realistic) as our non-linear cut-off wavenumbers. The corresponding 1-$\sigma$ sensitivity of our MCMC forecasts can be seen in table \ref{table:lensing_sensitivity}.}
	\label{fig:Chi2ofL_euclid}
\end{figure}

\begin{table}
\caption{Planck (see section \ref{sec:datasets}) plus Euclid cosmic shear 1-$\sigma$ sensitivity (normalized by corresponding Planck-only values) of MCMC forecasts for the non-linear cut-off values used in Figure \ref{fig:Chi2ofL_euclid}. We see that most sensitivities do not depend strongly on the choice of a given $k_{\text{NL}}(0)$. Only $n_s$ and $M_{\nu}$ show a non-negligible improvement in sensitivity, despite the large changes in the cut-off. Therefore, we find that the non-linear cut-off scheme is appropriate for our analysis.}
\label{table:lensing_sensitivity}
\centering
\begin{tabular}{c|ccccccc}
	\toprule
	$k_{\text{max}}$ & $100\omega_b$ & $\omega_{\text{cdm}}$ & $\theta_s$ & $\ln(10^{10}A_s)$ & $n_s$ & $\tau_{\text{reio}}$ & $M_{\nu}$ [eV]\\
	\hline
	0.5 h/Mpc & 0.77 & 0.27 & 0.97 & 0.94 & 0.72 & 0.96 & 0.50 \\
	1.0 h/Mpc & 0.76 & 0.27 & 0.94 & 0.95 & 0.70 & 0.98 & 0.41 \\
	2.0 h/Mpc & 0.76 & 0.25 & 0.97 & 0.94 & 0.65 & 0.97 & 0.36 \\
	$l_{\text{max}}=5000$ & 0.74 & 0.24 & 0.94 & 0.94 & 0.58 & 0.96 & 0.30 \\
	\hline
	Planck only & 1.00 & 1.00 & 1.00 & 1.00 & 1.00 & 1.00 & 1.00 \\
\end{tabular}
\end{table}

In Figure~\ref{fig:Chi2ofL_euclid} we see the $\Delta \chi^2_\ell$ contributions to the Euclid cosmic shear likelihood for different choices of $k_{NL}(0)$. Whenever $\ell$ reaches a value where an additional redshift bin has to be discarded according to the cut-off scheme described above, the $\Delta \chi^2$ drops sharply. A comparison of forecasts for Planck + Euclid cosmic shear for the same values of $k_{\text{NL}}(0)$ is shown in Table~\ref{table:lensing_sensitivity}. We see that the sensitivity does not differ by a large amount despite great changes in the non-linear cut-off, with only $n_s$ and $M_{\nu}$ showing non-negligible improvement in sensitivity with increasing cut-off values. Since the results do not depend strongly on the choice of $k_{\text{NL}}$, the cut-off approximation is accurate enough.

For our analysis we will adopt two values: a ``conservative'' cut-off $k_{\text{NL}}(0)=0.5 h/\text{Mpc}$, and a ``realistic'' cut-off $k_{\text{NL}}(0)=2.0 h/\text{Mpc}$. The realistic case is supposed to reflect improvements in the modelling of non-linear scales in the analysis of future data. Previous analyses like that of Ref.~\cite{Kohlinger:2017sxk} using a sharp bin-independent cut-off at $l=1310$ used an amount of information somewhere ``in between'' our  conservative and realistic assumptions.

\begin{figure}
	\makebox[\textwidth]{
		\input{figures/PaperChi2ofL_opt.pgf}
		\input{figures/PaperChi2ofL_con.pgf}
	}
	\caption{Sensitivity distribution for all cosmic shear likelihoods. The left panel shows the realistic approach and the right panel the conservative one. The $\Delta \chi^2$-values are contributions for each multipole $l$ obtained by setting $\Delta P = 0.001 P$ for all $k$. We find that SKA1 is not competitive, but that SKA2 will out-perform Euclid.}
	\label{Chi2_contr_Plots2}
\end{figure}
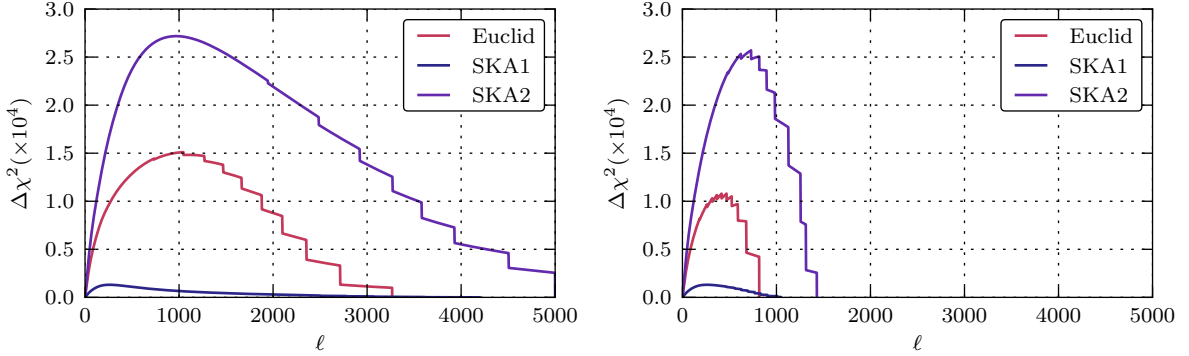

In Figure~\ref{Chi2_contr_Plots2} we show the sensitivity distribution for Euclid, SKA1 and SKA2 for the realistic (left) and conservative (right) non-linear cut-offs. We see that SKA1 is not competitive and that SKA2 will be more constraining than Euclid, because of the better accuracy of redshift measurements and of the greater sky coverage.

\subsection{Power spectrum error modelling (galaxy clustering + intensity mapping)}

\label{sec:GC_IM_error}
For the case of the three-dimensional galaxy power spectrum $P_g(\bm{k},z)=P_g(k,\mu,z)$, things are a bit more complicated. After binning in $z$-space, instead of dealing with a discrete expansion parameter $\ell$ in each bin of mean redshift $\bar{z}$, we have two continuous variables $(k, \mu)$. The traditional way to build a likelihood is recalled in Appendix~\ref{Apendix:Galaxy clustering likelihood} . The contribution of one redshift bin and of the interval $(k \pm \frac{\dd k}{2}, \mu \pm \frac{\dd \mu}{2})$ to $-2 \ln {\cal L}$ (i.e.~to the $\chi^2$) can be written in a differential form:
\begin{equation}
\frac{\dd \chi^2}{\dd k \dd \mu} = k^2\frac{V_r(\bar{z})}{2(2\pi)^2} \times \left[ \frac{\Delta P_g(k,\mu,\bar{z})}{\sigma_\mathrm{obs}(k,\mu,\bar{z})}\right]^2 \ ,
\end{equation}
where $\Delta P_g(k,\mu,\bar{z})$ is the difference between the predicted and observed galaxy power spectrum, the prefactor proportional to $k^2$ accounts for the density of independent Fourier modes, and $V_r(\bar{z})$ is the volume of one redshift bin given by Eq.~\ref{eq:Vr}.
The observational error is $\sigma_{\text{obs}}(k,\mu,\bar{z}) = P_g(k,\mu,\bar{z}) + P_N$ where $P_N$ is some constant noise. 
If we want to understand how the experimental sensitivity depends on different scales, it is worth looking at the effective error:
\begin{equation}
\frac{\dd \chi^2}{\dd k \dd \mu} \equiv   \left[ \frac{\Delta P_g(k,\mu,\bar{z})}{\sigma_\mathrm{eff}(k,\mu,\bar{z})} \right]^2 \ \Rightarrow \qquad         \sigma_{\text{eff}}(k,\mu,\bar{z}) = \sigma_{\text{obs}}(k,\mu,\bar{z}) \, \left[k^2\frac{V_r(\bar{z})}{2(2\pi)^2}\right]^{-1/2} \ .
\label{effErrDef}
\end{equation}
The power spectrum decreases when $k$ increases, such that the effective error decreases approximately like $\propto k^{-2}$. This in turn means that the amount of accessible information grows to infinity. If a cut-off $k_{\text{NL}}$ is used to prevent this, the region directly below this cut-off will be the one with the biggest weight in the likelihood, making the results very sensitive to the choice of $k_{\text{NL}}$. Thus, a more realistic way to account for the theoretical error is needed.

We first review the approach of Ref.~\cite{Audren:2012vy} to this problem. It starts from the assumption that for each $(k,\mu,z)$, we can reasonably estimate the 1$\sigma$ uncertainty $\delta P_g (k, \mu, z)$ on the theoretical prediction for the galaxy power spectrum $P_g(k,\mu,z)$ (e.g.\,by comparing the output of various simulations) and define a 1$\sigma$ envelope function $\alpha$ for the relative error:
\begin{equation}
\alpha(k,\mu,z)  =  \delta P_g (k, \mu, z)/P_g(k, \mu, z)~.
\end{equation}
The concrete implementation of this error in the likelihood is not trivial, because the errors made on $P_g(k,\mu,z)$ at different values of $(k,\mu,z)$ should in principle be correlated.
If we assume that the whole $(k, \mu,z)$ volume probed by the experiment can be split in bins in which the errors are uncorrelated, we can introduce one independent nuisance parameter per bin, and marginalise over it. This approach is actually numerically expensive, but in good approximation, the marginalisation can be replaced by an analytic minimisation. This leads to a simple expression for the contribution of each uncorrelated bin $(k \pm \frac{\dd k}{2}, \mu \pm \frac{\dd \mu}{2}, \bar{z} \pm \frac{\Delta \bar{z}}{2})$:
\begin{equation}
\frac{\dd \chi^2}{\dd k \dd \mu} = k^2\frac{V_r(\bar{z})}{2(2\pi)^2} \times \frac{\left[ \Delta P_g(k,\mu,\bar{z}) \right]^2}{\sigma_{\text{obs}}^2(k,\mu,\bar{z})+\sigma_{\text{th}}^2(k,\mu,\bar{z})} \,
\end{equation}
where  $\sigma_{\text{th}}(k,\mu,z)=\alpha(k,\mu,z)P_g(k,\mu,z)$. Note that the bin width in $(k,\mu)$-space appears explicitly in the differential expression on the left-hand side, while the bin width in redshift space appears implicitly in the expression of $V_r(\bar{z})$ given in Equation~(\ref{eq:Vr}). Equivalently one can take an arbitrary binning in $(k, \mu, z)$-space, provided that the theoretical error is rescaled self-consistently: this is the approach followed in Ref.~\cite{Audren:2012vy}. 

The difficulty is then to evaluate the correlation length in $(k , \mu, z)$-space. The authors of Ref.~\cite{Audren:2012vy} choose a method that compares the effect of the theoretical error to a reference $\Delta \chi^2 = 1$ obtained by varying a single Gaussian nuisance parameter. 
However, this method makes the amplitude of the error dependent on the range of the integrals $[k_{\text{min}}, k_{\text{max}}]$, and on the number of redshift bins. As a result, the error depends on the survey specifications and cannot be used in our combined forecast of the future sensitivity of various experiments. It is thus necessary to take a closer look at the correlation of the theoretical error.

The authors of Ref.~\cite{Baldauf:2016sjb} address this problem by introducing the full correlation matrix of the theoretical error. They write the contribution of each redshift bin to their log-likelihood as
\begin{equation}
\chi^2 = \sum_{i,j} \left[ (\Delta P(k_i) - Q(k_i)) \frac{\delta_{ij}}{\sigma_\mathrm{obs}(k_i)\sigma_\mathrm{obs}(k_j)} (\Delta P(k_j) - Q(k_i)) + Q(k_i) C^{-1}_{ij} Q(k_j) \right] \ ,
\label{eq:baldauf}
\end{equation}
where the wavenumber range has been discretised arbitrarily, redshift-space distortions are neglected (thus quantitites do not depend on $\mu$), $Q(k_i)$ is a single realization of the theoretical error, and $C_{ij} = \langle Q(k_i) Q(k_j) \rangle$ is the error correlation matrix, for which one needs to make some assumption. Ref.~\cite{Baldauf:2016sjb} parametrises $C_{ij}$ in terms of an error amplitude for each given $k_i$ (equivalent to $\delta P(k_i) = \alpha(k_i) P(k_i)$ in our notations), and a correlation length $\Delta k$ such that the correlation is exponentially suppressed for $|k_i-k_j|>\Delta k$:
\begin{equation}
C_{ij} = \alpha(k_i) P(k_i) \exp\left[\frac{-(k_i-k_j)^2}{2\Delta k^2}\right] \alpha(k_j) P(k_j) ~.
\end{equation}
The forecasts can then be performed with a marginalisation over each nuisance parameter $Q(k_i)$. Compared to the previous method, this approach relies on one more assumption: one needs to postulate not only an error amplitude function $\alpha(k)$, but also a correlation length $\Delta k$, accounting for the minimum typical scale over which we allow the theoretical error to fluctuate randomly. This enlarged parametrisation compared to Ref.~\cite{Audren:2012vy} is a good thing, because it makes more clear and explicit assumptions on the theoretical error. The authors of Ref.~\cite{Baldauf:2016sjb} argue that $\Delta k$ can be matched with the Baryon Acoustic Oscillation (BAO) scale.

Unfortunately, there are several reasons for which we cannot transpose directly this approach in our forecast. First, the marginalisation over one nuisance parameter for each $k_i$ is still tractable for the authors of Ref.~\cite{Baldauf:2016sjb} because their forecast is Fisher matrix based. In a full MCMC forecast like ours, it would introduce too many varying parameters and the convergence of the MCMC chains would be prohibitively slow. Additionally, Ref.~\cite{Baldauf:2016sjb} neglected RSD corrections and assumed no correlations between the error in different redshift bins. In the present work, we wish to incorporate RSD effects and to address the issue of error correlations in redshift space. In principle this would imply a generalisation of Eq.~\ref{eq:baldauf} to one nuisance parameter for each argument $Q(k_i, \mu_j, z_k)$, with a six-dimensional correlation matrix. 
This problem is too heavy to be solved with an MCMC approach. Note that we cannot simply approximate the marginalisation over each $Q$ by an analytic minimization, because there is no simple analytic solution in presence of an exponential correlation function. 

Thus we chose to stick to the idea of Ref.~\cite{Baldauf:2016sjb} of formulating the problem in terms of correlation lengths, but, to make it computationally tractable, we assume that the $(k, \mu, z)$ space can be split into approximately uncorrelated bins. In other words, we replace the non-diagonal  covariance matrix by a diagonal one with a bigger spacing $(\Delta k, \Delta \mu, \Delta z)$ between adjacent bin centers $(k_{i\pm1}, \mu_{j\pm1}, z_{k\pm1})$, such that the $Q(k_i, \mu_j, z_k)$ are statistically independent:
\begin{equation}
\chi^2 = \sum_{m, n} \left[ \frac{(\Delta P_g({\bf k}_m, \bar{z}_n) - Q({\bf k}_m, \bar{z}_n))^2}{\sigma^2_\mathrm{obs}({\bf k}_m, \bar{z}_n)} \right] + \sum_{i,j,k} \frac{Q^2(k_i,\mu_j,z_k)}{(\alpha(k_i,\mu_j,z_k)P_g(k_i,\mu_j,z_k))^2} \ .
\label{eq:chi2baldauf}
\end{equation}
As usual in the expression of a galaxy survey likelihood, the first sum runs over all independent Fourier modes ${\bf k}_m$ (see Appendix~\ref{Apendix:Galaxy clustering likelihood} or seminal papers like Ref.~\cite{Feldman:1993ky}) and over redshift bins of mean $\bar{z}_n$ separated by $\Delta \bar{z}$. Instead, the second sum runs over the centers of the larger bins with uncorrelated theoretical errors, spaced by steps $(\Delta k, \Delta \mu, \Delta z)$ that play the role of correlation lengths.
The function $Q$ is assumed to be continuous, with a few nodes $Q(k_i,\mu_j,z_k)$ which are treated as nuisance parameters. In principle, the value of $Q$ in an arbitrary point $(k,\mu,z)$ or equivalently $({\bf k},z)$ which appears in the first sum could be obtained by performing a smooth interpolation of the $Q$ function between the node values. However, in practise, we do not need to perform any such interpolation in our numerical implementation: we shall see below that $Q$ can be eliminated analytically and does not appear in our final expression (\ref{eq:finalGCchi2}).

It is conventional to assume that in the first sum, the volume of each independent mode is so small that the discrete sum can be represented as an integral (see Appendix~\ref{Apendix:Galaxy clustering likelihood}):
\begin{equation}
\sum_{m,n} \rightarrow \sum_{n} \int_{k_\mathrm{min}}^{k_\mathrm{max}} \!\!\! \dd k\cdot k^2  \int_{-1}^1 \!\!\!\dd \mu \,\, \frac{V_r(\bar{z}_n)}{2(2\pi)^2} \ .
\end{equation}
Actually, the second sum can also be replaced by an integral, provided that we rescale the steps of integration by the correlation lengths, in order to avoid multiple countings of the contribution of each independent nuisance parameter:
\begin{equation}
\sum_{i,j,k} \rightarrow \sum_{\bar{z}} \frac{\Delta\bar{z}}{\Delta z} \int_{k_\mathrm{min}}^{k_\mathrm{max}} \frac{\dd k}{\Delta k} \int_{-1}^1 \frac{\dd \mu}{\Delta \mu} \ .
\end{equation}
In other words, this is equivalent to increasing the number of nuisance parameters by some factor, while dividing their weight by the same factor in order to keep a fully equivalent expression.
Then the $\chi^2$ can be written with a single integral 
\begin{equation}
\chi^2 = \sum_{n} \int_{k_\mathrm{min}}^{k_\mathrm{max}} \!\!\! \dd k\cdot k^2  \int_{-1}^1 \!\!\! \dd \mu \,\, \frac{V_r(\bar{z}_n)}{2(2\pi)^2} \left[ \frac{(\Delta P_g(k,\mu, \bar{z}_n) - Q(k,\mu, \bar{z}_n))^2}{\sigma^2_\mathrm{obs}(k,\mu, \bar{z}_n)}  +  \frac{Q^2(k,\mu,\bar{z}_n)}{\sigma_\mathrm{th}^2(k,\mu, \bar{z}_n)} \right] 
\end{equation}
where we defined
\begin{equation}
\sigma_\mathrm{th} (k,\mu , z) = \left[ \frac{V_r(z)}{2(2\pi)^2} k^2 \Delta k \Delta \mu \frac{\Delta z}{\Delta\bar{z}}\right]^{1/2}  \alpha(k,\mu,z) P_g(k,\mu,z)~.
\label{eq:sigmath}
\end{equation}
Finally we can approximate the marginalisation over nuisance parameters by an analytic minimisation\footnote{Like in Ref.\cite{Audren:2012vy}, we perform the analytic minimization under a small approximation. Since $\sigma_\mathrm{obs}$ contains $P_g$, it should also contain a term $Q$ added to it. This small dependence of the standard deviation on the theoretical error has an extremely small impact. We neglect it and stick to the definition $\sigma_\mathrm{obs} = P_g+P_N$.} and obtain a computationally tractable expression:
\begin{equation}
\chi^2 = \sum_{n} \int_{k_\mathrm{min}}^{k_\mathrm{max}} \!\!\! \dd k\cdot k^2  \int_{-1}^1 \!\!\! \dd \mu \,\, \frac{V_r(\bar{z}_n)}{2(2\pi)^2} \left[ \frac{(\Delta P_g(k,\mu, \bar{z}_n))^2}{\sigma^2_\mathrm{obs}(k,\mu, \bar{z}_n)+\sigma_\mathrm{th}^2(k,\mu, \bar{z}_n)}   \right] ~.
\label{eq:finalGCchi2}
\end{equation}
This expression differs from the usual likelihood derived in absence of a theoretical error only through the presence of the term $\sigma_\mathrm{th}^2$  in the denominator. 
The $\chi^2$ defined in Eq.~\ref{eq:finalGCchi2} has the same form as the one used in Ref.~\cite{Audren:2012vy}, but the new ingredient is the more rigorous definition of the quantity 
$\sigma_\mathrm{th}$ in Eq.~\ref{eq:sigmath}, which comes from a precise discussion of the role of correlation lengths $(\Delta k, \Delta \mu, \Delta z)$. We must now specify this term and motivate some choices for the correlation lengths and for the error envelope function $\alpha(k,\mu,z)$. These choices should be guided by the types of errors which are expected to be made on theoretical predictions for non-linear corrections, and by the shape and amplitude of these errors.

The choice of $\Delta \mu$ relates to the question: for a given bin $(k_i,z_k)$, how many independent nuisance parameters should describe the error for different $\mu_j$ values? A most reasonable answer is one. Indeed, the error made on the prediction of the non-linear power spectrum $P_m(\hat{k},z)$ and on the (possibly non-linear) bias $b(\hat{k},z)$ is isotropic in 3D Fourier space. When it propagates to $\alpha(k,z,\mu)$ (the relative error on $P_g(k,\mu,z)$), it only gives a small $\mu$-dependence through the projection from $\hat{k}$ to $k$, with no further error introduced in this projection. On the other hand, the departure from our possibly too simplistic ansatz for RSD corrections, fingers-of-God corrections and instrumental resolution effects could potentially be strongly $\mu$-dependent and motivate the introduction of more than one independent nuisance parameter per $(k_i,z_k)$ bin.

However, there are some arguments in favor of neglecting the theoretical error from RSD modeling compared to other theoretical error sources. They will be discussed later in the context of a quantification of the total theoretical error. 
We conclude that for a given bin $(k_i,z_k)$, the errors made on $P_g(k,\mu,z)$ for different $\mu$ values can be considered as fully correlated with each other in very good approximation. Taking a single independent nuisance parameter per $(k_i,z_k)$ bin is mathematically equivalent to setting $\Delta \mu= \mu_\mathrm{max} - \mu_\mathrm{min}=2$ in Eq.~\ref{eq:sigmath}. Thus the theoretical error expression gets reduced to
\begin{equation}
\sigma_\mathrm{th} (k,\mu , z) = \left[ \frac{V_r(z)}{(2\pi)^2} k^2 \Delta k \frac{\Delta z}{\Delta\bar{z}}\right]^{1/2}  \alpha(k,\mu,z) P_g(k,\mu,z)~.
\label{eq:sigmath2}
\end{equation}

For the correlation length in wavenumber space, we will use $\Delta k = 0.05 \, h/$Mpc like in Ref.~\cite{Baldauf:2016sjb}. This is chosen to be close to the BAO scale,  which is the smallest inherent scale in the matter power spectrum and a conservative guess for the correlation length in $k$-space. 

The correlation length in redshift space is harder to guess. Theoretical errors are not necessarily correlated throughout the whole redshift range probed by the experiment. A value close to but slightly smaller than the total redshift range probed by the experiment (1.3 for Euclid, 1.9 for SKA2) should be a conservative guess. We assume $\Delta z=1$, which is equivalent to assuming between one and two $z$-bins with independent  theoretical error for each $k_i$ bin. Note that Ref.~\cite{Baldauf:2016sjb} did not discuss the issue of correlations in redshift space and used the $\chi^2$ formula~(\ref{eq:chi2baldauf}) in each redshift bin, each time with a new bunch of independent nuisance parameters $Q(k_i)$. This is equivalent to setting a correlation length $\Delta z$ implicitly equal to the size of individual redshift bins $\Delta \bar{z}$. We do not adopt this approach, since the errors made by $N$-body simulations in the prediction of $P_m(\hat{k},z)$ at a given scale $\hat{k}$ but for two very nearby redshifts $\bar{z}_n$ and $\bar{z}_{n+1}$ should not be statistically independent. Moreover the impact of the theoretical error should not directly depend on the number of bins in which one chooses to split the data.

The relative error envelope function $\alpha(k,\mu,z)$ should model uncertainties on three types of non-linear corrections: the prediction of the matter power spectrum itself, of the bias and of redshift-space distortions. 
Here we will neglect the third contribution (i.e.\,uncertainties on the RSD correction term) for several reasons. First, the fingers-of-God correction term already leads to a strong suppression of the power spectrum which results in a big relative observational error when the power spectrum becomes smaller than the noise power. At this point any extra theoretical error term becomes irrelevant. Furthermore, RSD corrections can be modeled up to higher order than used here~\cite{Heavens:1998es} and their modeling is continuously improving (see e.g.\,Refs.~\cite{Bianchi:2014kba} and \cite{Song:2018afp} and references therein).
Thus we can focus on the theoretical error on the non-linear matter power spectrum and bias predictions.
The bias is usually assumed to be linear up to scales $k<0.2$\,h/Mpc. Beyond, a non-linear treatment would be more realistic (see e.g.\,\cite{Jennings:2015lea}), but non-linear bias can be be predicted by future simulations up to some residual uncertainty. Hence, the theoretical uncertainty should account mainly for inaccuracies in matter power spectrum and bias predictions from simulations.

The \texttt{HALOFIT} semi-analytic formula \cite{Takahashi:2012em,Bird:2011rb}, which we use for the present forecasts, only reaches accuracies of 5\% at $k<1$\,h/Mpc and 10\% at $k<10$\,h/Mpc according to Ref.~\cite{Takahashi:2012em}. This error was estimated from a comparison with N-body simulations. The more recent HMcode \cite{Mead:2015yca} achieves better precision than \texttt{HALOFIT} for k values larger than the BAO scale ($k > 1$\,h/Mpc), while the precision on BAO scales is a little worse. Overall, the effect is an error of ~5\% on all scales \cite{Mead:2015yca}. Finally Ref.~\cite{Schneider:2015yka} found that present-day N-body codes (Ramses \cite{Teyssier:2001cp}, Pkdgrav3 \cite{Potter:2016ttn}, Gadget3 \cite{Springel:2005mi}) agree to within 1\% at $k=1$\,h/Mpc and 3\% at $k=10$\,h/Mpc. Following the plots in Ref.~\cite{Schneider:2015yka}, we assume an exponential growth of the uncertainty in the decimal logarithm of $k$ crossing 0.33\% at $k=0.01$\,h/Mpc. However, these are pure dark matter simulations and the effect of baryonic feedback, as well as the $k$-dependence of galaxy-to-mass bias on nonlinear scales will increase the error.
Different hydrodynamical simulations find different results for the suppression of power induced by baryonic feedback~\cite{vanDaalen:2013ita, Chisari:2018prw}, mainly depending on the implementation of AGN feedback. According to the OWLS simulations of Ref.~\cite{vanDaalen:2011xb} the suppression starts at $k \lesssim 1 \,h / \mathrm{Mpc}$ and reaches $\sim 30$ \% at $k=10\,h / \mathrm{Mpc}$.
To account for uncertainties in the future modeling of baryonic feedback, which will of course be smaller than the effect itself, and to allow for remaining additional uncertainty from bias and RSD modeling at small scales, we increase the theoretical error to 1\% at $k=0.3$\,h/Mpc and 10\% at $k=10$\,h/Mpc.
Consequently, we define a relative error function passing through these three fixed points:
\begin{enumerate}
\item 0.33\% error at $k=0.01$\,h/Mpc
\item 1\% error at $k=0.3$\,h/Mpc
\item 10\% error at $k=10$\,h/Mpc
\end{enumerate}
This can be achieved with the following ansatz:
\begin{equation}
\label{eq:th_error}
\alpha(\bm{k},z) = \begin{cases}
a_1 \exp\left(c_1\cdot\log_{10}\frac{k}{k_1(z)}\right)~,&\frac{k}{k_1(z)}<0.3\\[10pt]
a_2 \exp\left(c_2\cdot\log_{10}\frac{k}{k_1(z)}\right)~,&\frac{k}{k_1(z)}>0.3 \ ,
\end{cases}
\end{equation}
where the wavenumber $k_1$ coincides with 1~h/Mpc at redhsift zero and scales with redshift like
\begin{equation}
k_1(z) = \frac{1h}{\text{Mpc}} \cdot(1+z)^{2/(2+n_s)}  \ .
\end{equation} 
The four free factors are fixed by the three fixed points defined above and the condition of continuity:
\begin{align*}
a_1 = 1.4806\,\% \ &, \ c_1 = 0.75056 \ ,\\
a_2 = 2.2047\,\% \ &, \ c_2 = 1.5120 \ .
\end{align*}
What we will call later the ``realistic case'' amounts in trusting this error function up to large wavenumbers. Then the information coming from small scales is suppressed gradually by the increasing relative error function, and the actual value of the cut-off $k_\mathrm{max}$ becomes effectively totally irrelevant. What we will call instead the ``conservative case'' is an analysis using this error function and additionally introducing a sharp cut-off at $k=0.2$\,h/Mpc (i.e., the error is effectively infinite above this value), following the scaling in redshift as defined Eq.~\ref{eq:k_NL_z}.

\begin{figure}
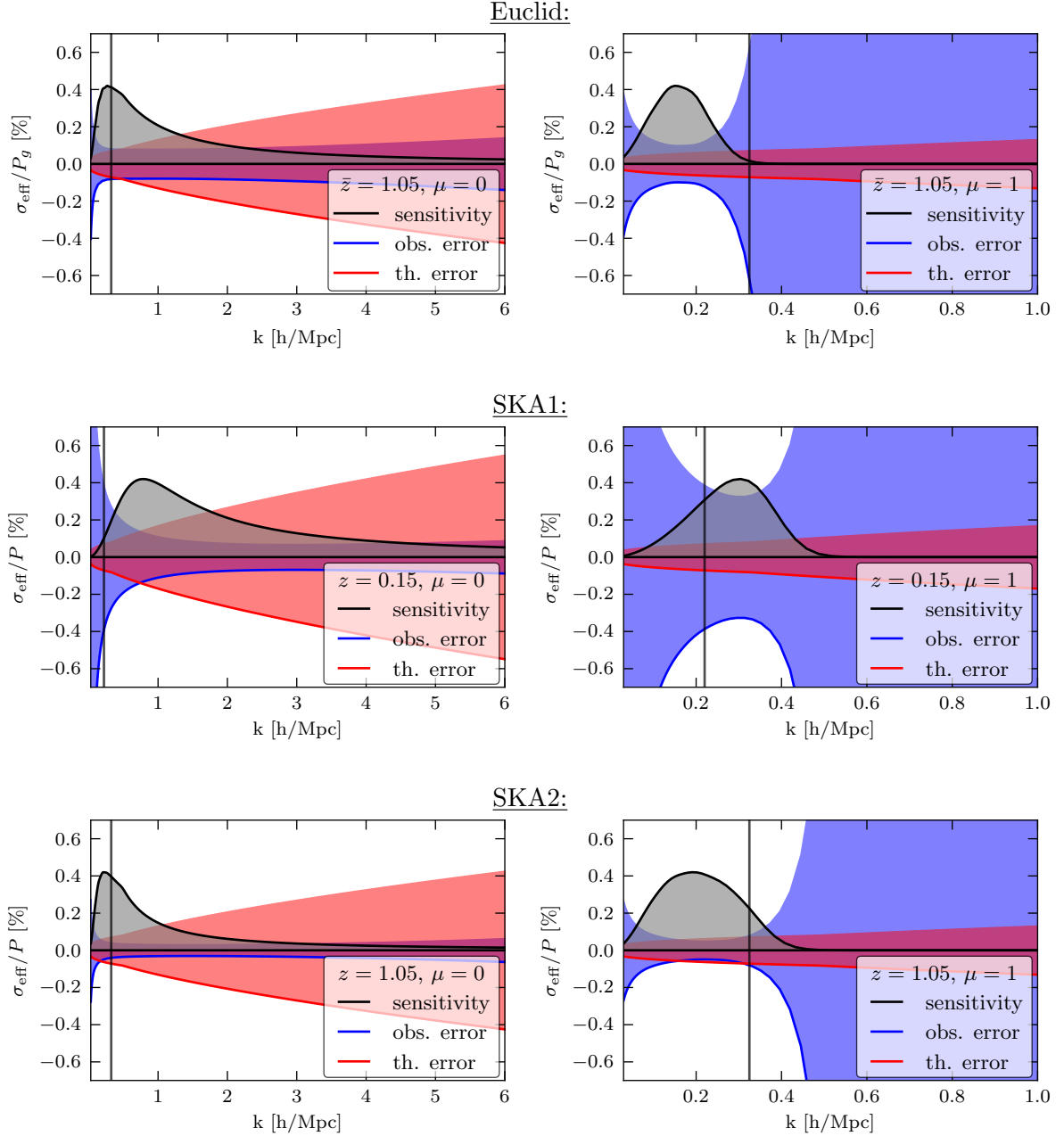

	\begin{center}
		\underline{Euclid:}
		\makebox[\textwidth]{
			\input{figures/euclid_pk_z1mu0.pgf}
			\input{figures/euclid_pk_z1mu1.pgf}
		}
		\text{}\\
		\underline{SKA1:}
		\makebox[\textwidth]{
			\input{figures/ska1_pk_z015mu0.pgf}
			\input{figures/ska1_pk_z015mu1.pgf}
		}
		\text{}\\
		\underline{SKA2:}
		\makebox[\textwidth]{
			\input{figures/ska2_pk_z105mu0.pgf}
			\input{figures/ska2_pk_z105mu1.pgf}
		}
	\end{center}
	\caption{Galaxy clustering: Examples for the relative effective errors $\sigma_{\rm eff}/P_g$ in selected redshift bins, decomposed into the contribution from the observational error (blue) and theoretical error (red).
	To show to which scale the experiment is most sensitive (taking these errors into account), we also show in grey the function $\sim \dd \chi^2/(\dd k \dd \mu)$ arbitrarily normalised to a constant relative difference between the theoretical and observed spectra ($\Delta P_g = \epsilon P_g$). The vertical line marks $k_{\text{NL}}(\bar{z})$, used as a sharp cut-off for the conservative setting.}
	\label{ErrorPlots_Pk}
\end{figure}
\begin{figure}
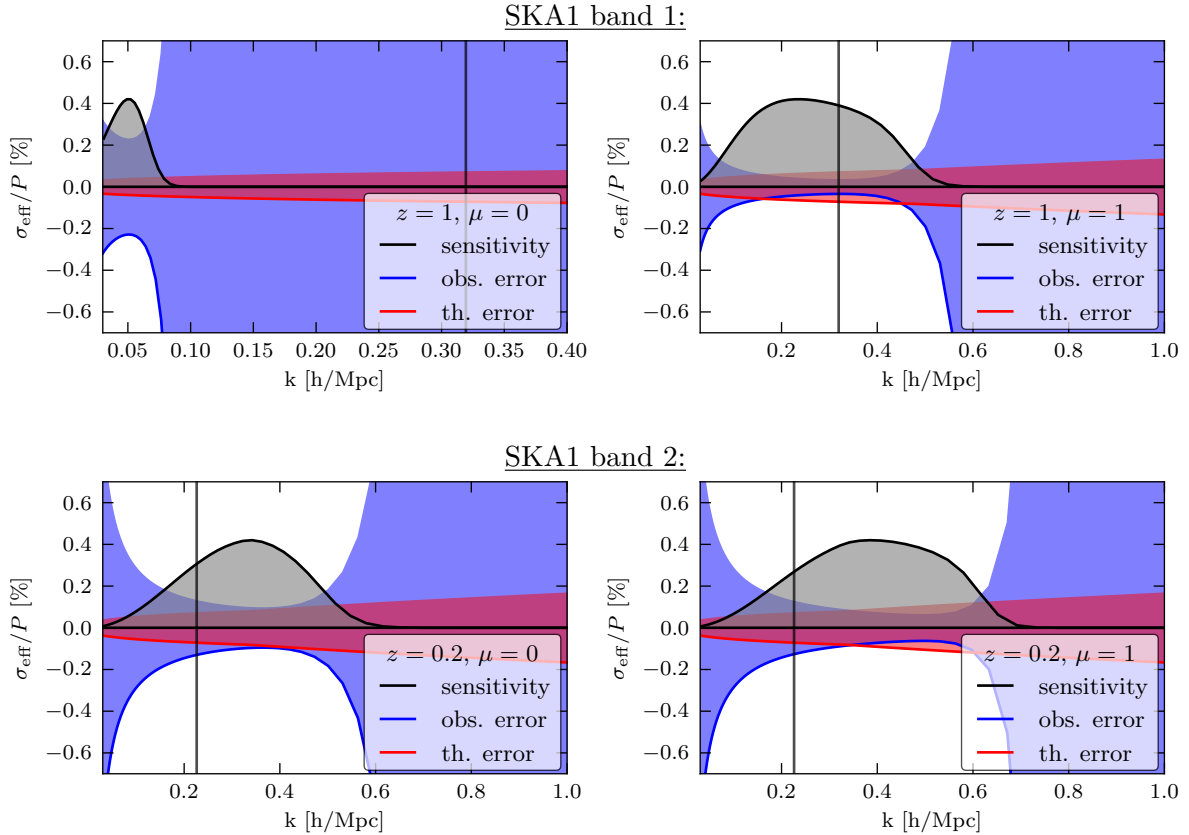

	\centering
	\underline{SKA1 band 1:}
	\makebox[\textwidth]{
		\input{figures/ska1_im1_z1mu0.pgf}
		\input{figures/ska1_im1_z1mu1.pgf}
	}
	\text{}\\
	\underline{SKA1 band 2:}
	\makebox[\textwidth]{
		\input{figures/ska1_im2_z02mu0.pgf}
		\input{figures/ska1_im2_z02mu1.pgf}
	}
	\caption{Intensity mapping: Examples for relative effective errors $\sigma_{\rm eff}/P_{21}$ and sensitivity contributions $\sim \dd \chi^2/(\dd k \dd \mu)$ arbitrarily normalised to $\Delta P_{21}=\epsilon P_{21}$. The vertical line marks $k_{\text{NL}}(\bar{z})$, used as a sharp cut-off for the conservative setting. Both the realistic and conservative setting make use of the theoretical error (red). Note the different $k_{\text{NL}}(z)$ values, corresponding to the mean redshift of each bin.}
	\label{ErrorPlots_IM}
\end{figure}
\begin{figure}
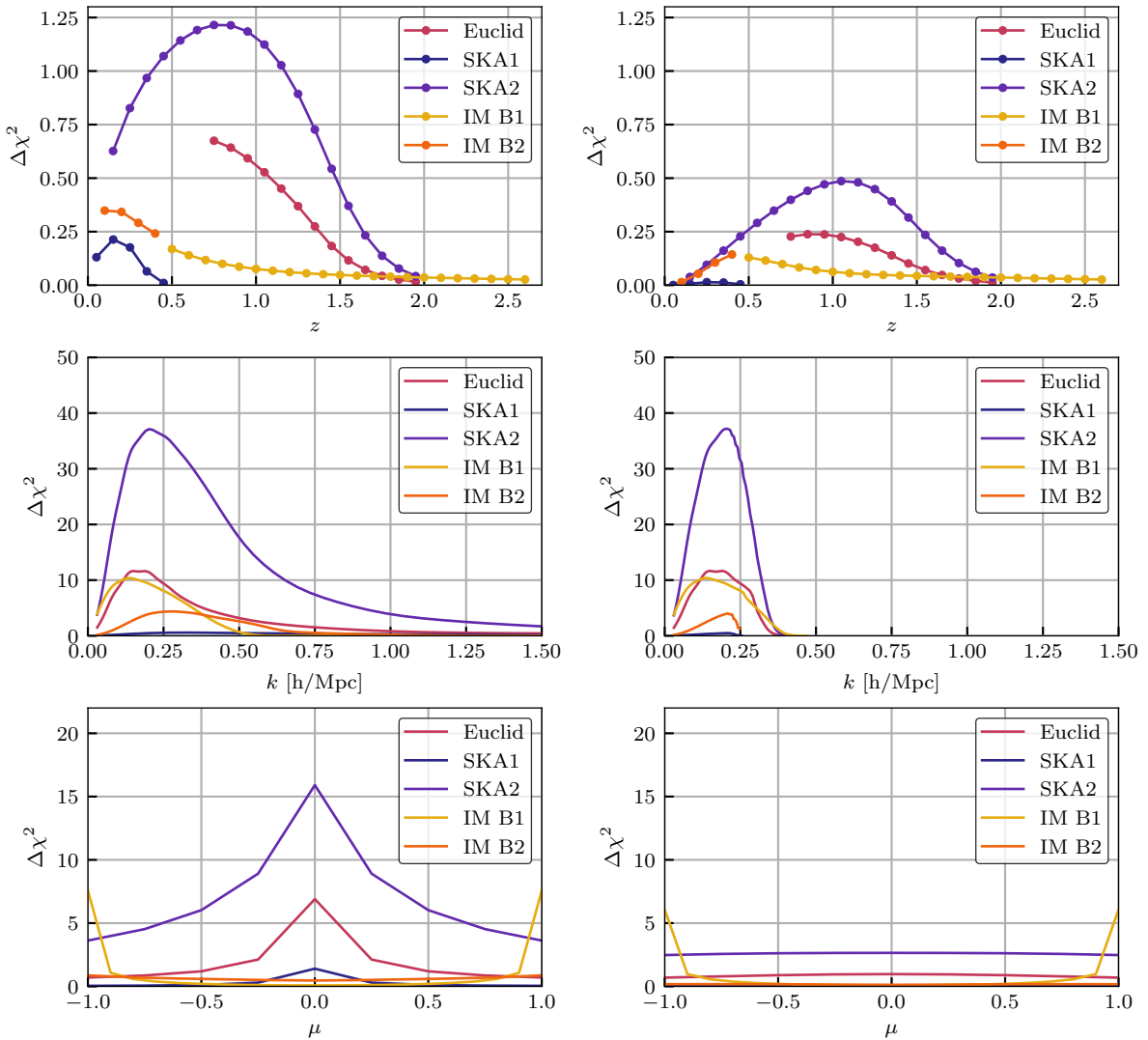

	\makebox[\textwidth]{
		\input{figures/PaperChi2ofZ_opt.pgf}
		\input{figures/PaperChi2ofZ_con.pgf}
	}
	\makebox[\textwidth]{
		\input{figures/PaperChi2ofK_opt.pgf}
		\input{figures/PaperChi2ofK_con.pgf}
	}
	\makebox[\textwidth]{
		\input{figures/PaperChi2ofMu_opt.pgf}
		\input{figures/PaperChi2ofMu_con.pgf}
	}
	\caption{Sensitivity distribution for all three-dimensional power spectrum likelihoods. The left panel shows the realistic approach and the right panel the conservative one. The $\Delta \chi^2$-values are contributions when two of the quantities $k$, $\mu$ and $z$ are integrated or summed over, and $\Delta P = 0.001 P$ everywhere. For intensity mapping (IM), band 1 and 2 of SKA1 are considered.}
	\label{Chi2_contr_Plots}
\end{figure}

Examples of effective errors\footnote{The theoretical effective error is defined analogously to the observational one (Eq.~\ref{effErrDef}).} for each likelihood and for a few selected redshift bins are shown in Figure \ref{ErrorPlots_Pk} for galaxy clustering and Figure \ref{ErrorPlots_IM} for intensity mapping. On the same plot, we show to which scales the experiments are most sensitive by just plotting  $\frac{\dd \chi^2}{\dd k \dd \mu}$ arbitrarily normalised to a constant relative difference between the theoretical and observed spectra ($\Delta P_g = \epsilon P_g$). 
The vertical line marks $k_{\text{NL}}(\bar{z})$, used as a sharp cut-off for the conservative setting.
Note the different $k_{\text{NL}}(\bar{z})$ values, corresponding to the mean redshift of each bin.  Both the realistic and conservative setting make use of the theoretical error (red). 
For galaxy clustering,
the observational errors (blue) dominate the error in the radial direction (right panel) due to redshift space distortions, whereas the closer we get to $\mu=0$ the observational error diminishes and the theoretical error takes over, illustrating the necessity for introducing a measure of the error on non-linear modelling. This effect is also seen as a function of $\mu$ in the bottom panel of Figure \ref{Chi2_contr_Plots}.
For intensity mapping, we notice that the trend is different from the galaxy clustering case (Figure \ref{ErrorPlots_Pk}). Due to the poor angular resolution, the observational error (blue) is larger in the transverse direction compared to the line-of-sight, especially for band 1, and dominates on most scales.

Just like we did for cosmic shear, we can illustrate the sensitivity of the likelihood to different regions in the data set by computing the contribution to the $\chi^2$ projected on the parameters $\bar{z}$, $k$ or $\mu$. The likelihood for galaxy surveys and intensity mapping can be written as
\begin{equation}
\chi^2 = \sum_{\bar{z}} \int_{k_{\text{min}}}^{k_{\text{max}}(\bar{z})} \dd k \int_{-1}^{+1} \dd \mu ~ k^2\frac{V_r(\bar{z})}{2(2\pi)^2} \times\frac{\Delta P^2}{\sigma_{\text{obs}}^2+\sigma_{\text{th}}^2} \ ,
\end{equation}
where $\Delta P$ is the difference between the fiducial and sampled power spectrum. Again, by omitting one of the sums or integrals, we obtain the desired projection. Similar to Figure \ref{Chi2_contr_Plots2} for cosmic shear, we see in Figure \ref{Chi2_contr_Plots} the result where the power spectrum differs everywhere by $\Delta P = \Delta P(\bm{k},\bar{z}) = P-\hat{P} = 0.001 P$, but now projected onto $\bar{z}$ (top row), $k$ (middle row) or $\mu$ (bottom row) for galaxy clustering and intensity mapping. The left panel shows the realistic case and the right panel our conservative case.

From Figure \ref{Chi2_contr_Plots}, we see again that the sensitivity of galaxy survey forecasts for SKA1 are not competitive compared to the ones for Euclid and SKA2. This is because of the small sky coverage and low number of galaxies, which result in a large observational error. SKA1 will cover a much smaller redshift range than Euclid and SKA2. At low redshifts the constraining power of galaxy clustering data strongly depends on the amount of non-linear information available, i.e. on the non-linear uncertainty and cut-off. Since an SKA1 galaxy survey is limited to low redshifts, the difference between the constraints obtained from the conservative scheme compared to the realistic one are expected to be quite large. We can see this effect in Figure \ref{ErrorPlots_Pk} (middle row), where the conservative sharp cut-off at $k_{\text{NL}}(z)$ removes the $k$-range in which galaxy clustering with SKA1 is sensitive.

Intensity mapping with SKA1 is more promising than SKA1 galaxy clustering. In contrast to galaxy surveys, the observational error of intensity mapping dominates for high $k$ for both $\mu=0$ and $\mu=1$ (Figure \ref{ErrorPlots_IM}). As a result, there is less non-linear information available. The poor angular resolution limits the information gain to radial directions $\mu \simeq 1$ for large redshifts, i.e. for band 1 (Figure \ref{Chi2_contr_Plots}, bottom row). For the realistic case, band 1 provides a greater amount of information on linear and on mildly non-linear scales, whereas band 2 is better at probing highly non-linear scales (Figure \ref{ErrorPlots_IM}, left panel, with the cumulative effect summarized in Figure \ref{Chi2_contr_Plots}, middle row, left panel). In the conservative case, band 1 accesses more information than band 2, since the non-linear scales are largely removed by the sharp cut-off (Figure \ref{ErrorPlots_IM}, with the cumulative effect in Figure \ref{Chi2_contr_Plots}, middle row, right panel).

Galaxy surveys with Euclid or SKA2 are much more sensitive. The excellent angular resolution, high sky coverage and low shot noise levels strongly suppress observational errors. For $\mu=0$ the information gain is almost entirely limited by theoretical accuracy, especially on non- and quasi-linear scales, so there is a lot of potential for increasing the sensitivity of predictions with improved modelling of non-linear effects on structure formation (Figure \ref{ErrorPlots_Pk}, left panel). At $\mu=1$ redshift-space distortions suppress the power spectrum at high $k$ to a level where shot noise again dominates (Figure \ref{ErrorPlots_Pk}, right panel). In comparison, SKA2 is more sensitive than Euclid, because of the greater sky coverage achieved due to the transparency of the Milky-Way to 21cm radiation, and the enormous amount of available accurate redshift measurements (Figure \ref{Chi2_contr_Plots}, top row).

%% file: figures/PaperChi2ofL_opt.pgf
\begingroup%
\makeatletter%
\begin{pgfpicture}%
\pgfpathrectangle{\pgfpointorigin}{\pgfqpoint{3.044140in}{1.881382in}}%
\pgfusepath{use as bounding box}%
\begin{pgfscope}%
\pgfsetrectcap%
\pgfsetroundjoin%
\definecolor{currentfill}{rgb}{1.000000,1.000000,1.000000}%
\pgfsetfillcolor{currentfill}%
\pgfsetlinewidth{0.000000pt}%
\definecolor{currentstroke}{rgb}{1.000000,1.000000,1.000000}%
\pgfsetstrokecolor{currentstroke}%
\pgfsetdash{}{0pt}%
\pgfpathmoveto{\pgfqpoint{0.000000in}{0.000000in}}%
\pgfpathlineto{\pgfqpoint{3.044140in}{0.000000in}}%
\pgfpathlineto{\pgfqpoint{3.044140in}{1.881382in}}%
\pgfpathlineto{\pgfqpoint{0.000000in}{1.881382in}}%
\pgfpathclose%
\pgfusepath{fill}%
\end{pgfscope}%
\begin{pgfscope}%
\pgfsetrectcap%
\pgfsetroundjoin%
\definecolor{currentfill}{rgb}{1.000000,1.000000,1.000000}%
\pgfsetfillcolor{currentfill}%
\pgfsetlinewidth{0.000000pt}%
\definecolor{currentstroke}{rgb}{0.000000,0.000000,0.000000}%
\pgfsetstrokecolor{currentstroke}%
\pgfsetdash{}{0pt}%
\pgfpathmoveto{\pgfqpoint{0.456621in}{0.329242in}}%
\pgfpathlineto{\pgfqpoint{2.891933in}{0.329242in}}%
\pgfpathlineto{\pgfqpoint{2.891933in}{1.834347in}}%
\pgfpathlineto{\pgfqpoint{0.456621in}{1.834347in}}%
\pgfpathclose%
\pgfusepath{fill}%
\end{pgfscope}%
\begin{pgfscope}%
\pgfpathrectangle{\pgfqpoint{0.456621in}{0.329242in}}{\pgfqpoint{2.435312in}{1.505106in}} %
\pgfusepath{clip}%
\pgfsetrectcap%
\pgfsetroundjoin%
\pgfsetlinewidth{1.003750pt}%
\definecolor{currentstroke}{rgb}{0.767843,0.220980,0.353137}%
\pgfsetstrokecolor{currentstroke}%
\pgfsetdash{}{0pt}%
\pgfpathmoveto{\pgfqpoint{0.459056in}{0.347619in}}%
\pgfpathlineto{\pgfqpoint{0.475129in}{0.464510in}}%
\pgfpathlineto{\pgfqpoint{0.485845in}{0.525315in}}%
\pgfpathlineto{\pgfqpoint{0.497534in}{0.579386in}}%
\pgfpathlineto{\pgfqpoint{0.512633in}{0.638257in}}%
\pgfpathlineto{\pgfqpoint{0.529193in}{0.694143in}}%
\pgfpathlineto{\pgfqpoint{0.543318in}{0.734258in}}%
\pgfpathlineto{\pgfqpoint{0.557930in}{0.768957in}}%
\pgfpathlineto{\pgfqpoint{0.575951in}{0.805758in}}%
\pgfpathlineto{\pgfqpoint{0.598843in}{0.847457in}}%
\pgfpathlineto{\pgfqpoint{0.618813in}{0.879243in}}%
\pgfpathlineto{\pgfqpoint{0.639269in}{0.907725in}}%
\pgfpathlineto{\pgfqpoint{0.662161in}{0.935586in}}%
\pgfpathlineto{\pgfqpoint{0.686514in}{0.961751in}}%
\pgfpathlineto{\pgfqpoint{0.712816in}{0.986691in}}%
\pgfpathlineto{\pgfqpoint{0.736195in}{1.005964in}}%
\pgfpathlineto{\pgfqpoint{0.762009in}{1.024423in}}%
\pgfpathlineto{\pgfqpoint{0.788311in}{1.040527in}}%
\pgfpathlineto{\pgfqpoint{0.813638in}{1.053576in}}%
\pgfpathlineto{\pgfqpoint{0.814125in}{1.049168in}}%
\pgfpathlineto{\pgfqpoint{0.815099in}{1.049631in}}%
\pgfpathlineto{\pgfqpoint{0.843349in}{1.061783in}}%
\pgfpathlineto{\pgfqpoint{0.869650in}{1.070741in}}%
\pgfpathlineto{\pgfqpoint{0.898874in}{1.078296in}}%
\pgfpathlineto{\pgfqpoint{0.929072in}{1.083827in}}%
\pgfpathlineto{\pgfqpoint{0.958782in}{1.087007in}}%
\pgfpathlineto{\pgfqpoint{0.966088in}{1.087466in}}%
\pgfpathlineto{\pgfqpoint{0.966575in}{1.072572in}}%
\pgfpathlineto{\pgfqpoint{0.967549in}{1.072622in}}%
\pgfpathlineto{\pgfqpoint{1.001644in}{1.073164in}}%
\pgfpathlineto{\pgfqpoint{1.038174in}{1.071423in}}%
\pgfpathlineto{\pgfqpoint{1.074216in}{1.067408in}}%
\pgfpathlineto{\pgfqpoint{1.075677in}{1.041206in}}%
\pgfpathlineto{\pgfqpoint{1.118539in}{1.033593in}}%
\pgfpathlineto{\pgfqpoint{1.167245in}{1.022527in}}%
\pgfpathlineto{\pgfqpoint{1.171629in}{1.021415in}}%
\pgfpathlineto{\pgfqpoint{1.173090in}{0.981529in}}%
\pgfpathlineto{\pgfqpoint{1.225205in}{0.966610in}}%
\pgfpathlineto{\pgfqpoint{1.267093in}{0.953384in}}%
\pgfpathlineto{\pgfqpoint{1.268554in}{0.897512in}}%
\pgfpathlineto{\pgfqpoint{1.362557in}{0.865765in}}%
\pgfpathlineto{\pgfqpoint{1.371324in}{0.862691in}}%
\pgfpathlineto{\pgfqpoint{1.372785in}{0.788783in}}%
\pgfpathlineto{\pgfqpoint{1.477991in}{0.753897in}}%
\pgfpathlineto{\pgfqpoint{1.479452in}{0.662904in}}%
\pgfpathlineto{\pgfqpoint{1.598782in}{0.630083in}}%
\pgfpathlineto{\pgfqpoint{1.602679in}{0.629051in}}%
\pgfpathlineto{\pgfqpoint{1.604140in}{0.525921in}}%
\pgfpathlineto{\pgfqpoint{1.710320in}{0.506533in}}%
\pgfpathlineto{\pgfqpoint{1.777534in}{0.495359in}}%
\pgfpathlineto{\pgfqpoint{1.778995in}{0.395458in}}%
\pgfpathlineto{\pgfqpoint{1.939239in}{0.385071in}}%
\pgfpathlineto{\pgfqpoint{2.047367in}{0.379201in}}%
\pgfpathlineto{\pgfqpoint{2.047854in}{0.329242in}}%
\pgfpathlineto{\pgfqpoint{2.047854in}{0.329242in}}%
\pgfusepath{stroke}%
\end{pgfscope}%
\begin{pgfscope}%
\pgfpathrectangle{\pgfqpoint{0.456621in}{0.329242in}}{\pgfqpoint{2.435312in}{1.505106in}} %
\pgfusepath{clip}%
\pgfsetrectcap%
\pgfsetroundjoin%
\pgfsetlinewidth{1.003750pt}%
\definecolor{currentstroke}{rgb}{0.169412,0.150000,0.532353}%
\pgfsetstrokecolor{currentstroke}%
\pgfsetdash{}{0pt}%
\pgfpathmoveto{\pgfqpoint{0.459056in}{0.332469in}}%
\pgfpathlineto{\pgfqpoint{0.475616in}{0.352328in}}%
\pgfpathlineto{\pgfqpoint{0.488767in}{0.364331in}}%
\pgfpathlineto{\pgfqpoint{0.503379in}{0.374628in}}%
\pgfpathlineto{\pgfqpoint{0.519452in}{0.383210in}}%
\pgfpathlineto{\pgfqpoint{0.535038in}{0.389054in}}%
\pgfpathlineto{\pgfqpoint{0.551598in}{0.392836in}}%
\pgfpathlineto{\pgfqpoint{0.570594in}{0.394783in}}%
\pgfpathlineto{\pgfqpoint{0.594460in}{0.394895in}}%
\pgfpathlineto{\pgfqpoint{0.626606in}{0.392753in}}%
\pgfpathlineto{\pgfqpoint{0.678721in}{0.386835in}}%
\pgfpathlineto{\pgfqpoint{0.785875in}{0.374684in}}%
\pgfpathlineto{\pgfqpoint{0.876956in}{0.366791in}}%
\pgfpathlineto{\pgfqpoint{0.982161in}{0.359989in}}%
\pgfpathlineto{\pgfqpoint{1.109285in}{0.353953in}}%
\pgfpathlineto{\pgfqpoint{1.301187in}{0.347366in}}%
\pgfpathlineto{\pgfqpoint{1.669893in}{0.338757in}}%
\pgfpathlineto{\pgfqpoint{1.674277in}{0.338705in}}%
\pgfpathlineto{\pgfqpoint{1.675738in}{0.337418in}}%
\pgfpathlineto{\pgfqpoint{1.810654in}{0.336191in}}%
\pgfpathlineto{\pgfqpoint{1.812116in}{0.334877in}}%
\pgfpathlineto{\pgfqpoint{1.950928in}{0.334045in}}%
\pgfpathlineto{\pgfqpoint{1.952390in}{0.332811in}}%
\pgfpathlineto{\pgfqpoint{2.214916in}{0.331007in}}%
\pgfpathlineto{\pgfqpoint{2.216377in}{0.330162in}}%
\pgfpathlineto{\pgfqpoint{2.499848in}{0.329242in}}%
\pgfpathlineto{\pgfqpoint{2.499848in}{0.329242in}}%
\pgfusepath{stroke}%
\end{pgfscope}%
\begin{pgfscope}%
\pgfpathrectangle{\pgfqpoint{0.456621in}{0.329242in}}{\pgfqpoint{2.435312in}{1.505106in}} %
\pgfusepath{clip}%
\pgfsetrectcap%
\pgfsetroundjoin%
\pgfsetlinewidth{1.003750pt}%
\definecolor{currentstroke}{rgb}{0.387059,0.164510,0.677451}%
\pgfsetstrokecolor{currentstroke}%
\pgfsetdash{}{0pt}%
\pgfpathmoveto{\pgfqpoint{0.459056in}{0.355727in}}%
\pgfpathlineto{\pgfqpoint{0.477565in}{0.555360in}}%
\pgfpathlineto{\pgfqpoint{0.489741in}{0.660924in}}%
\pgfpathlineto{\pgfqpoint{0.502405in}{0.752314in}}%
\pgfpathlineto{\pgfqpoint{0.519452in}{0.858627in}}%
\pgfpathlineto{\pgfqpoint{0.542831in}{0.989548in}}%
\pgfpathlineto{\pgfqpoint{0.562314in}{1.087053in}}%
\pgfpathlineto{\pgfqpoint{0.580822in}{1.168597in}}%
\pgfpathlineto{\pgfqpoint{0.600791in}{1.246279in}}%
\pgfpathlineto{\pgfqpoint{0.621735in}{1.318892in}}%
\pgfpathlineto{\pgfqpoint{0.641705in}{1.380562in}}%
\pgfpathlineto{\pgfqpoint{0.661187in}{1.433734in}}%
\pgfpathlineto{\pgfqpoint{0.679209in}{1.476884in}}%
\pgfpathlineto{\pgfqpoint{0.697717in}{1.515361in}}%
\pgfpathlineto{\pgfqpoint{0.716225in}{1.548788in}}%
\pgfpathlineto{\pgfqpoint{0.734247in}{1.576778in}}%
\pgfpathlineto{\pgfqpoint{0.753242in}{1.602097in}}%
\pgfpathlineto{\pgfqpoint{0.771750in}{1.623191in}}%
\pgfpathlineto{\pgfqpoint{0.788798in}{1.639504in}}%
\pgfpathlineto{\pgfqpoint{0.806819in}{1.653794in}}%
\pgfpathlineto{\pgfqpoint{0.825814in}{1.666121in}}%
\pgfpathlineto{\pgfqpoint{0.845784in}{1.676353in}}%
\pgfpathlineto{\pgfqpoint{0.863805in}{1.683162in}}%
\pgfpathlineto{\pgfqpoint{0.883775in}{1.688356in}}%
\pgfpathlineto{\pgfqpoint{0.905205in}{1.691685in}}%
\pgfpathlineto{\pgfqpoint{0.927123in}{1.693000in}}%
\pgfpathlineto{\pgfqpoint{0.950502in}{1.692054in}}%
\pgfpathlineto{\pgfqpoint{0.976804in}{1.688388in}}%
\pgfpathlineto{\pgfqpoint{1.005540in}{1.681983in}}%
\pgfpathlineto{\pgfqpoint{1.036712in}{1.672681in}}%
\pgfpathlineto{\pgfqpoint{1.065936in}{1.661703in}}%
\pgfpathlineto{\pgfqpoint{1.099056in}{1.647045in}}%
\pgfpathlineto{\pgfqpoint{1.147763in}{1.622554in}}%
\pgfpathlineto{\pgfqpoint{1.191111in}{1.598301in}}%
\pgfpathlineto{\pgfqpoint{1.251020in}{1.561475in}}%
\pgfpathlineto{\pgfqpoint{1.340639in}{1.502917in}}%
\pgfpathlineto{\pgfqpoint{1.402983in}{1.460313in}}%
\pgfpathlineto{\pgfqpoint{1.404444in}{1.446814in}}%
\pgfpathlineto{\pgfqpoint{1.584658in}{1.324454in}}%
\pgfpathlineto{\pgfqpoint{1.666484in}{1.270347in}}%
\pgfpathlineto{\pgfqpoint{1.667945in}{1.229001in}}%
\pgfpathlineto{\pgfqpoint{1.728341in}{1.191664in}}%
\pgfpathlineto{\pgfqpoint{1.797017in}{1.150667in}}%
\pgfpathlineto{\pgfqpoint{1.878843in}{1.103681in}}%
\pgfpathlineto{\pgfqpoint{1.880304in}{1.041781in}}%
\pgfpathlineto{\pgfqpoint{1.977230in}{0.992664in}}%
\pgfpathlineto{\pgfqpoint{2.048828in}{0.958624in}}%
\pgfpathlineto{\pgfqpoint{2.050289in}{0.884269in}}%
\pgfpathlineto{\pgfqpoint{2.138935in}{0.848309in}}%
\pgfpathlineto{\pgfqpoint{2.199817in}{0.824975in}}%
\pgfpathlineto{\pgfqpoint{2.201279in}{0.743877in}}%
\pgfpathlineto{\pgfqpoint{2.277260in}{0.720354in}}%
\pgfpathlineto{\pgfqpoint{2.369802in}{0.694012in}}%
\pgfpathlineto{\pgfqpoint{2.371263in}{0.612941in}}%
\pgfpathlineto{\pgfqpoint{2.463318in}{0.593925in}}%
\pgfpathlineto{\pgfqpoint{2.570959in}{0.573874in}}%
\pgfpathlineto{\pgfqpoint{2.650350in}{0.560300in}}%
\pgfpathlineto{\pgfqpoint{2.651811in}{0.482976in}}%
\pgfpathlineto{\pgfqpoint{2.863683in}{0.460215in}}%
\pgfpathlineto{\pgfqpoint{2.891446in}{0.457327in}}%
\pgfpathlineto{\pgfqpoint{2.891933in}{0.329242in}}%
\pgfpathlineto{\pgfqpoint{2.891933in}{0.329242in}}%
\pgfusepath{stroke}%
\end{pgfscope}%
\begin{pgfscope}%
\pgfpathrectangle{\pgfqpoint{0.456621in}{0.329242in}}{\pgfqpoint{2.435312in}{1.505106in}} %
\pgfusepath{clip}%
\pgfsetbuttcap%
\pgfsetroundjoin%
\pgfsetlinewidth{0.501875pt}%
\definecolor{currentstroke}{rgb}{0.000000,0.000000,0.000000}%
\pgfsetstrokecolor{currentstroke}%
\pgfsetdash{{1.000000pt}{3.000000pt}}{0.000000pt}%
\pgfpathmoveto{\pgfqpoint{0.456621in}{0.329242in}}%
\pgfpathlineto{\pgfqpoint{0.456621in}{1.834347in}}%
\pgfusepath{stroke}%
\end{pgfscope}%
\begin{pgfscope}%
\pgfsetbuttcap%
\pgfsetroundjoin%
\definecolor{currentfill}{rgb}{0.000000,0.000000,0.000000}%
\pgfsetfillcolor{currentfill}%
\pgfsetlinewidth{0.501875pt}%
\definecolor{currentstroke}{rgb}{0.000000,0.000000,0.000000}%
\pgfsetstrokecolor{currentstroke}%
\pgfsetdash{}{0pt}%
\pgfsys@defobject{currentmarker}{\pgfqpoint{0.000000in}{0.000000in}}{\pgfqpoint{0.000000in}{0.055556in}}{%
\pgfpathmoveto{\pgfqpoint{0.000000in}{0.000000in}}%
\pgfpathlineto{\pgfqpoint{0.000000in}{0.055556in}}%
\pgfusepath{stroke,fill}%
}%
\begin{pgfscope}%
\pgfsys@transformshift{0.456621in}{0.329242in}%
\pgfsys@useobject{currentmarker}{}%
\end{pgfscope}%
\end{pgfscope}%
\begin{pgfscope}%
\pgfsetbuttcap%
\pgfsetroundjoin%
\definecolor{currentfill}{rgb}{0.000000,0.000000,0.000000}%
\pgfsetfillcolor{currentfill}%
\pgfsetlinewidth{0.501875pt}%
\definecolor{currentstroke}{rgb}{0.000000,0.000000,0.000000}%
\pgfsetstrokecolor{currentstroke}%
\pgfsetdash{}{0pt}%
\pgfsys@defobject{currentmarker}{\pgfqpoint{0.000000in}{-0.055556in}}{\pgfqpoint{0.000000in}{0.000000in}}{%
\pgfpathmoveto{\pgfqpoint{0.000000in}{0.000000in}}%
\pgfpathlineto{\pgfqpoint{0.000000in}{-0.055556in}}%
\pgfusepath{stroke,fill}%
}%
\begin{pgfscope}%
\pgfsys@transformshift{0.456621in}{1.834347in}%
\pgfsys@useobject{currentmarker}{}%
\end{pgfscope}%
\end{pgfscope}%
\begin{pgfscope}%
\pgftext[left,bottom,x=0.427107in,y=0.202081in,rotate=0.000000]{{\rmfamily\fontsize{8.000000}{9.600000}\selectfont \(\displaystyle 0\)}}
\end{pgfscope}%
\begin{pgfscope}%
\pgfpathrectangle{\pgfqpoint{0.456621in}{0.329242in}}{\pgfqpoint{2.435312in}{1.505106in}} %
\pgfusepath{clip}%
\pgfsetbuttcap%
\pgfsetroundjoin%
\pgfsetlinewidth{0.501875pt}%
\definecolor{currentstroke}{rgb}{0.000000,0.000000,0.000000}%
\pgfsetstrokecolor{currentstroke}%
\pgfsetdash{{1.000000pt}{3.000000pt}}{0.000000pt}%
\pgfpathmoveto{\pgfqpoint{0.943683in}{0.329242in}}%
\pgfpathlineto{\pgfqpoint{0.943683in}{1.834347in}}%
\pgfusepath{stroke}%
\end{pgfscope}%
\begin{pgfscope}%
\pgfsetbuttcap%
\pgfsetroundjoin%
\definecolor{currentfill}{rgb}{0.000000,0.000000,0.000000}%
\pgfsetfillcolor{currentfill}%
\pgfsetlinewidth{0.501875pt}%
\definecolor{currentstroke}{rgb}{0.000000,0.000000,0.000000}%
\pgfsetstrokecolor{currentstroke}%
\pgfsetdash{}{0pt}%
\pgfsys@defobject{currentmarker}{\pgfqpoint{0.000000in}{0.000000in}}{\pgfqpoint{0.000000in}{0.055556in}}{%
\pgfpathmoveto{\pgfqpoint{0.000000in}{0.000000in}}%
\pgfpathlineto{\pgfqpoint{0.000000in}{0.055556in}}%
\pgfusepath{stroke,fill}%
}%
\begin{pgfscope}%
\pgfsys@transformshift{0.943683in}{0.329242in}%
\pgfsys@useobject{currentmarker}{}%
\end{pgfscope}%
\end{pgfscope}%
\begin{pgfscope}%
\pgfsetbuttcap%
\pgfsetroundjoin%
\definecolor{currentfill}{rgb}{0.000000,0.000000,0.000000}%
\pgfsetfillcolor{currentfill}%
\pgfsetlinewidth{0.501875pt}%
\definecolor{currentstroke}{rgb}{0.000000,0.000000,0.000000}%
\pgfsetstrokecolor{currentstroke}%
\pgfsetdash{}{0pt}%
\pgfsys@defobject{currentmarker}{\pgfqpoint{0.000000in}{-0.055556in}}{\pgfqpoint{0.000000in}{0.000000in}}{%
\pgfpathmoveto{\pgfqpoint{0.000000in}{0.000000in}}%
\pgfpathlineto{\pgfqpoint{0.000000in}{-0.055556in}}%
\pgfusepath{stroke,fill}%
}%
\begin{pgfscope}%
\pgfsys@transformshift{0.943683in}{1.834347in}%
\pgfsys@useobject{currentmarker}{}%
\end{pgfscope}%
\end{pgfscope}%
\begin{pgfscope}%
\pgftext[left,bottom,x=0.825626in,y=0.202081in,rotate=0.000000]{{\rmfamily\fontsize{8.000000}{9.600000}\selectfont \(\displaystyle 1000\)}}
\end{pgfscope}%
\begin{pgfscope}%
\pgfpathrectangle{\pgfqpoint{0.456621in}{0.329242in}}{\pgfqpoint{2.435312in}{1.505106in}} %
\pgfusepath{clip}%
\pgfsetbuttcap%
\pgfsetroundjoin%
\pgfsetlinewidth{0.501875pt}%
\definecolor{currentstroke}{rgb}{0.000000,0.000000,0.000000}%
\pgfsetstrokecolor{currentstroke}%
\pgfsetdash{{1.000000pt}{3.000000pt}}{0.000000pt}%
\pgfpathmoveto{\pgfqpoint{1.430746in}{0.329242in}}%
\pgfpathlineto{\pgfqpoint{1.430746in}{1.834347in}}%
\pgfusepath{stroke}%
\end{pgfscope}%
\begin{pgfscope}%
\pgfsetbuttcap%
\pgfsetroundjoin%
\definecolor{currentfill}{rgb}{0.000000,0.000000,0.000000}%
\pgfsetfillcolor{currentfill}%
\pgfsetlinewidth{0.501875pt}%
\definecolor{currentstroke}{rgb}{0.000000,0.000000,0.000000}%
\pgfsetstrokecolor{currentstroke}%
\pgfsetdash{}{0pt}%
\pgfsys@defobject{currentmarker}{\pgfqpoint{0.000000in}{0.000000in}}{\pgfqpoint{0.000000in}{0.055556in}}{%
\pgfpathmoveto{\pgfqpoint{0.000000in}{0.000000in}}%
\pgfpathlineto{\pgfqpoint{0.000000in}{0.055556in}}%
\pgfusepath{stroke,fill}%
}%
\begin{pgfscope}%
\pgfsys@transformshift{1.430746in}{0.329242in}%
\pgfsys@useobject{currentmarker}{}%
\end{pgfscope}%
\end{pgfscope}%
\begin{pgfscope}%
\pgfsetbuttcap%
\pgfsetroundjoin%
\definecolor{currentfill}{rgb}{0.000000,0.000000,0.000000}%
\pgfsetfillcolor{currentfill}%
\pgfsetlinewidth{0.501875pt}%
\definecolor{currentstroke}{rgb}{0.000000,0.000000,0.000000}%
\pgfsetstrokecolor{currentstroke}%
\pgfsetdash{}{0pt}%
\pgfsys@defobject{currentmarker}{\pgfqpoint{0.000000in}{-0.055556in}}{\pgfqpoint{0.000000in}{0.000000in}}{%
\pgfpathmoveto{\pgfqpoint{0.000000in}{0.000000in}}%
\pgfpathlineto{\pgfqpoint{0.000000in}{-0.055556in}}%
\pgfusepath{stroke,fill}%
}%
\begin{pgfscope}%
\pgfsys@transformshift{1.430746in}{1.834347in}%
\pgfsys@useobject{currentmarker}{}%
\end{pgfscope}%
\end{pgfscope}%
\begin{pgfscope}%
\pgftext[left,bottom,x=1.312689in,y=0.202081in,rotate=0.000000]{{\rmfamily\fontsize{8.000000}{9.600000}\selectfont \(\displaystyle 2000\)}}
\end{pgfscope}%
\begin{pgfscope}%
\pgfpathrectangle{\pgfqpoint{0.456621in}{0.329242in}}{\pgfqpoint{2.435312in}{1.505106in}} %
\pgfusepath{clip}%
\pgfsetbuttcap%
\pgfsetroundjoin%
\pgfsetlinewidth{0.501875pt}%
\definecolor{currentstroke}{rgb}{0.000000,0.000000,0.000000}%
\pgfsetstrokecolor{currentstroke}%
\pgfsetdash{{1.000000pt}{3.000000pt}}{0.000000pt}%
\pgfpathmoveto{\pgfqpoint{1.917808in}{0.329242in}}%
\pgfpathlineto{\pgfqpoint{1.917808in}{1.834347in}}%
\pgfusepath{stroke}%
\end{pgfscope}%
\begin{pgfscope}%
\pgfsetbuttcap%
\pgfsetroundjoin%
\definecolor{currentfill}{rgb}{0.000000,0.000000,0.000000}%
\pgfsetfillcolor{currentfill}%
\pgfsetlinewidth{0.501875pt}%
\definecolor{currentstroke}{rgb}{0.000000,0.000000,0.000000}%
\pgfsetstrokecolor{currentstroke}%
\pgfsetdash{}{0pt}%
\pgfsys@defobject{currentmarker}{\pgfqpoint{0.000000in}{0.000000in}}{\pgfqpoint{0.000000in}{0.055556in}}{%
\pgfpathmoveto{\pgfqpoint{0.000000in}{0.000000in}}%
\pgfpathlineto{\pgfqpoint{0.000000in}{0.055556in}}%
\pgfusepath{stroke,fill}%
}%
\begin{pgfscope}%
\pgfsys@transformshift{1.917808in}{0.329242in}%
\pgfsys@useobject{currentmarker}{}%
\end{pgfscope}%
\end{pgfscope}%
\begin{pgfscope}%
\pgfsetbuttcap%
\pgfsetroundjoin%
\definecolor{currentfill}{rgb}{0.000000,0.000000,0.000000}%
\pgfsetfillcolor{currentfill}%
\pgfsetlinewidth{0.501875pt}%
\definecolor{currentstroke}{rgb}{0.000000,0.000000,0.000000}%
\pgfsetstrokecolor{currentstroke}%
\pgfsetdash{}{0pt}%
\pgfsys@defobject{currentmarker}{\pgfqpoint{0.000000in}{-0.055556in}}{\pgfqpoint{0.000000in}{0.000000in}}{%
\pgfpathmoveto{\pgfqpoint{0.000000in}{0.000000in}}%
\pgfpathlineto{\pgfqpoint{0.000000in}{-0.055556in}}%
\pgfusepath{stroke,fill}%
}%
\begin{pgfscope}%
\pgfsys@transformshift{1.917808in}{1.834347in}%
\pgfsys@useobject{currentmarker}{}%
\end{pgfscope}%
\end{pgfscope}%
\begin{pgfscope}%
\pgftext[left,bottom,x=1.799751in,y=0.202081in,rotate=0.000000]{{\rmfamily\fontsize{8.000000}{9.600000}\selectfont \(\displaystyle 3000\)}}
\end{pgfscope}%
\begin{pgfscope}%
\pgfpathrectangle{\pgfqpoint{0.456621in}{0.329242in}}{\pgfqpoint{2.435312in}{1.505106in}} %
\pgfusepath{clip}%
\pgfsetbuttcap%
\pgfsetroundjoin%
\pgfsetlinewidth{0.501875pt}%
\definecolor{currentstroke}{rgb}{0.000000,0.000000,0.000000}%
\pgfsetstrokecolor{currentstroke}%
\pgfsetdash{{1.000000pt}{3.000000pt}}{0.000000pt}%
\pgfpathmoveto{\pgfqpoint{2.404871in}{0.329242in}}%
\pgfpathlineto{\pgfqpoint{2.404871in}{1.834347in}}%
\pgfusepath{stroke}%
\end{pgfscope}%
\begin{pgfscope}%
\pgfsetbuttcap%
\pgfsetroundjoin%
\definecolor{currentfill}{rgb}{0.000000,0.000000,0.000000}%
\pgfsetfillcolor{currentfill}%
\pgfsetlinewidth{0.501875pt}%
\definecolor{currentstroke}{rgb}{0.000000,0.000000,0.000000}%
\pgfsetstrokecolor{currentstroke}%
\pgfsetdash{}{0pt}%
\pgfsys@defobject{currentmarker}{\pgfqpoint{0.000000in}{0.000000in}}{\pgfqpoint{0.000000in}{0.055556in}}{%
\pgfpathmoveto{\pgfqpoint{0.000000in}{0.000000in}}%
\pgfpathlineto{\pgfqpoint{0.000000in}{0.055556in}}%
\pgfusepath{stroke,fill}%
}%
\begin{pgfscope}%
\pgfsys@transformshift{2.404871in}{0.329242in}%
\pgfsys@useobject{currentmarker}{}%
\end{pgfscope}%
\end{pgfscope}%
\begin{pgfscope}%
\pgfsetbuttcap%
\pgfsetroundjoin%
\definecolor{currentfill}{rgb}{0.000000,0.000000,0.000000}%
\pgfsetfillcolor{currentfill}%
\pgfsetlinewidth{0.501875pt}%
\definecolor{currentstroke}{rgb}{0.000000,0.000000,0.000000}%
\pgfsetstrokecolor{currentstroke}%
\pgfsetdash{}{0pt}%
\pgfsys@defobject{currentmarker}{\pgfqpoint{0.000000in}{-0.055556in}}{\pgfqpoint{0.000000in}{0.000000in}}{%
\pgfpathmoveto{\pgfqpoint{0.000000in}{0.000000in}}%
\pgfpathlineto{\pgfqpoint{0.000000in}{-0.055556in}}%
\pgfusepath{stroke,fill}%
}%
\begin{pgfscope}%
\pgfsys@transformshift{2.404871in}{1.834347in}%
\pgfsys@useobject{currentmarker}{}%
\end{pgfscope}%
\end{pgfscope}%
\begin{pgfscope}%
\pgftext[left,bottom,x=2.286813in,y=0.202081in,rotate=0.000000]{{\rmfamily\fontsize{8.000000}{9.600000}\selectfont \(\displaystyle 4000\)}}
\end{pgfscope}%
\begin{pgfscope}%
\pgfpathrectangle{\pgfqpoint{0.456621in}{0.329242in}}{\pgfqpoint{2.435312in}{1.505106in}} %
\pgfusepath{clip}%
\pgfsetbuttcap%
\pgfsetroundjoin%
\pgfsetlinewidth{0.501875pt}%
\definecolor{currentstroke}{rgb}{0.000000,0.000000,0.000000}%
\pgfsetstrokecolor{currentstroke}%
\pgfsetdash{{1.000000pt}{3.000000pt}}{0.000000pt}%
\pgfpathmoveto{\pgfqpoint{2.891933in}{0.329242in}}%
\pgfpathlineto{\pgfqpoint{2.891933in}{1.834347in}}%
\pgfusepath{stroke}%
\end{pgfscope}%
\begin{pgfscope}%
\pgfsetbuttcap%
\pgfsetroundjoin%
\definecolor{currentfill}{rgb}{0.000000,0.000000,0.000000}%
\pgfsetfillcolor{currentfill}%
\pgfsetlinewidth{0.501875pt}%
\definecolor{currentstroke}{rgb}{0.000000,0.000000,0.000000}%
\pgfsetstrokecolor{currentstroke}%
\pgfsetdash{}{0pt}%
\pgfsys@defobject{currentmarker}{\pgfqpoint{0.000000in}{0.000000in}}{\pgfqpoint{0.000000in}{0.055556in}}{%
\pgfpathmoveto{\pgfqpoint{0.000000in}{0.000000in}}%
\pgfpathlineto{\pgfqpoint{0.000000in}{0.055556in}}%
\pgfusepath{stroke,fill}%
}%
\begin{pgfscope}%
\pgfsys@transformshift{2.891933in}{0.329242in}%
\pgfsys@useobject{currentmarker}{}%
\end{pgfscope}%
\end{pgfscope}%
\begin{pgfscope}%
\pgfsetbuttcap%
\pgfsetroundjoin%
\definecolor{currentfill}{rgb}{0.000000,0.000000,0.000000}%
\pgfsetfillcolor{currentfill}%
\pgfsetlinewidth{0.501875pt}%
\definecolor{currentstroke}{rgb}{0.000000,0.000000,0.000000}%
\pgfsetstrokecolor{currentstroke}%
\pgfsetdash{}{0pt}%
\pgfsys@defobject{currentmarker}{\pgfqpoint{0.000000in}{-0.055556in}}{\pgfqpoint{0.000000in}{0.000000in}}{%
\pgfpathmoveto{\pgfqpoint{0.000000in}{0.000000in}}%
\pgfpathlineto{\pgfqpoint{0.000000in}{-0.055556in}}%
\pgfusepath{stroke,fill}%
}%
\begin{pgfscope}%
\pgfsys@transformshift{2.891933in}{1.834347in}%
\pgfsys@useobject{currentmarker}{}%
\end{pgfscope}%
\end{pgfscope}%
\begin{pgfscope}%
\pgftext[left,bottom,x=2.773876in,y=0.202081in,rotate=0.000000]{{\rmfamily\fontsize{8.000000}{9.600000}\selectfont \(\displaystyle 5000\)}}
\end{pgfscope}%
\begin{pgfscope}%
\pgftext[left,bottom,x=1.650164in,y=0.055476in,rotate=0.000000]{{\rmfamily\fontsize{8.000000}{9.600000}\selectfont \(\displaystyle \ell\)}}
\end{pgfscope}%
\begin{pgfscope}%
\pgfpathrectangle{\pgfqpoint{0.456621in}{0.329242in}}{\pgfqpoint{2.435312in}{1.505106in}} %
\pgfusepath{clip}%
\pgfsetbuttcap%
\pgfsetroundjoin%
\pgfsetlinewidth{0.501875pt}%
\definecolor{currentstroke}{rgb}{0.000000,0.000000,0.000000}%
\pgfsetstrokecolor{currentstroke}%
\pgfsetdash{{1.000000pt}{3.000000pt}}{0.000000pt}%
\pgfpathmoveto{\pgfqpoint{0.456621in}{0.329242in}}%
\pgfpathlineto{\pgfqpoint{2.891933in}{0.329242in}}%
\pgfusepath{stroke}%
\end{pgfscope}%
\begin{pgfscope}%
\pgfsetbuttcap%
\pgfsetroundjoin%
\definecolor{currentfill}{rgb}{0.000000,0.000000,0.000000}%
\pgfsetfillcolor{currentfill}%
\pgfsetlinewidth{0.501875pt}%
\definecolor{currentstroke}{rgb}{0.000000,0.000000,0.000000}%
\pgfsetstrokecolor{currentstroke}%
\pgfsetdash{}{0pt}%
\pgfsys@defobject{currentmarker}{\pgfqpoint{0.000000in}{0.000000in}}{\pgfqpoint{0.055556in}{0.000000in}}{%
\pgfpathmoveto{\pgfqpoint{0.000000in}{0.000000in}}%
\pgfpathlineto{\pgfqpoint{0.055556in}{0.000000in}}%
\pgfusepath{stroke,fill}%
}%
\begin{pgfscope}%
\pgfsys@transformshift{0.456621in}{0.329242in}%
\pgfsys@useobject{currentmarker}{}%
\end{pgfscope}%
\end{pgfscope}%
\begin{pgfscope}%
\pgfsetbuttcap%
\pgfsetroundjoin%
\definecolor{currentfill}{rgb}{0.000000,0.000000,0.000000}%
\pgfsetfillcolor{currentfill}%
\pgfsetlinewidth{0.501875pt}%
\definecolor{currentstroke}{rgb}{0.000000,0.000000,0.000000}%
\pgfsetstrokecolor{currentstroke}%
\pgfsetdash{}{0pt}%
\pgfsys@defobject{currentmarker}{\pgfqpoint{-0.055556in}{0.000000in}}{\pgfqpoint{0.000000in}{0.000000in}}{%
\pgfpathmoveto{\pgfqpoint{0.000000in}{0.000000in}}%
\pgfpathlineto{\pgfqpoint{-0.055556in}{0.000000in}}%
\pgfusepath{stroke,fill}%
}%
\begin{pgfscope}%
\pgfsys@transformshift{2.891933in}{0.329242in}%
\pgfsys@useobject{currentmarker}{}%
\end{pgfscope}%
\end{pgfscope}%
\begin{pgfscope}%
\pgftext[left,bottom,x=0.250214in,y=0.293439in,rotate=0.000000]{{\rmfamily\fontsize{8.000000}{9.600000}\selectfont \(\displaystyle 0.0\)}}
\end{pgfscope}%
\begin{pgfscope}%
\pgfpathrectangle{\pgfqpoint{0.456621in}{0.329242in}}{\pgfqpoint{2.435312in}{1.505106in}} %
\pgfusepath{clip}%
\pgfsetbuttcap%
\pgfsetroundjoin%
\pgfsetlinewidth{0.501875pt}%
\definecolor{currentstroke}{rgb}{0.000000,0.000000,0.000000}%
\pgfsetstrokecolor{currentstroke}%
\pgfsetdash{{1.000000pt}{3.000000pt}}{0.000000pt}%
\pgfpathmoveto{\pgfqpoint{0.456621in}{0.580093in}}%
\pgfpathlineto{\pgfqpoint{2.891933in}{0.580093in}}%
\pgfusepath{stroke}%
\end{pgfscope}%
\begin{pgfscope}%
\pgfsetbuttcap%
\pgfsetroundjoin%
\definecolor{currentfill}{rgb}{0.000000,0.000000,0.000000}%
\pgfsetfillcolor{currentfill}%
\pgfsetlinewidth{0.501875pt}%
\definecolor{currentstroke}{rgb}{0.000000,0.000000,0.000000}%
\pgfsetstrokecolor{currentstroke}%
\pgfsetdash{}{0pt}%
\pgfsys@defobject{currentmarker}{\pgfqpoint{0.000000in}{0.000000in}}{\pgfqpoint{0.055556in}{0.000000in}}{%
\pgfpathmoveto{\pgfqpoint{0.000000in}{0.000000in}}%
\pgfpathlineto{\pgfqpoint{0.055556in}{0.000000in}}%
\pgfusepath{stroke,fill}%
}%
\begin{pgfscope}%
\pgfsys@transformshift{0.456621in}{0.580093in}%
\pgfsys@useobject{currentmarker}{}%
\end{pgfscope}%
\end{pgfscope}%
\begin{pgfscope}%
\pgfsetbuttcap%
\pgfsetroundjoin%
\definecolor{currentfill}{rgb}{0.000000,0.000000,0.000000}%
\pgfsetfillcolor{currentfill}%
\pgfsetlinewidth{0.501875pt}%
\definecolor{currentstroke}{rgb}{0.000000,0.000000,0.000000}%
\pgfsetstrokecolor{currentstroke}%
\pgfsetdash{}{0pt}%
\pgfsys@defobject{currentmarker}{\pgfqpoint{-0.055556in}{0.000000in}}{\pgfqpoint{0.000000in}{0.000000in}}{%
\pgfpathmoveto{\pgfqpoint{0.000000in}{0.000000in}}%
\pgfpathlineto{\pgfqpoint{-0.055556in}{0.000000in}}%
\pgfusepath{stroke,fill}%
}%
\begin{pgfscope}%
\pgfsys@transformshift{2.891933in}{0.580093in}%
\pgfsys@useobject{currentmarker}{}%
\end{pgfscope}%
\end{pgfscope}%
\begin{pgfscope}%
\pgftext[left,bottom,x=0.250214in,y=0.544290in,rotate=0.000000]{{\rmfamily\fontsize{8.000000}{9.600000}\selectfont \(\displaystyle 0.5\)}}
\end{pgfscope}%
\begin{pgfscope}%
\pgfpathrectangle{\pgfqpoint{0.456621in}{0.329242in}}{\pgfqpoint{2.435312in}{1.505106in}} %
\pgfusepath{clip}%
\pgfsetbuttcap%
\pgfsetroundjoin%
\pgfsetlinewidth{0.501875pt}%
\definecolor{currentstroke}{rgb}{0.000000,0.000000,0.000000}%
\pgfsetstrokecolor{currentstroke}%
\pgfsetdash{{1.000000pt}{3.000000pt}}{0.000000pt}%
\pgfpathmoveto{\pgfqpoint{0.456621in}{0.830944in}}%
\pgfpathlineto{\pgfqpoint{2.891933in}{0.830944in}}%
\pgfusepath{stroke}%
\end{pgfscope}%
\begin{pgfscope}%
\pgfsetbuttcap%
\pgfsetroundjoin%
\definecolor{currentfill}{rgb}{0.000000,0.000000,0.000000}%
\pgfsetfillcolor{currentfill}%
\pgfsetlinewidth{0.501875pt}%
\definecolor{currentstroke}{rgb}{0.000000,0.000000,0.000000}%
\pgfsetstrokecolor{currentstroke}%
\pgfsetdash{}{0pt}%
\pgfsys@defobject{currentmarker}{\pgfqpoint{0.000000in}{0.000000in}}{\pgfqpoint{0.055556in}{0.000000in}}{%
\pgfpathmoveto{\pgfqpoint{0.000000in}{0.000000in}}%
\pgfpathlineto{\pgfqpoint{0.055556in}{0.000000in}}%
\pgfusepath{stroke,fill}%
}%
\begin{pgfscope}%
\pgfsys@transformshift{0.456621in}{0.830944in}%
\pgfsys@useobject{currentmarker}{}%
\end{pgfscope}%
\end{pgfscope}%
\begin{pgfscope}%
\pgfsetbuttcap%
\pgfsetroundjoin%
\definecolor{currentfill}{rgb}{0.000000,0.000000,0.000000}%
\pgfsetfillcolor{currentfill}%
\pgfsetlinewidth{0.501875pt}%
\definecolor{currentstroke}{rgb}{0.000000,0.000000,0.000000}%
\pgfsetstrokecolor{currentstroke}%
\pgfsetdash{}{0pt}%
\pgfsys@defobject{currentmarker}{\pgfqpoint{-0.055556in}{0.000000in}}{\pgfqpoint{0.000000in}{0.000000in}}{%
\pgfpathmoveto{\pgfqpoint{0.000000in}{0.000000in}}%
\pgfpathlineto{\pgfqpoint{-0.055556in}{0.000000in}}%
\pgfusepath{stroke,fill}%
}%
\begin{pgfscope}%
\pgfsys@transformshift{2.891933in}{0.830944in}%
\pgfsys@useobject{currentmarker}{}%
\end{pgfscope}%
\end{pgfscope}%
\begin{pgfscope}%
\pgftext[left,bottom,x=0.250214in,y=0.795141in,rotate=0.000000]{{\rmfamily\fontsize{8.000000}{9.600000}\selectfont \(\displaystyle 1.0\)}}
\end{pgfscope}%
\begin{pgfscope}%
\pgfpathrectangle{\pgfqpoint{0.456621in}{0.329242in}}{\pgfqpoint{2.435312in}{1.505106in}} %
\pgfusepath{clip}%
\pgfsetbuttcap%
\pgfsetroundjoin%
\pgfsetlinewidth{0.501875pt}%
\definecolor{currentstroke}{rgb}{0.000000,0.000000,0.000000}%
\pgfsetstrokecolor{currentstroke}%
\pgfsetdash{{1.000000pt}{3.000000pt}}{0.000000pt}%
\pgfpathmoveto{\pgfqpoint{0.456621in}{1.081795in}}%
\pgfpathlineto{\pgfqpoint{2.891933in}{1.081795in}}%
\pgfusepath{stroke}%
\end{pgfscope}%
\begin{pgfscope}%
\pgfsetbuttcap%
\pgfsetroundjoin%
\definecolor{currentfill}{rgb}{0.000000,0.000000,0.000000}%
\pgfsetfillcolor{currentfill}%
\pgfsetlinewidth{0.501875pt}%
\definecolor{currentstroke}{rgb}{0.000000,0.000000,0.000000}%
\pgfsetstrokecolor{currentstroke}%
\pgfsetdash{}{0pt}%
\pgfsys@defobject{currentmarker}{\pgfqpoint{0.000000in}{0.000000in}}{\pgfqpoint{0.055556in}{0.000000in}}{%
\pgfpathmoveto{\pgfqpoint{0.000000in}{0.000000in}}%
\pgfpathlineto{\pgfqpoint{0.055556in}{0.000000in}}%
\pgfusepath{stroke,fill}%
}%
\begin{pgfscope}%
\pgfsys@transformshift{0.456621in}{1.081795in}%
\pgfsys@useobject{currentmarker}{}%
\end{pgfscope}%
\end{pgfscope}%
\begin{pgfscope}%
\pgfsetbuttcap%
\pgfsetroundjoin%
\definecolor{currentfill}{rgb}{0.000000,0.000000,0.000000}%
\pgfsetfillcolor{currentfill}%
\pgfsetlinewidth{0.501875pt}%
\definecolor{currentstroke}{rgb}{0.000000,0.000000,0.000000}%
\pgfsetstrokecolor{currentstroke}%
\pgfsetdash{}{0pt}%
\pgfsys@defobject{currentmarker}{\pgfqpoint{-0.055556in}{0.000000in}}{\pgfqpoint{0.000000in}{0.000000in}}{%
\pgfpathmoveto{\pgfqpoint{0.000000in}{0.000000in}}%
\pgfpathlineto{\pgfqpoint{-0.055556in}{0.000000in}}%
\pgfusepath{stroke,fill}%
}%
\begin{pgfscope}%
\pgfsys@transformshift{2.891933in}{1.081795in}%
\pgfsys@useobject{currentmarker}{}%
\end{pgfscope}%
\end{pgfscope}%
\begin{pgfscope}%
\pgftext[left,bottom,x=0.250214in,y=1.045992in,rotate=0.000000]{{\rmfamily\fontsize{8.000000}{9.600000}\selectfont \(\displaystyle 1.5\)}}
\end{pgfscope}%
\begin{pgfscope}%
\pgfpathrectangle{\pgfqpoint{0.456621in}{0.329242in}}{\pgfqpoint{2.435312in}{1.505106in}} %
\pgfusepath{clip}%
\pgfsetbuttcap%
\pgfsetroundjoin%
\pgfsetlinewidth{0.501875pt}%
\definecolor{currentstroke}{rgb}{0.000000,0.000000,0.000000}%
\pgfsetstrokecolor{currentstroke}%
\pgfsetdash{{1.000000pt}{3.000000pt}}{0.000000pt}%
\pgfpathmoveto{\pgfqpoint{0.456621in}{1.332646in}}%
\pgfpathlineto{\pgfqpoint{2.891933in}{1.332646in}}%
\pgfusepath{stroke}%
\end{pgfscope}%
\begin{pgfscope}%
\pgfsetbuttcap%
\pgfsetroundjoin%
\definecolor{currentfill}{rgb}{0.000000,0.000000,0.000000}%
\pgfsetfillcolor{currentfill}%
\pgfsetlinewidth{0.501875pt}%
\definecolor{currentstroke}{rgb}{0.000000,0.000000,0.000000}%
\pgfsetstrokecolor{currentstroke}%
\pgfsetdash{}{0pt}%
\pgfsys@defobject{currentmarker}{\pgfqpoint{0.000000in}{0.000000in}}{\pgfqpoint{0.055556in}{0.000000in}}{%
\pgfpathmoveto{\pgfqpoint{0.000000in}{0.000000in}}%
\pgfpathlineto{\pgfqpoint{0.055556in}{0.000000in}}%
\pgfusepath{stroke,fill}%
}%
\begin{pgfscope}%
\pgfsys@transformshift{0.456621in}{1.332646in}%
\pgfsys@useobject{currentmarker}{}%
\end{pgfscope}%
\end{pgfscope}%
\begin{pgfscope}%
\pgfsetbuttcap%
\pgfsetroundjoin%
\definecolor{currentfill}{rgb}{0.000000,0.000000,0.000000}%
\pgfsetfillcolor{currentfill}%
\pgfsetlinewidth{0.501875pt}%
\definecolor{currentstroke}{rgb}{0.000000,0.000000,0.000000}%
\pgfsetstrokecolor{currentstroke}%
\pgfsetdash{}{0pt}%
\pgfsys@defobject{currentmarker}{\pgfqpoint{-0.055556in}{0.000000in}}{\pgfqpoint{0.000000in}{0.000000in}}{%
\pgfpathmoveto{\pgfqpoint{0.000000in}{0.000000in}}%
\pgfpathlineto{\pgfqpoint{-0.055556in}{0.000000in}}%
\pgfusepath{stroke,fill}%
}%
\begin{pgfscope}%
\pgfsys@transformshift{2.891933in}{1.332646in}%
\pgfsys@useobject{currentmarker}{}%
\end{pgfscope}%
\end{pgfscope}%
\begin{pgfscope}%
\pgftext[left,bottom,x=0.250214in,y=1.296843in,rotate=0.000000]{{\rmfamily\fontsize{8.000000}{9.600000}\selectfont \(\displaystyle 2.0\)}}
\end{pgfscope}%
\begin{pgfscope}%
\pgfpathrectangle{\pgfqpoint{0.456621in}{0.329242in}}{\pgfqpoint{2.435312in}{1.505106in}} %
\pgfusepath{clip}%
\pgfsetbuttcap%
\pgfsetroundjoin%
\pgfsetlinewidth{0.501875pt}%
\definecolor{currentstroke}{rgb}{0.000000,0.000000,0.000000}%
\pgfsetstrokecolor{currentstroke}%
\pgfsetdash{{1.000000pt}{3.000000pt}}{0.000000pt}%
\pgfpathmoveto{\pgfqpoint{0.456621in}{1.583497in}}%
\pgfpathlineto{\pgfqpoint{2.891933in}{1.583497in}}%
\pgfusepath{stroke}%
\end{pgfscope}%
\begin{pgfscope}%
\pgfsetbuttcap%
\pgfsetroundjoin%
\definecolor{currentfill}{rgb}{0.000000,0.000000,0.000000}%
\pgfsetfillcolor{currentfill}%
\pgfsetlinewidth{0.501875pt}%
\definecolor{currentstroke}{rgb}{0.000000,0.000000,0.000000}%
\pgfsetstrokecolor{currentstroke}%
\pgfsetdash{}{0pt}%
\pgfsys@defobject{currentmarker}{\pgfqpoint{0.000000in}{0.000000in}}{\pgfqpoint{0.055556in}{0.000000in}}{%
\pgfpathmoveto{\pgfqpoint{0.000000in}{0.000000in}}%
\pgfpathlineto{\pgfqpoint{0.055556in}{0.000000in}}%
\pgfusepath{stroke,fill}%
}%
\begin{pgfscope}%
\pgfsys@transformshift{0.456621in}{1.583497in}%
\pgfsys@useobject{currentmarker}{}%
\end{pgfscope}%
\end{pgfscope}%
\begin{pgfscope}%
\pgfsetbuttcap%
\pgfsetroundjoin%
\definecolor{currentfill}{rgb}{0.000000,0.000000,0.000000}%
\pgfsetfillcolor{currentfill}%
\pgfsetlinewidth{0.501875pt}%
\definecolor{currentstroke}{rgb}{0.000000,0.000000,0.000000}%
\pgfsetstrokecolor{currentstroke}%
\pgfsetdash{}{0pt}%
\pgfsys@defobject{currentmarker}{\pgfqpoint{-0.055556in}{0.000000in}}{\pgfqpoint{0.000000in}{0.000000in}}{%
\pgfpathmoveto{\pgfqpoint{0.000000in}{0.000000in}}%
\pgfpathlineto{\pgfqpoint{-0.055556in}{0.000000in}}%
\pgfusepath{stroke,fill}%
}%
\begin{pgfscope}%
\pgfsys@transformshift{2.891933in}{1.583497in}%
\pgfsys@useobject{currentmarker}{}%
\end{pgfscope}%
\end{pgfscope}%
\begin{pgfscope}%
\pgftext[left,bottom,x=0.250214in,y=1.547694in,rotate=0.000000]{{\rmfamily\fontsize{8.000000}{9.600000}\selectfont \(\displaystyle 2.5\)}}
\end{pgfscope}%
\begin{pgfscope}%
\pgfpathrectangle{\pgfqpoint{0.456621in}{0.329242in}}{\pgfqpoint{2.435312in}{1.505106in}} %
\pgfusepath{clip}%
\pgfsetbuttcap%
\pgfsetroundjoin%
\pgfsetlinewidth{0.501875pt}%
\definecolor{currentstroke}{rgb}{0.000000,0.000000,0.000000}%
\pgfsetstrokecolor{currentstroke}%
\pgfsetdash{{1.000000pt}{3.000000pt}}{0.000000pt}%
\pgfpathmoveto{\pgfqpoint{0.456621in}{1.834347in}}%
\pgfpathlineto{\pgfqpoint{2.891933in}{1.834347in}}%
\pgfusepath{stroke}%
\end{pgfscope}%
\begin{pgfscope}%
\pgfsetbuttcap%
\pgfsetroundjoin%
\definecolor{currentfill}{rgb}{0.000000,0.000000,0.000000}%
\pgfsetfillcolor{currentfill}%
\pgfsetlinewidth{0.501875pt}%
\definecolor{currentstroke}{rgb}{0.000000,0.000000,0.000000}%
\pgfsetstrokecolor{currentstroke}%
\pgfsetdash{}{0pt}%
\pgfsys@defobject{currentmarker}{\pgfqpoint{0.000000in}{0.000000in}}{\pgfqpoint{0.055556in}{0.000000in}}{%
\pgfpathmoveto{\pgfqpoint{0.000000in}{0.000000in}}%
\pgfpathlineto{\pgfqpoint{0.055556in}{0.000000in}}%
\pgfusepath{stroke,fill}%
}%
\begin{pgfscope}%
\pgfsys@transformshift{0.456621in}{1.834347in}%
\pgfsys@useobject{currentmarker}{}%
\end{pgfscope}%
\end{pgfscope}%
\begin{pgfscope}%
\pgfsetbuttcap%
\pgfsetroundjoin%
\definecolor{currentfill}{rgb}{0.000000,0.000000,0.000000}%
\pgfsetfillcolor{currentfill}%
\pgfsetlinewidth{0.501875pt}%
\definecolor{currentstroke}{rgb}{0.000000,0.000000,0.000000}%
\pgfsetstrokecolor{currentstroke}%
\pgfsetdash{}{0pt}%
\pgfsys@defobject{currentmarker}{\pgfqpoint{-0.055556in}{0.000000in}}{\pgfqpoint{0.000000in}{0.000000in}}{%
\pgfpathmoveto{\pgfqpoint{0.000000in}{0.000000in}}%
\pgfpathlineto{\pgfqpoint{-0.055556in}{0.000000in}}%
\pgfusepath{stroke,fill}%
}%
\begin{pgfscope}%
\pgfsys@transformshift{2.891933in}{1.834347in}%
\pgfsys@useobject{currentmarker}{}%
\end{pgfscope}%
\end{pgfscope}%
\begin{pgfscope}%
\pgftext[left,bottom,x=0.250214in,y=1.798545in,rotate=0.000000]{{\rmfamily\fontsize{8.000000}{9.600000}\selectfont \(\displaystyle 3.0\)}}
\end{pgfscope}%
\begin{pgfscope}%
\pgftext[left,bottom,x=0.180770in,y=0.787232in,rotate=90.000000]{{\rmfamily\fontsize{8.000000}{9.600000}\selectfont \(\displaystyle \Delta\chi^2 (\times 10^4)\)}}
\end{pgfscope}%
\begin{pgfscope}%
\pgfsetrectcap%
\pgfsetroundjoin%
\pgfsetlinewidth{0.501875pt}%
\definecolor{currentstroke}{rgb}{0.000000,0.000000,0.000000}%
\pgfsetstrokecolor{currentstroke}%
\pgfsetdash{}{0pt}%
\pgfpathmoveto{\pgfqpoint{0.456621in}{1.834347in}}%
\pgfpathlineto{\pgfqpoint{2.891933in}{1.834347in}}%
\pgfusepath{stroke}%
\end{pgfscope}%
\begin{pgfscope}%
\pgfsetrectcap%
\pgfsetroundjoin%
\pgfsetlinewidth{0.501875pt}%
\definecolor{currentstroke}{rgb}{0.000000,0.000000,0.000000}%
\pgfsetstrokecolor{currentstroke}%
\pgfsetdash{}{0pt}%
\pgfpathmoveto{\pgfqpoint{2.891933in}{0.329242in}}%
\pgfpathlineto{\pgfqpoint{2.891933in}{1.834347in}}%
\pgfusepath{stroke}%
\end{pgfscope}%
\begin{pgfscope}%
\pgfsetrectcap%
\pgfsetroundjoin%
\pgfsetlinewidth{0.501875pt}%
\definecolor{currentstroke}{rgb}{0.000000,0.000000,0.000000}%
\pgfsetstrokecolor{currentstroke}%
\pgfsetdash{}{0pt}%
\pgfpathmoveto{\pgfqpoint{0.456621in}{0.329242in}}%
\pgfpathlineto{\pgfqpoint{2.891933in}{0.329242in}}%
\pgfusepath{stroke}%
\end{pgfscope}%
\begin{pgfscope}%
\pgfsetrectcap%
\pgfsetroundjoin%
\pgfsetlinewidth{0.501875pt}%
\definecolor{currentstroke}{rgb}{0.000000,0.000000,0.000000}%
\pgfsetstrokecolor{currentstroke}%
\pgfsetdash{}{0pt}%
\pgfpathmoveto{\pgfqpoint{0.456621in}{0.329242in}}%
\pgfpathlineto{\pgfqpoint{0.456621in}{1.834347in}}%
\pgfusepath{stroke}%
\end{pgfscope}%
\begin{pgfscope}%
\pgfsetrectcap%
\pgfsetroundjoin%
\definecolor{currentfill}{rgb}{1.000000,1.000000,1.000000}%
\pgfsetfillcolor{currentfill}%
\pgfsetlinewidth{0.501875pt}%
\definecolor{currentstroke}{rgb}{0.000000,0.000000,0.000000}%
\pgfsetstrokecolor{currentstroke}%
\pgfsetdash{}{0pt}%
\pgfpathmoveto{\pgfqpoint{2.129081in}{1.302249in}}%
\pgfpathlineto{\pgfqpoint{2.814155in}{1.302249in}}%
\pgfpathquadraticcurveto{\pgfqpoint{2.836377in}{1.302249in}}{\pgfqpoint{2.836377in}{1.324471in}}%
\pgfpathlineto{\pgfqpoint{2.836377in}{1.756570in}}%
\pgfpathquadraticcurveto{\pgfqpoint{2.836377in}{1.778792in}}{\pgfqpoint{2.814155in}{1.778792in}}%
\pgfpathlineto{\pgfqpoint{2.129081in}{1.778792in}}%
\pgfpathquadraticcurveto{\pgfqpoint{2.106859in}{1.778792in}}{\pgfqpoint{2.106859in}{1.756570in}}%
\pgfpathlineto{\pgfqpoint{2.106859in}{1.324471in}}%
\pgfpathquadraticcurveto{\pgfqpoint{2.106859in}{1.302249in}}{\pgfqpoint{2.129081in}{1.302249in}}%
\pgfpathclose%
\pgfusepath{stroke,fill}%
\end{pgfscope}%
\begin{pgfscope}%
\pgfsetrectcap%
\pgfsetroundjoin%
\pgfsetlinewidth{1.003750pt}%
\definecolor{currentstroke}{rgb}{0.767843,0.220980,0.353137}%
\pgfsetstrokecolor{currentstroke}%
\pgfsetdash{}{0pt}%
\pgfpathmoveto{\pgfqpoint{2.184637in}{1.695459in}}%
\pgfpathlineto{\pgfqpoint{2.340192in}{1.695459in}}%
\pgfusepath{stroke}%
\end{pgfscope}%
\begin{pgfscope}%
\pgftext[left,bottom,x=2.462414in,y=1.656570in,rotate=0.000000]{{\rmfamily\fontsize{8.000000}{9.600000}\selectfont Euclid}}
\end{pgfscope}%
\begin{pgfscope}%
\pgfsetrectcap%
\pgfsetroundjoin%
\pgfsetlinewidth{1.003750pt}%
\definecolor{currentstroke}{rgb}{0.169412,0.150000,0.532353}%
\pgfsetstrokecolor{currentstroke}%
\pgfsetdash{}{0pt}%
\pgfpathmoveto{\pgfqpoint{2.184637in}{1.540520in}}%
\pgfpathlineto{\pgfqpoint{2.340192in}{1.540520in}}%
\pgfusepath{stroke}%
\end{pgfscope}%
\begin{pgfscope}%
\pgftext[left,bottom,x=2.462414in,y=1.501631in,rotate=0.000000]{{\rmfamily\fontsize{8.000000}{9.600000}\selectfont SKA1}}
\end{pgfscope}%
\begin{pgfscope}%
\pgfsetrectcap%
\pgfsetroundjoin%
\pgfsetlinewidth{1.003750pt}%
\definecolor{currentstroke}{rgb}{0.387059,0.164510,0.677451}%
\pgfsetstrokecolor{currentstroke}%
\pgfsetdash{}{0pt}%
\pgfpathmoveto{\pgfqpoint{2.184637in}{1.385582in}}%
\pgfpathlineto{\pgfqpoint{2.340192in}{1.385582in}}%
\pgfusepath{stroke}%
\end{pgfscope}%
\begin{pgfscope}%
\pgftext[left,bottom,x=2.462414in,y=1.346693in,rotate=0.000000]{{\rmfamily\fontsize{8.000000}{9.600000}\selectfont SKA2}}
\end{pgfscope}%
\end{pgfpicture}%
\makeatother%
\endgroup%

%% file: figures/PaperChi2ofL_con.pgf
\begingroup%
\makeatletter%
\begin{pgfpicture}%
\pgfpathrectangle{\pgfpointorigin}{\pgfqpoint{3.044140in}{1.881382in}}%
\pgfusepath{use as bounding box}%
\begin{pgfscope}%
\pgfsetrectcap%
\pgfsetroundjoin%
\definecolor{currentfill}{rgb}{1.000000,1.000000,1.000000}%
\pgfsetfillcolor{currentfill}%
\pgfsetlinewidth{0.000000pt}%
\definecolor{currentstroke}{rgb}{1.000000,1.000000,1.000000}%
\pgfsetstrokecolor{currentstroke}%
\pgfsetdash{}{0pt}%
\pgfpathmoveto{\pgfqpoint{0.000000in}{0.000000in}}%
\pgfpathlineto{\pgfqpoint{3.044140in}{0.000000in}}%
\pgfpathlineto{\pgfqpoint{3.044140in}{1.881382in}}%
\pgfpathlineto{\pgfqpoint{0.000000in}{1.881382in}}%
\pgfpathclose%
\pgfusepath{fill}%
\end{pgfscope}%
\begin{pgfscope}%
\pgfsetrectcap%
\pgfsetroundjoin%
\definecolor{currentfill}{rgb}{1.000000,1.000000,1.000000}%
\pgfsetfillcolor{currentfill}%
\pgfsetlinewidth{0.000000pt}%
\definecolor{currentstroke}{rgb}{0.000000,0.000000,0.000000}%
\pgfsetstrokecolor{currentstroke}%
\pgfsetdash{}{0pt}%
\pgfpathmoveto{\pgfqpoint{0.456621in}{0.329242in}}%
\pgfpathlineto{\pgfqpoint{2.891933in}{0.329242in}}%
\pgfpathlineto{\pgfqpoint{2.891933in}{1.834347in}}%
\pgfpathlineto{\pgfqpoint{0.456621in}{1.834347in}}%
\pgfpathclose%
\pgfusepath{fill}%
\end{pgfscope}%
\begin{pgfscope}%
\pgfpathrectangle{\pgfqpoint{0.456621in}{0.329242in}}{\pgfqpoint{2.435312in}{1.505106in}} %
\pgfusepath{clip}%
\pgfsetrectcap%
\pgfsetroundjoin%
\pgfsetlinewidth{1.003750pt}%
\definecolor{currentstroke}{rgb}{0.767843,0.220980,0.353137}%
\pgfsetstrokecolor{currentstroke}%
\pgfsetdash{}{0pt}%
\pgfpathmoveto{\pgfqpoint{0.459056in}{0.347619in}}%
\pgfpathlineto{\pgfqpoint{0.475129in}{0.464510in}}%
\pgfpathlineto{\pgfqpoint{0.485845in}{0.525315in}}%
\pgfpathlineto{\pgfqpoint{0.497534in}{0.579386in}}%
\pgfpathlineto{\pgfqpoint{0.512633in}{0.638257in}}%
\pgfpathlineto{\pgfqpoint{0.529193in}{0.694143in}}%
\pgfpathlineto{\pgfqpoint{0.543318in}{0.734258in}}%
\pgfpathlineto{\pgfqpoint{0.545753in}{0.740466in}}%
\pgfpathlineto{\pgfqpoint{0.546240in}{0.735488in}}%
\pgfpathlineto{\pgfqpoint{0.546728in}{0.736698in}}%
\pgfpathlineto{\pgfqpoint{0.561826in}{0.771104in}}%
\pgfpathlineto{\pgfqpoint{0.581796in}{0.810654in}}%
\pgfpathlineto{\pgfqpoint{0.583744in}{0.814292in}}%
\pgfpathlineto{\pgfqpoint{0.584231in}{0.800512in}}%
\pgfpathlineto{\pgfqpoint{0.585205in}{0.802310in}}%
\pgfpathlineto{\pgfqpoint{0.607123in}{0.840371in}}%
\pgfpathlineto{\pgfqpoint{0.611020in}{0.846586in}}%
\pgfpathlineto{\pgfqpoint{0.611507in}{0.828533in}}%
\pgfpathlineto{\pgfqpoint{0.612481in}{0.830056in}}%
\pgfpathlineto{\pgfqpoint{0.633425in}{0.860626in}}%
\pgfpathlineto{\pgfqpoint{0.635373in}{0.863258in}}%
\pgfpathlineto{\pgfqpoint{0.635860in}{0.842143in}}%
\pgfpathlineto{\pgfqpoint{0.636834in}{0.843427in}}%
\pgfpathlineto{\pgfqpoint{0.658265in}{0.869728in}}%
\pgfpathlineto{\pgfqpoint{0.659239in}{0.870846in}}%
\pgfpathlineto{\pgfqpoint{0.659726in}{0.845774in}}%
\pgfpathlineto{\pgfqpoint{0.660700in}{0.846824in}}%
\pgfpathlineto{\pgfqpoint{0.682618in}{0.868764in}}%
\pgfpathlineto{\pgfqpoint{0.685053in}{0.871012in}}%
\pgfpathlineto{\pgfqpoint{0.685540in}{0.837734in}}%
\pgfpathlineto{\pgfqpoint{0.686514in}{0.838495in}}%
\pgfpathlineto{\pgfqpoint{0.708432in}{0.854159in}}%
\pgfpathlineto{\pgfqpoint{0.711842in}{0.856348in}}%
\pgfpathlineto{\pgfqpoint{0.712329in}{0.805789in}}%
\pgfpathlineto{\pgfqpoint{0.713303in}{0.806193in}}%
\pgfpathlineto{\pgfqpoint{0.733760in}{0.813442in}}%
\pgfpathlineto{\pgfqpoint{0.743014in}{0.816043in}}%
\pgfpathlineto{\pgfqpoint{0.744475in}{0.729569in}}%
\pgfpathlineto{\pgfqpoint{0.773212in}{0.727800in}}%
\pgfpathlineto{\pgfqpoint{0.786849in}{0.726238in}}%
\pgfpathlineto{\pgfqpoint{0.788311in}{0.562235in}}%
\pgfpathlineto{\pgfqpoint{0.854064in}{0.541150in}}%
\pgfpathlineto{\pgfqpoint{0.854551in}{0.329242in}}%
\pgfpathlineto{\pgfqpoint{0.854551in}{0.329242in}}%
\pgfusepath{stroke}%
\end{pgfscope}%
\begin{pgfscope}%
\pgfpathrectangle{\pgfqpoint{0.456621in}{0.329242in}}{\pgfqpoint{2.435312in}{1.505106in}} %
\pgfusepath{clip}%
\pgfsetrectcap%
\pgfsetroundjoin%
\pgfsetlinewidth{1.003750pt}%
\definecolor{currentstroke}{rgb}{0.169412,0.150000,0.532353}%
\pgfsetstrokecolor{currentstroke}%
\pgfsetdash{}{0pt}%
\pgfpathmoveto{\pgfqpoint{0.459056in}{0.332469in}}%
\pgfpathlineto{\pgfqpoint{0.475616in}{0.352328in}}%
\pgfpathlineto{\pgfqpoint{0.488767in}{0.364331in}}%
\pgfpathlineto{\pgfqpoint{0.503379in}{0.374628in}}%
\pgfpathlineto{\pgfqpoint{0.519452in}{0.383210in}}%
\pgfpathlineto{\pgfqpoint{0.535038in}{0.389054in}}%
\pgfpathlineto{\pgfqpoint{0.551598in}{0.392836in}}%
\pgfpathlineto{\pgfqpoint{0.570594in}{0.394783in}}%
\pgfpathlineto{\pgfqpoint{0.594460in}{0.394895in}}%
\pgfpathlineto{\pgfqpoint{0.626606in}{0.392753in}}%
\pgfpathlineto{\pgfqpoint{0.660700in}{0.388506in}}%
\pgfpathlineto{\pgfqpoint{0.683592in}{0.385694in}}%
\pgfpathlineto{\pgfqpoint{0.685053in}{0.383728in}}%
\pgfpathlineto{\pgfqpoint{0.725479in}{0.378823in}}%
\pgfpathlineto{\pgfqpoint{0.726941in}{0.375613in}}%
\pgfpathlineto{\pgfqpoint{0.761035in}{0.371857in}}%
\pgfpathlineto{\pgfqpoint{0.762496in}{0.367595in}}%
\pgfpathlineto{\pgfqpoint{0.795129in}{0.364511in}}%
\pgfpathlineto{\pgfqpoint{0.796591in}{0.359302in}}%
\pgfpathlineto{\pgfqpoint{0.830198in}{0.356739in}}%
\pgfpathlineto{\pgfqpoint{0.831659in}{0.350901in}}%
\pgfpathlineto{\pgfqpoint{0.861370in}{0.349217in}}%
\pgfpathlineto{\pgfqpoint{0.862831in}{0.343229in}}%
\pgfpathlineto{\pgfqpoint{0.895951in}{0.341984in}}%
\pgfpathlineto{\pgfqpoint{0.897412in}{0.336398in}}%
\pgfpathlineto{\pgfqpoint{0.947580in}{0.335455in}}%
\pgfpathlineto{\pgfqpoint{0.949041in}{0.331061in}}%
\pgfpathlineto{\pgfqpoint{0.967062in}{0.330969in}}%
\pgfpathlineto{\pgfqpoint{0.967549in}{0.329242in}}%
\pgfpathlineto{\pgfqpoint{0.967549in}{0.329242in}}%
\pgfusepath{stroke}%
\end{pgfscope}%
\begin{pgfscope}%
\pgfpathrectangle{\pgfqpoint{0.456621in}{0.329242in}}{\pgfqpoint{2.435312in}{1.505106in}} %
\pgfusepath{clip}%
\pgfsetrectcap%
\pgfsetroundjoin%
\pgfsetlinewidth{1.003750pt}%
\definecolor{currentstroke}{rgb}{0.387059,0.164510,0.677451}%
\pgfsetstrokecolor{currentstroke}%
\pgfsetdash{}{0pt}%
\pgfpathmoveto{\pgfqpoint{0.459056in}{0.355727in}}%
\pgfpathlineto{\pgfqpoint{0.477565in}{0.555360in}}%
\pgfpathlineto{\pgfqpoint{0.489741in}{0.660924in}}%
\pgfpathlineto{\pgfqpoint{0.502405in}{0.752314in}}%
\pgfpathlineto{\pgfqpoint{0.519452in}{0.858627in}}%
\pgfpathlineto{\pgfqpoint{0.542831in}{0.989548in}}%
\pgfpathlineto{\pgfqpoint{0.562314in}{1.087053in}}%
\pgfpathlineto{\pgfqpoint{0.580822in}{1.168597in}}%
\pgfpathlineto{\pgfqpoint{0.600791in}{1.246279in}}%
\pgfpathlineto{\pgfqpoint{0.621735in}{1.318892in}}%
\pgfpathlineto{\pgfqpoint{0.641705in}{1.380562in}}%
\pgfpathlineto{\pgfqpoint{0.661187in}{1.433734in}}%
\pgfpathlineto{\pgfqpoint{0.679209in}{1.476884in}}%
\pgfpathlineto{\pgfqpoint{0.692846in}{1.505751in}}%
\pgfpathlineto{\pgfqpoint{0.693333in}{1.498640in}}%
\pgfpathlineto{\pgfqpoint{0.694307in}{1.500573in}}%
\pgfpathlineto{\pgfqpoint{0.712816in}{1.534687in}}%
\pgfpathlineto{\pgfqpoint{0.730350in}{1.562545in}}%
\pgfpathlineto{\pgfqpoint{0.748858in}{1.587793in}}%
\pgfpathlineto{\pgfqpoint{0.759087in}{1.600170in}}%
\pgfpathlineto{\pgfqpoint{0.759574in}{1.575202in}}%
\pgfpathlineto{\pgfqpoint{0.760548in}{1.576255in}}%
\pgfpathlineto{\pgfqpoint{0.778082in}{1.593559in}}%
\pgfpathlineto{\pgfqpoint{0.795129in}{1.607355in}}%
\pgfpathlineto{\pgfqpoint{0.812177in}{1.618603in}}%
\pgfpathlineto{\pgfqpoint{0.812664in}{1.572869in}}%
\pgfpathlineto{\pgfqpoint{0.813638in}{1.573319in}}%
\pgfpathlineto{\pgfqpoint{0.832633in}{1.580798in}}%
\pgfpathlineto{\pgfqpoint{0.851629in}{1.585870in}}%
\pgfpathlineto{\pgfqpoint{0.854551in}{1.586432in}}%
\pgfpathlineto{\pgfqpoint{0.855038in}{1.516353in}}%
\pgfpathlineto{\pgfqpoint{0.856012in}{1.516369in}}%
\pgfpathlineto{\pgfqpoint{0.876469in}{1.515504in}}%
\pgfpathlineto{\pgfqpoint{0.892055in}{1.513545in}}%
\pgfpathlineto{\pgfqpoint{0.893516in}{1.413755in}}%
\pgfpathlineto{\pgfqpoint{0.923227in}{1.402630in}}%
\pgfpathlineto{\pgfqpoint{0.934916in}{1.397608in}}%
\pgfpathlineto{\pgfqpoint{0.936377in}{1.261401in}}%
\pgfpathlineto{\pgfqpoint{0.977778in}{1.235385in}}%
\pgfpathlineto{\pgfqpoint{1.005053in}{1.217517in}}%
\pgfpathlineto{\pgfqpoint{1.006514in}{1.017725in}}%
\pgfpathlineto{\pgfqpoint{1.067884in}{0.975092in}}%
\pgfpathlineto{\pgfqpoint{1.069346in}{0.724229in}}%
\pgfpathlineto{\pgfqpoint{1.095647in}{0.709794in}}%
\pgfpathlineto{\pgfqpoint{1.097108in}{0.471395in}}%
\pgfpathlineto{\pgfqpoint{1.151659in}{0.457954in}}%
\pgfpathlineto{\pgfqpoint{1.152633in}{0.457733in}}%
\pgfpathlineto{\pgfqpoint{1.153120in}{0.329242in}}%
\pgfpathlineto{\pgfqpoint{1.153120in}{0.329242in}}%
\pgfusepath{stroke}%
\end{pgfscope}%
\begin{pgfscope}%
\pgfpathrectangle{\pgfqpoint{0.456621in}{0.329242in}}{\pgfqpoint{2.435312in}{1.505106in}} %
\pgfusepath{clip}%
\pgfsetbuttcap%
\pgfsetroundjoin%
\pgfsetlinewidth{0.501875pt}%
\definecolor{currentstroke}{rgb}{0.000000,0.000000,0.000000}%
\pgfsetstrokecolor{currentstroke}%
\pgfsetdash{{1.000000pt}{3.000000pt}}{0.000000pt}%
\pgfpathmoveto{\pgfqpoint{0.456621in}{0.329242in}}%
\pgfpathlineto{\pgfqpoint{0.456621in}{1.834347in}}%
\pgfusepath{stroke}%
\end{pgfscope}%
\begin{pgfscope}%
\pgfsetbuttcap%
\pgfsetroundjoin%
\definecolor{currentfill}{rgb}{0.000000,0.000000,0.000000}%
\pgfsetfillcolor{currentfill}%
\pgfsetlinewidth{0.501875pt}%
\definecolor{currentstroke}{rgb}{0.000000,0.000000,0.000000}%
\pgfsetstrokecolor{currentstroke}%
\pgfsetdash{}{0pt}%
\pgfsys@defobject{currentmarker}{\pgfqpoint{0.000000in}{0.000000in}}{\pgfqpoint{0.000000in}{0.055556in}}{%
\pgfpathmoveto{\pgfqpoint{0.000000in}{0.000000in}}%
\pgfpathlineto{\pgfqpoint{0.000000in}{0.055556in}}%
\pgfusepath{stroke,fill}%
}%
\begin{pgfscope}%
\pgfsys@transformshift{0.456621in}{0.329242in}%
\pgfsys@useobject{currentmarker}{}%
\end{pgfscope}%
\end{pgfscope}%
\begin{pgfscope}%
\pgfsetbuttcap%
\pgfsetroundjoin%
\definecolor{currentfill}{rgb}{0.000000,0.000000,0.000000}%
\pgfsetfillcolor{currentfill}%
\pgfsetlinewidth{0.501875pt}%
\definecolor{currentstroke}{rgb}{0.000000,0.000000,0.000000}%
\pgfsetstrokecolor{currentstroke}%
\pgfsetdash{}{0pt}%
\pgfsys@defobject{currentmarker}{\pgfqpoint{0.000000in}{-0.055556in}}{\pgfqpoint{0.000000in}{0.000000in}}{%
\pgfpathmoveto{\pgfqpoint{0.000000in}{0.000000in}}%
\pgfpathlineto{\pgfqpoint{0.000000in}{-0.055556in}}%
\pgfusepath{stroke,fill}%
}%
\begin{pgfscope}%
\pgfsys@transformshift{0.456621in}{1.834347in}%
\pgfsys@useobject{currentmarker}{}%
\end{pgfscope}%
\end{pgfscope}%
\begin{pgfscope}%
\pgftext[left,bottom,x=0.427107in,y=0.202081in,rotate=0.000000]{{\rmfamily\fontsize{8.000000}{9.600000}\selectfont \(\displaystyle 0\)}}
\end{pgfscope}%
\begin{pgfscope}%
\pgfpathrectangle{\pgfqpoint{0.456621in}{0.329242in}}{\pgfqpoint{2.435312in}{1.505106in}} %
\pgfusepath{clip}%
\pgfsetbuttcap%
\pgfsetroundjoin%
\pgfsetlinewidth{0.501875pt}%
\definecolor{currentstroke}{rgb}{0.000000,0.000000,0.000000}%
\pgfsetstrokecolor{currentstroke}%
\pgfsetdash{{1.000000pt}{3.000000pt}}{0.000000pt}%
\pgfpathmoveto{\pgfqpoint{0.943683in}{0.329242in}}%
\pgfpathlineto{\pgfqpoint{0.943683in}{1.834347in}}%
\pgfusepath{stroke}%
\end{pgfscope}%
\begin{pgfscope}%
\pgfsetbuttcap%
\pgfsetroundjoin%
\definecolor{currentfill}{rgb}{0.000000,0.000000,0.000000}%
\pgfsetfillcolor{currentfill}%
\pgfsetlinewidth{0.501875pt}%
\definecolor{currentstroke}{rgb}{0.000000,0.000000,0.000000}%
\pgfsetstrokecolor{currentstroke}%
\pgfsetdash{}{0pt}%
\pgfsys@defobject{currentmarker}{\pgfqpoint{0.000000in}{0.000000in}}{\pgfqpoint{0.000000in}{0.055556in}}{%
\pgfpathmoveto{\pgfqpoint{0.000000in}{0.000000in}}%
\pgfpathlineto{\pgfqpoint{0.000000in}{0.055556in}}%
\pgfusepath{stroke,fill}%
}%
\begin{pgfscope}%
\pgfsys@transformshift{0.943683in}{0.329242in}%
\pgfsys@useobject{currentmarker}{}%
\end{pgfscope}%
\end{pgfscope}%
\begin{pgfscope}%
\pgfsetbuttcap%
\pgfsetroundjoin%
\definecolor{currentfill}{rgb}{0.000000,0.000000,0.000000}%
\pgfsetfillcolor{currentfill}%
\pgfsetlinewidth{0.501875pt}%
\definecolor{currentstroke}{rgb}{0.000000,0.000000,0.000000}%
\pgfsetstrokecolor{currentstroke}%
\pgfsetdash{}{0pt}%
\pgfsys@defobject{currentmarker}{\pgfqpoint{0.000000in}{-0.055556in}}{\pgfqpoint{0.000000in}{0.000000in}}{%
\pgfpathmoveto{\pgfqpoint{0.000000in}{0.000000in}}%
\pgfpathlineto{\pgfqpoint{0.000000in}{-0.055556in}}%
\pgfusepath{stroke,fill}%
}%
\begin{pgfscope}%
\pgfsys@transformshift{0.943683in}{1.834347in}%
\pgfsys@useobject{currentmarker}{}%
\end{pgfscope}%
\end{pgfscope}%
\begin{pgfscope}%
\pgftext[left,bottom,x=0.825626in,y=0.202081in,rotate=0.000000]{{\rmfamily\fontsize{8.000000}{9.600000}\selectfont \(\displaystyle 1000\)}}
\end{pgfscope}%
\begin{pgfscope}%
\pgfpathrectangle{\pgfqpoint{0.456621in}{0.329242in}}{\pgfqpoint{2.435312in}{1.505106in}} %
\pgfusepath{clip}%
\pgfsetbuttcap%
\pgfsetroundjoin%
\pgfsetlinewidth{0.501875pt}%
\definecolor{currentstroke}{rgb}{0.000000,0.000000,0.000000}%
\pgfsetstrokecolor{currentstroke}%
\pgfsetdash{{1.000000pt}{3.000000pt}}{0.000000pt}%
\pgfpathmoveto{\pgfqpoint{1.430746in}{0.329242in}}%
\pgfpathlineto{\pgfqpoint{1.430746in}{1.834347in}}%
\pgfusepath{stroke}%
\end{pgfscope}%
\begin{pgfscope}%
\pgfsetbuttcap%
\pgfsetroundjoin%
\definecolor{currentfill}{rgb}{0.000000,0.000000,0.000000}%
\pgfsetfillcolor{currentfill}%
\pgfsetlinewidth{0.501875pt}%
\definecolor{currentstroke}{rgb}{0.000000,0.000000,0.000000}%
\pgfsetstrokecolor{currentstroke}%
\pgfsetdash{}{0pt}%
\pgfsys@defobject{currentmarker}{\pgfqpoint{0.000000in}{0.000000in}}{\pgfqpoint{0.000000in}{0.055556in}}{%
\pgfpathmoveto{\pgfqpoint{0.000000in}{0.000000in}}%
\pgfpathlineto{\pgfqpoint{0.000000in}{0.055556in}}%
\pgfusepath{stroke,fill}%
}%
\begin{pgfscope}%
\pgfsys@transformshift{1.430746in}{0.329242in}%
\pgfsys@useobject{currentmarker}{}%
\end{pgfscope}%
\end{pgfscope}%
\begin{pgfscope}%
\pgfsetbuttcap%
\pgfsetroundjoin%
\definecolor{currentfill}{rgb}{0.000000,0.000000,0.000000}%
\pgfsetfillcolor{currentfill}%
\pgfsetlinewidth{0.501875pt}%
\definecolor{currentstroke}{rgb}{0.000000,0.000000,0.000000}%
\pgfsetstrokecolor{currentstroke}%
\pgfsetdash{}{0pt}%
\pgfsys@defobject{currentmarker}{\pgfqpoint{0.000000in}{-0.055556in}}{\pgfqpoint{0.000000in}{0.000000in}}{%
\pgfpathmoveto{\pgfqpoint{0.000000in}{0.000000in}}%
\pgfpathlineto{\pgfqpoint{0.000000in}{-0.055556in}}%
\pgfusepath{stroke,fill}%
}%
\begin{pgfscope}%
\pgfsys@transformshift{1.430746in}{1.834347in}%
\pgfsys@useobject{currentmarker}{}%
\end{pgfscope}%
\end{pgfscope}%
\begin{pgfscope}%
\pgftext[left,bottom,x=1.312689in,y=0.202081in,rotate=0.000000]{{\rmfamily\fontsize{8.000000}{9.600000}\selectfont \(\displaystyle 2000\)}}
\end{pgfscope}%
\begin{pgfscope}%
\pgfpathrectangle{\pgfqpoint{0.456621in}{0.329242in}}{\pgfqpoint{2.435312in}{1.505106in}} %
\pgfusepath{clip}%
\pgfsetbuttcap%
\pgfsetroundjoin%
\pgfsetlinewidth{0.501875pt}%
\definecolor{currentstroke}{rgb}{0.000000,0.000000,0.000000}%
\pgfsetstrokecolor{currentstroke}%
\pgfsetdash{{1.000000pt}{3.000000pt}}{0.000000pt}%
\pgfpathmoveto{\pgfqpoint{1.917808in}{0.329242in}}%
\pgfpathlineto{\pgfqpoint{1.917808in}{1.834347in}}%
\pgfusepath{stroke}%
\end{pgfscope}%
\begin{pgfscope}%
\pgfsetbuttcap%
\pgfsetroundjoin%
\definecolor{currentfill}{rgb}{0.000000,0.000000,0.000000}%
\pgfsetfillcolor{currentfill}%
\pgfsetlinewidth{0.501875pt}%
\definecolor{currentstroke}{rgb}{0.000000,0.000000,0.000000}%
\pgfsetstrokecolor{currentstroke}%
\pgfsetdash{}{0pt}%
\pgfsys@defobject{currentmarker}{\pgfqpoint{0.000000in}{0.000000in}}{\pgfqpoint{0.000000in}{0.055556in}}{%
\pgfpathmoveto{\pgfqpoint{0.000000in}{0.000000in}}%
\pgfpathlineto{\pgfqpoint{0.000000in}{0.055556in}}%
\pgfusepath{stroke,fill}%
}%
\begin{pgfscope}%
\pgfsys@transformshift{1.917808in}{0.329242in}%
\pgfsys@useobject{currentmarker}{}%
\end{pgfscope}%
\end{pgfscope}%
\begin{pgfscope}%
\pgfsetbuttcap%
\pgfsetroundjoin%
\definecolor{currentfill}{rgb}{0.000000,0.000000,0.000000}%
\pgfsetfillcolor{currentfill}%
\pgfsetlinewidth{0.501875pt}%
\definecolor{currentstroke}{rgb}{0.000000,0.000000,0.000000}%
\pgfsetstrokecolor{currentstroke}%
\pgfsetdash{}{0pt}%
\pgfsys@defobject{currentmarker}{\pgfqpoint{0.000000in}{-0.055556in}}{\pgfqpoint{0.000000in}{0.000000in}}{%
\pgfpathmoveto{\pgfqpoint{0.000000in}{0.000000in}}%
\pgfpathlineto{\pgfqpoint{0.000000in}{-0.055556in}}%
\pgfusepath{stroke,fill}%
}%
\begin{pgfscope}%
\pgfsys@transformshift{1.917808in}{1.834347in}%
\pgfsys@useobject{currentmarker}{}%
\end{pgfscope}%
\end{pgfscope}%
\begin{pgfscope}%
\pgftext[left,bottom,x=1.799751in,y=0.202081in,rotate=0.000000]{{\rmfamily\fontsize{8.000000}{9.600000}\selectfont \(\displaystyle 3000\)}}
\end{pgfscope}%
\begin{pgfscope}%
\pgfpathrectangle{\pgfqpoint{0.456621in}{0.329242in}}{\pgfqpoint{2.435312in}{1.505106in}} %
\pgfusepath{clip}%
\pgfsetbuttcap%
\pgfsetroundjoin%
\pgfsetlinewidth{0.501875pt}%
\definecolor{currentstroke}{rgb}{0.000000,0.000000,0.000000}%
\pgfsetstrokecolor{currentstroke}%
\pgfsetdash{{1.000000pt}{3.000000pt}}{0.000000pt}%
\pgfpathmoveto{\pgfqpoint{2.404871in}{0.329242in}}%
\pgfpathlineto{\pgfqpoint{2.404871in}{1.834347in}}%
\pgfusepath{stroke}%
\end{pgfscope}%
\begin{pgfscope}%
\pgfsetbuttcap%
\pgfsetroundjoin%
\definecolor{currentfill}{rgb}{0.000000,0.000000,0.000000}%
\pgfsetfillcolor{currentfill}%
\pgfsetlinewidth{0.501875pt}%
\definecolor{currentstroke}{rgb}{0.000000,0.000000,0.000000}%
\pgfsetstrokecolor{currentstroke}%
\pgfsetdash{}{0pt}%
\pgfsys@defobject{currentmarker}{\pgfqpoint{0.000000in}{0.000000in}}{\pgfqpoint{0.000000in}{0.055556in}}{%
\pgfpathmoveto{\pgfqpoint{0.000000in}{0.000000in}}%
\pgfpathlineto{\pgfqpoint{0.000000in}{0.055556in}}%
\pgfusepath{stroke,fill}%
}%
\begin{pgfscope}%
\pgfsys@transformshift{2.404871in}{0.329242in}%
\pgfsys@useobject{currentmarker}{}%
\end{pgfscope}%
\end{pgfscope}%
\begin{pgfscope}%
\pgfsetbuttcap%
\pgfsetroundjoin%
\definecolor{currentfill}{rgb}{0.000000,0.000000,0.000000}%
\pgfsetfillcolor{currentfill}%
\pgfsetlinewidth{0.501875pt}%
\definecolor{currentstroke}{rgb}{0.000000,0.000000,0.000000}%
\pgfsetstrokecolor{currentstroke}%
\pgfsetdash{}{0pt}%
\pgfsys@defobject{currentmarker}{\pgfqpoint{0.000000in}{-0.055556in}}{\pgfqpoint{0.000000in}{0.000000in}}{%
\pgfpathmoveto{\pgfqpoint{0.000000in}{0.000000in}}%
\pgfpathlineto{\pgfqpoint{0.000000in}{-0.055556in}}%
\pgfusepath{stroke,fill}%
}%
\begin{pgfscope}%
\pgfsys@transformshift{2.404871in}{1.834347in}%
\pgfsys@useobject{currentmarker}{}%
\end{pgfscope}%
\end{pgfscope}%
\begin{pgfscope}%
\pgftext[left,bottom,x=2.286813in,y=0.202081in,rotate=0.000000]{{\rmfamily\fontsize{8.000000}{9.600000}\selectfont \(\displaystyle 4000\)}}
\end{pgfscope}%
\begin{pgfscope}%
\pgfpathrectangle{\pgfqpoint{0.456621in}{0.329242in}}{\pgfqpoint{2.435312in}{1.505106in}} %
\pgfusepath{clip}%
\pgfsetbuttcap%
\pgfsetroundjoin%
\pgfsetlinewidth{0.501875pt}%
\definecolor{currentstroke}{rgb}{0.000000,0.000000,0.000000}%
\pgfsetstrokecolor{currentstroke}%
\pgfsetdash{{1.000000pt}{3.000000pt}}{0.000000pt}%
\pgfpathmoveto{\pgfqpoint{2.891933in}{0.329242in}}%
\pgfpathlineto{\pgfqpoint{2.891933in}{1.834347in}}%
\pgfusepath{stroke}%
\end{pgfscope}%
\begin{pgfscope}%
\pgfsetbuttcap%
\pgfsetroundjoin%
\definecolor{currentfill}{rgb}{0.000000,0.000000,0.000000}%
\pgfsetfillcolor{currentfill}%
\pgfsetlinewidth{0.501875pt}%
\definecolor{currentstroke}{rgb}{0.000000,0.000000,0.000000}%
\pgfsetstrokecolor{currentstroke}%
\pgfsetdash{}{0pt}%
\pgfsys@defobject{currentmarker}{\pgfqpoint{0.000000in}{0.000000in}}{\pgfqpoint{0.000000in}{0.055556in}}{%
\pgfpathmoveto{\pgfqpoint{0.000000in}{0.000000in}}%
\pgfpathlineto{\pgfqpoint{0.000000in}{0.055556in}}%
\pgfusepath{stroke,fill}%
}%
\begin{pgfscope}%
\pgfsys@transformshift{2.891933in}{0.329242in}%
\pgfsys@useobject{currentmarker}{}%
\end{pgfscope}%
\end{pgfscope}%
\begin{pgfscope}%
\pgfsetbuttcap%
\pgfsetroundjoin%
\definecolor{currentfill}{rgb}{0.000000,0.000000,0.000000}%
\pgfsetfillcolor{currentfill}%
\pgfsetlinewidth{0.501875pt}%
\definecolor{currentstroke}{rgb}{0.000000,0.000000,0.000000}%
\pgfsetstrokecolor{currentstroke}%
\pgfsetdash{}{0pt}%
\pgfsys@defobject{currentmarker}{\pgfqpoint{0.000000in}{-0.055556in}}{\pgfqpoint{0.000000in}{0.000000in}}{%
\pgfpathmoveto{\pgfqpoint{0.000000in}{0.000000in}}%
\pgfpathlineto{\pgfqpoint{0.000000in}{-0.055556in}}%
\pgfusepath{stroke,fill}%
}%
\begin{pgfscope}%
\pgfsys@transformshift{2.891933in}{1.834347in}%
\pgfsys@useobject{currentmarker}{}%
\end{pgfscope}%
\end{pgfscope}%
\begin{pgfscope}%
\pgftext[left,bottom,x=2.773876in,y=0.202081in,rotate=0.000000]{{\rmfamily\fontsize{8.000000}{9.600000}\selectfont \(\displaystyle 5000\)}}
\end{pgfscope}%
\begin{pgfscope}%
\pgftext[left,bottom,x=1.650164in,y=0.055476in,rotate=0.000000]{{\rmfamily\fontsize{8.000000}{9.600000}\selectfont \(\displaystyle \ell\)}}
\end{pgfscope}%
\begin{pgfscope}%
\pgfpathrectangle{\pgfqpoint{0.456621in}{0.329242in}}{\pgfqpoint{2.435312in}{1.505106in}} %
\pgfusepath{clip}%
\pgfsetbuttcap%
\pgfsetroundjoin%
\pgfsetlinewidth{0.501875pt}%
\definecolor{currentstroke}{rgb}{0.000000,0.000000,0.000000}%
\pgfsetstrokecolor{currentstroke}%
\pgfsetdash{{1.000000pt}{3.000000pt}}{0.000000pt}%
\pgfpathmoveto{\pgfqpoint{0.456621in}{0.329242in}}%
\pgfpathlineto{\pgfqpoint{2.891933in}{0.329242in}}%
\pgfusepath{stroke}%
\end{pgfscope}%
\begin{pgfscope}%
\pgfsetbuttcap%
\pgfsetroundjoin%
\definecolor{currentfill}{rgb}{0.000000,0.000000,0.000000}%
\pgfsetfillcolor{currentfill}%
\pgfsetlinewidth{0.501875pt}%
\definecolor{currentstroke}{rgb}{0.000000,0.000000,0.000000}%
\pgfsetstrokecolor{currentstroke}%
\pgfsetdash{}{0pt}%
\pgfsys@defobject{currentmarker}{\pgfqpoint{0.000000in}{0.000000in}}{\pgfqpoint{0.055556in}{0.000000in}}{%
\pgfpathmoveto{\pgfqpoint{0.000000in}{0.000000in}}%
\pgfpathlineto{\pgfqpoint{0.055556in}{0.000000in}}%
\pgfusepath{stroke,fill}%
}%
\begin{pgfscope}%
\pgfsys@transformshift{0.456621in}{0.329242in}%
\pgfsys@useobject{currentmarker}{}%
\end{pgfscope}%
\end{pgfscope}%
\begin{pgfscope}%
\pgfsetbuttcap%
\pgfsetroundjoin%
\definecolor{currentfill}{rgb}{0.000000,0.000000,0.000000}%
\pgfsetfillcolor{currentfill}%
\pgfsetlinewidth{0.501875pt}%
\definecolor{currentstroke}{rgb}{0.000000,0.000000,0.000000}%
\pgfsetstrokecolor{currentstroke}%
\pgfsetdash{}{0pt}%
\pgfsys@defobject{currentmarker}{\pgfqpoint{-0.055556in}{0.000000in}}{\pgfqpoint{0.000000in}{0.000000in}}{%
\pgfpathmoveto{\pgfqpoint{0.000000in}{0.000000in}}%
\pgfpathlineto{\pgfqpoint{-0.055556in}{0.000000in}}%
\pgfusepath{stroke,fill}%
}%
\begin{pgfscope}%
\pgfsys@transformshift{2.891933in}{0.329242in}%
\pgfsys@useobject{currentmarker}{}%
\end{pgfscope}%
\end{pgfscope}%
\begin{pgfscope}%
\pgftext[left,bottom,x=0.250214in,y=0.293439in,rotate=0.000000]{{\rmfamily\fontsize{8.000000}{9.600000}\selectfont \(\displaystyle 0.0\)}}
\end{pgfscope}%
\begin{pgfscope}%
\pgfpathrectangle{\pgfqpoint{0.456621in}{0.329242in}}{\pgfqpoint{2.435312in}{1.505106in}} %
\pgfusepath{clip}%
\pgfsetbuttcap%
\pgfsetroundjoin%
\pgfsetlinewidth{0.501875pt}%
\definecolor{currentstroke}{rgb}{0.000000,0.000000,0.000000}%
\pgfsetstrokecolor{currentstroke}%
\pgfsetdash{{1.000000pt}{3.000000pt}}{0.000000pt}%
\pgfpathmoveto{\pgfqpoint{0.456621in}{0.580093in}}%
\pgfpathlineto{\pgfqpoint{2.891933in}{0.580093in}}%
\pgfusepath{stroke}%
\end{pgfscope}%
\begin{pgfscope}%
\pgfsetbuttcap%
\pgfsetroundjoin%
\definecolor{currentfill}{rgb}{0.000000,0.000000,0.000000}%
\pgfsetfillcolor{currentfill}%
\pgfsetlinewidth{0.501875pt}%
\definecolor{currentstroke}{rgb}{0.000000,0.000000,0.000000}%
\pgfsetstrokecolor{currentstroke}%
\pgfsetdash{}{0pt}%
\pgfsys@defobject{currentmarker}{\pgfqpoint{0.000000in}{0.000000in}}{\pgfqpoint{0.055556in}{0.000000in}}{%
\pgfpathmoveto{\pgfqpoint{0.000000in}{0.000000in}}%
\pgfpathlineto{\pgfqpoint{0.055556in}{0.000000in}}%
\pgfusepath{stroke,fill}%
}%
\begin{pgfscope}%
\pgfsys@transformshift{0.456621in}{0.580093in}%
\pgfsys@useobject{currentmarker}{}%
\end{pgfscope}%
\end{pgfscope}%
\begin{pgfscope}%
\pgfsetbuttcap%
\pgfsetroundjoin%
\definecolor{currentfill}{rgb}{0.000000,0.000000,0.000000}%
\pgfsetfillcolor{currentfill}%
\pgfsetlinewidth{0.501875pt}%
\definecolor{currentstroke}{rgb}{0.000000,0.000000,0.000000}%
\pgfsetstrokecolor{currentstroke}%
\pgfsetdash{}{0pt}%
\pgfsys@defobject{currentmarker}{\pgfqpoint{-0.055556in}{0.000000in}}{\pgfqpoint{0.000000in}{0.000000in}}{%
\pgfpathmoveto{\pgfqpoint{0.000000in}{0.000000in}}%
\pgfpathlineto{\pgfqpoint{-0.055556in}{0.000000in}}%
\pgfusepath{stroke,fill}%
}%
\begin{pgfscope}%
\pgfsys@transformshift{2.891933in}{0.580093in}%
\pgfsys@useobject{currentmarker}{}%
\end{pgfscope}%
\end{pgfscope}%
\begin{pgfscope}%
\pgftext[left,bottom,x=0.250214in,y=0.544290in,rotate=0.000000]{{\rmfamily\fontsize{8.000000}{9.600000}\selectfont \(\displaystyle 0.5\)}}
\end{pgfscope}%
\begin{pgfscope}%
\pgfpathrectangle{\pgfqpoint{0.456621in}{0.329242in}}{\pgfqpoint{2.435312in}{1.505106in}} %
\pgfusepath{clip}%
\pgfsetbuttcap%
\pgfsetroundjoin%
\pgfsetlinewidth{0.501875pt}%
\definecolor{currentstroke}{rgb}{0.000000,0.000000,0.000000}%
\pgfsetstrokecolor{currentstroke}%
\pgfsetdash{{1.000000pt}{3.000000pt}}{0.000000pt}%
\pgfpathmoveto{\pgfqpoint{0.456621in}{0.830944in}}%
\pgfpathlineto{\pgfqpoint{2.891933in}{0.830944in}}%
\pgfusepath{stroke}%
\end{pgfscope}%
\begin{pgfscope}%
\pgfsetbuttcap%
\pgfsetroundjoin%
\definecolor{currentfill}{rgb}{0.000000,0.000000,0.000000}%
\pgfsetfillcolor{currentfill}%
\pgfsetlinewidth{0.501875pt}%
\definecolor{currentstroke}{rgb}{0.000000,0.000000,0.000000}%
\pgfsetstrokecolor{currentstroke}%
\pgfsetdash{}{0pt}%
\pgfsys@defobject{currentmarker}{\pgfqpoint{0.000000in}{0.000000in}}{\pgfqpoint{0.055556in}{0.000000in}}{%
\pgfpathmoveto{\pgfqpoint{0.000000in}{0.000000in}}%
\pgfpathlineto{\pgfqpoint{0.055556in}{0.000000in}}%
\pgfusepath{stroke,fill}%
}%
\begin{pgfscope}%
\pgfsys@transformshift{0.456621in}{0.830944in}%
\pgfsys@useobject{currentmarker}{}%
\end{pgfscope}%
\end{pgfscope}%
\begin{pgfscope}%
\pgfsetbuttcap%
\pgfsetroundjoin%
\definecolor{currentfill}{rgb}{0.000000,0.000000,0.000000}%
\pgfsetfillcolor{currentfill}%
\pgfsetlinewidth{0.501875pt}%
\definecolor{currentstroke}{rgb}{0.000000,0.000000,0.000000}%
\pgfsetstrokecolor{currentstroke}%
\pgfsetdash{}{0pt}%
\pgfsys@defobject{currentmarker}{\pgfqpoint{-0.055556in}{0.000000in}}{\pgfqpoint{0.000000in}{0.000000in}}{%
\pgfpathmoveto{\pgfqpoint{0.000000in}{0.000000in}}%
\pgfpathlineto{\pgfqpoint{-0.055556in}{0.000000in}}%
\pgfusepath{stroke,fill}%
}%
\begin{pgfscope}%
\pgfsys@transformshift{2.891933in}{0.830944in}%
\pgfsys@useobject{currentmarker}{}%
\end{pgfscope}%
\end{pgfscope}%
\begin{pgfscope}%
\pgftext[left,bottom,x=0.250214in,y=0.795141in,rotate=0.000000]{{\rmfamily\fontsize{8.000000}{9.600000}\selectfont \(\displaystyle 1.0\)}}
\end{pgfscope}%
\begin{pgfscope}%
\pgfpathrectangle{\pgfqpoint{0.456621in}{0.329242in}}{\pgfqpoint{2.435312in}{1.505106in}} %
\pgfusepath{clip}%
\pgfsetbuttcap%
\pgfsetroundjoin%
\pgfsetlinewidth{0.501875pt}%
\definecolor{currentstroke}{rgb}{0.000000,0.000000,0.000000}%
\pgfsetstrokecolor{currentstroke}%
\pgfsetdash{{1.000000pt}{3.000000pt}}{0.000000pt}%
\pgfpathmoveto{\pgfqpoint{0.456621in}{1.081795in}}%
\pgfpathlineto{\pgfqpoint{2.891933in}{1.081795in}}%
\pgfusepath{stroke}%
\end{pgfscope}%
\begin{pgfscope}%
\pgfsetbuttcap%
\pgfsetroundjoin%
\definecolor{currentfill}{rgb}{0.000000,0.000000,0.000000}%
\pgfsetfillcolor{currentfill}%
\pgfsetlinewidth{0.501875pt}%
\definecolor{currentstroke}{rgb}{0.000000,0.000000,0.000000}%
\pgfsetstrokecolor{currentstroke}%
\pgfsetdash{}{0pt}%
\pgfsys@defobject{currentmarker}{\pgfqpoint{0.000000in}{0.000000in}}{\pgfqpoint{0.055556in}{0.000000in}}{%
\pgfpathmoveto{\pgfqpoint{0.000000in}{0.000000in}}%
\pgfpathlineto{\pgfqpoint{0.055556in}{0.000000in}}%
\pgfusepath{stroke,fill}%
}%
\begin{pgfscope}%
\pgfsys@transformshift{0.456621in}{1.081795in}%
\pgfsys@useobject{currentmarker}{}%
\end{pgfscope}%
\end{pgfscope}%
\begin{pgfscope}%
\pgfsetbuttcap%
\pgfsetroundjoin%
\definecolor{currentfill}{rgb}{0.000000,0.000000,0.000000}%
\pgfsetfillcolor{currentfill}%
\pgfsetlinewidth{0.501875pt}%
\definecolor{currentstroke}{rgb}{0.000000,0.000000,0.000000}%
\pgfsetstrokecolor{currentstroke}%
\pgfsetdash{}{0pt}%
\pgfsys@defobject{currentmarker}{\pgfqpoint{-0.055556in}{0.000000in}}{\pgfqpoint{0.000000in}{0.000000in}}{%
\pgfpathmoveto{\pgfqpoint{0.000000in}{0.000000in}}%
\pgfpathlineto{\pgfqpoint{-0.055556in}{0.000000in}}%
\pgfusepath{stroke,fill}%
}%
\begin{pgfscope}%
\pgfsys@transformshift{2.891933in}{1.081795in}%
\pgfsys@useobject{currentmarker}{}%
\end{pgfscope}%
\end{pgfscope}%
\begin{pgfscope}%
\pgftext[left,bottom,x=0.250214in,y=1.045992in,rotate=0.000000]{{\rmfamily\fontsize{8.000000}{9.600000}\selectfont \(\displaystyle 1.5\)}}
\end{pgfscope}%
\begin{pgfscope}%
\pgfpathrectangle{\pgfqpoint{0.456621in}{0.329242in}}{\pgfqpoint{2.435312in}{1.505106in}} %
\pgfusepath{clip}%
\pgfsetbuttcap%
\pgfsetroundjoin%
\pgfsetlinewidth{0.501875pt}%
\definecolor{currentstroke}{rgb}{0.000000,0.000000,0.000000}%
\pgfsetstrokecolor{currentstroke}%
\pgfsetdash{{1.000000pt}{3.000000pt}}{0.000000pt}%
\pgfpathmoveto{\pgfqpoint{0.456621in}{1.332646in}}%
\pgfpathlineto{\pgfqpoint{2.891933in}{1.332646in}}%
\pgfusepath{stroke}%
\end{pgfscope}%
\begin{pgfscope}%
\pgfsetbuttcap%
\pgfsetroundjoin%
\definecolor{currentfill}{rgb}{0.000000,0.000000,0.000000}%
\pgfsetfillcolor{currentfill}%
\pgfsetlinewidth{0.501875pt}%
\definecolor{currentstroke}{rgb}{0.000000,0.000000,0.000000}%
\pgfsetstrokecolor{currentstroke}%
\pgfsetdash{}{0pt}%
\pgfsys@defobject{currentmarker}{\pgfqpoint{0.000000in}{0.000000in}}{\pgfqpoint{0.055556in}{0.000000in}}{%
\pgfpathmoveto{\pgfqpoint{0.000000in}{0.000000in}}%
\pgfpathlineto{\pgfqpoint{0.055556in}{0.000000in}}%
\pgfusepath{stroke,fill}%
}%
\begin{pgfscope}%
\pgfsys@transformshift{0.456621in}{1.332646in}%
\pgfsys@useobject{currentmarker}{}%
\end{pgfscope}%
\end{pgfscope}%
\begin{pgfscope}%
\pgfsetbuttcap%
\pgfsetroundjoin%
\definecolor{currentfill}{rgb}{0.000000,0.000000,0.000000}%
\pgfsetfillcolor{currentfill}%
\pgfsetlinewidth{0.501875pt}%
\definecolor{currentstroke}{rgb}{0.000000,0.000000,0.000000}%
\pgfsetstrokecolor{currentstroke}%
\pgfsetdash{}{0pt}%
\pgfsys@defobject{currentmarker}{\pgfqpoint{-0.055556in}{0.000000in}}{\pgfqpoint{0.000000in}{0.000000in}}{%
\pgfpathmoveto{\pgfqpoint{0.000000in}{0.000000in}}%
\pgfpathlineto{\pgfqpoint{-0.055556in}{0.000000in}}%
\pgfusepath{stroke,fill}%
}%
\begin{pgfscope}%
\pgfsys@transformshift{2.891933in}{1.332646in}%
\pgfsys@useobject{currentmarker}{}%
\end{pgfscope}%
\end{pgfscope}%
\begin{pgfscope}%
\pgftext[left,bottom,x=0.250214in,y=1.296843in,rotate=0.000000]{{\rmfamily\fontsize{8.000000}{9.600000}\selectfont \(\displaystyle 2.0\)}}
\end{pgfscope}%
\begin{pgfscope}%
\pgfpathrectangle{\pgfqpoint{0.456621in}{0.329242in}}{\pgfqpoint{2.435312in}{1.505106in}} %
\pgfusepath{clip}%
\pgfsetbuttcap%
\pgfsetroundjoin%
\pgfsetlinewidth{0.501875pt}%
\definecolor{currentstroke}{rgb}{0.000000,0.000000,0.000000}%
\pgfsetstrokecolor{currentstroke}%
\pgfsetdash{{1.000000pt}{3.000000pt}}{0.000000pt}%
\pgfpathmoveto{\pgfqpoint{0.456621in}{1.583497in}}%
\pgfpathlineto{\pgfqpoint{2.891933in}{1.583497in}}%
\pgfusepath{stroke}%
\end{pgfscope}%
\begin{pgfscope}%
\pgfsetbuttcap%
\pgfsetroundjoin%
\definecolor{currentfill}{rgb}{0.000000,0.000000,0.000000}%
\pgfsetfillcolor{currentfill}%
\pgfsetlinewidth{0.501875pt}%
\definecolor{currentstroke}{rgb}{0.000000,0.000000,0.000000}%
\pgfsetstrokecolor{currentstroke}%
\pgfsetdash{}{0pt}%
\pgfsys@defobject{currentmarker}{\pgfqpoint{0.000000in}{0.000000in}}{\pgfqpoint{0.055556in}{0.000000in}}{%
\pgfpathmoveto{\pgfqpoint{0.000000in}{0.000000in}}%
\pgfpathlineto{\pgfqpoint{0.055556in}{0.000000in}}%
\pgfusepath{stroke,fill}%
}%
\begin{pgfscope}%
\pgfsys@transformshift{0.456621in}{1.583497in}%
\pgfsys@useobject{currentmarker}{}%
\end{pgfscope}%
\end{pgfscope}%
\begin{pgfscope}%
\pgfsetbuttcap%
\pgfsetroundjoin%
\definecolor{currentfill}{rgb}{0.000000,0.000000,0.000000}%
\pgfsetfillcolor{currentfill}%
\pgfsetlinewidth{0.501875pt}%
\definecolor{currentstroke}{rgb}{0.000000,0.000000,0.000000}%
\pgfsetstrokecolor{currentstroke}%
\pgfsetdash{}{0pt}%
\pgfsys@defobject{currentmarker}{\pgfqpoint{-0.055556in}{0.000000in}}{\pgfqpoint{0.000000in}{0.000000in}}{%
\pgfpathmoveto{\pgfqpoint{0.000000in}{0.000000in}}%
\pgfpathlineto{\pgfqpoint{-0.055556in}{0.000000in}}%
\pgfusepath{stroke,fill}%
}%
\begin{pgfscope}%
\pgfsys@transformshift{2.891933in}{1.583497in}%
\pgfsys@useobject{currentmarker}{}%
\end{pgfscope}%
\end{pgfscope}%
\begin{pgfscope}%
\pgftext[left,bottom,x=0.250214in,y=1.547694in,rotate=0.000000]{{\rmfamily\fontsize{8.000000}{9.600000}\selectfont \(\displaystyle 2.5\)}}
\end{pgfscope}%
\begin{pgfscope}%
\pgfpathrectangle{\pgfqpoint{0.456621in}{0.329242in}}{\pgfqpoint{2.435312in}{1.505106in}} %
\pgfusepath{clip}%
\pgfsetbuttcap%
\pgfsetroundjoin%
\pgfsetlinewidth{0.501875pt}%
\definecolor{currentstroke}{rgb}{0.000000,0.000000,0.000000}%
\pgfsetstrokecolor{currentstroke}%
\pgfsetdash{{1.000000pt}{3.000000pt}}{0.000000pt}%
\pgfpathmoveto{\pgfqpoint{0.456621in}{1.834347in}}%
\pgfpathlineto{\pgfqpoint{2.891933in}{1.834347in}}%
\pgfusepath{stroke}%
\end{pgfscope}%
\begin{pgfscope}%
\pgfsetbuttcap%
\pgfsetroundjoin%
\definecolor{currentfill}{rgb}{0.000000,0.000000,0.000000}%
\pgfsetfillcolor{currentfill}%
\pgfsetlinewidth{0.501875pt}%
\definecolor{currentstroke}{rgb}{0.000000,0.000000,0.000000}%
\pgfsetstrokecolor{currentstroke}%
\pgfsetdash{}{0pt}%
\pgfsys@defobject{currentmarker}{\pgfqpoint{0.000000in}{0.000000in}}{\pgfqpoint{0.055556in}{0.000000in}}{%
\pgfpathmoveto{\pgfqpoint{0.000000in}{0.000000in}}%
\pgfpathlineto{\pgfqpoint{0.055556in}{0.000000in}}%
\pgfusepath{stroke,fill}%
}%
\begin{pgfscope}%
\pgfsys@transformshift{0.456621in}{1.834347in}%
\pgfsys@useobject{currentmarker}{}%
\end{pgfscope}%
\end{pgfscope}%
\begin{pgfscope}%
\pgfsetbuttcap%
\pgfsetroundjoin%
\definecolor{currentfill}{rgb}{0.000000,0.000000,0.000000}%
\pgfsetfillcolor{currentfill}%
\pgfsetlinewidth{0.501875pt}%
\definecolor{currentstroke}{rgb}{0.000000,0.000000,0.000000}%
\pgfsetstrokecolor{currentstroke}%
\pgfsetdash{}{0pt}%
\pgfsys@defobject{currentmarker}{\pgfqpoint{-0.055556in}{0.000000in}}{\pgfqpoint{0.000000in}{0.000000in}}{%
\pgfpathmoveto{\pgfqpoint{0.000000in}{0.000000in}}%
\pgfpathlineto{\pgfqpoint{-0.055556in}{0.000000in}}%
\pgfusepath{stroke,fill}%
}%
\begin{pgfscope}%
\pgfsys@transformshift{2.891933in}{1.834347in}%
\pgfsys@useobject{currentmarker}{}%
\end{pgfscope}%
\end{pgfscope}%
\begin{pgfscope}%
\pgftext[left,bottom,x=0.250214in,y=1.798545in,rotate=0.000000]{{\rmfamily\fontsize{8.000000}{9.600000}\selectfont \(\displaystyle 3.0\)}}
\end{pgfscope}%
\begin{pgfscope}%
\pgftext[left,bottom,x=0.180770in,y=0.787232in,rotate=90.000000]{{\rmfamily\fontsize{8.000000}{9.600000}\selectfont \(\displaystyle \Delta\chi^2 (\times 10^4)\)}}
\end{pgfscope}%
\begin{pgfscope}%
\pgfsetrectcap%
\pgfsetroundjoin%
\pgfsetlinewidth{0.501875pt}%
\definecolor{currentstroke}{rgb}{0.000000,0.000000,0.000000}%
\pgfsetstrokecolor{currentstroke}%
\pgfsetdash{}{0pt}%
\pgfpathmoveto{\pgfqpoint{0.456621in}{1.834347in}}%
\pgfpathlineto{\pgfqpoint{2.891933in}{1.834347in}}%
\pgfusepath{stroke}%
\end{pgfscope}%
\begin{pgfscope}%
\pgfsetrectcap%
\pgfsetroundjoin%
\pgfsetlinewidth{0.501875pt}%
\definecolor{currentstroke}{rgb}{0.000000,0.000000,0.000000}%
\pgfsetstrokecolor{currentstroke}%
\pgfsetdash{}{0pt}%
\pgfpathmoveto{\pgfqpoint{2.891933in}{0.329242in}}%
\pgfpathlineto{\pgfqpoint{2.891933in}{1.834347in}}%
\pgfusepath{stroke}%
\end{pgfscope}%
\begin{pgfscope}%
\pgfsetrectcap%
\pgfsetroundjoin%
\pgfsetlinewidth{0.501875pt}%
\definecolor{currentstroke}{rgb}{0.000000,0.000000,0.000000}%
\pgfsetstrokecolor{currentstroke}%
\pgfsetdash{}{0pt}%
\pgfpathmoveto{\pgfqpoint{0.456621in}{0.329242in}}%
\pgfpathlineto{\pgfqpoint{2.891933in}{0.329242in}}%
\pgfusepath{stroke}%
\end{pgfscope}%
\begin{pgfscope}%
\pgfsetrectcap%
\pgfsetroundjoin%
\pgfsetlinewidth{0.501875pt}%
\definecolor{currentstroke}{rgb}{0.000000,0.000000,0.000000}%
\pgfsetstrokecolor{currentstroke}%
\pgfsetdash{}{0pt}%
\pgfpathmoveto{\pgfqpoint{0.456621in}{0.329242in}}%
\pgfpathlineto{\pgfqpoint{0.456621in}{1.834347in}}%
\pgfusepath{stroke}%
\end{pgfscope}%
\begin{pgfscope}%
\pgfsetrectcap%
\pgfsetroundjoin%
\definecolor{currentfill}{rgb}{1.000000,1.000000,1.000000}%
\pgfsetfillcolor{currentfill}%
\pgfsetlinewidth{0.501875pt}%
\definecolor{currentstroke}{rgb}{0.000000,0.000000,0.000000}%
\pgfsetstrokecolor{currentstroke}%
\pgfsetdash{}{0pt}%
\pgfpathmoveto{\pgfqpoint{2.129081in}{1.302249in}}%
\pgfpathlineto{\pgfqpoint{2.814155in}{1.302249in}}%
\pgfpathquadraticcurveto{\pgfqpoint{2.836377in}{1.302249in}}{\pgfqpoint{2.836377in}{1.324471in}}%
\pgfpathlineto{\pgfqpoint{2.836377in}{1.756570in}}%
\pgfpathquadraticcurveto{\pgfqpoint{2.836377in}{1.778792in}}{\pgfqpoint{2.814155in}{1.778792in}}%
\pgfpathlineto{\pgfqpoint{2.129081in}{1.778792in}}%
\pgfpathquadraticcurveto{\pgfqpoint{2.106859in}{1.778792in}}{\pgfqpoint{2.106859in}{1.756570in}}%
\pgfpathlineto{\pgfqpoint{2.106859in}{1.324471in}}%
\pgfpathquadraticcurveto{\pgfqpoint{2.106859in}{1.302249in}}{\pgfqpoint{2.129081in}{1.302249in}}%
\pgfpathclose%
\pgfusepath{stroke,fill}%
\end{pgfscope}%
\begin{pgfscope}%
\pgfsetrectcap%
\pgfsetroundjoin%
\pgfsetlinewidth{1.003750pt}%
\definecolor{currentstroke}{rgb}{0.767843,0.220980,0.353137}%
\pgfsetstrokecolor{currentstroke}%
\pgfsetdash{}{0pt}%
\pgfpathmoveto{\pgfqpoint{2.184637in}{1.695459in}}%
\pgfpathlineto{\pgfqpoint{2.340192in}{1.695459in}}%
\pgfusepath{stroke}%
\end{pgfscope}%
\begin{pgfscope}%
\pgftext[left,bottom,x=2.462414in,y=1.656570in,rotate=0.000000]{{\rmfamily\fontsize{8.000000}{9.600000}\selectfont Euclid}}
\end{pgfscope}%
\begin{pgfscope}%
\pgfsetrectcap%
\pgfsetroundjoin%
\pgfsetlinewidth{1.003750pt}%
\definecolor{currentstroke}{rgb}{0.169412,0.150000,0.532353}%
\pgfsetstrokecolor{currentstroke}%
\pgfsetdash{}{0pt}%
\pgfpathmoveto{\pgfqpoint{2.184637in}{1.540520in}}%
\pgfpathlineto{\pgfqpoint{2.340192in}{1.540520in}}%
\pgfusepath{stroke}%
\end{pgfscope}%
\begin{pgfscope}%
\pgftext[left,bottom,x=2.462414in,y=1.501631in,rotate=0.000000]{{\rmfamily\fontsize{8.000000}{9.600000}\selectfont SKA1}}
\end{pgfscope}%
\begin{pgfscope}%
\pgfsetrectcap%
\pgfsetroundjoin%
\pgfsetlinewidth{1.003750pt}%
\definecolor{currentstroke}{rgb}{0.387059,0.164510,0.677451}%
\pgfsetstrokecolor{currentstroke}%
\pgfsetdash{}{0pt}%
\pgfpathmoveto{\pgfqpoint{2.184637in}{1.385582in}}%
\pgfpathlineto{\pgfqpoint{2.340192in}{1.385582in}}%
\pgfusepath{stroke}%
\end{pgfscope}%
\begin{pgfscope}%
\pgftext[left,bottom,x=2.462414in,y=1.346693in,rotate=0.000000]{{\rmfamily\fontsize{8.000000}{9.600000}\selectfont SKA2}}
\end{pgfscope}%
\end{pgfpicture}%
\makeatother%
\endgroup%

%% file: figures/PaperChi2ofK_opt.pgf
\begingroup%
\makeatletter%
\begin{pgfpicture}%
\pgfpathrectangle{\pgfpointorigin}{\pgfqpoint{3.044140in}{1.881382in}}%
\pgfusepath{use as bounding box, clip}%
\begin{pgfscope}%
\pgfsetbuttcap%
\pgfsetmiterjoin%
\definecolor{currentfill}{rgb}{1.000000,1.000000,1.000000}%
\pgfsetfillcolor{currentfill}%
\pgfsetlinewidth{0.000000pt}%
\definecolor{currentstroke}{rgb}{1.000000,1.000000,1.000000}%
\pgfsetstrokecolor{currentstroke}%
\pgfsetdash{}{0pt}%
\pgfpathmoveto{\pgfqpoint{0.000000in}{0.000000in}}%
\pgfpathlineto{\pgfqpoint{3.044140in}{0.000000in}}%
\pgfpathlineto{\pgfqpoint{3.044140in}{1.881382in}}%
\pgfpathlineto{\pgfqpoint{0.000000in}{1.881382in}}%
\pgfpathclose%
\pgfusepath{fill}%
\end{pgfscope}%
\begin{pgfscope}%
\pgfsetbuttcap%
\pgfsetmiterjoin%
\definecolor{currentfill}{rgb}{1.000000,1.000000,1.000000}%
\pgfsetfillcolor{currentfill}%
\pgfsetlinewidth{0.000000pt}%
\definecolor{currentstroke}{rgb}{0.000000,0.000000,0.000000}%
\pgfsetstrokecolor{currentstroke}%
\pgfsetstrokeopacity{0.000000}%
\pgfsetdash{}{0pt}%
\pgfpathmoveto{\pgfqpoint{0.456621in}{0.329242in}}%
\pgfpathlineto{\pgfqpoint{2.891933in}{0.329242in}}%
\pgfpathlineto{\pgfqpoint{2.891933in}{1.834347in}}%
\pgfpathlineto{\pgfqpoint{0.456621in}{1.834347in}}%
\pgfpathclose%
\pgfusepath{fill}%
\end{pgfscope}%
\begin{pgfscope}%
\pgfpathrectangle{\pgfqpoint{0.456621in}{0.329242in}}{\pgfqpoint{2.435312in}{1.505106in}} %
\pgfusepath{clip}%
\pgfsetrectcap%
\pgfsetroundjoin%
\pgfsetlinewidth{0.803000pt}%
\definecolor{currentstroke}{rgb}{0.690196,0.690196,0.690196}%
\pgfsetstrokecolor{currentstroke}%
\pgfsetdash{}{0pt}%
\pgfpathmoveto{\pgfqpoint{0.456621in}{0.329242in}}%
\pgfpathlineto{\pgfqpoint{0.456621in}{1.834347in}}%
\pgfusepath{stroke}%
\end{pgfscope}%
\begin{pgfscope}%
\pgfsetbuttcap%
\pgfsetroundjoin%
\definecolor{currentfill}{rgb}{0.000000,0.000000,0.000000}%
\pgfsetfillcolor{currentfill}%
\pgfsetlinewidth{0.803000pt}%
\definecolor{currentstroke}{rgb}{0.000000,0.000000,0.000000}%
\pgfsetstrokecolor{currentstroke}%
\pgfsetdash{}{0pt}%
\pgfsys@defobject{currentmarker}{\pgfqpoint{0.000000in}{0.000000in}}{\pgfqpoint{0.000000in}{0.048611in}}{%
\pgfpathmoveto{\pgfqpoint{0.000000in}{0.000000in}}%
\pgfpathlineto{\pgfqpoint{0.000000in}{0.048611in}}%
\pgfusepath{stroke,fill}%
}%
\begin{pgfscope}%
\pgfsys@transformshift{0.456621in}{0.329242in}%
\pgfsys@useobject{currentmarker}{}%
\end{pgfscope}%
\end{pgfscope}%
\begin{pgfscope}%
\pgftext[x=0.456621in,y=0.280631in,,top]{\rmfamily\fontsize{8.000000}{9.600000}\selectfont \(\displaystyle 0.00\)}%
\end{pgfscope}%
\begin{pgfscope}%
\pgfpathrectangle{\pgfqpoint{0.456621in}{0.329242in}}{\pgfqpoint{2.435312in}{1.505106in}} %
\pgfusepath{clip}%
\pgfsetrectcap%
\pgfsetroundjoin%
\pgfsetlinewidth{0.803000pt}%
\definecolor{currentstroke}{rgb}{0.690196,0.690196,0.690196}%
\pgfsetstrokecolor{currentstroke}%
\pgfsetdash{}{0pt}%
\pgfpathmoveto{\pgfqpoint{0.862506in}{0.329242in}}%
\pgfpathlineto{\pgfqpoint{0.862506in}{1.834347in}}%
\pgfusepath{stroke}%
\end{pgfscope}%
\begin{pgfscope}%
\pgfsetbuttcap%
\pgfsetroundjoin%
\definecolor{currentfill}{rgb}{0.000000,0.000000,0.000000}%
\pgfsetfillcolor{currentfill}%
\pgfsetlinewidth{0.803000pt}%
\definecolor{currentstroke}{rgb}{0.000000,0.000000,0.000000}%
\pgfsetstrokecolor{currentstroke}%
\pgfsetdash{}{0pt}%
\pgfsys@defobject{currentmarker}{\pgfqpoint{0.000000in}{0.000000in}}{\pgfqpoint{0.000000in}{0.048611in}}{%
\pgfpathmoveto{\pgfqpoint{0.000000in}{0.000000in}}%
\pgfpathlineto{\pgfqpoint{0.000000in}{0.048611in}}%
\pgfusepath{stroke,fill}%
}%
\begin{pgfscope}%
\pgfsys@transformshift{0.862506in}{0.329242in}%
\pgfsys@useobject{currentmarker}{}%
\end{pgfscope}%
\end{pgfscope}%
\begin{pgfscope}%
\pgftext[x=0.862506in,y=0.280631in,,top]{\rmfamily\fontsize{8.000000}{9.600000}\selectfont \(\displaystyle 0.25\)}%
\end{pgfscope}%
\begin{pgfscope}%
\pgfpathrectangle{\pgfqpoint{0.456621in}{0.329242in}}{\pgfqpoint{2.435312in}{1.505106in}} %
\pgfusepath{clip}%
\pgfsetrectcap%
\pgfsetroundjoin%
\pgfsetlinewidth{0.803000pt}%
\definecolor{currentstroke}{rgb}{0.690196,0.690196,0.690196}%
\pgfsetstrokecolor{currentstroke}%
\pgfsetdash{}{0pt}%
\pgfpathmoveto{\pgfqpoint{1.268392in}{0.329242in}}%
\pgfpathlineto{\pgfqpoint{1.268392in}{1.834347in}}%
\pgfusepath{stroke}%
\end{pgfscope}%
\begin{pgfscope}%
\pgfsetbuttcap%
\pgfsetroundjoin%
\definecolor{currentfill}{rgb}{0.000000,0.000000,0.000000}%
\pgfsetfillcolor{currentfill}%
\pgfsetlinewidth{0.803000pt}%
\definecolor{currentstroke}{rgb}{0.000000,0.000000,0.000000}%
\pgfsetstrokecolor{currentstroke}%
\pgfsetdash{}{0pt}%
\pgfsys@defobject{currentmarker}{\pgfqpoint{0.000000in}{0.000000in}}{\pgfqpoint{0.000000in}{0.048611in}}{%
\pgfpathmoveto{\pgfqpoint{0.000000in}{0.000000in}}%
\pgfpathlineto{\pgfqpoint{0.000000in}{0.048611in}}%
\pgfusepath{stroke,fill}%
}%
\begin{pgfscope}%
\pgfsys@transformshift{1.268392in}{0.329242in}%
\pgfsys@useobject{currentmarker}{}%
\end{pgfscope}%
\end{pgfscope}%
\begin{pgfscope}%
\pgftext[x=1.268392in,y=0.280631in,,top]{\rmfamily\fontsize{8.000000}{9.600000}\selectfont \(\displaystyle 0.50\)}%
\end{pgfscope}%
\begin{pgfscope}%
\pgfpathrectangle{\pgfqpoint{0.456621in}{0.329242in}}{\pgfqpoint{2.435312in}{1.505106in}} %
\pgfusepath{clip}%
\pgfsetrectcap%
\pgfsetroundjoin%
\pgfsetlinewidth{0.803000pt}%
\definecolor{currentstroke}{rgb}{0.690196,0.690196,0.690196}%
\pgfsetstrokecolor{currentstroke}%
\pgfsetdash{}{0pt}%
\pgfpathmoveto{\pgfqpoint{1.674277in}{0.329242in}}%
\pgfpathlineto{\pgfqpoint{1.674277in}{1.834347in}}%
\pgfusepath{stroke}%
\end{pgfscope}%
\begin{pgfscope}%
\pgfsetbuttcap%
\pgfsetroundjoin%
\definecolor{currentfill}{rgb}{0.000000,0.000000,0.000000}%
\pgfsetfillcolor{currentfill}%
\pgfsetlinewidth{0.803000pt}%
\definecolor{currentstroke}{rgb}{0.000000,0.000000,0.000000}%
\pgfsetstrokecolor{currentstroke}%
\pgfsetdash{}{0pt}%
\pgfsys@defobject{currentmarker}{\pgfqpoint{0.000000in}{0.000000in}}{\pgfqpoint{0.000000in}{0.048611in}}{%
\pgfpathmoveto{\pgfqpoint{0.000000in}{0.000000in}}%
\pgfpathlineto{\pgfqpoint{0.000000in}{0.048611in}}%
\pgfusepath{stroke,fill}%
}%
\begin{pgfscope}%
\pgfsys@transformshift{1.674277in}{0.329242in}%
\pgfsys@useobject{currentmarker}{}%
\end{pgfscope}%
\end{pgfscope}%
\begin{pgfscope}%
\pgftext[x=1.674277in,y=0.280631in,,top]{\rmfamily\fontsize{8.000000}{9.600000}\selectfont \(\displaystyle 0.75\)}%
\end{pgfscope}%
\begin{pgfscope}%
\pgfpathrectangle{\pgfqpoint{0.456621in}{0.329242in}}{\pgfqpoint{2.435312in}{1.505106in}} %
\pgfusepath{clip}%
\pgfsetrectcap%
\pgfsetroundjoin%
\pgfsetlinewidth{0.803000pt}%
\definecolor{currentstroke}{rgb}{0.690196,0.690196,0.690196}%
\pgfsetstrokecolor{currentstroke}%
\pgfsetdash{}{0pt}%
\pgfpathmoveto{\pgfqpoint{2.080162in}{0.329242in}}%
\pgfpathlineto{\pgfqpoint{2.080162in}{1.834347in}}%
\pgfusepath{stroke}%
\end{pgfscope}%
\begin{pgfscope}%
\pgfsetbuttcap%
\pgfsetroundjoin%
\definecolor{currentfill}{rgb}{0.000000,0.000000,0.000000}%
\pgfsetfillcolor{currentfill}%
\pgfsetlinewidth{0.803000pt}%
\definecolor{currentstroke}{rgb}{0.000000,0.000000,0.000000}%
\pgfsetstrokecolor{currentstroke}%
\pgfsetdash{}{0pt}%
\pgfsys@defobject{currentmarker}{\pgfqpoint{0.000000in}{0.000000in}}{\pgfqpoint{0.000000in}{0.048611in}}{%
\pgfpathmoveto{\pgfqpoint{0.000000in}{0.000000in}}%
\pgfpathlineto{\pgfqpoint{0.000000in}{0.048611in}}%
\pgfusepath{stroke,fill}%
}%
\begin{pgfscope}%
\pgfsys@transformshift{2.080162in}{0.329242in}%
\pgfsys@useobject{currentmarker}{}%
\end{pgfscope}%
\end{pgfscope}%
\begin{pgfscope}%
\pgftext[x=2.080162in,y=0.280631in,,top]{\rmfamily\fontsize{8.000000}{9.600000}\selectfont \(\displaystyle 1.00\)}%
\end{pgfscope}%
\begin{pgfscope}%
\pgfpathrectangle{\pgfqpoint{0.456621in}{0.329242in}}{\pgfqpoint{2.435312in}{1.505106in}} %
\pgfusepath{clip}%
\pgfsetrectcap%
\pgfsetroundjoin%
\pgfsetlinewidth{0.803000pt}%
\definecolor{currentstroke}{rgb}{0.690196,0.690196,0.690196}%
\pgfsetstrokecolor{currentstroke}%
\pgfsetdash{}{0pt}%
\pgfpathmoveto{\pgfqpoint{2.486048in}{0.329242in}}%
\pgfpathlineto{\pgfqpoint{2.486048in}{1.834347in}}%
\pgfusepath{stroke}%
\end{pgfscope}%
\begin{pgfscope}%
\pgfsetbuttcap%
\pgfsetroundjoin%
\definecolor{currentfill}{rgb}{0.000000,0.000000,0.000000}%
\pgfsetfillcolor{currentfill}%
\pgfsetlinewidth{0.803000pt}%
\definecolor{currentstroke}{rgb}{0.000000,0.000000,0.000000}%
\pgfsetstrokecolor{currentstroke}%
\pgfsetdash{}{0pt}%
\pgfsys@defobject{currentmarker}{\pgfqpoint{0.000000in}{0.000000in}}{\pgfqpoint{0.000000in}{0.048611in}}{%
\pgfpathmoveto{\pgfqpoint{0.000000in}{0.000000in}}%
\pgfpathlineto{\pgfqpoint{0.000000in}{0.048611in}}%
\pgfusepath{stroke,fill}%
}%
\begin{pgfscope}%
\pgfsys@transformshift{2.486048in}{0.329242in}%
\pgfsys@useobject{currentmarker}{}%
\end{pgfscope}%
\end{pgfscope}%
\begin{pgfscope}%
\pgftext[x=2.486048in,y=0.280631in,,top]{\rmfamily\fontsize{8.000000}{9.600000}\selectfont \(\displaystyle 1.25\)}%
\end{pgfscope}%
\begin{pgfscope}%
\pgfpathrectangle{\pgfqpoint{0.456621in}{0.329242in}}{\pgfqpoint{2.435312in}{1.505106in}} %
\pgfusepath{clip}%
\pgfsetrectcap%
\pgfsetroundjoin%
\pgfsetlinewidth{0.803000pt}%
\definecolor{currentstroke}{rgb}{0.690196,0.690196,0.690196}%
\pgfsetstrokecolor{currentstroke}%
\pgfsetdash{}{0pt}%
\pgfpathmoveto{\pgfqpoint{2.891933in}{0.329242in}}%
\pgfpathlineto{\pgfqpoint{2.891933in}{1.834347in}}%
\pgfusepath{stroke}%
\end{pgfscope}%
\begin{pgfscope}%
\pgfsetbuttcap%
\pgfsetroundjoin%
\definecolor{currentfill}{rgb}{0.000000,0.000000,0.000000}%
\pgfsetfillcolor{currentfill}%
\pgfsetlinewidth{0.803000pt}%
\definecolor{currentstroke}{rgb}{0.000000,0.000000,0.000000}%
\pgfsetstrokecolor{currentstroke}%
\pgfsetdash{}{0pt}%
\pgfsys@defobject{currentmarker}{\pgfqpoint{0.000000in}{0.000000in}}{\pgfqpoint{0.000000in}{0.048611in}}{%
\pgfpathmoveto{\pgfqpoint{0.000000in}{0.000000in}}%
\pgfpathlineto{\pgfqpoint{0.000000in}{0.048611in}}%
\pgfusepath{stroke,fill}%
}%
\begin{pgfscope}%
\pgfsys@transformshift{2.891933in}{0.329242in}%
\pgfsys@useobject{currentmarker}{}%
\end{pgfscope}%
\end{pgfscope}%
\begin{pgfscope}%
\pgftext[x=2.891933in,y=0.280631in,,top]{\rmfamily\fontsize{8.000000}{9.600000}\selectfont \(\displaystyle 1.50\)}%
\end{pgfscope}%
\begin{pgfscope}%
\pgftext[x=1.674277in,y=0.126310in,,top]{\rmfamily\fontsize{8.000000}{9.600000}\selectfont \(\displaystyle k\) [h/Mpc]}%
\end{pgfscope}%
\begin{pgfscope}%
\pgfpathrectangle{\pgfqpoint{0.456621in}{0.329242in}}{\pgfqpoint{2.435312in}{1.505106in}} %
\pgfusepath{clip}%
\pgfsetrectcap%
\pgfsetroundjoin%
\pgfsetlinewidth{0.803000pt}%
\definecolor{currentstroke}{rgb}{0.690196,0.690196,0.690196}%
\pgfsetstrokecolor{currentstroke}%
\pgfsetdash{}{0pt}%
\pgfpathmoveto{\pgfqpoint{0.456621in}{0.329242in}}%
\pgfpathlineto{\pgfqpoint{2.891933in}{0.329242in}}%
\pgfusepath{stroke}%
\end{pgfscope}%
\begin{pgfscope}%
\pgfsetbuttcap%
\pgfsetroundjoin%
\definecolor{currentfill}{rgb}{0.000000,0.000000,0.000000}%
\pgfsetfillcolor{currentfill}%
\pgfsetlinewidth{0.803000pt}%
\definecolor{currentstroke}{rgb}{0.000000,0.000000,0.000000}%
\pgfsetstrokecolor{currentstroke}%
\pgfsetdash{}{0pt}%
\pgfsys@defobject{currentmarker}{\pgfqpoint{0.000000in}{0.000000in}}{\pgfqpoint{0.048611in}{0.000000in}}{%
\pgfpathmoveto{\pgfqpoint{0.000000in}{0.000000in}}%
\pgfpathlineto{\pgfqpoint{0.048611in}{0.000000in}}%
\pgfusepath{stroke,fill}%
}%
\begin{pgfscope}%
\pgfsys@transformshift{0.456621in}{0.329242in}%
\pgfsys@useobject{currentmarker}{}%
\end{pgfscope}%
\end{pgfscope}%
\begin{pgfscope}%
\pgftext[x=0.348981in,y=0.290662in,left,base]{\rmfamily\fontsize{8.000000}{9.600000}\selectfont \(\displaystyle 0\)}%
\end{pgfscope}%
\begin{pgfscope}%
\pgfpathrectangle{\pgfqpoint{0.456621in}{0.329242in}}{\pgfqpoint{2.435312in}{1.505106in}} %
\pgfusepath{clip}%
\pgfsetrectcap%
\pgfsetroundjoin%
\pgfsetlinewidth{0.803000pt}%
\definecolor{currentstroke}{rgb}{0.690196,0.690196,0.690196}%
\pgfsetstrokecolor{currentstroke}%
\pgfsetdash{}{0pt}%
\pgfpathmoveto{\pgfqpoint{0.456621in}{0.630263in}}%
\pgfpathlineto{\pgfqpoint{2.891933in}{0.630263in}}%
\pgfusepath{stroke}%
\end{pgfscope}%
\begin{pgfscope}%
\pgfsetbuttcap%
\pgfsetroundjoin%
\definecolor{currentfill}{rgb}{0.000000,0.000000,0.000000}%
\pgfsetfillcolor{currentfill}%
\pgfsetlinewidth{0.803000pt}%
\definecolor{currentstroke}{rgb}{0.000000,0.000000,0.000000}%
\pgfsetstrokecolor{currentstroke}%
\pgfsetdash{}{0pt}%
\pgfsys@defobject{currentmarker}{\pgfqpoint{0.000000in}{0.000000in}}{\pgfqpoint{0.048611in}{0.000000in}}{%
\pgfpathmoveto{\pgfqpoint{0.000000in}{0.000000in}}%
\pgfpathlineto{\pgfqpoint{0.048611in}{0.000000in}}%
\pgfusepath{stroke,fill}%
}%
\begin{pgfscope}%
\pgfsys@transformshift{0.456621in}{0.630263in}%
\pgfsys@useobject{currentmarker}{}%
\end{pgfscope}%
\end{pgfscope}%
\begin{pgfscope}%
\pgftext[x=0.289953in,y=0.591683in,left,base]{\rmfamily\fontsize{8.000000}{9.600000}\selectfont \(\displaystyle 10\)}%
\end{pgfscope}%
\begin{pgfscope}%
\pgfpathrectangle{\pgfqpoint{0.456621in}{0.329242in}}{\pgfqpoint{2.435312in}{1.505106in}} %
\pgfusepath{clip}%
\pgfsetrectcap%
\pgfsetroundjoin%
\pgfsetlinewidth{0.803000pt}%
\definecolor{currentstroke}{rgb}{0.690196,0.690196,0.690196}%
\pgfsetstrokecolor{currentstroke}%
\pgfsetdash{}{0pt}%
\pgfpathmoveto{\pgfqpoint{0.456621in}{0.931284in}}%
\pgfpathlineto{\pgfqpoint{2.891933in}{0.931284in}}%
\pgfusepath{stroke}%
\end{pgfscope}%
\begin{pgfscope}%
\pgfsetbuttcap%
\pgfsetroundjoin%
\definecolor{currentfill}{rgb}{0.000000,0.000000,0.000000}%
\pgfsetfillcolor{currentfill}%
\pgfsetlinewidth{0.803000pt}%
\definecolor{currentstroke}{rgb}{0.000000,0.000000,0.000000}%
\pgfsetstrokecolor{currentstroke}%
\pgfsetdash{}{0pt}%
\pgfsys@defobject{currentmarker}{\pgfqpoint{0.000000in}{0.000000in}}{\pgfqpoint{0.048611in}{0.000000in}}{%
\pgfpathmoveto{\pgfqpoint{0.000000in}{0.000000in}}%
\pgfpathlineto{\pgfqpoint{0.048611in}{0.000000in}}%
\pgfusepath{stroke,fill}%
}%
\begin{pgfscope}%
\pgfsys@transformshift{0.456621in}{0.931284in}%
\pgfsys@useobject{currentmarker}{}%
\end{pgfscope}%
\end{pgfscope}%
\begin{pgfscope}%
\pgftext[x=0.289953in,y=0.892704in,left,base]{\rmfamily\fontsize{8.000000}{9.600000}\selectfont \(\displaystyle 20\)}%
\end{pgfscope}%
\begin{pgfscope}%
\pgfpathrectangle{\pgfqpoint{0.456621in}{0.329242in}}{\pgfqpoint{2.435312in}{1.505106in}} %
\pgfusepath{clip}%
\pgfsetrectcap%
\pgfsetroundjoin%
\pgfsetlinewidth{0.803000pt}%
\definecolor{currentstroke}{rgb}{0.690196,0.690196,0.690196}%
\pgfsetstrokecolor{currentstroke}%
\pgfsetdash{}{0pt}%
\pgfpathmoveto{\pgfqpoint{0.456621in}{1.232305in}}%
\pgfpathlineto{\pgfqpoint{2.891933in}{1.232305in}}%
\pgfusepath{stroke}%
\end{pgfscope}%
\begin{pgfscope}%
\pgfsetbuttcap%
\pgfsetroundjoin%
\definecolor{currentfill}{rgb}{0.000000,0.000000,0.000000}%
\pgfsetfillcolor{currentfill}%
\pgfsetlinewidth{0.803000pt}%
\definecolor{currentstroke}{rgb}{0.000000,0.000000,0.000000}%
\pgfsetstrokecolor{currentstroke}%
\pgfsetdash{}{0pt}%
\pgfsys@defobject{currentmarker}{\pgfqpoint{0.000000in}{0.000000in}}{\pgfqpoint{0.048611in}{0.000000in}}{%
\pgfpathmoveto{\pgfqpoint{0.000000in}{0.000000in}}%
\pgfpathlineto{\pgfqpoint{0.048611in}{0.000000in}}%
\pgfusepath{stroke,fill}%
}%
\begin{pgfscope}%
\pgfsys@transformshift{0.456621in}{1.232305in}%
\pgfsys@useobject{currentmarker}{}%
\end{pgfscope}%
\end{pgfscope}%
\begin{pgfscope}%
\pgftext[x=0.289953in,y=1.193725in,left,base]{\rmfamily\fontsize{8.000000}{9.600000}\selectfont \(\displaystyle 30\)}%
\end{pgfscope}%
\begin{pgfscope}%
\pgfpathrectangle{\pgfqpoint{0.456621in}{0.329242in}}{\pgfqpoint{2.435312in}{1.505106in}} %
\pgfusepath{clip}%
\pgfsetrectcap%
\pgfsetroundjoin%
\pgfsetlinewidth{0.803000pt}%
\definecolor{currentstroke}{rgb}{0.690196,0.690196,0.690196}%
\pgfsetstrokecolor{currentstroke}%
\pgfsetdash{}{0pt}%
\pgfpathmoveto{\pgfqpoint{0.456621in}{1.533326in}}%
\pgfpathlineto{\pgfqpoint{2.891933in}{1.533326in}}%
\pgfusepath{stroke}%
\end{pgfscope}%
\begin{pgfscope}%
\pgfsetbuttcap%
\pgfsetroundjoin%
\definecolor{currentfill}{rgb}{0.000000,0.000000,0.000000}%
\pgfsetfillcolor{currentfill}%
\pgfsetlinewidth{0.803000pt}%
\definecolor{currentstroke}{rgb}{0.000000,0.000000,0.000000}%
\pgfsetstrokecolor{currentstroke}%
\pgfsetdash{}{0pt}%
\pgfsys@defobject{currentmarker}{\pgfqpoint{0.000000in}{0.000000in}}{\pgfqpoint{0.048611in}{0.000000in}}{%
\pgfpathmoveto{\pgfqpoint{0.000000in}{0.000000in}}%
\pgfpathlineto{\pgfqpoint{0.048611in}{0.000000in}}%
\pgfusepath{stroke,fill}%
}%
\begin{pgfscope}%
\pgfsys@transformshift{0.456621in}{1.533326in}%
\pgfsys@useobject{currentmarker}{}%
\end{pgfscope}%
\end{pgfscope}%
\begin{pgfscope}%
\pgftext[x=0.289953in,y=1.494746in,left,base]{\rmfamily\fontsize{8.000000}{9.600000}\selectfont \(\displaystyle 40\)}%
\end{pgfscope}%
\begin{pgfscope}%
\pgfpathrectangle{\pgfqpoint{0.456621in}{0.329242in}}{\pgfqpoint{2.435312in}{1.505106in}} %
\pgfusepath{clip}%
\pgfsetrectcap%
\pgfsetroundjoin%
\pgfsetlinewidth{0.803000pt}%
\definecolor{currentstroke}{rgb}{0.690196,0.690196,0.690196}%
\pgfsetstrokecolor{currentstroke}%
\pgfsetdash{}{0pt}%
\pgfpathmoveto{\pgfqpoint{0.456621in}{1.834347in}}%
\pgfpathlineto{\pgfqpoint{2.891933in}{1.834347in}}%
\pgfusepath{stroke}%
\end{pgfscope}%
\begin{pgfscope}%
\pgfsetbuttcap%
\pgfsetroundjoin%
\definecolor{currentfill}{rgb}{0.000000,0.000000,0.000000}%
\pgfsetfillcolor{currentfill}%
\pgfsetlinewidth{0.803000pt}%
\definecolor{currentstroke}{rgb}{0.000000,0.000000,0.000000}%
\pgfsetstrokecolor{currentstroke}%
\pgfsetdash{}{0pt}%
\pgfsys@defobject{currentmarker}{\pgfqpoint{0.000000in}{0.000000in}}{\pgfqpoint{0.048611in}{0.000000in}}{%
\pgfpathmoveto{\pgfqpoint{0.000000in}{0.000000in}}%
\pgfpathlineto{\pgfqpoint{0.048611in}{0.000000in}}%
\pgfusepath{stroke,fill}%
}%
\begin{pgfscope}%
\pgfsys@transformshift{0.456621in}{1.834347in}%
\pgfsys@useobject{currentmarker}{}%
\end{pgfscope}%
\end{pgfscope}%
\begin{pgfscope}%
\pgftext[x=0.289953in,y=1.795767in,left,base]{\rmfamily\fontsize{8.000000}{9.600000}\selectfont \(\displaystyle 50\)}%
\end{pgfscope}%
\begin{pgfscope}%
\pgftext[x=0.234397in,y=1.081795in,,bottom,rotate=90.000000]{\rmfamily\fontsize{8.000000}{9.600000}\selectfont \(\displaystyle \Delta\chi^2\)}%
\end{pgfscope}%
\begin{pgfscope}%
\pgfpathrectangle{\pgfqpoint{0.456621in}{0.329242in}}{\pgfqpoint{2.435312in}{1.505106in}} %
\pgfusepath{clip}%
\pgfsetrectcap%
\pgfsetroundjoin%
\pgfsetlinewidth{1.003750pt}%
\definecolor{currentstroke}{rgb}{0.767843,0.220980,0.353137}%
\pgfsetstrokecolor{currentstroke}%
\pgfsetdash{}{0pt}%
\pgfpathmoveto{\pgfqpoint{0.506625in}{0.374112in}}%
\pgfpathlineto{\pgfqpoint{0.509649in}{0.378968in}}%
\pgfpathlineto{\pgfqpoint{0.512856in}{0.384263in}}%
\pgfpathlineto{\pgfqpoint{0.516257in}{0.390031in}}%
\pgfpathlineto{\pgfqpoint{0.519863in}{0.396319in}}%
\pgfpathlineto{\pgfqpoint{0.523688in}{0.403187in}}%
\pgfpathlineto{\pgfqpoint{0.527744in}{0.410709in}}%
\pgfpathlineto{\pgfqpoint{0.532045in}{0.418978in}}%
\pgfpathlineto{\pgfqpoint{0.536607in}{0.428103in}}%
\pgfpathlineto{\pgfqpoint{0.541444in}{0.438194in}}%
\pgfpathlineto{\pgfqpoint{0.546574in}{0.449349in}}%
\pgfpathlineto{\pgfqpoint{0.552014in}{0.461621in}}%
\pgfpathlineto{\pgfqpoint{0.557783in}{0.474991in}}%
\pgfpathlineto{\pgfqpoint{0.563901in}{0.489324in}}%
\pgfpathlineto{\pgfqpoint{0.570388in}{0.504353in}}%
\pgfpathlineto{\pgfqpoint{0.577269in}{0.519673in}}%
\pgfpathlineto{\pgfqpoint{0.584565in}{0.534793in}}%
\pgfpathlineto{\pgfqpoint{0.592302in}{0.549240in}}%
\pgfpathlineto{\pgfqpoint{0.600508in}{0.562750in}}%
\pgfpathlineto{\pgfqpoint{0.609210in}{0.575483in}}%
\pgfpathlineto{\pgfqpoint{0.618438in}{0.588148in}}%
\pgfpathlineto{\pgfqpoint{0.628224in}{0.601848in}}%
\pgfpathlineto{\pgfqpoint{0.638602in}{0.617502in}}%
\pgfpathlineto{\pgfqpoint{0.649607in}{0.635027in}}%
\pgfpathlineto{\pgfqpoint{0.661278in}{0.652746in}}%
\pgfpathlineto{\pgfqpoint{0.673655in}{0.667659in}}%
\pgfpathlineto{\pgfqpoint{0.686780in}{0.676822in}}%
\pgfpathlineto{\pgfqpoint{0.700700in}{0.679371in}}%
\pgfpathlineto{\pgfqpoint{0.715460in}{0.677775in}}%
\pgfpathlineto{\pgfqpoint{0.731114in}{0.676476in}}%
\pgfpathlineto{\pgfqpoint{0.747714in}{0.677823in}}%
\pgfpathlineto{\pgfqpoint{0.765319in}{0.678898in}}%
\pgfpathlineto{\pgfqpoint{0.783987in}{0.673494in}}%
\pgfpathlineto{\pgfqpoint{0.803785in}{0.658971in}}%
\pgfpathlineto{\pgfqpoint{0.824781in}{0.639957in}}%
\pgfpathlineto{\pgfqpoint{0.847045in}{0.622629in}}%
\pgfpathlineto{\pgfqpoint{0.870657in}{0.606729in}}%
\pgfpathlineto{\pgfqpoint{0.895696in}{0.587372in}}%
\pgfpathlineto{\pgfqpoint{0.922250in}{0.564504in}}%
\pgfpathlineto{\pgfqpoint{0.950409in}{0.543359in}}%
\pgfpathlineto{\pgfqpoint{0.980272in}{0.525373in}}%
\pgfpathlineto{\pgfqpoint{1.011940in}{0.507931in}}%
\pgfpathlineto{\pgfqpoint{1.045523in}{0.491300in}}%
\pgfpathlineto{\pgfqpoint{1.081138in}{0.476962in}}%
\pgfpathlineto{\pgfqpoint{1.118906in}{0.464021in}}%
\pgfpathlineto{\pgfqpoint{1.158959in}{0.451947in}}%
\pgfpathlineto{\pgfqpoint{1.201434in}{0.440736in}}%
\pgfpathlineto{\pgfqpoint{1.246477in}{0.429916in}}%
\pgfpathlineto{\pgfqpoint{1.294245in}{0.419756in}}%
\pgfpathlineto{\pgfqpoint{1.344901in}{0.410583in}}%
\pgfpathlineto{\pgfqpoint{1.398621in}{0.402398in}}%
\pgfpathlineto{\pgfqpoint{1.455589in}{0.395142in}}%
\pgfpathlineto{\pgfqpoint{1.516003in}{0.388628in}}%
\pgfpathlineto{\pgfqpoint{1.580070in}{0.382687in}}%
\pgfpathlineto{\pgfqpoint{1.648012in}{0.377175in}}%
\pgfpathlineto{\pgfqpoint{1.720063in}{0.372019in}}%
\pgfpathlineto{\pgfqpoint{1.796471in}{0.367193in}}%
\pgfpathlineto{\pgfqpoint{1.877500in}{0.362717in}}%
\pgfpathlineto{\pgfqpoint{1.963429in}{0.358636in}}%
\pgfpathlineto{\pgfqpoint{2.054555in}{0.355010in}}%
\pgfpathlineto{\pgfqpoint{2.151192in}{0.351892in}}%
\pgfpathlineto{\pgfqpoint{2.253674in}{0.349310in}}%
\pgfpathlineto{\pgfqpoint{2.362353in}{0.347242in}}%
\pgfpathlineto{\pgfqpoint{2.477604in}{0.345619in}}%
\pgfpathlineto{\pgfqpoint{2.599825in}{0.344336in}}%
\pgfpathlineto{\pgfqpoint{2.729438in}{0.343288in}}%
\pgfpathlineto{\pgfqpoint{2.866889in}{0.342388in}}%
\pgfpathlineto{\pgfqpoint{2.901933in}{0.342193in}}%
\pgfusepath{stroke}%
\end{pgfscope}%
\begin{pgfscope}%
\pgfpathrectangle{\pgfqpoint{0.456621in}{0.329242in}}{\pgfqpoint{2.435312in}{1.505106in}} %
\pgfusepath{clip}%
\pgfsetrectcap%
\pgfsetroundjoin%
\pgfsetlinewidth{1.003750pt}%
\definecolor{currentstroke}{rgb}{0.169412,0.150000,0.532353}%
\pgfsetstrokecolor{currentstroke}%
\pgfsetdash{}{0pt}%
\pgfpathmoveto{\pgfqpoint{0.506625in}{0.330167in}}%
\pgfpathlineto{\pgfqpoint{0.509649in}{0.330270in}}%
\pgfpathlineto{\pgfqpoint{0.512856in}{0.330383in}}%
\pgfpathlineto{\pgfqpoint{0.516257in}{0.330508in}}%
\pgfpathlineto{\pgfqpoint{0.519863in}{0.330645in}}%
\pgfpathlineto{\pgfqpoint{0.523688in}{0.330795in}}%
\pgfpathlineto{\pgfqpoint{0.527744in}{0.330962in}}%
\pgfpathlineto{\pgfqpoint{0.532045in}{0.331146in}}%
\pgfpathlineto{\pgfqpoint{0.536607in}{0.331351in}}%
\pgfpathlineto{\pgfqpoint{0.541444in}{0.331579in}}%
\pgfpathlineto{\pgfqpoint{0.546574in}{0.331834in}}%
\pgfpathlineto{\pgfqpoint{0.552014in}{0.332118in}}%
\pgfpathlineto{\pgfqpoint{0.557783in}{0.332432in}}%
\pgfpathlineto{\pgfqpoint{0.563901in}{0.332776in}}%
\pgfpathlineto{\pgfqpoint{0.570388in}{0.333145in}}%
\pgfpathlineto{\pgfqpoint{0.577269in}{0.333535in}}%
\pgfpathlineto{\pgfqpoint{0.584565in}{0.333938in}}%
\pgfpathlineto{\pgfqpoint{0.592302in}{0.334348in}}%
\pgfpathlineto{\pgfqpoint{0.600508in}{0.334760in}}%
\pgfpathlineto{\pgfqpoint{0.609210in}{0.335183in}}%
\pgfpathlineto{\pgfqpoint{0.618438in}{0.335632in}}%
\pgfpathlineto{\pgfqpoint{0.628224in}{0.336137in}}%
\pgfpathlineto{\pgfqpoint{0.638602in}{0.336723in}}%
\pgfpathlineto{\pgfqpoint{0.649607in}{0.337398in}}%
\pgfpathlineto{\pgfqpoint{0.661278in}{0.338132in}}%
\pgfpathlineto{\pgfqpoint{0.673655in}{0.338865in}}%
\pgfpathlineto{\pgfqpoint{0.686780in}{0.339529in}}%
\pgfpathlineto{\pgfqpoint{0.700700in}{0.340101in}}%
\pgfpathlineto{\pgfqpoint{0.715460in}{0.340637in}}%
\pgfpathlineto{\pgfqpoint{0.731114in}{0.341250in}}%
\pgfpathlineto{\pgfqpoint{0.747714in}{0.342010in}}%
\pgfpathlineto{\pgfqpoint{0.765319in}{0.342848in}}%
\pgfpathlineto{\pgfqpoint{0.783987in}{0.343584in}}%
\pgfpathlineto{\pgfqpoint{0.803785in}{0.344112in}}%
\pgfpathlineto{\pgfqpoint{0.824781in}{0.344541in}}%
\pgfpathlineto{\pgfqpoint{0.847045in}{0.345053in}}%
\pgfpathlineto{\pgfqpoint{0.870657in}{0.345625in}}%
\pgfpathlineto{\pgfqpoint{0.895696in}{0.346042in}}%
\pgfpathlineto{\pgfqpoint{0.922250in}{0.346234in}}%
\pgfpathlineto{\pgfqpoint{0.950409in}{0.346371in}}%
\pgfpathlineto{\pgfqpoint{0.980272in}{0.346492in}}%
\pgfpathlineto{\pgfqpoint{1.011940in}{0.346441in}}%
\pgfpathlineto{\pgfqpoint{1.045523in}{0.346214in}}%
\pgfpathlineto{\pgfqpoint{1.081138in}{0.345923in}}%
\pgfpathlineto{\pgfqpoint{1.118906in}{0.345550in}}%
\pgfpathlineto{\pgfqpoint{1.158959in}{0.345105in}}%
\pgfpathlineto{\pgfqpoint{1.201434in}{0.344647in}}%
\pgfpathlineto{\pgfqpoint{1.246477in}{0.344177in}}%
\pgfpathlineto{\pgfqpoint{1.294245in}{0.343708in}}%
\pgfpathlineto{\pgfqpoint{1.344901in}{0.343245in}}%
\pgfpathlineto{\pgfqpoint{1.398621in}{0.342779in}}%
\pgfpathlineto{\pgfqpoint{1.455589in}{0.342304in}}%
\pgfpathlineto{\pgfqpoint{1.516003in}{0.341818in}}%
\pgfpathlineto{\pgfqpoint{1.580070in}{0.341316in}}%
\pgfpathlineto{\pgfqpoint{1.648012in}{0.340794in}}%
\pgfpathlineto{\pgfqpoint{1.720063in}{0.340255in}}%
\pgfpathlineto{\pgfqpoint{1.796471in}{0.339713in}}%
\pgfpathlineto{\pgfqpoint{1.877500in}{0.339181in}}%
\pgfpathlineto{\pgfqpoint{1.963429in}{0.338661in}}%
\pgfpathlineto{\pgfqpoint{2.054555in}{0.338140in}}%
\pgfpathlineto{\pgfqpoint{2.151192in}{0.337599in}}%
\pgfpathlineto{\pgfqpoint{2.253674in}{0.337020in}}%
\pgfpathlineto{\pgfqpoint{2.362353in}{0.336401in}}%
\pgfpathlineto{\pgfqpoint{2.477604in}{0.335750in}}%
\pgfpathlineto{\pgfqpoint{2.599825in}{0.335084in}}%
\pgfpathlineto{\pgfqpoint{2.729438in}{0.334429in}}%
\pgfpathlineto{\pgfqpoint{2.866889in}{0.333815in}}%
\pgfpathlineto{\pgfqpoint{2.901933in}{0.333686in}}%
\pgfusepath{stroke}%
\end{pgfscope}%
\begin{pgfscope}%
\pgfpathrectangle{\pgfqpoint{0.456621in}{0.329242in}}{\pgfqpoint{2.435312in}{1.505106in}} %
\pgfusepath{clip}%
\pgfsetrectcap%
\pgfsetroundjoin%
\pgfsetlinewidth{1.003750pt}%
\definecolor{currentstroke}{rgb}{0.387059,0.164510,0.677451}%
\pgfsetstrokecolor{currentstroke}%
\pgfsetdash{}{0pt}%
\pgfpathmoveto{\pgfqpoint{0.506625in}{0.443473in}}%
\pgfpathlineto{\pgfqpoint{0.509649in}{0.456294in}}%
\pgfpathlineto{\pgfqpoint{0.512856in}{0.470379in}}%
\pgfpathlineto{\pgfqpoint{0.516257in}{0.485839in}}%
\pgfpathlineto{\pgfqpoint{0.519863in}{0.502807in}}%
\pgfpathlineto{\pgfqpoint{0.523688in}{0.521436in}}%
\pgfpathlineto{\pgfqpoint{0.527744in}{0.541913in}}%
\pgfpathlineto{\pgfqpoint{0.532045in}{0.564452in}}%
\pgfpathlineto{\pgfqpoint{0.536607in}{0.589290in}}%
\pgfpathlineto{\pgfqpoint{0.541444in}{0.616667in}}%
\pgfpathlineto{\pgfqpoint{0.546574in}{0.646793in}}%
\pgfpathlineto{\pgfqpoint{0.552014in}{0.679792in}}%
\pgfpathlineto{\pgfqpoint{0.557783in}{0.715651in}}%
\pgfpathlineto{\pgfqpoint{0.563901in}{0.754155in}}%
\pgfpathlineto{\pgfqpoint{0.570388in}{0.794857in}}%
\pgfpathlineto{\pgfqpoint{0.577269in}{0.837077in}}%
\pgfpathlineto{\pgfqpoint{0.584565in}{0.879985in}}%
\pgfpathlineto{\pgfqpoint{0.592302in}{0.922782in}}%
\pgfpathlineto{\pgfqpoint{0.600508in}{0.964989in}}%
\pgfpathlineto{\pgfqpoint{0.609210in}{1.006789in}}%
\pgfpathlineto{\pgfqpoint{0.618438in}{1.049197in}}%
\pgfpathlineto{\pgfqpoint{0.628224in}{1.093774in}}%
\pgfpathlineto{\pgfqpoint{0.638602in}{1.141694in}}%
\pgfpathlineto{\pgfqpoint{0.649607in}{1.192481in}}%
\pgfpathlineto{\pgfqpoint{0.661278in}{1.243254in}}%
\pgfpathlineto{\pgfqpoint{0.673655in}{1.289337in}}%
\pgfpathlineto{\pgfqpoint{0.686780in}{1.326411in}}%
\pgfpathlineto{\pgfqpoint{0.700700in}{1.353315in}}%
\pgfpathlineto{\pgfqpoint{0.715460in}{1.373582in}}%
\pgfpathlineto{\pgfqpoint{0.731114in}{1.393367in}}%
\pgfpathlineto{\pgfqpoint{0.747714in}{1.415702in}}%
\pgfpathlineto{\pgfqpoint{0.765319in}{1.436286in}}%
\pgfpathlineto{\pgfqpoint{0.783987in}{1.446537in}}%
\pgfpathlineto{\pgfqpoint{0.803785in}{1.442808in}}%
\pgfpathlineto{\pgfqpoint{0.824781in}{1.431085in}}%
\pgfpathlineto{\pgfqpoint{0.847045in}{1.419393in}}%
\pgfpathlineto{\pgfqpoint{0.870657in}{1.406919in}}%
\pgfpathlineto{\pgfqpoint{0.895696in}{1.385981in}}%
\pgfpathlineto{\pgfqpoint{0.922250in}{1.355229in}}%
\pgfpathlineto{\pgfqpoint{0.950409in}{1.321371in}}%
\pgfpathlineto{\pgfqpoint{0.980272in}{1.286130in}}%
\pgfpathlineto{\pgfqpoint{1.011940in}{1.244695in}}%
\pgfpathlineto{\pgfqpoint{1.045523in}{1.196704in}}%
\pgfpathlineto{\pgfqpoint{1.081138in}{1.144352in}}%
\pgfpathlineto{\pgfqpoint{1.118906in}{1.086252in}}%
\pgfpathlineto{\pgfqpoint{1.158959in}{1.022575in}}%
\pgfpathlineto{\pgfqpoint{1.201434in}{0.956234in}}%
\pgfpathlineto{\pgfqpoint{1.246477in}{0.889481in}}%
\pgfpathlineto{\pgfqpoint{1.294245in}{0.825529in}}%
\pgfpathlineto{\pgfqpoint{1.344901in}{0.767592in}}%
\pgfpathlineto{\pgfqpoint{1.398621in}{0.716271in}}%
\pgfpathlineto{\pgfqpoint{1.455589in}{0.670332in}}%
\pgfpathlineto{\pgfqpoint{1.516003in}{0.628942in}}%
\pgfpathlineto{\pgfqpoint{1.580070in}{0.592692in}}%
\pgfpathlineto{\pgfqpoint{1.648012in}{0.561866in}}%
\pgfpathlineto{\pgfqpoint{1.720063in}{0.535278in}}%
\pgfpathlineto{\pgfqpoint{1.796471in}{0.511294in}}%
\pgfpathlineto{\pgfqpoint{1.877500in}{0.489071in}}%
\pgfpathlineto{\pgfqpoint{1.963429in}{0.468598in}}%
\pgfpathlineto{\pgfqpoint{2.054555in}{0.450486in}}%
\pgfpathlineto{\pgfqpoint{2.151192in}{0.435500in}}%
\pgfpathlineto{\pgfqpoint{2.253674in}{0.423562in}}%
\pgfpathlineto{\pgfqpoint{2.362353in}{0.413649in}}%
\pgfpathlineto{\pgfqpoint{2.477604in}{0.404799in}}%
\pgfpathlineto{\pgfqpoint{2.599825in}{0.396578in}}%
\pgfpathlineto{\pgfqpoint{2.729438in}{0.388848in}}%
\pgfpathlineto{\pgfqpoint{2.866889in}{0.381563in}}%
\pgfpathlineto{\pgfqpoint{2.901933in}{0.379912in}}%
\pgfusepath{stroke}%
\end{pgfscope}%
\begin{pgfscope}%
\pgfpathrectangle{\pgfqpoint{0.456621in}{0.329242in}}{\pgfqpoint{2.435312in}{1.505106in}} %
\pgfusepath{clip}%
\pgfsetrectcap%
\pgfsetroundjoin%
\pgfsetlinewidth{1.003750pt}%
\definecolor{currentstroke}{rgb}{0.900000,0.677451,0.070980}%
\pgfsetstrokecolor{currentstroke}%
\pgfsetdash{}{0pt}%
\pgfpathmoveto{\pgfqpoint{0.506625in}{0.446034in}}%
\pgfpathlineto{\pgfqpoint{0.509649in}{0.454160in}}%
\pgfpathlineto{\pgfqpoint{0.512856in}{0.462474in}}%
\pgfpathlineto{\pgfqpoint{0.516257in}{0.470966in}}%
\pgfpathlineto{\pgfqpoint{0.519863in}{0.479634in}}%
\pgfpathlineto{\pgfqpoint{0.523688in}{0.488486in}}%
\pgfpathlineto{\pgfqpoint{0.527744in}{0.497525in}}%
\pgfpathlineto{\pgfqpoint{0.532045in}{0.506745in}}%
\pgfpathlineto{\pgfqpoint{0.536607in}{0.516136in}}%
\pgfpathlineto{\pgfqpoint{0.541444in}{0.525702in}}%
\pgfpathlineto{\pgfqpoint{0.546574in}{0.535451in}}%
\pgfpathlineto{\pgfqpoint{0.552014in}{0.545371in}}%
\pgfpathlineto{\pgfqpoint{0.557783in}{0.555396in}}%
\pgfpathlineto{\pgfqpoint{0.563901in}{0.565386in}}%
\pgfpathlineto{\pgfqpoint{0.570388in}{0.575140in}}%
\pgfpathlineto{\pgfqpoint{0.577269in}{0.584424in}}%
\pgfpathlineto{\pgfqpoint{0.584565in}{0.593017in}}%
\pgfpathlineto{\pgfqpoint{0.592302in}{0.600761in}}%
\pgfpathlineto{\pgfqpoint{0.600508in}{0.607619in}}%
\pgfpathlineto{\pgfqpoint{0.609210in}{0.613716in}}%
\pgfpathlineto{\pgfqpoint{0.618438in}{0.619317in}}%
\pgfpathlineto{\pgfqpoint{0.628224in}{0.624723in}}%
\pgfpathlineto{\pgfqpoint{0.638602in}{0.630056in}}%
\pgfpathlineto{\pgfqpoint{0.649607in}{0.635069in}}%
\pgfpathlineto{\pgfqpoint{0.661278in}{0.639098in}}%
\pgfpathlineto{\pgfqpoint{0.673655in}{0.641303in}}%
\pgfpathlineto{\pgfqpoint{0.686780in}{0.641085in}}%
\pgfpathlineto{\pgfqpoint{0.700700in}{0.638445in}}%
\pgfpathlineto{\pgfqpoint{0.715460in}{0.634137in}}%
\pgfpathlineto{\pgfqpoint{0.731114in}{0.629208in}}%
\pgfpathlineto{\pgfqpoint{0.747714in}{0.624148in}}%
\pgfpathlineto{\pgfqpoint{0.765319in}{0.618493in}}%
\pgfpathlineto{\pgfqpoint{0.783987in}{0.611168in}}%
\pgfpathlineto{\pgfqpoint{0.803785in}{0.601773in}}%
\pgfpathlineto{\pgfqpoint{0.824781in}{0.591166in}}%
\pgfpathlineto{\pgfqpoint{0.847045in}{0.580272in}}%
\pgfpathlineto{\pgfqpoint{0.870657in}{0.568878in}}%
\pgfpathlineto{\pgfqpoint{0.895696in}{0.555828in}}%
\pgfpathlineto{\pgfqpoint{0.922250in}{0.540517in}}%
\pgfpathlineto{\pgfqpoint{0.950409in}{0.523522in}}%
\pgfpathlineto{\pgfqpoint{0.980272in}{0.505085in}}%
\pgfpathlineto{\pgfqpoint{1.011940in}{0.484473in}}%
\pgfpathlineto{\pgfqpoint{1.045523in}{0.461879in}}%
\pgfpathlineto{\pgfqpoint{1.081138in}{0.438668in}}%
\pgfpathlineto{\pgfqpoint{1.118906in}{0.415674in}}%
\pgfpathlineto{\pgfqpoint{1.158959in}{0.393627in}}%
\pgfpathlineto{\pgfqpoint{1.201434in}{0.373513in}}%
\pgfpathlineto{\pgfqpoint{1.246477in}{0.356333in}}%
\pgfpathlineto{\pgfqpoint{1.294245in}{0.343231in}}%
\pgfpathlineto{\pgfqpoint{1.344901in}{0.334877in}}%
\pgfpathlineto{\pgfqpoint{1.398621in}{0.330863in}}%
\pgfpathlineto{\pgfqpoint{1.455589in}{0.329565in}}%
\pgfpathlineto{\pgfqpoint{1.516003in}{0.329289in}}%
\pgfpathlineto{\pgfqpoint{1.580070in}{0.329247in}}%
\pgfpathlineto{\pgfqpoint{1.648012in}{0.329242in}}%
\pgfpathlineto{\pgfqpoint{1.720063in}{0.329242in}}%
\pgfpathlineto{\pgfqpoint{1.796471in}{0.329242in}}%
\pgfpathlineto{\pgfqpoint{1.877500in}{0.329242in}}%
\pgfpathlineto{\pgfqpoint{1.963429in}{0.329242in}}%
\pgfpathlineto{\pgfqpoint{2.054555in}{0.329242in}}%
\pgfpathlineto{\pgfqpoint{2.151192in}{0.329242in}}%
\pgfpathlineto{\pgfqpoint{2.253674in}{0.329242in}}%
\pgfpathlineto{\pgfqpoint{2.362353in}{0.329242in}}%
\pgfpathlineto{\pgfqpoint{2.477604in}{0.329242in}}%
\pgfpathlineto{\pgfqpoint{2.599825in}{0.329242in}}%
\pgfpathlineto{\pgfqpoint{2.729438in}{0.329242in}}%
\pgfpathlineto{\pgfqpoint{2.866889in}{0.329242in}}%
\pgfpathlineto{\pgfqpoint{2.901933in}{0.329242in}}%
\pgfusepath{stroke}%
\end{pgfscope}%
\begin{pgfscope}%
\pgfpathrectangle{\pgfqpoint{0.456621in}{0.329242in}}{\pgfqpoint{2.435312in}{1.505106in}} %
\pgfusepath{clip}%
\pgfsetrectcap%
\pgfsetroundjoin%
\pgfsetlinewidth{1.003750pt}%
\definecolor{currentstroke}{rgb}{0.941961,0.395098,0.062941}%
\pgfsetstrokecolor{currentstroke}%
\pgfsetdash{}{0pt}%
\pgfpathmoveto{\pgfqpoint{0.506625in}{0.333724in}}%
\pgfpathlineto{\pgfqpoint{0.509649in}{0.334279in}}%
\pgfpathlineto{\pgfqpoint{0.512856in}{0.334902in}}%
\pgfpathlineto{\pgfqpoint{0.516257in}{0.335601in}}%
\pgfpathlineto{\pgfqpoint{0.519863in}{0.336386in}}%
\pgfpathlineto{\pgfqpoint{0.523688in}{0.337266in}}%
\pgfpathlineto{\pgfqpoint{0.527744in}{0.338252in}}%
\pgfpathlineto{\pgfqpoint{0.532045in}{0.339358in}}%
\pgfpathlineto{\pgfqpoint{0.536607in}{0.340596in}}%
\pgfpathlineto{\pgfqpoint{0.541444in}{0.341982in}}%
\pgfpathlineto{\pgfqpoint{0.546574in}{0.343533in}}%
\pgfpathlineto{\pgfqpoint{0.552014in}{0.345265in}}%
\pgfpathlineto{\pgfqpoint{0.557783in}{0.347199in}}%
\pgfpathlineto{\pgfqpoint{0.563901in}{0.349355in}}%
\pgfpathlineto{\pgfqpoint{0.570388in}{0.351754in}}%
\pgfpathlineto{\pgfqpoint{0.577269in}{0.354420in}}%
\pgfpathlineto{\pgfqpoint{0.584565in}{0.357374in}}%
\pgfpathlineto{\pgfqpoint{0.592302in}{0.360640in}}%
\pgfpathlineto{\pgfqpoint{0.600508in}{0.364241in}}%
\pgfpathlineto{\pgfqpoint{0.609210in}{0.368198in}}%
\pgfpathlineto{\pgfqpoint{0.618438in}{0.372533in}}%
\pgfpathlineto{\pgfqpoint{0.628224in}{0.377264in}}%
\pgfpathlineto{\pgfqpoint{0.638602in}{0.382410in}}%
\pgfpathlineto{\pgfqpoint{0.649607in}{0.387977in}}%
\pgfpathlineto{\pgfqpoint{0.661278in}{0.393955in}}%
\pgfpathlineto{\pgfqpoint{0.673655in}{0.400308in}}%
\pgfpathlineto{\pgfqpoint{0.686780in}{0.406968in}}%
\pgfpathlineto{\pgfqpoint{0.700700in}{0.413843in}}%
\pgfpathlineto{\pgfqpoint{0.715460in}{0.420827in}}%
\pgfpathlineto{\pgfqpoint{0.731114in}{0.427829in}}%
\pgfpathlineto{\pgfqpoint{0.747714in}{0.434726in}}%
\pgfpathlineto{\pgfqpoint{0.765319in}{0.441268in}}%
\pgfpathlineto{\pgfqpoint{0.783987in}{0.447040in}}%
\pgfpathlineto{\pgfqpoint{0.803785in}{0.451670in}}%
\pgfpathlineto{\pgfqpoint{0.824781in}{0.455176in}}%
\pgfpathlineto{\pgfqpoint{0.847045in}{0.457864in}}%
\pgfpathlineto{\pgfqpoint{0.870657in}{0.459806in}}%
\pgfpathlineto{\pgfqpoint{0.895696in}{0.460744in}}%
\pgfpathlineto{\pgfqpoint{0.922250in}{0.460502in}}%
\pgfpathlineto{\pgfqpoint{0.950409in}{0.459085in}}%
\pgfpathlineto{\pgfqpoint{0.980272in}{0.456656in}}%
\pgfpathlineto{\pgfqpoint{1.011940in}{0.453254in}}%
\pgfpathlineto{\pgfqpoint{1.045523in}{0.448310in}}%
\pgfpathlineto{\pgfqpoint{1.081138in}{0.441412in}}%
\pgfpathlineto{\pgfqpoint{1.118906in}{0.433558in}}%
\pgfpathlineto{\pgfqpoint{1.158959in}{0.426137in}}%
\pgfpathlineto{\pgfqpoint{1.201434in}{0.419170in}}%
\pgfpathlineto{\pgfqpoint{1.246477in}{0.411593in}}%
\pgfpathlineto{\pgfqpoint{1.294245in}{0.402857in}}%
\pgfpathlineto{\pgfqpoint{1.344901in}{0.392812in}}%
\pgfpathlineto{\pgfqpoint{1.398621in}{0.380503in}}%
\pgfpathlineto{\pgfqpoint{1.455589in}{0.366715in}}%
\pgfpathlineto{\pgfqpoint{1.516003in}{0.355580in}}%
\pgfpathlineto{\pgfqpoint{1.580070in}{0.349460in}}%
\pgfpathlineto{\pgfqpoint{1.648012in}{0.346300in}}%
\pgfpathlineto{\pgfqpoint{1.720063in}{0.343813in}}%
\pgfpathlineto{\pgfqpoint{1.796471in}{0.341611in}}%
\pgfpathlineto{\pgfqpoint{1.877500in}{0.339535in}}%
\pgfpathlineto{\pgfqpoint{1.963429in}{0.337513in}}%
\pgfpathlineto{\pgfqpoint{2.054555in}{0.335710in}}%
\pgfpathlineto{\pgfqpoint{2.151192in}{0.333964in}}%
\pgfpathlineto{\pgfqpoint{2.253674in}{0.332313in}}%
\pgfpathlineto{\pgfqpoint{2.362353in}{0.330941in}}%
\pgfpathlineto{\pgfqpoint{2.477604in}{0.329964in}}%
\pgfpathlineto{\pgfqpoint{2.599825in}{0.329461in}}%
\pgfpathlineto{\pgfqpoint{2.729438in}{0.329290in}}%
\pgfpathlineto{\pgfqpoint{2.866889in}{0.329250in}}%
\pgfpathlineto{\pgfqpoint{2.901933in}{0.329248in}}%
\pgfusepath{stroke}%
\end{pgfscope}%
\begin{pgfscope}%
\pgfsetrectcap%
\pgfsetmiterjoin%
\pgfsetlinewidth{0.501875pt}%
\definecolor{currentstroke}{rgb}{0.000000,0.000000,0.000000}%
\pgfsetstrokecolor{currentstroke}%
\pgfsetdash{}{0pt}%
\pgfpathmoveto{\pgfqpoint{0.456621in}{0.329242in}}%
\pgfpathlineto{\pgfqpoint{0.456621in}{1.834347in}}%
\pgfusepath{stroke}%
\end{pgfscope}%
\begin{pgfscope}%
\pgfsetrectcap%
\pgfsetmiterjoin%
\pgfsetlinewidth{0.501875pt}%
\definecolor{currentstroke}{rgb}{0.000000,0.000000,0.000000}%
\pgfsetstrokecolor{currentstroke}%
\pgfsetdash{}{0pt}%
\pgfpathmoveto{\pgfqpoint{2.891933in}{0.329242in}}%
\pgfpathlineto{\pgfqpoint{2.891933in}{1.834347in}}%
\pgfusepath{stroke}%
\end{pgfscope}%
\begin{pgfscope}%
\pgfsetrectcap%
\pgfsetmiterjoin%
\pgfsetlinewidth{0.501875pt}%
\definecolor{currentstroke}{rgb}{0.000000,0.000000,0.000000}%
\pgfsetstrokecolor{currentstroke}%
\pgfsetdash{}{0pt}%
\pgfpathmoveto{\pgfqpoint{0.456621in}{0.329242in}}%
\pgfpathlineto{\pgfqpoint{2.891933in}{0.329242in}}%
\pgfusepath{stroke}%
\end{pgfscope}%
\begin{pgfscope}%
\pgfsetrectcap%
\pgfsetmiterjoin%
\pgfsetlinewidth{0.501875pt}%
\definecolor{currentstroke}{rgb}{0.000000,0.000000,0.000000}%
\pgfsetstrokecolor{currentstroke}%
\pgfsetdash{}{0pt}%
\pgfpathmoveto{\pgfqpoint{0.456621in}{1.834347in}}%
\pgfpathlineto{\pgfqpoint{2.891933in}{1.834347in}}%
\pgfusepath{stroke}%
\end{pgfscope}%
\begin{pgfscope}%
\pgfsetbuttcap%
\pgfsetmiterjoin%
\definecolor{currentfill}{rgb}{1.000000,1.000000,1.000000}%
\pgfsetfillcolor{currentfill}%
\pgfsetfillopacity{0.800000}%
\pgfsetlinewidth{0.501875pt}%
\definecolor{currentstroke}{rgb}{0.000000,0.000000,0.000000}%
\pgfsetstrokecolor{currentstroke}%
\pgfsetstrokeopacity{0.800000}%
\pgfsetdash{}{0pt}%
\pgfpathmoveto{\pgfqpoint{2.148255in}{0.992989in}}%
\pgfpathlineto{\pgfqpoint{2.814155in}{0.992989in}}%
\pgfpathquadraticcurveto{\pgfqpoint{2.836377in}{0.992989in}}{\pgfqpoint{2.836377in}{1.015211in}}%
\pgfpathlineto{\pgfqpoint{2.836377in}{1.756570in}}%
\pgfpathquadraticcurveto{\pgfqpoint{2.836377in}{1.778792in}}{\pgfqpoint{2.814155in}{1.778792in}}%
\pgfpathlineto{\pgfqpoint{2.148255in}{1.778792in}}%
\pgfpathquadraticcurveto{\pgfqpoint{2.126033in}{1.778792in}}{\pgfqpoint{2.126033in}{1.756570in}}%
\pgfpathlineto{\pgfqpoint{2.126033in}{1.015211in}}%
\pgfpathquadraticcurveto{\pgfqpoint{2.126033in}{0.992989in}}{\pgfqpoint{2.148255in}{0.992989in}}%
\pgfpathclose%
\pgfusepath{stroke,fill}%
\end{pgfscope}%
\begin{pgfscope}%
\pgfsetrectcap%
\pgfsetroundjoin%
\pgfsetlinewidth{1.003750pt}%
\definecolor{currentstroke}{rgb}{0.767843,0.220980,0.353137}%
\pgfsetstrokecolor{currentstroke}%
\pgfsetdash{}{0pt}%
\pgfpathmoveto{\pgfqpoint{2.159366in}{1.706570in}}%
\pgfpathlineto{\pgfqpoint{2.381589in}{1.706570in}}%
\pgfusepath{stroke}%
\end{pgfscope}%
\begin{pgfscope}%
\pgftext[x=2.470477in,y=1.667681in,left,base]{\rmfamily\fontsize{8.000000}{9.600000}\selectfont Euclid}%
\end{pgfscope}%
\begin{pgfscope}%
\pgfsetrectcap%
\pgfsetroundjoin%
\pgfsetlinewidth{1.003750pt}%
\definecolor{currentstroke}{rgb}{0.169412,0.150000,0.532353}%
\pgfsetstrokecolor{currentstroke}%
\pgfsetdash{}{0pt}%
\pgfpathmoveto{\pgfqpoint{2.159366in}{1.551631in}}%
\pgfpathlineto{\pgfqpoint{2.381589in}{1.551631in}}%
\pgfusepath{stroke}%
\end{pgfscope}%
\begin{pgfscope}%
\pgftext[x=2.470477in,y=1.512742in,left,base]{\rmfamily\fontsize{8.000000}{9.600000}\selectfont SKA1}%
\end{pgfscope}%
\begin{pgfscope}%
\pgfsetrectcap%
\pgfsetroundjoin%
\pgfsetlinewidth{1.003750pt}%
\definecolor{currentstroke}{rgb}{0.387059,0.164510,0.677451}%
\pgfsetstrokecolor{currentstroke}%
\pgfsetdash{}{0pt}%
\pgfpathmoveto{\pgfqpoint{2.159366in}{1.396693in}}%
\pgfpathlineto{\pgfqpoint{2.381589in}{1.396693in}}%
\pgfusepath{stroke}%
\end{pgfscope}%
\begin{pgfscope}%
\pgftext[x=2.470477in,y=1.357804in,left,base]{\rmfamily\fontsize{8.000000}{9.600000}\selectfont SKA2}%
\end{pgfscope}%
\begin{pgfscope}%
\pgfsetrectcap%
\pgfsetroundjoin%
\pgfsetlinewidth{1.003750pt}%
\definecolor{currentstroke}{rgb}{0.900000,0.677451,0.070980}%
\pgfsetstrokecolor{currentstroke}%
\pgfsetdash{}{0pt}%
\pgfpathmoveto{\pgfqpoint{2.159366in}{1.241755in}}%
\pgfpathlineto{\pgfqpoint{2.381589in}{1.241755in}}%
\pgfusepath{stroke}%
\end{pgfscope}%
\begin{pgfscope}%
\pgftext[x=2.470477in,y=1.202866in,left,base]{\rmfamily\fontsize{8.000000}{9.600000}\selectfont IM B1}%
\end{pgfscope}%
\begin{pgfscope}%
\pgfsetrectcap%
\pgfsetroundjoin%
\pgfsetlinewidth{1.003750pt}%
\definecolor{currentstroke}{rgb}{0.941961,0.395098,0.062941}%
\pgfsetstrokecolor{currentstroke}%
\pgfsetdash{}{0pt}%
\pgfpathmoveto{\pgfqpoint{2.159366in}{1.086816in}}%
\pgfpathlineto{\pgfqpoint{2.381589in}{1.086816in}}%
\pgfusepath{stroke}%
\end{pgfscope}%
\begin{pgfscope}%
\pgftext[x=2.470477in,y=1.047927in,left,base]{\rmfamily\fontsize{8.000000}{9.600000}\selectfont IM B2}%
\end{pgfscope}%
\end{pgfpicture}%
\makeatother%
\endgroup%

%% file: figures/PaperChi2ofMu_opt.pgf
\begingroup%
\makeatletter%
\begin{pgfpicture}%
\pgfpathrectangle{\pgfpointorigin}{\pgfqpoint{3.044140in}{1.881382in}}%
\pgfusepath{use as bounding box, clip}%
\begin{pgfscope}%
\pgfsetbuttcap%
\pgfsetmiterjoin%
\definecolor{currentfill}{rgb}{1.000000,1.000000,1.000000}%
\pgfsetfillcolor{currentfill}%
\pgfsetlinewidth{0.000000pt}%
\definecolor{currentstroke}{rgb}{1.000000,1.000000,1.000000}%
\pgfsetstrokecolor{currentstroke}%
\pgfsetdash{}{0pt}%
\pgfpathmoveto{\pgfqpoint{0.000000in}{0.000000in}}%
\pgfpathlineto{\pgfqpoint{3.044140in}{0.000000in}}%
\pgfpathlineto{\pgfqpoint{3.044140in}{1.881382in}}%
\pgfpathlineto{\pgfqpoint{0.000000in}{1.881382in}}%
\pgfpathclose%
\pgfusepath{fill}%
\end{pgfscope}%
\begin{pgfscope}%
\pgfsetbuttcap%
\pgfsetmiterjoin%
\definecolor{currentfill}{rgb}{1.000000,1.000000,1.000000}%
\pgfsetfillcolor{currentfill}%
\pgfsetlinewidth{0.000000pt}%
\definecolor{currentstroke}{rgb}{0.000000,0.000000,0.000000}%
\pgfsetstrokecolor{currentstroke}%
\pgfsetstrokeopacity{0.000000}%
\pgfsetdash{}{0pt}%
\pgfpathmoveto{\pgfqpoint{0.456621in}{0.329242in}}%
\pgfpathlineto{\pgfqpoint{2.891933in}{0.329242in}}%
\pgfpathlineto{\pgfqpoint{2.891933in}{1.834347in}}%
\pgfpathlineto{\pgfqpoint{0.456621in}{1.834347in}}%
\pgfpathclose%
\pgfusepath{fill}%
\end{pgfscope}%
\begin{pgfscope}%
\pgfpathrectangle{\pgfqpoint{0.456621in}{0.329242in}}{\pgfqpoint{2.435312in}{1.505106in}} %
\pgfusepath{clip}%
\pgfsetrectcap%
\pgfsetroundjoin%
\pgfsetlinewidth{0.803000pt}%
\definecolor{currentstroke}{rgb}{0.690196,0.690196,0.690196}%
\pgfsetstrokecolor{currentstroke}%
\pgfsetdash{}{0pt}%
\pgfpathmoveto{\pgfqpoint{0.456621in}{0.329242in}}%
\pgfpathlineto{\pgfqpoint{0.456621in}{1.834347in}}%
\pgfusepath{stroke}%
\end{pgfscope}%
\begin{pgfscope}%
\pgfsetbuttcap%
\pgfsetroundjoin%
\definecolor{currentfill}{rgb}{0.000000,0.000000,0.000000}%
\pgfsetfillcolor{currentfill}%
\pgfsetlinewidth{0.803000pt}%
\definecolor{currentstroke}{rgb}{0.000000,0.000000,0.000000}%
\pgfsetstrokecolor{currentstroke}%
\pgfsetdash{}{0pt}%
\pgfsys@defobject{currentmarker}{\pgfqpoint{0.000000in}{0.000000in}}{\pgfqpoint{0.000000in}{0.048611in}}{%
\pgfpathmoveto{\pgfqpoint{0.000000in}{0.000000in}}%
\pgfpathlineto{\pgfqpoint{0.000000in}{0.048611in}}%
\pgfusepath{stroke,fill}%
}%
\begin{pgfscope}%
\pgfsys@transformshift{0.456621in}{0.329242in}%
\pgfsys@useobject{currentmarker}{}%
\end{pgfscope}%
\end{pgfscope}%
\begin{pgfscope}%
\pgftext[x=0.456621in,y=0.280631in,,top]{\rmfamily\fontsize{8.000000}{9.600000}\selectfont \(\displaystyle -1.0\)}%
\end{pgfscope}%
\begin{pgfscope}%
\pgfpathrectangle{\pgfqpoint{0.456621in}{0.329242in}}{\pgfqpoint{2.435312in}{1.505106in}} %
\pgfusepath{clip}%
\pgfsetrectcap%
\pgfsetroundjoin%
\pgfsetlinewidth{0.803000pt}%
\definecolor{currentstroke}{rgb}{0.690196,0.690196,0.690196}%
\pgfsetstrokecolor{currentstroke}%
\pgfsetdash{}{0pt}%
\pgfpathmoveto{\pgfqpoint{1.065449in}{0.329242in}}%
\pgfpathlineto{\pgfqpoint{1.065449in}{1.834347in}}%
\pgfusepath{stroke}%
\end{pgfscope}%
\begin{pgfscope}%
\pgfsetbuttcap%
\pgfsetroundjoin%
\definecolor{currentfill}{rgb}{0.000000,0.000000,0.000000}%
\pgfsetfillcolor{currentfill}%
\pgfsetlinewidth{0.803000pt}%
\definecolor{currentstroke}{rgb}{0.000000,0.000000,0.000000}%
\pgfsetstrokecolor{currentstroke}%
\pgfsetdash{}{0pt}%
\pgfsys@defobject{currentmarker}{\pgfqpoint{0.000000in}{0.000000in}}{\pgfqpoint{0.000000in}{0.048611in}}{%
\pgfpathmoveto{\pgfqpoint{0.000000in}{0.000000in}}%
\pgfpathlineto{\pgfqpoint{0.000000in}{0.048611in}}%
\pgfusepath{stroke,fill}%
}%
\begin{pgfscope}%
\pgfsys@transformshift{1.065449in}{0.329242in}%
\pgfsys@useobject{currentmarker}{}%
\end{pgfscope}%
\end{pgfscope}%
\begin{pgfscope}%
\pgftext[x=1.065449in,y=0.280631in,,top]{\rmfamily\fontsize{8.000000}{9.600000}\selectfont \(\displaystyle -0.5\)}%
\end{pgfscope}%
\begin{pgfscope}%
\pgfpathrectangle{\pgfqpoint{0.456621in}{0.329242in}}{\pgfqpoint{2.435312in}{1.505106in}} %
\pgfusepath{clip}%
\pgfsetrectcap%
\pgfsetroundjoin%
\pgfsetlinewidth{0.803000pt}%
\definecolor{currentstroke}{rgb}{0.690196,0.690196,0.690196}%
\pgfsetstrokecolor{currentstroke}%
\pgfsetdash{}{0pt}%
\pgfpathmoveto{\pgfqpoint{1.674277in}{0.329242in}}%
\pgfpathlineto{\pgfqpoint{1.674277in}{1.834347in}}%
\pgfusepath{stroke}%
\end{pgfscope}%
\begin{pgfscope}%
\pgfsetbuttcap%
\pgfsetroundjoin%
\definecolor{currentfill}{rgb}{0.000000,0.000000,0.000000}%
\pgfsetfillcolor{currentfill}%
\pgfsetlinewidth{0.803000pt}%
\definecolor{currentstroke}{rgb}{0.000000,0.000000,0.000000}%
\pgfsetstrokecolor{currentstroke}%
\pgfsetdash{}{0pt}%
\pgfsys@defobject{currentmarker}{\pgfqpoint{0.000000in}{0.000000in}}{\pgfqpoint{0.000000in}{0.048611in}}{%
\pgfpathmoveto{\pgfqpoint{0.000000in}{0.000000in}}%
\pgfpathlineto{\pgfqpoint{0.000000in}{0.048611in}}%
\pgfusepath{stroke,fill}%
}%
\begin{pgfscope}%
\pgfsys@transformshift{1.674277in}{0.329242in}%
\pgfsys@useobject{currentmarker}{}%
\end{pgfscope}%
\end{pgfscope}%
\begin{pgfscope}%
\pgftext[x=1.674277in,y=0.280631in,,top]{\rmfamily\fontsize{8.000000}{9.600000}\selectfont \(\displaystyle 0.0\)}%
\end{pgfscope}%
\begin{pgfscope}%
\pgfpathrectangle{\pgfqpoint{0.456621in}{0.329242in}}{\pgfqpoint{2.435312in}{1.505106in}} %
\pgfusepath{clip}%
\pgfsetrectcap%
\pgfsetroundjoin%
\pgfsetlinewidth{0.803000pt}%
\definecolor{currentstroke}{rgb}{0.690196,0.690196,0.690196}%
\pgfsetstrokecolor{currentstroke}%
\pgfsetdash{}{0pt}%
\pgfpathmoveto{\pgfqpoint{2.283105in}{0.329242in}}%
\pgfpathlineto{\pgfqpoint{2.283105in}{1.834347in}}%
\pgfusepath{stroke}%
\end{pgfscope}%
\begin{pgfscope}%
\pgfsetbuttcap%
\pgfsetroundjoin%
\definecolor{currentfill}{rgb}{0.000000,0.000000,0.000000}%
\pgfsetfillcolor{currentfill}%
\pgfsetlinewidth{0.803000pt}%
\definecolor{currentstroke}{rgb}{0.000000,0.000000,0.000000}%
\pgfsetstrokecolor{currentstroke}%
\pgfsetdash{}{0pt}%
\pgfsys@defobject{currentmarker}{\pgfqpoint{0.000000in}{0.000000in}}{\pgfqpoint{0.000000in}{0.048611in}}{%
\pgfpathmoveto{\pgfqpoint{0.000000in}{0.000000in}}%
\pgfpathlineto{\pgfqpoint{0.000000in}{0.048611in}}%
\pgfusepath{stroke,fill}%
}%
\begin{pgfscope}%
\pgfsys@transformshift{2.283105in}{0.329242in}%
\pgfsys@useobject{currentmarker}{}%
\end{pgfscope}%
\end{pgfscope}%
\begin{pgfscope}%
\pgftext[x=2.283105in,y=0.280631in,,top]{\rmfamily\fontsize{8.000000}{9.600000}\selectfont \(\displaystyle 0.5\)}%
\end{pgfscope}%
\begin{pgfscope}%
\pgfpathrectangle{\pgfqpoint{0.456621in}{0.329242in}}{\pgfqpoint{2.435312in}{1.505106in}} %
\pgfusepath{clip}%
\pgfsetrectcap%
\pgfsetroundjoin%
\pgfsetlinewidth{0.803000pt}%
\definecolor{currentstroke}{rgb}{0.690196,0.690196,0.690196}%
\pgfsetstrokecolor{currentstroke}%
\pgfsetdash{}{0pt}%
\pgfpathmoveto{\pgfqpoint{2.891933in}{0.329242in}}%
\pgfpathlineto{\pgfqpoint{2.891933in}{1.834347in}}%
\pgfusepath{stroke}%
\end{pgfscope}%
\begin{pgfscope}%
\pgfsetbuttcap%
\pgfsetroundjoin%
\definecolor{currentfill}{rgb}{0.000000,0.000000,0.000000}%
\pgfsetfillcolor{currentfill}%
\pgfsetlinewidth{0.803000pt}%
\definecolor{currentstroke}{rgb}{0.000000,0.000000,0.000000}%
\pgfsetstrokecolor{currentstroke}%
\pgfsetdash{}{0pt}%
\pgfsys@defobject{currentmarker}{\pgfqpoint{0.000000in}{0.000000in}}{\pgfqpoint{0.000000in}{0.048611in}}{%
\pgfpathmoveto{\pgfqpoint{0.000000in}{0.000000in}}%
\pgfpathlineto{\pgfqpoint{0.000000in}{0.048611in}}%
\pgfusepath{stroke,fill}%
}%
\begin{pgfscope}%
\pgfsys@transformshift{2.891933in}{0.329242in}%
\pgfsys@useobject{currentmarker}{}%
\end{pgfscope}%
\end{pgfscope}%
\begin{pgfscope}%
\pgftext[x=2.891933in,y=0.280631in,,top]{\rmfamily\fontsize{8.000000}{9.600000}\selectfont \(\displaystyle 1.0\)}%
\end{pgfscope}%
\begin{pgfscope}%
\pgftext[x=1.674277in,y=0.126310in,,top]{\rmfamily\fontsize{8.000000}{9.600000}\selectfont \(\displaystyle \mu\)}%
\end{pgfscope}%
\begin{pgfscope}%
\pgfpathrectangle{\pgfqpoint{0.456621in}{0.329242in}}{\pgfqpoint{2.435312in}{1.505106in}} %
\pgfusepath{clip}%
\pgfsetrectcap%
\pgfsetroundjoin%
\pgfsetlinewidth{0.803000pt}%
\definecolor{currentstroke}{rgb}{0.690196,0.690196,0.690196}%
\pgfsetstrokecolor{currentstroke}%
\pgfsetdash{}{0pt}%
\pgfpathmoveto{\pgfqpoint{0.456621in}{0.329242in}}%
\pgfpathlineto{\pgfqpoint{2.891933in}{0.329242in}}%
\pgfusepath{stroke}%
\end{pgfscope}%
\begin{pgfscope}%
\pgfsetbuttcap%
\pgfsetroundjoin%
\definecolor{currentfill}{rgb}{0.000000,0.000000,0.000000}%
\pgfsetfillcolor{currentfill}%
\pgfsetlinewidth{0.803000pt}%
\definecolor{currentstroke}{rgb}{0.000000,0.000000,0.000000}%
\pgfsetstrokecolor{currentstroke}%
\pgfsetdash{}{0pt}%
\pgfsys@defobject{currentmarker}{\pgfqpoint{0.000000in}{0.000000in}}{\pgfqpoint{0.048611in}{0.000000in}}{%
\pgfpathmoveto{\pgfqpoint{0.000000in}{0.000000in}}%
\pgfpathlineto{\pgfqpoint{0.048611in}{0.000000in}}%
\pgfusepath{stroke,fill}%
}%
\begin{pgfscope}%
\pgfsys@transformshift{0.456621in}{0.329242in}%
\pgfsys@useobject{currentmarker}{}%
\end{pgfscope}%
\end{pgfscope}%
\begin{pgfscope}%
\pgftext[x=0.348981in,y=0.290662in,left,base]{\rmfamily\fontsize{8.000000}{9.600000}\selectfont \(\displaystyle 0\)}%
\end{pgfscope}%
\begin{pgfscope}%
\pgfpathrectangle{\pgfqpoint{0.456621in}{0.329242in}}{\pgfqpoint{2.435312in}{1.505106in}} %
\pgfusepath{clip}%
\pgfsetrectcap%
\pgfsetroundjoin%
\pgfsetlinewidth{0.803000pt}%
\definecolor{currentstroke}{rgb}{0.690196,0.690196,0.690196}%
\pgfsetstrokecolor{currentstroke}%
\pgfsetdash{}{0pt}%
\pgfpathmoveto{\pgfqpoint{0.456621in}{0.671311in}}%
\pgfpathlineto{\pgfqpoint{2.891933in}{0.671311in}}%
\pgfusepath{stroke}%
\end{pgfscope}%
\begin{pgfscope}%
\pgfsetbuttcap%
\pgfsetroundjoin%
\definecolor{currentfill}{rgb}{0.000000,0.000000,0.000000}%
\pgfsetfillcolor{currentfill}%
\pgfsetlinewidth{0.803000pt}%
\definecolor{currentstroke}{rgb}{0.000000,0.000000,0.000000}%
\pgfsetstrokecolor{currentstroke}%
\pgfsetdash{}{0pt}%
\pgfsys@defobject{currentmarker}{\pgfqpoint{0.000000in}{0.000000in}}{\pgfqpoint{0.048611in}{0.000000in}}{%
\pgfpathmoveto{\pgfqpoint{0.000000in}{0.000000in}}%
\pgfpathlineto{\pgfqpoint{0.048611in}{0.000000in}}%
\pgfusepath{stroke,fill}%
}%
\begin{pgfscope}%
\pgfsys@transformshift{0.456621in}{0.671311in}%
\pgfsys@useobject{currentmarker}{}%
\end{pgfscope}%
\end{pgfscope}%
\begin{pgfscope}%
\pgftext[x=0.348981in,y=0.632731in,left,base]{\rmfamily\fontsize{8.000000}{9.600000}\selectfont \(\displaystyle 5\)}%
\end{pgfscope}%
\begin{pgfscope}%
\pgfpathrectangle{\pgfqpoint{0.456621in}{0.329242in}}{\pgfqpoint{2.435312in}{1.505106in}} %
\pgfusepath{clip}%
\pgfsetrectcap%
\pgfsetroundjoin%
\pgfsetlinewidth{0.803000pt}%
\definecolor{currentstroke}{rgb}{0.690196,0.690196,0.690196}%
\pgfsetstrokecolor{currentstroke}%
\pgfsetdash{}{0pt}%
\pgfpathmoveto{\pgfqpoint{0.456621in}{1.013381in}}%
\pgfpathlineto{\pgfqpoint{2.891933in}{1.013381in}}%
\pgfusepath{stroke}%
\end{pgfscope}%
\begin{pgfscope}%
\pgfsetbuttcap%
\pgfsetroundjoin%
\definecolor{currentfill}{rgb}{0.000000,0.000000,0.000000}%
\pgfsetfillcolor{currentfill}%
\pgfsetlinewidth{0.803000pt}%
\definecolor{currentstroke}{rgb}{0.000000,0.000000,0.000000}%
\pgfsetstrokecolor{currentstroke}%
\pgfsetdash{}{0pt}%
\pgfsys@defobject{currentmarker}{\pgfqpoint{0.000000in}{0.000000in}}{\pgfqpoint{0.048611in}{0.000000in}}{%
\pgfpathmoveto{\pgfqpoint{0.000000in}{0.000000in}}%
\pgfpathlineto{\pgfqpoint{0.048611in}{0.000000in}}%
\pgfusepath{stroke,fill}%
}%
\begin{pgfscope}%
\pgfsys@transformshift{0.456621in}{1.013381in}%
\pgfsys@useobject{currentmarker}{}%
\end{pgfscope}%
\end{pgfscope}%
\begin{pgfscope}%
\pgftext[x=0.289953in,y=0.974800in,left,base]{\rmfamily\fontsize{8.000000}{9.600000}\selectfont \(\displaystyle 10\)}%
\end{pgfscope}%
\begin{pgfscope}%
\pgfpathrectangle{\pgfqpoint{0.456621in}{0.329242in}}{\pgfqpoint{2.435312in}{1.505106in}} %
\pgfusepath{clip}%
\pgfsetrectcap%
\pgfsetroundjoin%
\pgfsetlinewidth{0.803000pt}%
\definecolor{currentstroke}{rgb}{0.690196,0.690196,0.690196}%
\pgfsetstrokecolor{currentstroke}%
\pgfsetdash{}{0pt}%
\pgfpathmoveto{\pgfqpoint{0.456621in}{1.355450in}}%
\pgfpathlineto{\pgfqpoint{2.891933in}{1.355450in}}%
\pgfusepath{stroke}%
\end{pgfscope}%
\begin{pgfscope}%
\pgfsetbuttcap%
\pgfsetroundjoin%
\definecolor{currentfill}{rgb}{0.000000,0.000000,0.000000}%
\pgfsetfillcolor{currentfill}%
\pgfsetlinewidth{0.803000pt}%
\definecolor{currentstroke}{rgb}{0.000000,0.000000,0.000000}%
\pgfsetstrokecolor{currentstroke}%
\pgfsetdash{}{0pt}%
\pgfsys@defobject{currentmarker}{\pgfqpoint{0.000000in}{0.000000in}}{\pgfqpoint{0.048611in}{0.000000in}}{%
\pgfpathmoveto{\pgfqpoint{0.000000in}{0.000000in}}%
\pgfpathlineto{\pgfqpoint{0.048611in}{0.000000in}}%
\pgfusepath{stroke,fill}%
}%
\begin{pgfscope}%
\pgfsys@transformshift{0.456621in}{1.355450in}%
\pgfsys@useobject{currentmarker}{}%
\end{pgfscope}%
\end{pgfscope}%
\begin{pgfscope}%
\pgftext[x=0.289953in,y=1.316870in,left,base]{\rmfamily\fontsize{8.000000}{9.600000}\selectfont \(\displaystyle 15\)}%
\end{pgfscope}%
\begin{pgfscope}%
\pgfpathrectangle{\pgfqpoint{0.456621in}{0.329242in}}{\pgfqpoint{2.435312in}{1.505106in}} %
\pgfusepath{clip}%
\pgfsetrectcap%
\pgfsetroundjoin%
\pgfsetlinewidth{0.803000pt}%
\definecolor{currentstroke}{rgb}{0.690196,0.690196,0.690196}%
\pgfsetstrokecolor{currentstroke}%
\pgfsetdash{}{0pt}%
\pgfpathmoveto{\pgfqpoint{0.456621in}{1.697520in}}%
\pgfpathlineto{\pgfqpoint{2.891933in}{1.697520in}}%
\pgfusepath{stroke}%
\end{pgfscope}%
\begin{pgfscope}%
\pgfsetbuttcap%
\pgfsetroundjoin%
\definecolor{currentfill}{rgb}{0.000000,0.000000,0.000000}%
\pgfsetfillcolor{currentfill}%
\pgfsetlinewidth{0.803000pt}%
\definecolor{currentstroke}{rgb}{0.000000,0.000000,0.000000}%
\pgfsetstrokecolor{currentstroke}%
\pgfsetdash{}{0pt}%
\pgfsys@defobject{currentmarker}{\pgfqpoint{0.000000in}{0.000000in}}{\pgfqpoint{0.048611in}{0.000000in}}{%
\pgfpathmoveto{\pgfqpoint{0.000000in}{0.000000in}}%
\pgfpathlineto{\pgfqpoint{0.048611in}{0.000000in}}%
\pgfusepath{stroke,fill}%
}%
\begin{pgfscope}%
\pgfsys@transformshift{0.456621in}{1.697520in}%
\pgfsys@useobject{currentmarker}{}%
\end{pgfscope}%
\end{pgfscope}%
\begin{pgfscope}%
\pgftext[x=0.289953in,y=1.658939in,left,base]{\rmfamily\fontsize{8.000000}{9.600000}\selectfont \(\displaystyle 20\)}%
\end{pgfscope}%
\begin{pgfscope}%
\pgftext[x=0.234397in,y=1.081795in,,bottom,rotate=90.000000]{\rmfamily\fontsize{8.000000}{9.600000}\selectfont \(\displaystyle \Delta\chi^2\)}%
\end{pgfscope}%
\begin{pgfscope}%
\pgfpathrectangle{\pgfqpoint{0.456621in}{0.329242in}}{\pgfqpoint{2.435312in}{1.505106in}} %
\pgfusepath{clip}%
\pgfsetrectcap%
\pgfsetroundjoin%
\pgfsetlinewidth{1.003750pt}%
\definecolor{currentstroke}{rgb}{0.767843,0.220980,0.353137}%
\pgfsetstrokecolor{currentstroke}%
\pgfsetdash{}{0pt}%
\pgfpathmoveto{\pgfqpoint{0.456621in}{0.377382in}}%
\pgfpathlineto{\pgfqpoint{0.761035in}{0.388338in}}%
\pgfpathlineto{\pgfqpoint{1.065449in}{0.410777in}}%
\pgfpathlineto{\pgfqpoint{1.369863in}{0.474337in}}%
\pgfpathlineto{\pgfqpoint{1.674277in}{0.801291in}}%
\pgfpathlineto{\pgfqpoint{1.978691in}{0.474337in}}%
\pgfpathlineto{\pgfqpoint{2.283105in}{0.410777in}}%
\pgfpathlineto{\pgfqpoint{2.587519in}{0.388338in}}%
\pgfpathlineto{\pgfqpoint{2.891933in}{0.377382in}}%
\pgfusepath{stroke}%
\end{pgfscope}%
\begin{pgfscope}%
\pgfpathrectangle{\pgfqpoint{0.456621in}{0.329242in}}{\pgfqpoint{2.435312in}{1.505106in}} %
\pgfusepath{clip}%
\pgfsetrectcap%
\pgfsetroundjoin%
\pgfsetlinewidth{1.003750pt}%
\definecolor{currentstroke}{rgb}{0.169412,0.150000,0.532353}%
\pgfsetstrokecolor{currentstroke}%
\pgfsetdash{}{0pt}%
\pgfpathmoveto{\pgfqpoint{0.456621in}{0.332303in}}%
\pgfpathlineto{\pgfqpoint{0.761035in}{0.333514in}}%
\pgfpathlineto{\pgfqpoint{1.065449in}{0.336640in}}%
\pgfpathlineto{\pgfqpoint{1.369863in}{0.349682in}}%
\pgfpathlineto{\pgfqpoint{1.674277in}{0.425074in}}%
\pgfpathlineto{\pgfqpoint{1.978691in}{0.349682in}}%
\pgfpathlineto{\pgfqpoint{2.283105in}{0.336640in}}%
\pgfpathlineto{\pgfqpoint{2.587519in}{0.333514in}}%
\pgfpathlineto{\pgfqpoint{2.891933in}{0.332303in}}%
\pgfusepath{stroke}%
\end{pgfscope}%
\begin{pgfscope}%
\pgfpathrectangle{\pgfqpoint{0.456621in}{0.329242in}}{\pgfqpoint{2.435312in}{1.505106in}} %
\pgfusepath{clip}%
\pgfsetrectcap%
\pgfsetroundjoin%
\pgfsetlinewidth{1.003750pt}%
\definecolor{currentstroke}{rgb}{0.387059,0.164510,0.677451}%
\pgfsetstrokecolor{currentstroke}%
\pgfsetdash{}{0pt}%
\pgfpathmoveto{\pgfqpoint{0.456621in}{0.576923in}}%
\pgfpathlineto{\pgfqpoint{0.761035in}{0.638912in}}%
\pgfpathlineto{\pgfqpoint{1.065449in}{0.741489in}}%
\pgfpathlineto{\pgfqpoint{1.369863in}{0.938737in}}%
\pgfpathlineto{\pgfqpoint{1.674277in}{1.417675in}}%
\pgfpathlineto{\pgfqpoint{1.978691in}{0.938737in}}%
\pgfpathlineto{\pgfqpoint{2.283105in}{0.741489in}}%
\pgfpathlineto{\pgfqpoint{2.587519in}{0.638912in}}%
\pgfpathlineto{\pgfqpoint{2.891933in}{0.576923in}}%
\pgfusepath{stroke}%
\end{pgfscope}%
\begin{pgfscope}%
\pgfpathrectangle{\pgfqpoint{0.456621in}{0.329242in}}{\pgfqpoint{2.435312in}{1.505106in}} %
\pgfusepath{clip}%
\pgfsetrectcap%
\pgfsetroundjoin%
\pgfsetlinewidth{1.003750pt}%
\definecolor{currentstroke}{rgb}{0.900000,0.677451,0.070980}%
\pgfsetstrokecolor{currentstroke}%
\pgfsetdash{}{0pt}%
\pgfpathmoveto{\pgfqpoint{0.456621in}{0.847824in}}%
\pgfpathlineto{\pgfqpoint{0.578387in}{0.402667in}}%
\pgfpathlineto{\pgfqpoint{0.700152in}{0.369103in}}%
\pgfpathlineto{\pgfqpoint{0.821918in}{0.355266in}}%
\pgfpathlineto{\pgfqpoint{0.943683in}{0.347680in}}%
\pgfpathlineto{\pgfqpoint{1.065449in}{0.342905in}}%
\pgfpathlineto{\pgfqpoint{1.187215in}{0.339652in}}%
\pgfpathlineto{\pgfqpoint{1.308980in}{0.337368in}}%
\pgfpathlineto{\pgfqpoint{1.430746in}{0.335809in}}%
\pgfpathlineto{\pgfqpoint{1.552511in}{0.334886in}}%
\pgfpathlineto{\pgfqpoint{1.674277in}{0.334578in}}%
\pgfpathlineto{\pgfqpoint{1.796043in}{0.334886in}}%
\pgfpathlineto{\pgfqpoint{1.917808in}{0.335809in}}%
\pgfpathlineto{\pgfqpoint{2.039574in}{0.337368in}}%
\pgfpathlineto{\pgfqpoint{2.161339in}{0.339652in}}%
\pgfpathlineto{\pgfqpoint{2.283105in}{0.342905in}}%
\pgfpathlineto{\pgfqpoint{2.404871in}{0.347680in}}%
\pgfpathlineto{\pgfqpoint{2.526636in}{0.355266in}}%
\pgfpathlineto{\pgfqpoint{2.648402in}{0.369103in}}%
\pgfpathlineto{\pgfqpoint{2.770167in}{0.402667in}}%
\pgfpathlineto{\pgfqpoint{2.891933in}{0.847824in}}%
\pgfusepath{stroke}%
\end{pgfscope}%
\begin{pgfscope}%
\pgfpathrectangle{\pgfqpoint{0.456621in}{0.329242in}}{\pgfqpoint{2.435312in}{1.505106in}} %
\pgfusepath{clip}%
\pgfsetrectcap%
\pgfsetroundjoin%
\pgfsetlinewidth{1.003750pt}%
\definecolor{currentstroke}{rgb}{0.941961,0.395098,0.062941}%
\pgfsetstrokecolor{currentstroke}%
\pgfsetdash{}{0pt}%
\pgfpathmoveto{\pgfqpoint{0.456621in}{0.388910in}}%
\pgfpathlineto{\pgfqpoint{0.700152in}{0.377752in}}%
\pgfpathlineto{\pgfqpoint{0.943683in}{0.371857in}}%
\pgfpathlineto{\pgfqpoint{1.187215in}{0.367521in}}%
\pgfpathlineto{\pgfqpoint{1.430746in}{0.363402in}}%
\pgfpathlineto{\pgfqpoint{1.674277in}{0.361088in}}%
\pgfpathlineto{\pgfqpoint{1.917808in}{0.363402in}}%
\pgfpathlineto{\pgfqpoint{2.161339in}{0.367521in}}%
\pgfpathlineto{\pgfqpoint{2.404871in}{0.371857in}}%
\pgfpathlineto{\pgfqpoint{2.648402in}{0.377752in}}%
\pgfpathlineto{\pgfqpoint{2.891933in}{0.388910in}}%
\pgfusepath{stroke}%
\end{pgfscope}%
\begin{pgfscope}%
\pgfsetrectcap%
\pgfsetmiterjoin%
\pgfsetlinewidth{0.501875pt}%
\definecolor{currentstroke}{rgb}{0.000000,0.000000,0.000000}%
\pgfsetstrokecolor{currentstroke}%
\pgfsetdash{}{0pt}%
\pgfpathmoveto{\pgfqpoint{0.456621in}{0.329242in}}%
\pgfpathlineto{\pgfqpoint{0.456621in}{1.834347in}}%
\pgfusepath{stroke}%
\end{pgfscope}%
\begin{pgfscope}%
\pgfsetrectcap%
\pgfsetmiterjoin%
\pgfsetlinewidth{0.501875pt}%
\definecolor{currentstroke}{rgb}{0.000000,0.000000,0.000000}%
\pgfsetstrokecolor{currentstroke}%
\pgfsetdash{}{0pt}%
\pgfpathmoveto{\pgfqpoint{2.891933in}{0.329242in}}%
\pgfpathlineto{\pgfqpoint{2.891933in}{1.834347in}}%
\pgfusepath{stroke}%
\end{pgfscope}%
\begin{pgfscope}%
\pgfsetrectcap%
\pgfsetmiterjoin%
\pgfsetlinewidth{0.501875pt}%
\definecolor{currentstroke}{rgb}{0.000000,0.000000,0.000000}%
\pgfsetstrokecolor{currentstroke}%
\pgfsetdash{}{0pt}%
\pgfpathmoveto{\pgfqpoint{0.456621in}{0.329242in}}%
\pgfpathlineto{\pgfqpoint{2.891933in}{0.329242in}}%
\pgfusepath{stroke}%
\end{pgfscope}%
\begin{pgfscope}%
\pgfsetrectcap%
\pgfsetmiterjoin%
\pgfsetlinewidth{0.501875pt}%
\definecolor{currentstroke}{rgb}{0.000000,0.000000,0.000000}%
\pgfsetstrokecolor{currentstroke}%
\pgfsetdash{}{0pt}%
\pgfpathmoveto{\pgfqpoint{0.456621in}{1.834347in}}%
\pgfpathlineto{\pgfqpoint{2.891933in}{1.834347in}}%
\pgfusepath{stroke}%
\end{pgfscope}%
\begin{pgfscope}%
\pgfsetbuttcap%
\pgfsetmiterjoin%
\definecolor{currentfill}{rgb}{1.000000,1.000000,1.000000}%
\pgfsetfillcolor{currentfill}%
\pgfsetfillopacity{0.800000}%
\pgfsetlinewidth{0.501875pt}%
\definecolor{currentstroke}{rgb}{0.000000,0.000000,0.000000}%
\pgfsetstrokecolor{currentstroke}%
\pgfsetstrokeopacity{0.800000}%
\pgfsetdash{}{0pt}%
\pgfpathmoveto{\pgfqpoint{2.148255in}{0.992989in}}%
\pgfpathlineto{\pgfqpoint{2.814155in}{0.992989in}}%
\pgfpathquadraticcurveto{\pgfqpoint{2.836377in}{0.992989in}}{\pgfqpoint{2.836377in}{1.015211in}}%
\pgfpathlineto{\pgfqpoint{2.836377in}{1.756570in}}%
\pgfpathquadraticcurveto{\pgfqpoint{2.836377in}{1.778792in}}{\pgfqpoint{2.814155in}{1.778792in}}%
\pgfpathlineto{\pgfqpoint{2.148255in}{1.778792in}}%
\pgfpathquadraticcurveto{\pgfqpoint{2.126033in}{1.778792in}}{\pgfqpoint{2.126033in}{1.756570in}}%
\pgfpathlineto{\pgfqpoint{2.126033in}{1.015211in}}%
\pgfpathquadraticcurveto{\pgfqpoint{2.126033in}{0.992989in}}{\pgfqpoint{2.148255in}{0.992989in}}%
\pgfpathclose%
\pgfusepath{stroke,fill}%
\end{pgfscope}%
\begin{pgfscope}%
\pgfsetrectcap%
\pgfsetroundjoin%
\pgfsetlinewidth{1.003750pt}%
\definecolor{currentstroke}{rgb}{0.767843,0.220980,0.353137}%
\pgfsetstrokecolor{currentstroke}%
\pgfsetdash{}{0pt}%
\pgfpathmoveto{\pgfqpoint{2.159366in}{1.706570in}}%
\pgfpathlineto{\pgfqpoint{2.381589in}{1.706570in}}%
\pgfusepath{stroke}%
\end{pgfscope}%
\begin{pgfscope}%
\pgftext[x=2.470477in,y=1.667681in,left,base]{\rmfamily\fontsize{8.000000}{9.600000}\selectfont Euclid}%
\end{pgfscope}%
\begin{pgfscope}%
\pgfsetrectcap%
\pgfsetroundjoin%
\pgfsetlinewidth{1.003750pt}%
\definecolor{currentstroke}{rgb}{0.169412,0.150000,0.532353}%
\pgfsetstrokecolor{currentstroke}%
\pgfsetdash{}{0pt}%
\pgfpathmoveto{\pgfqpoint{2.159366in}{1.551631in}}%
\pgfpathlineto{\pgfqpoint{2.381589in}{1.551631in}}%
\pgfusepath{stroke}%
\end{pgfscope}%
\begin{pgfscope}%
\pgftext[x=2.470477in,y=1.512742in,left,base]{\rmfamily\fontsize{8.000000}{9.600000}\selectfont SKA1}%
\end{pgfscope}%
\begin{pgfscope}%
\pgfsetrectcap%
\pgfsetroundjoin%
\pgfsetlinewidth{1.003750pt}%
\definecolor{currentstroke}{rgb}{0.387059,0.164510,0.677451}%
\pgfsetstrokecolor{currentstroke}%
\pgfsetdash{}{0pt}%
\pgfpathmoveto{\pgfqpoint{2.159366in}{1.396693in}}%
\pgfpathlineto{\pgfqpoint{2.381589in}{1.396693in}}%
\pgfusepath{stroke}%
\end{pgfscope}%
\begin{pgfscope}%
\pgftext[x=2.470477in,y=1.357804in,left,base]{\rmfamily\fontsize{8.000000}{9.600000}\selectfont SKA2}%
\end{pgfscope}%
\begin{pgfscope}%
\pgfsetrectcap%
\pgfsetroundjoin%
\pgfsetlinewidth{1.003750pt}%
\definecolor{currentstroke}{rgb}{0.900000,0.677451,0.070980}%
\pgfsetstrokecolor{currentstroke}%
\pgfsetdash{}{0pt}%
\pgfpathmoveto{\pgfqpoint{2.159366in}{1.241755in}}%
\pgfpathlineto{\pgfqpoint{2.381589in}{1.241755in}}%
\pgfusepath{stroke}%
\end{pgfscope}%
\begin{pgfscope}%
\pgftext[x=2.470477in,y=1.202866in,left,base]{\rmfamily\fontsize{8.000000}{9.600000}\selectfont IM B1}%
\end{pgfscope}%
\begin{pgfscope}%
\pgfsetrectcap%
\pgfsetroundjoin%
\pgfsetlinewidth{1.003750pt}%
\definecolor{currentstroke}{rgb}{0.941961,0.395098,0.062941}%
\pgfsetstrokecolor{currentstroke}%
\pgfsetdash{}{0pt}%
\pgfpathmoveto{\pgfqpoint{2.159366in}{1.086816in}}%
\pgfpathlineto{\pgfqpoint{2.381589in}{1.086816in}}%
\pgfusepath{stroke}%
\end{pgfscope}%
\begin{pgfscope}%
\pgftext[x=2.470477in,y=1.047927in,left,base]{\rmfamily\fontsize{8.000000}{9.600000}\selectfont IM B2}%
\end{pgfscope}%
\end{pgfpicture}%
\makeatother%
\endgroup%

%% file: figures/PaperChi2ofMu_con.pgf
\begingroup%
\makeatletter%
\begin{pgfpicture}%
\pgfpathrectangle{\pgfpointorigin}{\pgfqpoint{3.044140in}{1.881382in}}%
\pgfusepath{use as bounding box, clip}%
\begin{pgfscope}%
\pgfsetbuttcap%
\pgfsetmiterjoin%
\definecolor{currentfill}{rgb}{1.000000,1.000000,1.000000}%
\pgfsetfillcolor{currentfill}%
\pgfsetlinewidth{0.000000pt}%
\definecolor{currentstroke}{rgb}{1.000000,1.000000,1.000000}%
\pgfsetstrokecolor{currentstroke}%
\pgfsetdash{}{0pt}%
\pgfpathmoveto{\pgfqpoint{0.000000in}{0.000000in}}%
\pgfpathlineto{\pgfqpoint{3.044140in}{0.000000in}}%
\pgfpathlineto{\pgfqpoint{3.044140in}{1.881382in}}%
\pgfpathlineto{\pgfqpoint{0.000000in}{1.881382in}}%
\pgfpathclose%
\pgfusepath{fill}%
\end{pgfscope}%
\begin{pgfscope}%
\pgfsetbuttcap%
\pgfsetmiterjoin%
\definecolor{currentfill}{rgb}{1.000000,1.000000,1.000000}%
\pgfsetfillcolor{currentfill}%
\pgfsetlinewidth{0.000000pt}%
\definecolor{currentstroke}{rgb}{0.000000,0.000000,0.000000}%
\pgfsetstrokecolor{currentstroke}%
\pgfsetstrokeopacity{0.000000}%
\pgfsetdash{}{0pt}%
\pgfpathmoveto{\pgfqpoint{0.456621in}{0.329242in}}%
\pgfpathlineto{\pgfqpoint{2.891933in}{0.329242in}}%
\pgfpathlineto{\pgfqpoint{2.891933in}{1.834347in}}%
\pgfpathlineto{\pgfqpoint{0.456621in}{1.834347in}}%
\pgfpathclose%
\pgfusepath{fill}%
\end{pgfscope}%
\begin{pgfscope}%
\pgfpathrectangle{\pgfqpoint{0.456621in}{0.329242in}}{\pgfqpoint{2.435312in}{1.505106in}} %
\pgfusepath{clip}%
\pgfsetrectcap%
\pgfsetroundjoin%
\pgfsetlinewidth{0.803000pt}%
\definecolor{currentstroke}{rgb}{0.690196,0.690196,0.690196}%
\pgfsetstrokecolor{currentstroke}%
\pgfsetdash{}{0pt}%
\pgfpathmoveto{\pgfqpoint{0.456621in}{0.329242in}}%
\pgfpathlineto{\pgfqpoint{0.456621in}{1.834347in}}%
\pgfusepath{stroke}%
\end{pgfscope}%
\begin{pgfscope}%
\pgfsetbuttcap%
\pgfsetroundjoin%
\definecolor{currentfill}{rgb}{0.000000,0.000000,0.000000}%
\pgfsetfillcolor{currentfill}%
\pgfsetlinewidth{0.803000pt}%
\definecolor{currentstroke}{rgb}{0.000000,0.000000,0.000000}%
\pgfsetstrokecolor{currentstroke}%
\pgfsetdash{}{0pt}%
\pgfsys@defobject{currentmarker}{\pgfqpoint{0.000000in}{0.000000in}}{\pgfqpoint{0.000000in}{0.048611in}}{%
\pgfpathmoveto{\pgfqpoint{0.000000in}{0.000000in}}%
\pgfpathlineto{\pgfqpoint{0.000000in}{0.048611in}}%
\pgfusepath{stroke,fill}%
}%
\begin{pgfscope}%
\pgfsys@transformshift{0.456621in}{0.329242in}%
\pgfsys@useobject{currentmarker}{}%
\end{pgfscope}%
\end{pgfscope}%
\begin{pgfscope}%
\pgftext[x=0.456621in,y=0.280631in,,top]{\rmfamily\fontsize{8.000000}{9.600000}\selectfont \(\displaystyle -1.0\)}%
\end{pgfscope}%
\begin{pgfscope}%
\pgfpathrectangle{\pgfqpoint{0.456621in}{0.329242in}}{\pgfqpoint{2.435312in}{1.505106in}} %
\pgfusepath{clip}%
\pgfsetrectcap%
\pgfsetroundjoin%
\pgfsetlinewidth{0.803000pt}%
\definecolor{currentstroke}{rgb}{0.690196,0.690196,0.690196}%
\pgfsetstrokecolor{currentstroke}%
\pgfsetdash{}{0pt}%
\pgfpathmoveto{\pgfqpoint{1.065449in}{0.329242in}}%
\pgfpathlineto{\pgfqpoint{1.065449in}{1.834347in}}%
\pgfusepath{stroke}%
\end{pgfscope}%
\begin{pgfscope}%
\pgfsetbuttcap%
\pgfsetroundjoin%
\definecolor{currentfill}{rgb}{0.000000,0.000000,0.000000}%
\pgfsetfillcolor{currentfill}%
\pgfsetlinewidth{0.803000pt}%
\definecolor{currentstroke}{rgb}{0.000000,0.000000,0.000000}%
\pgfsetstrokecolor{currentstroke}%
\pgfsetdash{}{0pt}%
\pgfsys@defobject{currentmarker}{\pgfqpoint{0.000000in}{0.000000in}}{\pgfqpoint{0.000000in}{0.048611in}}{%
\pgfpathmoveto{\pgfqpoint{0.000000in}{0.000000in}}%
\pgfpathlineto{\pgfqpoint{0.000000in}{0.048611in}}%
\pgfusepath{stroke,fill}%
}%
\begin{pgfscope}%
\pgfsys@transformshift{1.065449in}{0.329242in}%
\pgfsys@useobject{currentmarker}{}%
\end{pgfscope}%
\end{pgfscope}%
\begin{pgfscope}%
\pgftext[x=1.065449in,y=0.280631in,,top]{\rmfamily\fontsize{8.000000}{9.600000}\selectfont \(\displaystyle -0.5\)}%
\end{pgfscope}%
\begin{pgfscope}%
\pgfpathrectangle{\pgfqpoint{0.456621in}{0.329242in}}{\pgfqpoint{2.435312in}{1.505106in}} %
\pgfusepath{clip}%
\pgfsetrectcap%
\pgfsetroundjoin%
\pgfsetlinewidth{0.803000pt}%
\definecolor{currentstroke}{rgb}{0.690196,0.690196,0.690196}%
\pgfsetstrokecolor{currentstroke}%
\pgfsetdash{}{0pt}%
\pgfpathmoveto{\pgfqpoint{1.674277in}{0.329242in}}%
\pgfpathlineto{\pgfqpoint{1.674277in}{1.834347in}}%
\pgfusepath{stroke}%
\end{pgfscope}%
\begin{pgfscope}%
\pgfsetbuttcap%
\pgfsetroundjoin%
\definecolor{currentfill}{rgb}{0.000000,0.000000,0.000000}%
\pgfsetfillcolor{currentfill}%
\pgfsetlinewidth{0.803000pt}%
\definecolor{currentstroke}{rgb}{0.000000,0.000000,0.000000}%
\pgfsetstrokecolor{currentstroke}%
\pgfsetdash{}{0pt}%
\pgfsys@defobject{currentmarker}{\pgfqpoint{0.000000in}{0.000000in}}{\pgfqpoint{0.000000in}{0.048611in}}{%
\pgfpathmoveto{\pgfqpoint{0.000000in}{0.000000in}}%
\pgfpathlineto{\pgfqpoint{0.000000in}{0.048611in}}%
\pgfusepath{stroke,fill}%
}%
\begin{pgfscope}%
\pgfsys@transformshift{1.674277in}{0.329242in}%
\pgfsys@useobject{currentmarker}{}%
\end{pgfscope}%
\end{pgfscope}%
\begin{pgfscope}%
\pgftext[x=1.674277in,y=0.280631in,,top]{\rmfamily\fontsize{8.000000}{9.600000}\selectfont \(\displaystyle 0.0\)}%
\end{pgfscope}%
\begin{pgfscope}%
\pgfpathrectangle{\pgfqpoint{0.456621in}{0.329242in}}{\pgfqpoint{2.435312in}{1.505106in}} %
\pgfusepath{clip}%
\pgfsetrectcap%
\pgfsetroundjoin%
\pgfsetlinewidth{0.803000pt}%
\definecolor{currentstroke}{rgb}{0.690196,0.690196,0.690196}%
\pgfsetstrokecolor{currentstroke}%
\pgfsetdash{}{0pt}%
\pgfpathmoveto{\pgfqpoint{2.283105in}{0.329242in}}%
\pgfpathlineto{\pgfqpoint{2.283105in}{1.834347in}}%
\pgfusepath{stroke}%
\end{pgfscope}%
\begin{pgfscope}%
\pgfsetbuttcap%
\pgfsetroundjoin%
\definecolor{currentfill}{rgb}{0.000000,0.000000,0.000000}%
\pgfsetfillcolor{currentfill}%
\pgfsetlinewidth{0.803000pt}%
\definecolor{currentstroke}{rgb}{0.000000,0.000000,0.000000}%
\pgfsetstrokecolor{currentstroke}%
\pgfsetdash{}{0pt}%
\pgfsys@defobject{currentmarker}{\pgfqpoint{0.000000in}{0.000000in}}{\pgfqpoint{0.000000in}{0.048611in}}{%
\pgfpathmoveto{\pgfqpoint{0.000000in}{0.000000in}}%
\pgfpathlineto{\pgfqpoint{0.000000in}{0.048611in}}%
\pgfusepath{stroke,fill}%
}%
\begin{pgfscope}%
\pgfsys@transformshift{2.283105in}{0.329242in}%
\pgfsys@useobject{currentmarker}{}%
\end{pgfscope}%
\end{pgfscope}%
\begin{pgfscope}%
\pgftext[x=2.283105in,y=0.280631in,,top]{\rmfamily\fontsize{8.000000}{9.600000}\selectfont \(\displaystyle 0.5\)}%
\end{pgfscope}%
\begin{pgfscope}%
\pgfpathrectangle{\pgfqpoint{0.456621in}{0.329242in}}{\pgfqpoint{2.435312in}{1.505106in}} %
\pgfusepath{clip}%
\pgfsetrectcap%
\pgfsetroundjoin%
\pgfsetlinewidth{0.803000pt}%
\definecolor{currentstroke}{rgb}{0.690196,0.690196,0.690196}%
\pgfsetstrokecolor{currentstroke}%
\pgfsetdash{}{0pt}%
\pgfpathmoveto{\pgfqpoint{2.891933in}{0.329242in}}%
\pgfpathlineto{\pgfqpoint{2.891933in}{1.834347in}}%
\pgfusepath{stroke}%
\end{pgfscope}%
\begin{pgfscope}%
\pgfsetbuttcap%
\pgfsetroundjoin%
\definecolor{currentfill}{rgb}{0.000000,0.000000,0.000000}%
\pgfsetfillcolor{currentfill}%
\pgfsetlinewidth{0.803000pt}%
\definecolor{currentstroke}{rgb}{0.000000,0.000000,0.000000}%
\pgfsetstrokecolor{currentstroke}%
\pgfsetdash{}{0pt}%
\pgfsys@defobject{currentmarker}{\pgfqpoint{0.000000in}{0.000000in}}{\pgfqpoint{0.000000in}{0.048611in}}{%
\pgfpathmoveto{\pgfqpoint{0.000000in}{0.000000in}}%
\pgfpathlineto{\pgfqpoint{0.000000in}{0.048611in}}%
\pgfusepath{stroke,fill}%
}%
\begin{pgfscope}%
\pgfsys@transformshift{2.891933in}{0.329242in}%
\pgfsys@useobject{currentmarker}{}%
\end{pgfscope}%
\end{pgfscope}%
\begin{pgfscope}%
\pgftext[x=2.891933in,y=0.280631in,,top]{\rmfamily\fontsize{8.000000}{9.600000}\selectfont \(\displaystyle 1.0\)}%
\end{pgfscope}%
\begin{pgfscope}%
\pgftext[x=1.674277in,y=0.126310in,,top]{\rmfamily\fontsize{8.000000}{9.600000}\selectfont \(\displaystyle \mu\)}%
\end{pgfscope}%
\begin{pgfscope}%
\pgfpathrectangle{\pgfqpoint{0.456621in}{0.329242in}}{\pgfqpoint{2.435312in}{1.505106in}} %
\pgfusepath{clip}%
\pgfsetrectcap%
\pgfsetroundjoin%
\pgfsetlinewidth{0.803000pt}%
\definecolor{currentstroke}{rgb}{0.690196,0.690196,0.690196}%
\pgfsetstrokecolor{currentstroke}%
\pgfsetdash{}{0pt}%
\pgfpathmoveto{\pgfqpoint{0.456621in}{0.329242in}}%
\pgfpathlineto{\pgfqpoint{2.891933in}{0.329242in}}%
\pgfusepath{stroke}%
\end{pgfscope}%
\begin{pgfscope}%
\pgfsetbuttcap%
\pgfsetroundjoin%
\definecolor{currentfill}{rgb}{0.000000,0.000000,0.000000}%
\pgfsetfillcolor{currentfill}%
\pgfsetlinewidth{0.803000pt}%
\definecolor{currentstroke}{rgb}{0.000000,0.000000,0.000000}%
\pgfsetstrokecolor{currentstroke}%
\pgfsetdash{}{0pt}%
\pgfsys@defobject{currentmarker}{\pgfqpoint{0.000000in}{0.000000in}}{\pgfqpoint{0.048611in}{0.000000in}}{%
\pgfpathmoveto{\pgfqpoint{0.000000in}{0.000000in}}%
\pgfpathlineto{\pgfqpoint{0.048611in}{0.000000in}}%
\pgfusepath{stroke,fill}%
}%
\begin{pgfscope}%
\pgfsys@transformshift{0.456621in}{0.329242in}%
\pgfsys@useobject{currentmarker}{}%
\end{pgfscope}%
\end{pgfscope}%
\begin{pgfscope}%
\pgftext[x=0.348981in,y=0.290662in,left,base]{\rmfamily\fontsize{8.000000}{9.600000}\selectfont \(\displaystyle 0\)}%
\end{pgfscope}%
\begin{pgfscope}%
\pgfpathrectangle{\pgfqpoint{0.456621in}{0.329242in}}{\pgfqpoint{2.435312in}{1.505106in}} %
\pgfusepath{clip}%
\pgfsetrectcap%
\pgfsetroundjoin%
\pgfsetlinewidth{0.803000pt}%
\definecolor{currentstroke}{rgb}{0.690196,0.690196,0.690196}%
\pgfsetstrokecolor{currentstroke}%
\pgfsetdash{}{0pt}%
\pgfpathmoveto{\pgfqpoint{0.456621in}{0.671311in}}%
\pgfpathlineto{\pgfqpoint{2.891933in}{0.671311in}}%
\pgfusepath{stroke}%
\end{pgfscope}%
\begin{pgfscope}%
\pgfsetbuttcap%
\pgfsetroundjoin%
\definecolor{currentfill}{rgb}{0.000000,0.000000,0.000000}%
\pgfsetfillcolor{currentfill}%
\pgfsetlinewidth{0.803000pt}%
\definecolor{currentstroke}{rgb}{0.000000,0.000000,0.000000}%
\pgfsetstrokecolor{currentstroke}%
\pgfsetdash{}{0pt}%
\pgfsys@defobject{currentmarker}{\pgfqpoint{0.000000in}{0.000000in}}{\pgfqpoint{0.048611in}{0.000000in}}{%
\pgfpathmoveto{\pgfqpoint{0.000000in}{0.000000in}}%
\pgfpathlineto{\pgfqpoint{0.048611in}{0.000000in}}%
\pgfusepath{stroke,fill}%
}%
\begin{pgfscope}%
\pgfsys@transformshift{0.456621in}{0.671311in}%
\pgfsys@useobject{currentmarker}{}%
\end{pgfscope}%
\end{pgfscope}%
\begin{pgfscope}%
\pgftext[x=0.348981in,y=0.632731in,left,base]{\rmfamily\fontsize{8.000000}{9.600000}\selectfont \(\displaystyle 5\)}%
\end{pgfscope}%
\begin{pgfscope}%
\pgfpathrectangle{\pgfqpoint{0.456621in}{0.329242in}}{\pgfqpoint{2.435312in}{1.505106in}} %
\pgfusepath{clip}%
\pgfsetrectcap%
\pgfsetroundjoin%
\pgfsetlinewidth{0.803000pt}%
\definecolor{currentstroke}{rgb}{0.690196,0.690196,0.690196}%
\pgfsetstrokecolor{currentstroke}%
\pgfsetdash{}{0pt}%
\pgfpathmoveto{\pgfqpoint{0.456621in}{1.013381in}}%
\pgfpathlineto{\pgfqpoint{2.891933in}{1.013381in}}%
\pgfusepath{stroke}%
\end{pgfscope}%
\begin{pgfscope}%
\pgfsetbuttcap%
\pgfsetroundjoin%
\definecolor{currentfill}{rgb}{0.000000,0.000000,0.000000}%
\pgfsetfillcolor{currentfill}%
\pgfsetlinewidth{0.803000pt}%
\definecolor{currentstroke}{rgb}{0.000000,0.000000,0.000000}%
\pgfsetstrokecolor{currentstroke}%
\pgfsetdash{}{0pt}%
\pgfsys@defobject{currentmarker}{\pgfqpoint{0.000000in}{0.000000in}}{\pgfqpoint{0.048611in}{0.000000in}}{%
\pgfpathmoveto{\pgfqpoint{0.000000in}{0.000000in}}%
\pgfpathlineto{\pgfqpoint{0.048611in}{0.000000in}}%
\pgfusepath{stroke,fill}%
}%
\begin{pgfscope}%
\pgfsys@transformshift{0.456621in}{1.013381in}%
\pgfsys@useobject{currentmarker}{}%
\end{pgfscope}%
\end{pgfscope}%
\begin{pgfscope}%
\pgftext[x=0.289953in,y=0.974800in,left,base]{\rmfamily\fontsize{8.000000}{9.600000}\selectfont \(\displaystyle 10\)}%
\end{pgfscope}%
\begin{pgfscope}%
\pgfpathrectangle{\pgfqpoint{0.456621in}{0.329242in}}{\pgfqpoint{2.435312in}{1.505106in}} %
\pgfusepath{clip}%
\pgfsetrectcap%
\pgfsetroundjoin%
\pgfsetlinewidth{0.803000pt}%
\definecolor{currentstroke}{rgb}{0.690196,0.690196,0.690196}%
\pgfsetstrokecolor{currentstroke}%
\pgfsetdash{}{0pt}%
\pgfpathmoveto{\pgfqpoint{0.456621in}{1.355450in}}%
\pgfpathlineto{\pgfqpoint{2.891933in}{1.355450in}}%
\pgfusepath{stroke}%
\end{pgfscope}%
\begin{pgfscope}%
\pgfsetbuttcap%
\pgfsetroundjoin%
\definecolor{currentfill}{rgb}{0.000000,0.000000,0.000000}%
\pgfsetfillcolor{currentfill}%
\pgfsetlinewidth{0.803000pt}%
\definecolor{currentstroke}{rgb}{0.000000,0.000000,0.000000}%
\pgfsetstrokecolor{currentstroke}%
\pgfsetdash{}{0pt}%
\pgfsys@defobject{currentmarker}{\pgfqpoint{0.000000in}{0.000000in}}{\pgfqpoint{0.048611in}{0.000000in}}{%
\pgfpathmoveto{\pgfqpoint{0.000000in}{0.000000in}}%
\pgfpathlineto{\pgfqpoint{0.048611in}{0.000000in}}%
\pgfusepath{stroke,fill}%
}%
\begin{pgfscope}%
\pgfsys@transformshift{0.456621in}{1.355450in}%
\pgfsys@useobject{currentmarker}{}%
\end{pgfscope}%
\end{pgfscope}%
\begin{pgfscope}%
\pgftext[x=0.289953in,y=1.316870in,left,base]{\rmfamily\fontsize{8.000000}{9.600000}\selectfont \(\displaystyle 15\)}%
\end{pgfscope}%
\begin{pgfscope}%
\pgfpathrectangle{\pgfqpoint{0.456621in}{0.329242in}}{\pgfqpoint{2.435312in}{1.505106in}} %
\pgfusepath{clip}%
\pgfsetrectcap%
\pgfsetroundjoin%
\pgfsetlinewidth{0.803000pt}%
\definecolor{currentstroke}{rgb}{0.690196,0.690196,0.690196}%
\pgfsetstrokecolor{currentstroke}%
\pgfsetdash{}{0pt}%
\pgfpathmoveto{\pgfqpoint{0.456621in}{1.697520in}}%
\pgfpathlineto{\pgfqpoint{2.891933in}{1.697520in}}%
\pgfusepath{stroke}%
\end{pgfscope}%
\begin{pgfscope}%
\pgfsetbuttcap%
\pgfsetroundjoin%
\definecolor{currentfill}{rgb}{0.000000,0.000000,0.000000}%
\pgfsetfillcolor{currentfill}%
\pgfsetlinewidth{0.803000pt}%
\definecolor{currentstroke}{rgb}{0.000000,0.000000,0.000000}%
\pgfsetstrokecolor{currentstroke}%
\pgfsetdash{}{0pt}%
\pgfsys@defobject{currentmarker}{\pgfqpoint{0.000000in}{0.000000in}}{\pgfqpoint{0.048611in}{0.000000in}}{%
\pgfpathmoveto{\pgfqpoint{0.000000in}{0.000000in}}%
\pgfpathlineto{\pgfqpoint{0.048611in}{0.000000in}}%
\pgfusepath{stroke,fill}%
}%
\begin{pgfscope}%
\pgfsys@transformshift{0.456621in}{1.697520in}%
\pgfsys@useobject{currentmarker}{}%
\end{pgfscope}%
\end{pgfscope}%
\begin{pgfscope}%
\pgftext[x=0.289953in,y=1.658939in,left,base]{\rmfamily\fontsize{8.000000}{9.600000}\selectfont \(\displaystyle 20\)}%
\end{pgfscope}%
\begin{pgfscope}%
\pgftext[x=0.234397in,y=1.081795in,,bottom,rotate=90.000000]{\rmfamily\fontsize{8.000000}{9.600000}\selectfont \(\displaystyle \Delta\chi^2\)}%
\end{pgfscope}%
\begin{pgfscope}%
\pgfpathrectangle{\pgfqpoint{0.456621in}{0.329242in}}{\pgfqpoint{2.435312in}{1.505106in}} %
\pgfusepath{clip}%
\pgfsetrectcap%
\pgfsetroundjoin%
\pgfsetlinewidth{1.003750pt}%
\definecolor{currentstroke}{rgb}{0.767843,0.220980,0.353137}%
\pgfsetstrokecolor{currentstroke}%
\pgfsetdash{}{0pt}%
\pgfpathmoveto{\pgfqpoint{0.456621in}{0.376661in}}%
\pgfpathlineto{\pgfqpoint{0.761035in}{0.383555in}}%
\pgfpathlineto{\pgfqpoint{1.065449in}{0.389852in}}%
\pgfpathlineto{\pgfqpoint{1.369863in}{0.394164in}}%
\pgfpathlineto{\pgfqpoint{1.674277in}{0.395678in}}%
\pgfpathlineto{\pgfqpoint{1.978691in}{0.394164in}}%
\pgfpathlineto{\pgfqpoint{2.283105in}{0.389852in}}%
\pgfpathlineto{\pgfqpoint{2.587519in}{0.383555in}}%
\pgfpathlineto{\pgfqpoint{2.891933in}{0.376661in}}%
\pgfusepath{stroke}%
\end{pgfscope}%
\begin{pgfscope}%
\pgfpathrectangle{\pgfqpoint{0.456621in}{0.329242in}}{\pgfqpoint{2.435312in}{1.505106in}} %
\pgfusepath{clip}%
\pgfsetrectcap%
\pgfsetroundjoin%
\pgfsetlinewidth{1.003750pt}%
\definecolor{currentstroke}{rgb}{0.169412,0.150000,0.532353}%
\pgfsetstrokecolor{currentstroke}%
\pgfsetdash{}{0pt}%
\pgfpathmoveto{\pgfqpoint{0.456621in}{0.330758in}}%
\pgfpathlineto{\pgfqpoint{0.761035in}{0.330671in}}%
\pgfpathlineto{\pgfqpoint{1.065449in}{0.330561in}}%
\pgfpathlineto{\pgfqpoint{1.369863in}{0.330469in}}%
\pgfpathlineto{\pgfqpoint{1.674277in}{0.330433in}}%
\pgfpathlineto{\pgfqpoint{1.978691in}{0.330469in}}%
\pgfpathlineto{\pgfqpoint{2.283105in}{0.330561in}}%
\pgfpathlineto{\pgfqpoint{2.587519in}{0.330671in}}%
\pgfpathlineto{\pgfqpoint{2.891933in}{0.330758in}}%
\pgfusepath{stroke}%
\end{pgfscope}%
\begin{pgfscope}%
\pgfpathrectangle{\pgfqpoint{0.456621in}{0.329242in}}{\pgfqpoint{2.435312in}{1.505106in}} %
\pgfusepath{clip}%
\pgfsetrectcap%
\pgfsetroundjoin%
\pgfsetlinewidth{1.003750pt}%
\definecolor{currentstroke}{rgb}{0.387059,0.164510,0.677451}%
\pgfsetstrokecolor{currentstroke}%
\pgfsetdash{}{0pt}%
\pgfpathmoveto{\pgfqpoint{0.456621in}{0.498659in}}%
\pgfpathlineto{\pgfqpoint{0.761035in}{0.504717in}}%
\pgfpathlineto{\pgfqpoint{1.065449in}{0.508364in}}%
\pgfpathlineto{\pgfqpoint{1.369863in}{0.510025in}}%
\pgfpathlineto{\pgfqpoint{1.674277in}{0.510444in}}%
\pgfpathlineto{\pgfqpoint{1.978691in}{0.510025in}}%
\pgfpathlineto{\pgfqpoint{2.283105in}{0.508364in}}%
\pgfpathlineto{\pgfqpoint{2.587519in}{0.504717in}}%
\pgfpathlineto{\pgfqpoint{2.891933in}{0.498659in}}%
\pgfusepath{stroke}%
\end{pgfscope}%
\begin{pgfscope}%
\pgfpathrectangle{\pgfqpoint{0.456621in}{0.329242in}}{\pgfqpoint{2.435312in}{1.505106in}} %
\pgfusepath{clip}%
\pgfsetrectcap%
\pgfsetroundjoin%
\pgfsetlinewidth{1.003750pt}%
\definecolor{currentstroke}{rgb}{0.900000,0.677451,0.070980}%
\pgfsetstrokecolor{currentstroke}%
\pgfsetdash{}{0pt}%
\pgfpathmoveto{\pgfqpoint{0.456621in}{0.745094in}}%
\pgfpathlineto{\pgfqpoint{0.578387in}{0.395773in}}%
\pgfpathlineto{\pgfqpoint{0.700152in}{0.367651in}}%
\pgfpathlineto{\pgfqpoint{0.821918in}{0.355068in}}%
\pgfpathlineto{\pgfqpoint{0.943683in}{0.347658in}}%
\pgfpathlineto{\pgfqpoint{1.065449in}{0.342897in}}%
\pgfpathlineto{\pgfqpoint{1.187215in}{0.339646in}}%
\pgfpathlineto{\pgfqpoint{1.308980in}{0.337363in}}%
\pgfpathlineto{\pgfqpoint{1.430746in}{0.335806in}}%
\pgfpathlineto{\pgfqpoint{1.552511in}{0.334883in}}%
\pgfpathlineto{\pgfqpoint{1.674277in}{0.334575in}}%
\pgfpathlineto{\pgfqpoint{1.796043in}{0.334883in}}%
\pgfpathlineto{\pgfqpoint{1.917808in}{0.335806in}}%
\pgfpathlineto{\pgfqpoint{2.039574in}{0.337363in}}%
\pgfpathlineto{\pgfqpoint{2.161339in}{0.339646in}}%
\pgfpathlineto{\pgfqpoint{2.283105in}{0.342897in}}%
\pgfpathlineto{\pgfqpoint{2.404871in}{0.347658in}}%
\pgfpathlineto{\pgfqpoint{2.526636in}{0.355068in}}%
\pgfpathlineto{\pgfqpoint{2.648402in}{0.367651in}}%
\pgfpathlineto{\pgfqpoint{2.770167in}{0.395773in}}%
\pgfpathlineto{\pgfqpoint{2.891933in}{0.745094in}}%
\pgfusepath{stroke}%
\end{pgfscope}%
\begin{pgfscope}%
\pgfpathrectangle{\pgfqpoint{0.456621in}{0.329242in}}{\pgfqpoint{2.435312in}{1.505106in}} %
\pgfusepath{clip}%
\pgfsetrectcap%
\pgfsetroundjoin%
\pgfsetlinewidth{1.003750pt}%
\definecolor{currentstroke}{rgb}{0.941961,0.395098,0.062941}%
\pgfsetstrokecolor{currentstroke}%
\pgfsetdash{}{0pt}%
\pgfpathmoveto{\pgfqpoint{0.456621in}{0.341193in}}%
\pgfpathlineto{\pgfqpoint{0.700152in}{0.341150in}}%
\pgfpathlineto{\pgfqpoint{0.943683in}{0.340940in}}%
\pgfpathlineto{\pgfqpoint{1.187215in}{0.340159in}}%
\pgfpathlineto{\pgfqpoint{1.430746in}{0.338767in}}%
\pgfpathlineto{\pgfqpoint{1.674277in}{0.337998in}}%
\pgfpathlineto{\pgfqpoint{1.917808in}{0.338767in}}%
\pgfpathlineto{\pgfqpoint{2.161339in}{0.340159in}}%
\pgfpathlineto{\pgfqpoint{2.404871in}{0.340940in}}%
\pgfpathlineto{\pgfqpoint{2.648402in}{0.341150in}}%
\pgfpathlineto{\pgfqpoint{2.891933in}{0.341193in}}%
\pgfusepath{stroke}%
\end{pgfscope}%
\begin{pgfscope}%
\pgfsetrectcap%
\pgfsetmiterjoin%
\pgfsetlinewidth{0.501875pt}%
\definecolor{currentstroke}{rgb}{0.000000,0.000000,0.000000}%
\pgfsetstrokecolor{currentstroke}%
\pgfsetdash{}{0pt}%
\pgfpathmoveto{\pgfqpoint{0.456621in}{0.329242in}}%
\pgfpathlineto{\pgfqpoint{0.456621in}{1.834347in}}%
\pgfusepath{stroke}%
\end{pgfscope}%
\begin{pgfscope}%
\pgfsetrectcap%
\pgfsetmiterjoin%
\pgfsetlinewidth{0.501875pt}%
\definecolor{currentstroke}{rgb}{0.000000,0.000000,0.000000}%
\pgfsetstrokecolor{currentstroke}%
\pgfsetdash{}{0pt}%
\pgfpathmoveto{\pgfqpoint{2.891933in}{0.329242in}}%
\pgfpathlineto{\pgfqpoint{2.891933in}{1.834347in}}%
\pgfusepath{stroke}%
\end{pgfscope}%
\begin{pgfscope}%
\pgfsetrectcap%
\pgfsetmiterjoin%
\pgfsetlinewidth{0.501875pt}%
\definecolor{currentstroke}{rgb}{0.000000,0.000000,0.000000}%
\pgfsetstrokecolor{currentstroke}%
\pgfsetdash{}{0pt}%
\pgfpathmoveto{\pgfqpoint{0.456621in}{0.329242in}}%
\pgfpathlineto{\pgfqpoint{2.891933in}{0.329242in}}%
\pgfusepath{stroke}%
\end{pgfscope}%
\begin{pgfscope}%
\pgfsetrectcap%
\pgfsetmiterjoin%
\pgfsetlinewidth{0.501875pt}%
\definecolor{currentstroke}{rgb}{0.000000,0.000000,0.000000}%
\pgfsetstrokecolor{currentstroke}%
\pgfsetdash{}{0pt}%
\pgfpathmoveto{\pgfqpoint{0.456621in}{1.834347in}}%
\pgfpathlineto{\pgfqpoint{2.891933in}{1.834347in}}%
\pgfusepath{stroke}%
\end{pgfscope}%
\begin{pgfscope}%
\pgfsetbuttcap%
\pgfsetmiterjoin%
\definecolor{currentfill}{rgb}{1.000000,1.000000,1.000000}%
\pgfsetfillcolor{currentfill}%
\pgfsetfillopacity{0.800000}%
\pgfsetlinewidth{0.501875pt}%
\definecolor{currentstroke}{rgb}{0.000000,0.000000,0.000000}%
\pgfsetstrokecolor{currentstroke}%
\pgfsetstrokeopacity{0.800000}%
\pgfsetdash{}{0pt}%
\pgfpathmoveto{\pgfqpoint{2.148255in}{0.992989in}}%
\pgfpathlineto{\pgfqpoint{2.814155in}{0.992989in}}%
\pgfpathquadraticcurveto{\pgfqpoint{2.836377in}{0.992989in}}{\pgfqpoint{2.836377in}{1.015211in}}%
\pgfpathlineto{\pgfqpoint{2.836377in}{1.756570in}}%
\pgfpathquadraticcurveto{\pgfqpoint{2.836377in}{1.778792in}}{\pgfqpoint{2.814155in}{1.778792in}}%
\pgfpathlineto{\pgfqpoint{2.148255in}{1.778792in}}%
\pgfpathquadraticcurveto{\pgfqpoint{2.126033in}{1.778792in}}{\pgfqpoint{2.126033in}{1.756570in}}%
\pgfpathlineto{\pgfqpoint{2.126033in}{1.015211in}}%
\pgfpathquadraticcurveto{\pgfqpoint{2.126033in}{0.992989in}}{\pgfqpoint{2.148255in}{0.992989in}}%
\pgfpathclose%
\pgfusepath{stroke,fill}%
\end{pgfscope}%
\begin{pgfscope}%
\pgfsetrectcap%
\pgfsetroundjoin%
\pgfsetlinewidth{1.003750pt}%
\definecolor{currentstroke}{rgb}{0.767843,0.220980,0.353137}%
\pgfsetstrokecolor{currentstroke}%
\pgfsetdash{}{0pt}%
\pgfpathmoveto{\pgfqpoint{2.159366in}{1.706570in}}%
\pgfpathlineto{\pgfqpoint{2.381589in}{1.706570in}}%
\pgfusepath{stroke}%
\end{pgfscope}%
\begin{pgfscope}%
\pgftext[x=2.470477in,y=1.667681in,left,base]{\rmfamily\fontsize{8.000000}{9.600000}\selectfont Euclid}%
\end{pgfscope}%
\begin{pgfscope}%
\pgfsetrectcap%
\pgfsetroundjoin%
\pgfsetlinewidth{1.003750pt}%
\definecolor{currentstroke}{rgb}{0.169412,0.150000,0.532353}%
\pgfsetstrokecolor{currentstroke}%
\pgfsetdash{}{0pt}%
\pgfpathmoveto{\pgfqpoint{2.159366in}{1.551631in}}%
\pgfpathlineto{\pgfqpoint{2.381589in}{1.551631in}}%
\pgfusepath{stroke}%
\end{pgfscope}%
\begin{pgfscope}%
\pgftext[x=2.470477in,y=1.512742in,left,base]{\rmfamily\fontsize{8.000000}{9.600000}\selectfont SKA1}%
\end{pgfscope}%
\begin{pgfscope}%
\pgfsetrectcap%
\pgfsetroundjoin%
\pgfsetlinewidth{1.003750pt}%
\definecolor{currentstroke}{rgb}{0.387059,0.164510,0.677451}%
\pgfsetstrokecolor{currentstroke}%
\pgfsetdash{}{0pt}%
\pgfpathmoveto{\pgfqpoint{2.159366in}{1.396693in}}%
\pgfpathlineto{\pgfqpoint{2.381589in}{1.396693in}}%
\pgfusepath{stroke}%
\end{pgfscope}%
\begin{pgfscope}%
\pgftext[x=2.470477in,y=1.357804in,left,base]{\rmfamily\fontsize{8.000000}{9.600000}\selectfont SKA2}%
\end{pgfscope}%
\begin{pgfscope}%
\pgfsetrectcap%
\pgfsetroundjoin%
\pgfsetlinewidth{1.003750pt}%
\definecolor{currentstroke}{rgb}{0.900000,0.677451,0.070980}%
\pgfsetstrokecolor{currentstroke}%
\pgfsetdash{}{0pt}%
\pgfpathmoveto{\pgfqpoint{2.159366in}{1.241755in}}%
\pgfpathlineto{\pgfqpoint{2.381589in}{1.241755in}}%
\pgfusepath{stroke}%
\end{pgfscope}%
\begin{pgfscope}%
\pgftext[x=2.470477in,y=1.202866in,left,base]{\rmfamily\fontsize{8.000000}{9.600000}\selectfont IM B1}%
\end{pgfscope}%
\begin{pgfscope}%
\pgfsetrectcap%
\pgfsetroundjoin%
\pgfsetlinewidth{1.003750pt}%
\definecolor{currentstroke}{rgb}{0.941961,0.395098,0.062941}%
\pgfsetstrokecolor{currentstroke}%
\pgfsetdash{}{0pt}%
\pgfpathmoveto{\pgfqpoint{2.159366in}{1.086816in}}%
\pgfpathlineto{\pgfqpoint{2.381589in}{1.086816in}}%
\pgfusepath{stroke}%
\end{pgfscope}%
\begin{pgfscope}%
\pgftext[x=2.470477in,y=1.047927in,left,base]{\rmfamily\fontsize{8.000000}{9.600000}\selectfont IM B2}%
\end{pgfscope}%
\end{pgfpicture}%
\makeatother%
\endgroup%

%% file: datasets.tex
\section{Datasets}
\label{sec:datasets}
In this section we summarize the mock data sets used in our forecasts, their shorthand names and relevant assumptions for non-linear modeling. GC stands for galaxy clustering, CS for Cosmic Shear and IM for intensity mapping. We consider several GC and CS mock data sets from Euclid and SKA, and IM data from SKA. These data sets can be combined in single forecasts. For this purpose we followed some rules to avoid double-counting of information:
\begin{itemize}
\item Only combine GC+GC, GC+IM or IM+IM if the redshift ranges do not overlap, since both take their information from the position of galaxies.
\item Do not combine CS+CS because they all use information down to redshift zero.
\end{itemize}

\noindent \textbf{Galaxy clustering} (\cref{sec:GC,sec:GC_IM_error})\\[5pt]
On large wavelengths we adopt a cut-off at $k_{\text{min}}=0.02 \mathrm{Mpc}^{-1}$ that removes scales that are bigger than the bin width or violate the small angle approximation.
On small wavelengths we compare two schemes: a
\textbf{non-linear cut-off} at $k_{\text{NL}}(z) = k_{\text{NL}}(0)\cdot(1+z)^{2/(2+n_s)}$ (see Eq.~\ref{eq:k_NL_z}), or a 
\textbf{theoretical uncertainty} growing with $k$ after $k = 0.01$\,$h/\text{Mpc}$ (see Eq.~\ref{eq:th_error}),
corresponding to relative errors of 0.33\% at $k \leq 0.01$\,$h/\text{Mpc}$, 1\% at $k=0.3$\,$h/\text{Mpc}$, and 10\% at $k=10$\,$h/\text{Mpc}$. Our data sets are:
\begin{itemize}
	\item \makebox[3.1cm][l]{\textbf{Euclid GC cons.}} \makebox[0.5cm][l]{} Euclid galaxy clustering, conservative \\
	\makebox[3.1cm][l]{} \makebox[0.5cm][l]{} Theoretical uncertainty, $k_{\text{NL}}(0)=0.2 $\,$h/\text{Mpc}$
	\item \makebox[3.1cm][l]{\textbf{Euclid GC real.}} \makebox[0.5cm][l]{} Euclid galaxy clustering, realistic \\
	\makebox[3.1cm][l]{} \makebox[0.5cm][l]{} Theoretical uncertainty, $k_{\text{max}}=10 $\,$h/\text{Mpc}$
	\item \makebox[3.1cm][l]{\textbf{SKA1 GC cons.}} \makebox[0.5cm][l]{} SKA1-MID band 2 galaxy clustering, conservative \\
	\makebox[3.1cm][l]{} \makebox[0.5cm][l]{} Theoretical uncertainty, $k_{\text{NL}}(0)=0.2 $\,$h/\text{Mpc}$
	\item \makebox[3.1cm][l]{\textbf{SKA1 GC real.}} \makebox[0.5cm][l]{} SKA1-MID band 2 galaxy clustering, realistic \\
	\makebox[3.1cm][l]{} \makebox[0.5cm][l]{} Theoretical uncertainty, $k_{\text{max}}=10 $\,$h/\text{Mpc}$
	\item \makebox[3.1cm][l]{\textbf{SKA2 GC cons.}} \makebox[0.5cm][l]{} SKA2-MID galaxy clustering, conservative \\
	\makebox[3.1cm][l]{} \makebox[0.5cm][l]{} Theoretical uncertainty, $k_{\text{NL}}(0)=0.2 $\,$h/\text{Mpc}$
	\item \makebox[3.1cm][l]{\textbf{SKA2 GC real.}} \makebox[0.5cm][l]{} SKA2-MID galaxy clustering, realistic \\
	\makebox[3.1cm][l]{} \makebox[0.5cm][l]{} Theoretical uncertainty, $k_{\text{max}}=10 $\,$h/\text{Mpc}$
\end{itemize}

\noindent \textbf{Intensity mapping} (\cref{sec:IM,sec:GC_IM_error})\\[5pt]
We adopt the same $k_{\text{min}}$, non-linear cut-off and theoretical uncertainty as for galaxy clustering. Our data sets are:
\begin{itemize}
	\item \makebox[3.1cm][l]{\textbf{SKA1 IM1 cons.}} \makebox[0.5cm][l]{} SKA1-MID band 1 intensity mapping, conservative \\
	\makebox[3.1cm][l]{} \makebox[0.5cm][l]{} Theoretical uncertainty, $k_{\text{NL}}(0)=0.2 $\,$h/\text{Mpc}$
	\item \makebox[3.1cm][l]{\textbf{SKA1 IM1 real.}} \makebox[0.5cm][l]{} SKA1-MID band 1 intensity mapping, realistic \\
	\makebox[3.1cm][l]{} \makebox[0.5cm][l]{} Theoretical uncertainty, $k_{\text{max}}=10 $\,$h/\text{Mpc}$
	\item \makebox[3.1cm][l]{\textbf{SKA1 IM2 cons.}} \makebox[0.5cm][l]{} SKA1-MID band 2 intensity mapping, conservative \\
	\makebox[3.1cm][l]{} \makebox[0.5cm][l]{} Theoretical uncertainty, $k_{\text{NL}}(0)=0.2 $\,$h/\text{Mpc}$
	\item \makebox[3.1cm][l]{\textbf{SKA1 IM2 real.}} \makebox[0.5cm][l]{} SKA1-MID band 2 intensity mapping, realistic \\
	\makebox[3.1cm][l]{} \makebox[0.5cm][l]{} Theoretical uncertainty, $k_{\text{max}}=10 $\,$h/\text{Mpc}$
\end{itemize}

\noindent \textbf{Cosmic shear} (\cref{sec:CS,sec:CS_error})\\[5pt]
We include multipoles from $\ell_{\text{min}}=5$ to a bin-dependent
\textbf{non-linear cut-off} at $\ell_{\text{max}}^i = k_{\text{NL}}(z) \cdot \bar{r}^{i}_{\text{peak}}$ (see \cref{eq:ell_max,eq:k_NL_z}). Our data sets are:
\begin{itemize}
	\item \makebox[3.1cm][l]{\textbf{Euclid CS cons.}} \makebox[0.5cm][l]{} Euclid cosmic shear, conservative \\
	\makebox[3.1cm][l]{} \makebox[0.5cm][l]{} $k_{\text{NL}}(0)=0.5 $\,$h/\text{Mpc}$
	\item \makebox[3.1cm][l]{\textbf{Euclid CS real.}} \makebox[0.5cm][l]{} Euclid cosmic shear, realistic \\
	\makebox[3.1cm][l]{} \makebox[0.5cm][l]{} $k_{\text{NL}}(0)=2.0 $\,$h/\text{Mpc}$
	\item \makebox[3.1cm][l]{\textbf{SKA1 CS cons.}} \makebox[0.5cm][l]{} SKA1-MID cosmic shear, conservative \\
	\makebox[3.1cm][l]{} \makebox[0.5cm][l]{} $k_{\text{NL}}(0)=0.5 $\,$h/\text{Mpc}$
	\item \makebox[3.1cm][l]{\textbf{SKA1 CS real.}} \makebox[0.5cm][l]{} SKA1-MID cosmic shear, realistic \\
	\makebox[3.1cm][l]{} \makebox[0.5cm][l]{} $k_{\text{NL}}(0)=2.0 $\,$h/\text{Mpc}$
	\item \makebox[3.1cm][l]{\textbf{SKA2 CS cons.}} \makebox[0.5cm][l]{} SKA2-MID cosmic shear, conservative \\
	\makebox[3.1cm][l]{} \makebox[0.5cm][l]{} $k_{\text{NL}}(0)=0.5 $\,$h/\text{Mpc}$
	\item \makebox[3.1cm][l]{\textbf{SKA2 CS real.}} \makebox[0.5cm][l]{} SKA2-MID cosmic shear, realistic \\
	\makebox[3.1cm][l]{} \makebox[0.5cm][l]{} $k_{\text{NL}}(0)=2.0 $\,$h/\text{Mpc}$
\end{itemize}

\noindent \textbf{Cosmic microwave background}
\begin{itemize}
	\item \textbf{Planck} \makebox[0.5cm][l]{}  Instead of the real Planck data, it is more convenient to run our forecasts with some mock temperature, polarization and CMB lensing data generated for the parameter values of our fiducial model. We go up to $\ell_{\rm max}=3000$ and we use noise spectra matching the expected sensitivity of the final Planck data release, in particular improving constraints from polarization.
	
\end{itemize}

%% file: Mnu.tex
\section{Baseline model results\label{sec:baseline}} 

It is well known from neutrino oscillation experiments that at least two neutrinos are massive.
However the absolute value of the mass has not been determined yet, neither by cosmology (for up-to-date upper limits see Refs.~\cite{Gariazzo:2018pei,Cuesta:2015iho,Vagnozzi:2017ovm}), nor by $\beta$-decay experiments (in this regard see the sensitivity of the forthcoming KATRIN experiment~\cite{Bonn:2007su}).
One of the biggest achievements of Euclid and SKA will be to pin down the neutrino mass sum $M_\nu$.
Therefore, our baseline model will be $\Lambda$CDM$+M_\nu$, parameterized as follows:
\begin{equation}
\{\omega_{\rm b},\,\omega_{\rm cdm},\,100\times\theta_{\rm s},\,\tau_\mathrm{reio},\,\ln(10^{10}A_{s}),\,n_{s},\,M_\nu\}.
\end{equation}
Our fiducial model assumes a minimal value of the total neutrino mass and some Planck inspired values for other parameters:
\begin{equation}
\{0.02218,\,0.1205,\,1.04156,\,0.0596,\,3.056,\,0.9619,\,0.06\mathrm{~eV}\}.
\end{equation}
We assume the total neutrino mass sum $M_\nu$ to be equally split among the three active neutrino species. This degenerate neutrino mass scheme is motivated by the fact that the deviation of its theoretical predictions both from the normal mass ordering and from the inverted mass ordering is negligible compared to the sensitivity of current and forthcoming cosmological data \cite{Lesgourgues:2006nd,Jimenez:2010ev,Giusarma:2016phn}.
For a detailed discussion of the physical effects involved in the measurements of $M_\nu$ with galaxy clustering and cosmic shear, and for the impact of 21-cm surveys, we refer the reader to Refs.~\cite{Archidiacono:2016lnv,Boyle:2017lzt}.
Here we just mention that low redshift measurements are sensitive to massive neutrinos because their free-streaming induces a relative suppression of the linear matter power spectrum on scales smaller than the free-streaming scale after the neutrino non-relativistic transition~\cite{Bashinsky:2003tk,Lesgourgues:2006nd,Hannestad:2010kz,Lesgourgues:1519137}. On top of the linear suppression, an additional dip appears at non-linear scales, caused by the delay of the onset of the non-linear growth in neutrino cosmologies~\cite{Bird:2011rb}. Euclid and SKA span a broad range of redshifts and scales, where the aforementioned effects on the shape of the matter power spectrum can be detected, as long as an accurate theoretical prediction is provided~\cite{Archidiacono:2015ota,Castorina:2015bma,Dupuy:2015ega,Fuhrer:2014zka,AliHaimoud:2012vj}.

In this regard, Refs.~\cite{Castorina:2013wga, Castorina:2015bma} have shown that in massive neutrino cosmologies the clustering properties of halos are determined by cold dark matter and baryons only (hereafter, {\it cb}), rather than the total matter field (i.e., cold dark matter + baryons + massive neutrinos). Therefore, the galaxy power spectrum must be reconstructed by taking into account only the {\it cb} field, ignoring the contribution of light massive neutrinos with a free-streaming length far larger than the typical size of a galaxy. Neglecting this effect can lead to sizeable errors \cite{Raccanelli:2017kht,Vagnozzi:2018pwo}. Following this approach, already used in Refs.~\cite{Obuljen:2017jiy,Villaescusa-Navarro:2017mfx}, we have modified Eq.~\ref{P_g-def} of the observed galaxy power spectrum as follows:
\begin{equation}
\label{P_g-def-cb}
P_g(k,\mu,z) =  f_{\text{AP}}(z) \times f_{\text{res}}(k,\mu,z) \times f_{\text{RSD}}(\hat{k},\hat{\mu},z) \times b^2(z) \times P_{cb}(\hat{k},z) \ .
\end{equation}
Moreover, the $\beta$ factor of the Kaiser formula, i.e. the ratio between the growth rate and the bias, embedded in the third term of Eq.~\ref{P_g-def-cb} ($f_{\text{RSD}}(\hat{k},\hat{\mu},z)$) and originally defined in Eq.~\ref{eq:beta} has to be rewritten as:
\begin{equation}
\beta(\hat{k},z) = -\frac{1+z}{2b(z)}\cdot\frac{\dd \ln P_{cb}(\hat{k},z)}{\dd z} \ ,
\end{equation}
where the bias is now rightfully assumed to be scale independent, being defined as $\delta_g = b(z) \times \delta_{cb}$.
The same considerations apply to the 21cm power spectrum, indeed the neutral hydrogen in low redshift galaxies is a biased tracer of the {\it cb} field only. Therefore, $P_m$ has to be replaced with $P_{cb}$ in Eq.~\ref{eq:P21_1} and Eq.~\ref{eq:P21}, and, as in the case of galaxy clustering, the Kaiser formula has to take into account the {\it cb} growth rate, rather than the total matter one.

\begin{figure}
\input{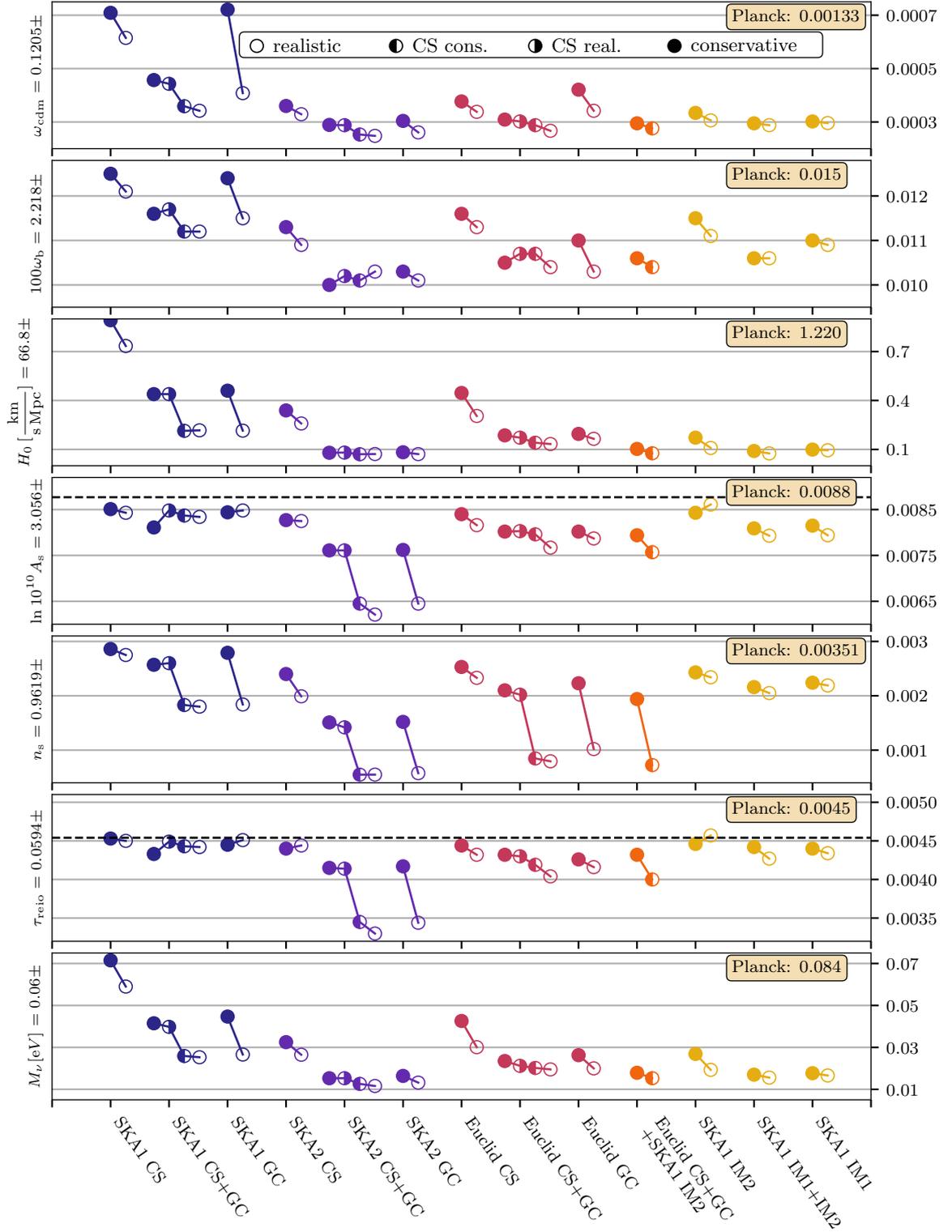}
\caption{$1 \, \sigma$ uncertainty for various combinations of experiments, all including Planck, to the baseline cosmological parameters.
The combinations of experiments are specified on the x axis at the bottom of the plot.
Empty circles denote realistic settings, while filled circles denote conservative settings. In the case of combinations of cosmic shear with other likelihoods, the left part of the circle describes the setting for the cosmic shear likelihood. The fiducial value is given on the left axis, while the scale of the $1 \, \sigma$ uncertainty is written on the right axis.
The Planck sensitivity is written inside the box in the upper right corner, and is also shown as a dashed line when within the range of the y axis on the right hand side.}
\label{fig:mnu_sigma}
\end{figure}

\begin{figure}
\input{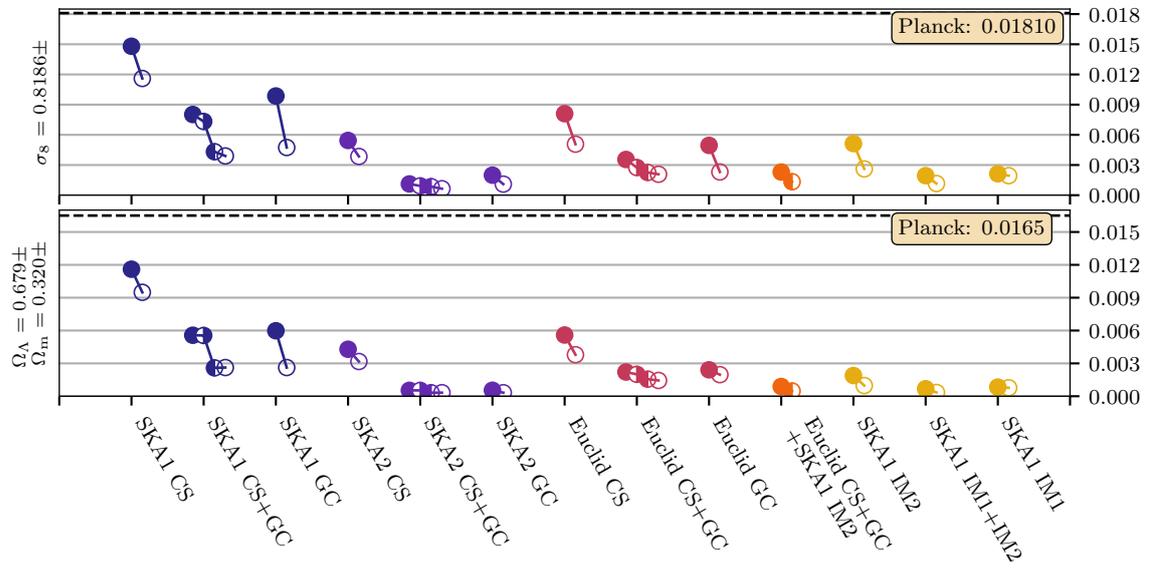}
\caption{Same as Figure~\ref{fig:mnu_sigma} but for the derived parameters $\Omega_\Lambda$, $\Omega_m$ and $\sigma_8$.
}
\label{fig:mnu_sigma2}
\end{figure}

We will now evaluate the sensitivity of Euclid and SKA (combined with Planck) to cosmological parameters by performing a Markov Chain Monte Carlo (MCMC) forecast, i.e. fitting the spectra of the fiducial model assuming the likelihood expressions discussed in the previous sections.
Our MCMC forecasts are obtained with the new version {\sc 3.0} of the {\sc MontePython} package\footnote{\tt https://github.com/brinckmann/montepython\_public}~\cite{Audren:2012wb,Brinckmann:2018cvx}, implementing our new Euclid and SKA likelihoods to fit to the mock data the theoretical spectra provided by the Boltzmann solver {\sc CLASS}~\cite{Blas:2011rf} {\tt v2.7}.
In Table~\ref{tab:mnu} we report the expected sensitivity of various probe combinations. 
In Figures~\ref{fig:mnu_sigma} and \ref{fig:mnu_sigma2} we depict the corresponding $1 \, \sigma$ uncertainty, in order to visualize the impact of different experiments, probe combinations and theoretical error prescriptions.

\begin{table}
\resizebox{1.0\textwidth}{!}{%
\input{Mnu_Table.tex}
}%
\caption{Expected $1\,\sigma$ sensitivity of Planck, Euclid and SKA to the cosmological parameters.
For each probe combination in the first and in the second column we indicate whether the cosmological probe is present or not, and whether the theoretical error is described with a conservative (c) or a realistic (r) approach.}
\label{tab:mnu}
\end{table}

We have already elaborated on our conservative and realistic approach to the implementation of the theoretical error in Section~\ref{sec:CS_error} and \ref{sec:GC_IM_error}. The sensitivity of galaxy clustering measurements is affected by the choice of the theoretical error more than cosmic shear and intensity mapping. As explained in Section~\ref{sec:GC_IM_error}, SKA1 is more sensitive than SKA2 and Euclid to the theoretical error prescription, because of its narrow redshift range. Indeed, Figure~\ref{fig:mnu_sigma} shows that only for SKA1 the uncertainty to every cosmological parameter shrinks in the realistic configuration with respect to the conservative one. In the case of SKA2 and Euclid, the improvement of the sensitivity due to the more optimistic theoretical error prescription mostly concerns the primordial power spectrum parameter $n_s$: 
In the absence of a sharp cut-off in the measured power spectrum, the extended lever arm in $k$ provides more constraining power. Finally, notice that the sensitivity of SKA1 band 2 intensity mapping to the derived parameters $\Omega_m$ and $\sigma_8$ greatly benefits from the realistic theoretical error configuration (Figure~\ref{fig:mnu_sigma2}). The improvement of the constraints on these parameters with respect to the conservative configuration is caused by low redshift probes breaking the degeneracy intrinsic in CMB data. This mechanism does not work for SKA1 IM2 if a conservative approach to the theoretical error is used, because of the inadequate sensitivity. However, the different correlation of SKA1 IM2 in this part of the parameter space appears if we apply a more optimistic approach to the theoretical error is assumed. 


Table~\ref{tab:mnu} and Figure~\ref{fig:mnu_sigma} show that both Euclid and SKA will greatly improve the sensitivity to the summed neutrino mass and to the other cosmological parameters with respect to currently available data sets.
For all error prescriptions, the combination Planck + SKA2 (CS + GC) achieves the best precision within our baseline $\Lambda$CDM$+M_\nu$ model, as expected from the previous sensitivity considerations. For experiments in a nearer future, Planck + Euclid (CS + GC) will give a very good precision for $n_{\text{s}}$, while the huge redshift range available to SKA1 intensity mapping enables a good measurement of $H_0$.

\begin{figure}
\includegraphics[width=1.0\linewidth]{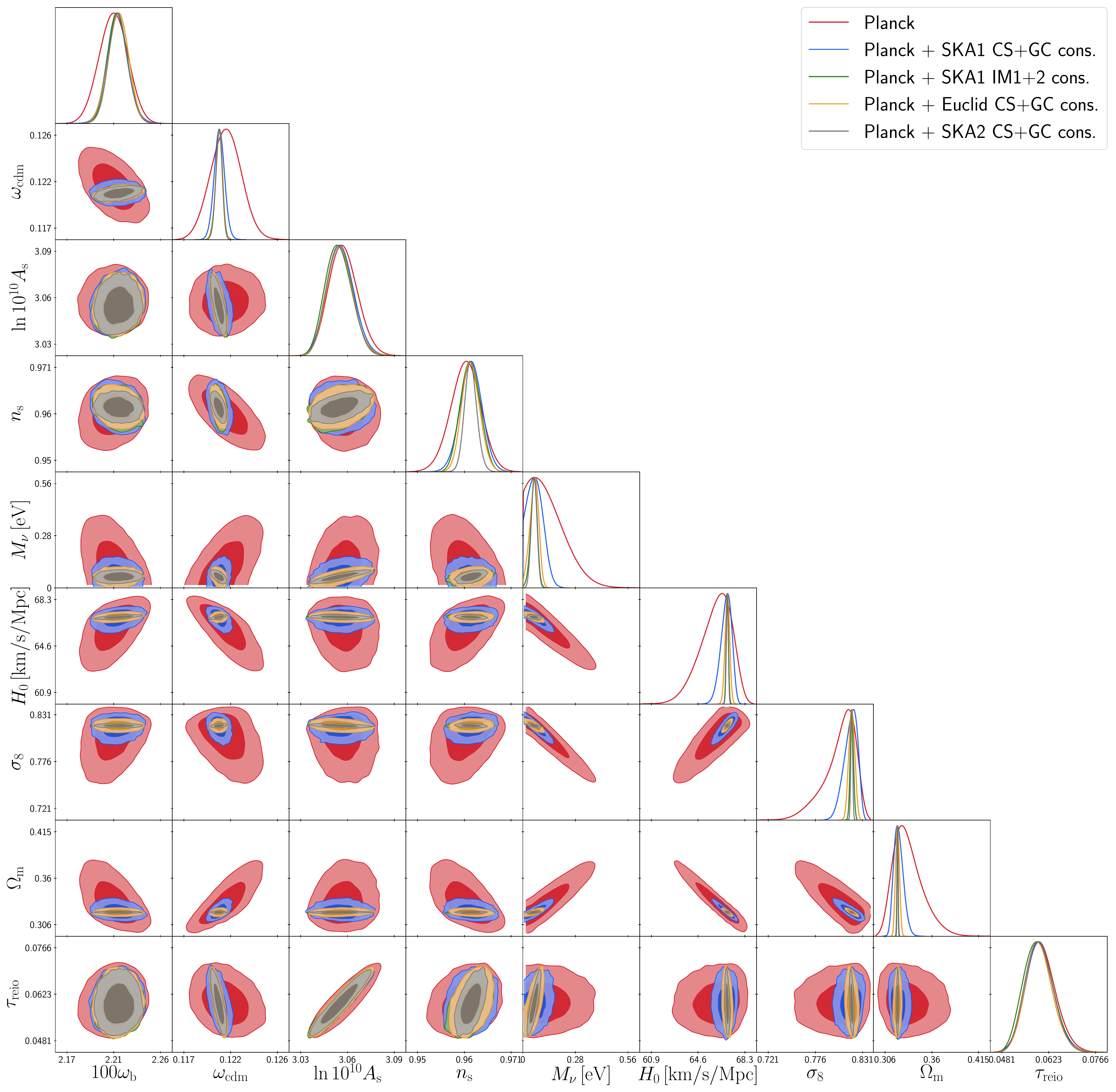}
\caption{Marginalized $1\, \sigma$ and $2\, \sigma$ contours and one-dimensional posteriors in the
$\left( \omega_b, \, \omega_\mathrm{cdm}, A_s, \, n_s, \, M_\nu, \, H_0, \, \tau_\mathrm{reio}, \, \sigma_8, \, \Omega_m \right)$
parameter space, showing the expected sensitivity of Planck-only, Planck + Euclid (CS + GC), Planck + SKA1 (CS + GC), Planck+SKA1-IM and Planck + SKA2 (CS + GC).
Here the analysis is performed following the conservative approach for the description of the theoretical error.}
\label{fig:mnu_triangle}
\end{figure}

\begin{figure}
\includegraphics[width=1.0\linewidth]{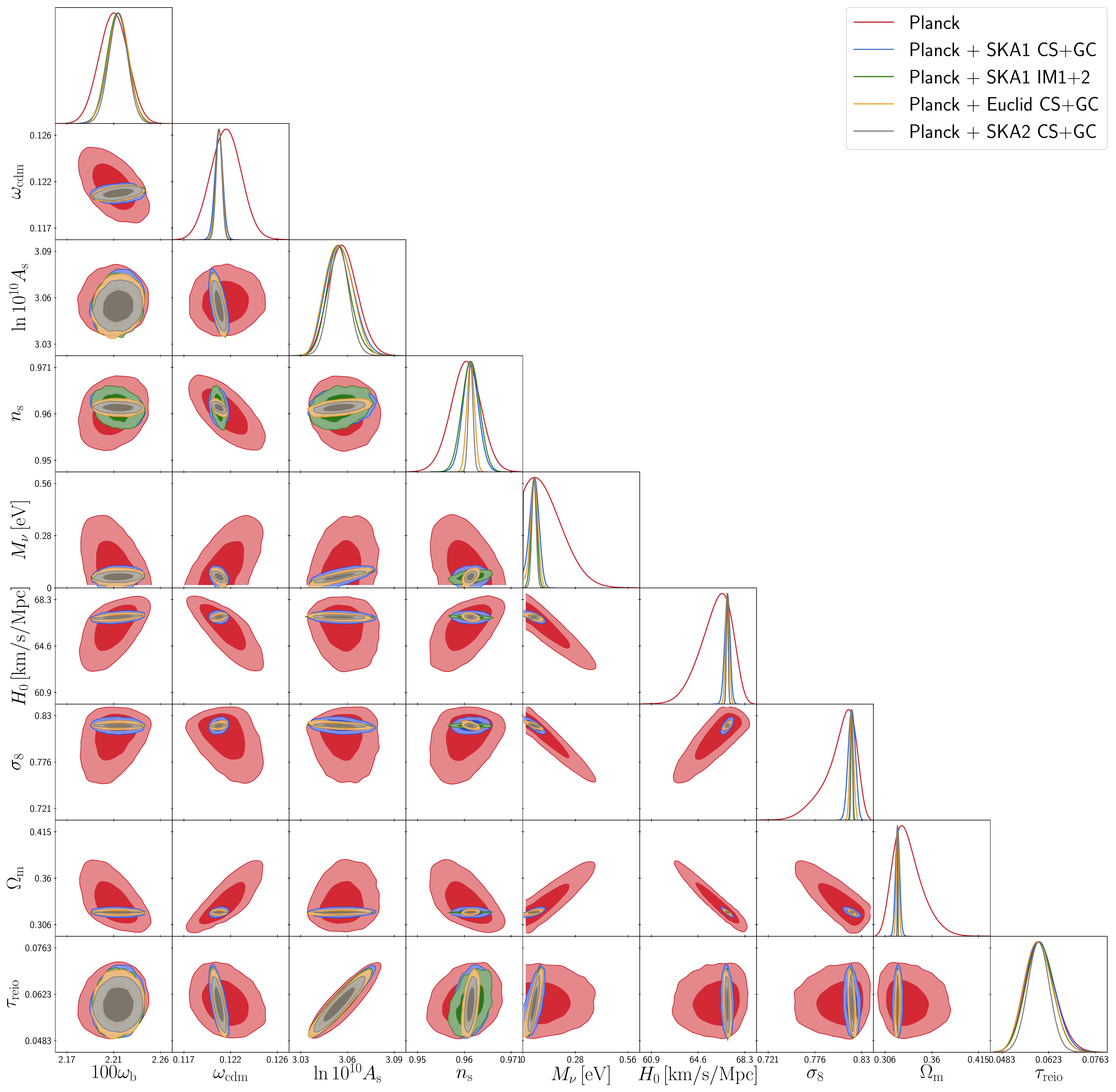}
\caption{Marginalized $1\, \sigma$ and $2\, \sigma$ contours and one-dimensional posteriors in the
$\left( \omega_b, \, \omega_\mathrm{cdm}, A_s, \, n_s, \, M_\nu, \, H_0, \, \tau_\mathrm{reio}, \, \sigma_8, \, \Omega_m \right)$
parameter space, showing the expected sensitivity of Planck-only, Planck + Euclid (CS + GC), Planck + SKA1 (CS + GC), Planck + SKA1 IM(1 + 2) and Planck + SKA2 (CS + GC).
Here the analysis is performed following the realistic modeling of the theoretical error for GC and IM and the conservative approach for CS. This is done to isolate the effect of a theoretical error approach without a cut-off from contributions of CS. }
\label{fig:mnu_triangle_real}
\end{figure}

Concerning the neutrino mass sum, with a realistic description of the theoretical error, the sensitivity of Planck + Euclid (CS+GC) + SKA1 (IM2) (resp. Planck + SKA2 (CS + GC)), to a fiducial neutrino mass of $60$~meV is $15$~meV (resp. $12$~meV). This fantastic accuracy implies a $4-5\,\sigma$ detection of a non-zero neutrino mass even in the minimal mass scenario.
Using a conservative prescription of the theoretical error degrades the sensitivity to $18$~meV (resp. $15$~meV), which still provides a more than $3\,\sigma$ detection and a factor $5$ improvement with respect to Planck only.
Our results on the sensitivity of Euclid and SKA to the neutrino mass sum are consistent with the previous literature~\cite{Audren:2012vy,Allison:2015qca,Liu:2015txa,Hamann:2012fe,Villaescusa-Navarro:2015cca}. Notice that Ref.~\cite{Allison:2015qca} finds
exactly the same uncertainty ($\sigma(M_\nu)=0.012$~eV) as in our most constraining data combinations (Planck + SKA2 (CS + GC)) by using Planck-polarization + CMB-Stage-IV + BAO-DESI + 21cm-HERA.
Figures~\ref{fig:mnu_triangle} and~\ref{fig:mnu_triangle_real} show the marginalized $1 \, \sigma$ and $2 \, \sigma$ contours and one dimensional posteriors in the
$\left( \omega_b, \, \omega_\mathrm{cdm}, A_s, \, n_s, \, M_\nu, \, H_0, \, \tau_\mathrm{reio}, \, \sigma_8, \, \Omega_m \right)$
parameter space for the most constraining data combinations with respect to Planck-only.
Figure~\ref{fig:mnu_triangle} assumes the conservative theoretical error setup, and Figure~\ref{fig:mnu_triangle_real} the realistic one.  

These figures show that combining high redshift CMB data with low redshift measurements breaks the degenerecies among the cosmological parameters that are present when using only CMB data (red lines and contours). For example, if we look at the contours in the plane $\left(M_\nu, \omega_\mathrm{cdm}\right)$ we can see that both Euclid and SKA (SKA1 intensity mapping, or SKA2 (CS + GC) lift the degeneracy and even slightly reverse it (a physical interpretation is given in~\cite{Archidiacono:2016lnv}). This propagates to other parameters, causing an overall shrinking of all contours and leading to an increase of the sensitivity.
From the triangle plots, we can verify once more that SKA1 (CS + GC) results (blue lines and contours) benefit from a realistic modelling of the theoretical error
more than the other experiments.
If we look at the SKA1 (CS + GC) one dimensional posterior on $H_0$, we can see that it is considerably narrower in the realistic configuration.
Given the correlation between $H_0$ and the other cosmological parameters, this effect leads to a remarkable increase of the sensitivity not only to $H_0$,
but also to the other cosmological parameters, e.g.\,the neutrino mass sum.
Indeed, only the realistic approach would allow for a significant detection of the minimal neutrino mass with SKA1 (CS + GC).
Interestingly, the combination and Planck and SKA1 intensity mapping (green lines and contours) has a sensitivity which is roughly comparable to that of Planck + Euclid (CS + GC) and Planck + SKA2 (CS + GC) for most parameters, with the notable exception of the primordial power spectrum parameters.

Finally, our analysis shows that, regardless of the theoretical error description, both Euclid and SKA (SKA1 intensity mapping, or SKA2 (CS + GC))
will provide a detection of a non-zero neutrino mass in the next few years. This result appears to be very robust at least as long as we assume a minimal 7-parameter cosmological model. One of the purposes of the next section is to assess to which extent this remains true in the presence of additional parameters.

%% file: Mnu_Table.tex
\begingroup
\setlength{\tabcolsep}{10pt}
\renewcommand{\arraystretch}{1.5}
\begin{tabular}{|l|c|c|c|c|c|c|c|c|c|c|}
\cline{2-11}
\multicolumn{1}{l|}{}&CS&GC&$\sigma(100 \times \omega_{\rm b})$&$\sigma(\omega_\mathrm{cdm})$&$\sigma( \ln \left[10^{10}A_s\right])$&$\sigma(n_s)$&$\sigma(H_0)/[\frac{\mathrm{km}}{\mathrm{s\,Mpc}}]$&$\sigma(\tau_\mathrm{reio})$&$\sigma(M_\nu)/[\mathrm{meV}]$&$\sigma(\sigma_8)$\\
\hline
Planck&-&-&0.015&0.00133&0.0088&0.00351&1.220&0.0045&84&0.01810\\
\hline
\multirow{8}{*}{Planck+SKA1}&c&\multirow{2}{*}{-}&0.013&0.00071&0.0085&0.00286&0.892&0.0045&72&0.01480\\
&r&&0.012&0.00061&0.0084&0.00275&0.734&0.0045&59&0.01160\\
\cdashline{2-11}
&-&\multirow{3}{*}{c}&0.012&0.00072&0.0084&0.00279&0.460&0.0044&45&0.00986\\
&c&&0.012&0.00046&0.0081&0.00257&0.439&0.0043&42&0.00802\\
&r&&0.012&0.00044&0.0085&0.00260&0.439&0.0045&40&0.00733\\
\cdashline{2-11}
&-&\multirow{3}{*}{r}&0.011&0.00041&0.0085&0.00184&0.215&0.0045&27&0.00474\\
&c&&0.011&0.00036&0.0084&0.00183&0.214&0.0044&26&0.00432\\
&r&&0.011&0.00034&0.0083&0.00180&0.217&0.0044&25&0.00390\\
\hline
\multirow{8}{*}{Planck+SKA2}&c&\multirow{2}{*}{-}&0.011&0.00036&0.0083&0.00240&0.339&0.0044&32&0.00544\\
&r&&0.011&0.00033&0.0083&0.00199&0.259&0.0044&26&0.00385\\
\cdashline{2-11}
&-&\multirow{3}{*}{c}&0.010&0.00030&0.0076&0.00152&0.083&0.0042&16&0.00199\\
&c&&0.010&0.00029&0.0076&0.00151&0.080&0.0042&15&0.00112\\
&r&&0.010&0.00029&0.0076&0.00142&0.081&0.0041&15&0.00092\\
\cdashline{2-11}
&-&\multirow{3}{*}{r}&0.010&0.00026&0.0065&0.00058&0.071&0.0034&13&0.00110\\
&c&&0.010&0.00025&0.0065&0.00055&0.070&0.0034&13&0.00085\\
&r&&0.010&0.00025&0.0062&0.00055&0.072&0.0033&12&0.00064\\
\hline
\multirow{8}{*}{Planck+Euclid}&c&\multirow{2}{*}{-}&0.012&0.00038&0.0084&0.00253&0.446&0.0044&43&0.00810\\
&r&&0.011&0.00034&0.0082&0.00233&0.305&0.0043&30&0.00507\\
\cdashline{2-11}
&-&\multirow{3}{*}{c}&0.011&0.00042&0.0080&0.00223&0.195&0.0043&26&0.00495\\
&c&&0.011&0.00031&0.0080&0.00210&0.186&0.0043&24&0.00355\\
&r&&0.011&0.00030&0.0080&0.00202&0.172&0.0043&21&0.00275\\
\cdashline{2-11}
&-&\multirow{3}{*}{r}&0.010&0.00034&0.0079&0.00102&0.165&0.0042&20&0.00231\\
&c&&0.011&0.00029&0.0080&0.00085&0.141&0.0042&20&0.00226\\
&r&&0.010&0.00027&0.0077&0.00080&0.133&0.0040&20&0.00208\\
\hline
\hline
Planck+Euclid&c&c&0.011&0.00030&0.0079&0.00194&0.103&0.0043&18&0.00230\\
\cdashline{2-2}
+SKA1&IM2&r&0.010&0.00028&0.0076&0.00072&0.076&0.0040&15&0.00134\\
\hline
\multirow{6}{*}{Planck+SKA1}&\multirow{2}{*}{IM1}&c&0.011&0.00030&0.0081&0.00224&0.099&0.0044&18&0.00213\\
&&r&0.011&0.00030&0.0079&0.00219&0.095&0.0043&17&0.00194\\
\cdashline{2-11}
&\multirow{2}{*}{IM2}&c&0.011&0.00033&0.0084&0.00243&0.172&0.0045&27&0.00513\\
&&r&0.011&0.00031&0.0086&0.00234&0.109&0.0046&19&0.00261\\
\cdashline{2-11}
&IM1&c&0.011&0.00030&0.0081&0.00216&0.090&0.0044&17&0.00194\\
&+IM2&r&0.011&0.00029&0.0079&0.00205&0.075&0.0043&16&0.00116\\
\hline
\end{tabular}
\endgroup

%% file: ext.tex
\section{Extended model forecasts} 

In this section, we describe the results obtained by fitting our mock data to three physically motivated extensions of the baseline model: (1) $\Lambda$CDM+$M_\nu$+$N_\mathrm{eff}$, (2) $\Lambda$CDM+$M_\nu$+$w_0$, (3) $\Lambda$CDM+$M_\nu$+$\left( w_0, w_a\right)$. These extensions have been discussed at various points in the past as potentially degenerate to some extent with neutrino masses, so the results of this section also aim at assessing the robustness of a future neutrino mass detection against extended cosmology assumptions. We leave for future work the study of models with a free spatial curvature parameter, which is also likely to degrade the sensitivity to the total neutrino mass~(see e.g.\,\cite{Boyle:2017lzt} for a recent discussion).

\begin{figure}
\includegraphics[width=0.5\textwidth]{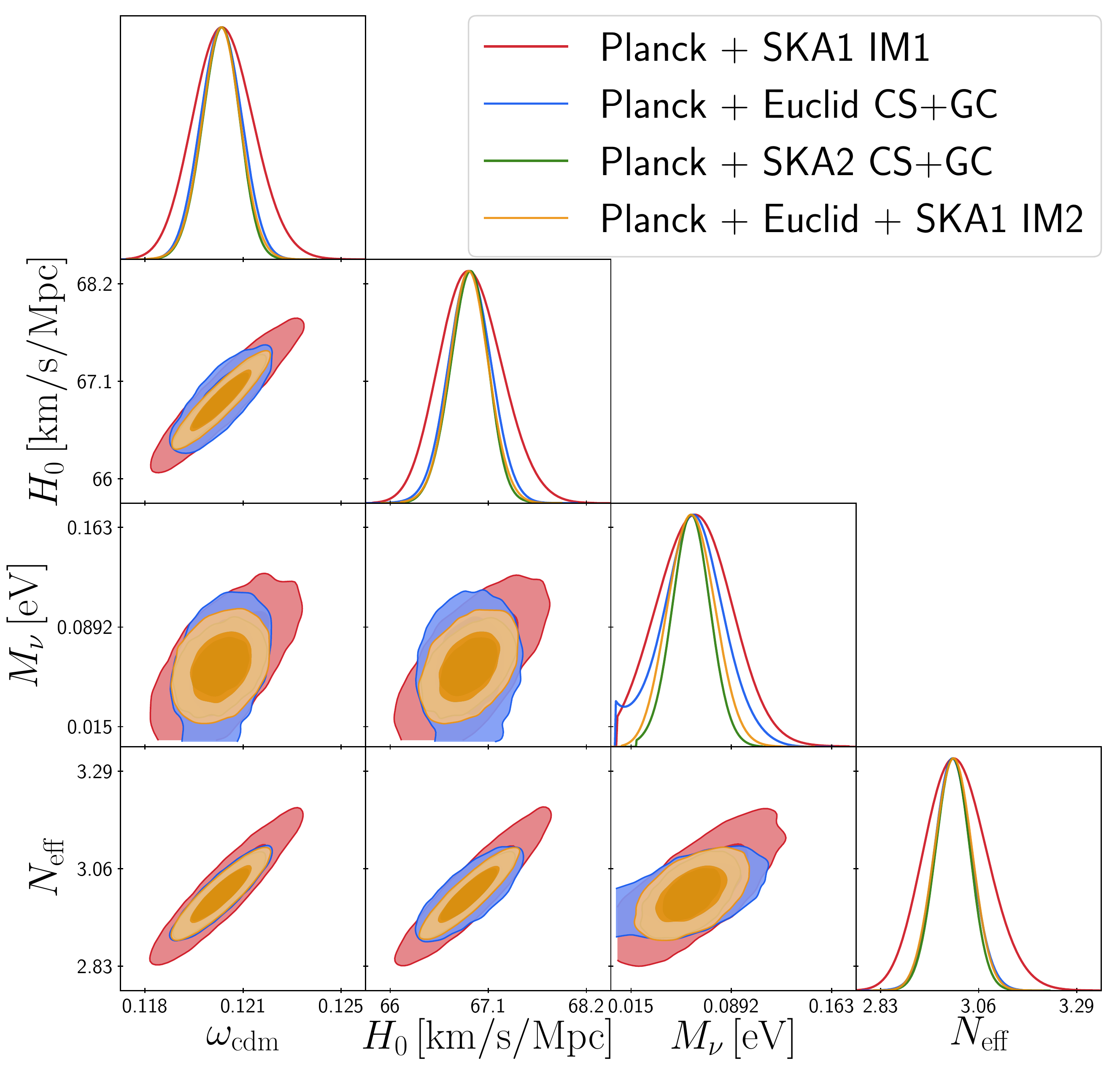}
\includegraphics[width=0.5\textwidth]{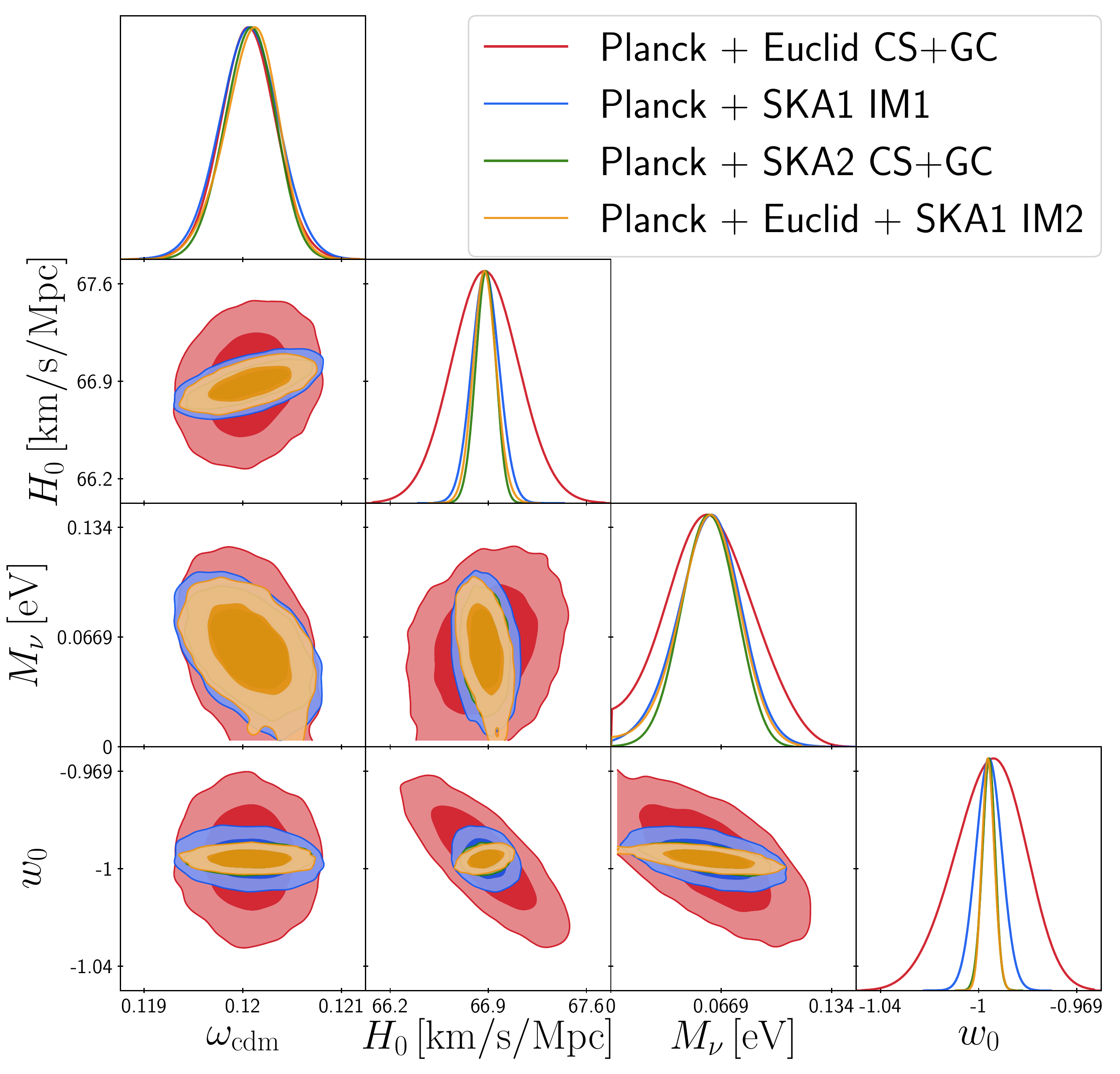}
\includegraphics[width=0.62\textwidth]{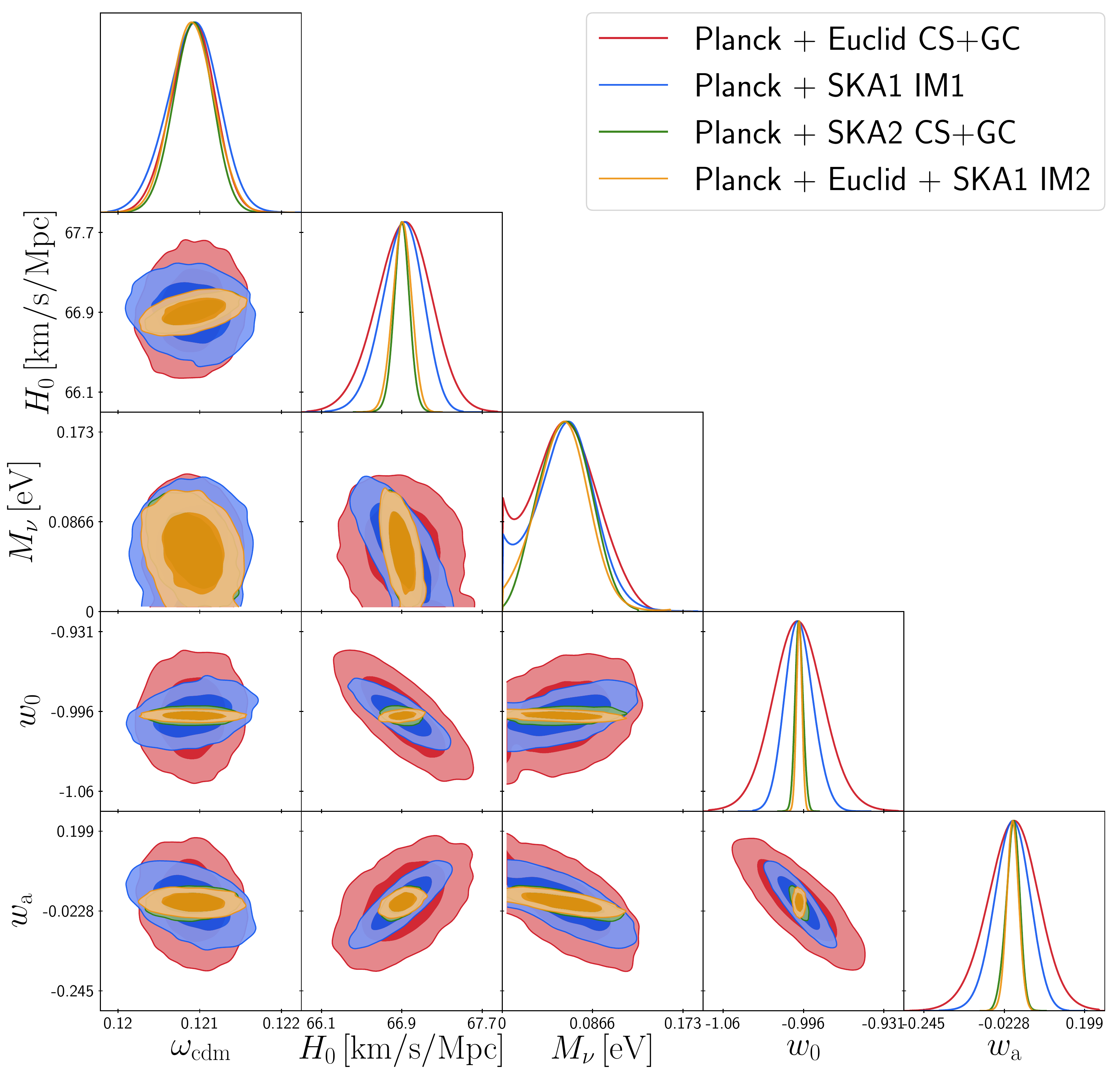}
\caption{Marginalized one- and two-$\sigma$ contours and one dimensional posteriors for the three extended models
$\Lambda$CDM+$M_\nu$+$N_\mathrm{eff}$ (top left), $\Lambda$CDM+$M_\nu$+$w_0$ (top right) and $\Lambda$CDM+$M_\nu$+$\left( w_0, w_a\right)$ (bottom left), in the realistic theoretical error scenario.}
\label{fig:ext_triangles}
\end{figure}

\begin{table}
\input{ext_Table.tex}
\caption{Expected $1\,\sigma$ sensitivity of Planck, Euclid and SKA to the cosmological parameters relevant in each extended model. For each probe combination the results in the top row are obtained with the conservative theoretical error approach applied to every observable; the results in the bottom row assume the realistic error prescription for GC and IM, while the conservative one is used for CS.}
\label{tab:ext}
\end{table}
\begin{figure}
\input{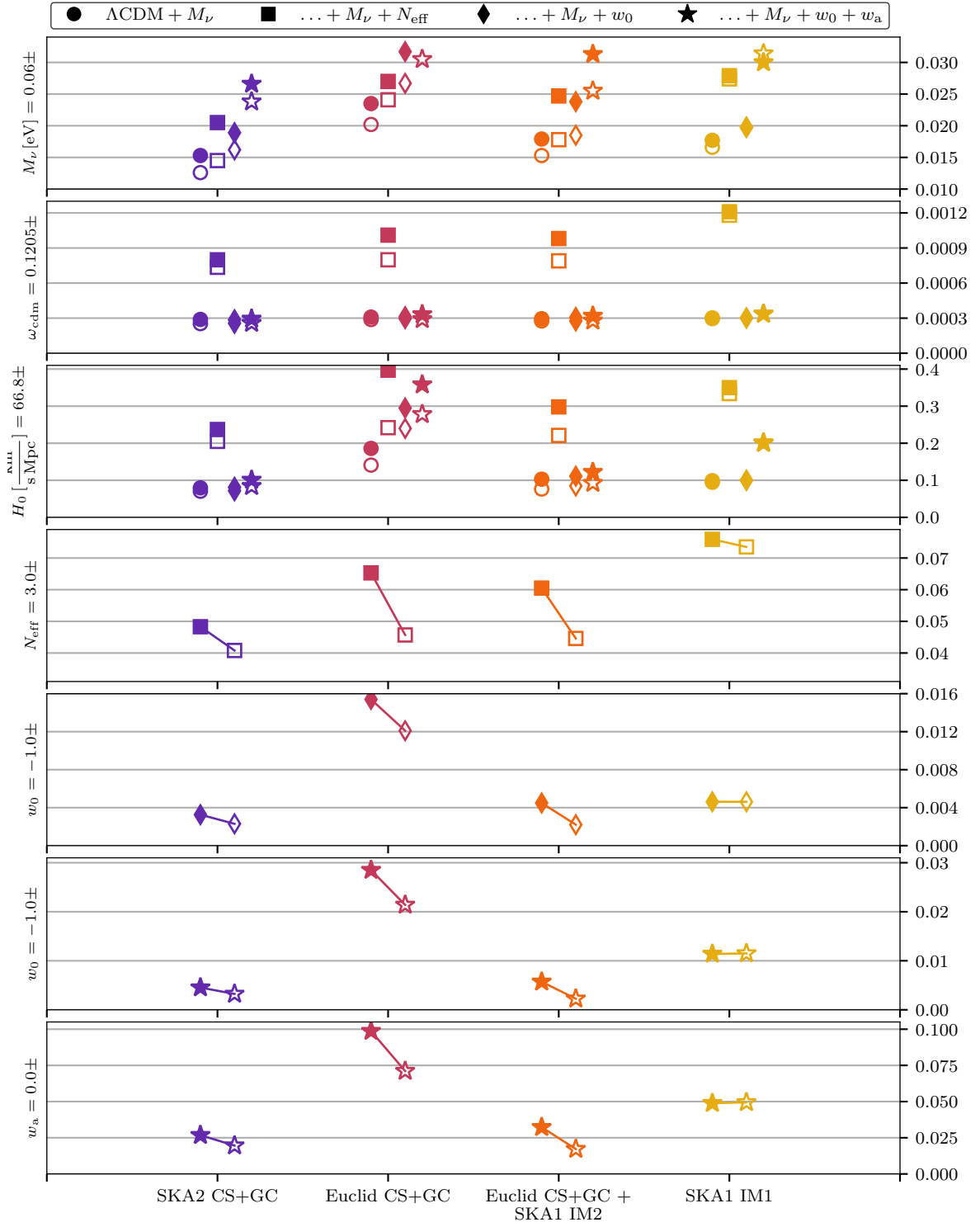}
\caption{$1 \, \sigma$ uncertainty for various combinations of experiments, all including Planck.
The different experiments, as well as the probe combinations, are specified on the x axis at the bottom of the plot.
The fiducial value is given on the left axis, while the $1 \, \sigma$ uncertainty is written on the right axis.
Finally, empty markers denote realistic settings, while filled markers denote conservative settings.
Different markers denote different extended models, according to the legend at the top of the plot.
}
\label{fig:ext_sigma}
\end{figure}
In Table~\ref{tab:ext} we report the expected sensitivity of various combinations of Planck, Euclid and SKA probes to the non-standard cosmological parameters and to those among the standard ones that show a relevant deviation from the baseline case.
In Figure~\ref{fig:ext_sigma} we depict the corresponding $1 \, \sigma$ uncertainty, to show how the sensitivity degrades in extended models. The decrease in the accuracy is caused by the degeneracies illustrated in the contour plots of Figure~\ref{fig:ext_triangles}.

We shall now proceed to discuss the results for each one of the three extended models. Notice that throughout this section, what we call the ``realistic'' case assumes the realistic theoretical error prescription for GC and IM, but still the conservative one for CS.

\subsection{$\Lambda$CDM+$M_\nu$+$N_\mathrm{eff}$}

We first promote the effective number of relativistic degrees of freedom $N_\mathrm{eff}$. This quantity parameterizes the radiation density ($\rho_r$) of the universe in the early universe beyond the photon density ($\rho_\gamma$),
\begin{equation*}
\rho_r=\rho_\gamma \left[1+ \frac{7}{8} \left( \frac{4}{11}\right)^{4/3} N_\mathrm{eff}\right].
\end{equation*}
In the cosmological standard model $N_\mathrm{eff}$ is equal to $3.045$~\cite{Mangano:2005cc,deSalas:2016ztq}, representing the three active neutrinos. We use this as our fiducial value. A deviation of $N_\mathrm{eff}$ from $3.045$ could be caused, for instance, by several plausible extensions of the Standard Model of particle physics, e.g.\,with sterile neutrinos~\cite{Archidiacono:2016kkh,Bridle:2016isd,Archidiacono:2012gv}.
Notice that a variation of $N_\mathrm{eff}$ implies a different expansion rate of the universe, with a profound impact on the cosmological observables at any 
redshift~\cite{Crotty:2004gm,Hou:2011ec,Archidiacono:2011gq,Lesgourgues:1519137,Archidiacono:2013fha}.

From Table~\ref{tab:ext} one can see that in the realistic theoretical error scenario the uncertainty is $\sigma(N_\mathrm{eff}) \leq 0.065$ for Planck + Euclid and Planck + SKA2. Such an accuracy  allows for a detection of more exotic models, e.g.\,with new bosons from new broken symmetries, leading to early decoupled or partially thermalized additional degrees of freedom~\cite{Baumann:2017gkg}.

In Figure~\ref{fig:ext_sigma} the predictions for this model appear as squared markers. We see that the sensitivity to $M_\nu$, $\omega_\mathrm{cdm}$ and $H_0$ degrades significantly with respect to the baseline model (circle markers) for every probe combination. In general, this degradation is caused by the strong degeneracies between $N_\mathrm{eff}$ and the aforementioned parameters. The importance of these correlations is illustrated in the top-left plot of Figure~\ref{fig:ext_triangles}. Observables that are more sensitive to the expansion rate of the Universe, such as cluster number counts, can break these degeneracies~\cite{Basse:2013zua}, and would bring back the sensitivity to all the baseline parameters close to the values obtained in the baseline model. Interestingly, the strongest correlations affect $\omega_\mathrm{cdm}$ and $H_0$, for which the reduction in sensitivity is more prominent than in the other dark energy extended models.
Instead, concerning $M_\nu$, adding SKA (IM2) to Euclid with a realistic theoretical error leads to $\sigma(M_\nu)=0.018$~eV, i.e.\,nearly the same sensitivity as in the baseline model, and a $3.3\sigma$ detection of a non-zero absolute neutrino mass. 
Therefore, the effect of $N_\mathrm{eff}$ on the cosmological probes can easily be disentangled from the effect of the neutrino mass sum with forthcoming galaxy and hydrogen surveys. 

\subsection{$\Lambda$CDM+$M_\nu$+$w_0$}

The second extended model includes a constant Dark Energy equation of state parameter $w_0$, with fiducial value $-1$ as in the $\Lambda$CDM model. From Table~\ref{tab:ext} and Figure~\ref{fig:ext_sigma}, one observes an important degradation (by almost a factor two) of the sensitivity to $M_\nu$ and $H_0$ for Euclid CS+GC and SKA2 CS+GC.   The importance of the parameter degeneracies with $w_0$ can be seen on the posterior distributions, displayed on Figure~\ref{fig:ext_triangles}.   But we also find that these degeneracies can be reduced, 1) by combining Euclid with SKA1-IM, 2) in the realistic theoretical error scenario, in such a way that the sensitivities to $M_\nu$ and $H_0$ are brought back to the ones of the baseline model.   

It is also found that the $1\, \sigma$ sensitivity to $w_0$ is improved almost by a factor two when considering the realistic theoretical error scenario, with $\sigma(w_0) = 0.0121 / 0.0023 / 0.0022 $ respectively for Planck+Euclid, Planck+SKA2 and Planck+Euclid+SKA1-IM.   It is therefore worth noticing that combining Euclid with the low-redshift SKA intensity mapping survey allows a very important improvement of the forecasted sensitivity to $w_0$. This is due to the intensity mapping of band 2 of SKA1 being sensitive down to very low redshift ($z_\mathrm{min}=0.05$).

\subsection{$\Lambda$CDM+$M_\nu$+$\left( w_0, w_a\right)$}

The third extended model allows for a variation of the Dark Energy equation of state with the CLP parameterization \cite{Chevallier:2000qy}
\begin{equation*}
w(a)=w_0+(1-a)w_a.
\end{equation*}
We adopt the $\Lambda$CDM values of these parameters $w_0 = -1$ and $w_a = 0$ in the fiducial model. 

In the conservative scenario, our results for Planck+Euclid are well compatible with those of Ref.~\cite{Amendola:2012ys}, and globally the sensitivities are significantly degraded with respect to the $\Lambda$CDM+$M_\nu$+$w_0$ model. In the conservative scenario, this is still true for the Planck+ SKA1-IM combination. However, as soon as we include CS+GC probes with a realistic error, the results are impressively stable and very mildly affected by the presence of an additional free parameter, as can be seen in Table~\ref{tab:ext} and Figure~\ref{fig:ext_sigma}. In Figure~\ref{fig:ext_triangles} we see the importance  of adding CS+GC information (green and yellow contours) in order to break the degeneracies of the Planck+SKA1-IM results (blue contours).

Ultimately, by combining Planck, Euclid CS+GC and SKA1-IM, one can expect sensitivities down to $\sigma(w_0) = 0.0023$ and $\sigma(w_a) = 0.017$, as well as $\sigma(M_\nu) = 0.026 \, {\rm eV}$.
By comparing with the results obtained for Planck  and Euclid CS+GC  only, we see that intensity mapping with SKA1 should lead to a useful increase of the sensitivity to the neutrino mass (allowing potentially for a 2.3$\sigma$- instead of 2.0$\sigma$-detection of the minimal mass), and to a very strong improvement in the sensitivity to the two DE parameters.

%% file: ext_Table.tex
\resizebox{1.0\textwidth}{!}{%
\begingroup
\setlength{\tabcolsep}{10pt}
\renewcommand{\arraystretch}{1.5}
\begin{tabular}{|c|c|c|c|c|c|}
\cline{3-6}
\multicolumn{2}{c|}{}&Planck+SKA2&Planck+Euclid&Planck+Euclid+SKA1&Planck+SKA1\\
\cline{2-2}
\multicolumn{1}{c|}{}&$\sigma$&(CS+GC)&(CS+GC)&(CS+GC)+(IM2)&(IM1)\\
\hline
\multirow{2}{*}{$\Lambda$CDM}&\multirow{2}{*}{$\omega_\mathrm{cdm}$}&0.00080&0.00101&0.00098&0.00121\\
&&0.00073&0.00080&0.00079&0.00118\\
\cdashline{2-6}
\multirow{2}{*}{+\,M$_{\nu}$}&\multirow{2}{*}{$H_0/[\frac{\mathrm{km}}{\mathrm{s\,Mpc}}]$}&0.237&0.397&0.298&0.350\\
&&0.205&0.242&0.221&0.334\\
\cdashline{2-6}
\multirow{2}{*}{+\,N$_{\mathrm{eff}}$}&\multirow{2}{*}{$M_\nu/[\mathrm{meV}]$}&20&27&25&28\\
&&14&24&18&27\\
\cdashline{2-6}
&\multirow{2}{*}{$N_{\mathrm{eff}}$}&0.048&0.065&0.060&0.076\\
&&0.041&0.046&0.045&0.073\\
\cdashline{2-6}
\hline
\multirow{2}{*}{$\Lambda$CDM}&\multirow{2}{*}{$\omega_\mathrm{cdm}$}&0.00029&0.00031&0.00031&0.00030\\
&&0.00026&0.00030&0.00028&0.00030\\
\cdashline{2-6}
\multirow{2}{*}{+\,M$_{\nu}$}&\multirow{2}{*}{$H_0/[\frac{\mathrm{km}}{\mathrm{s\,Mpc}}]$}&0.081&0.295&0.111&0.100\\
&&0.072&0.240&0.085&0.100\\
\cdashline{2-6}
\multirow{2}{*}{+\,w$_{0}$}&\multirow{2}{*}{$M_\nu/[\mathrm{meV}]$}&19&32&24&20\\
&&16&27&18&20\\
\cdashline{2-6}
&\multirow{2}{*}{$w_{0}$}&0.0033&0.0154&0.0045&0.0046\\
&&0.0023&0.0121&0.0022&0.0046\\
\cdashline{2-6}
\hline
\multirow{2}{*}{$\Lambda$CDM}&\multirow{2}{*}{$\omega_\mathrm{cdm}$}&0.00029&0.00033&0.00032&0.00034\\
&&0.00026&0.00029&0.00028&0.00034\\
\cdashline{2-6}
\multirow{2}{*}{+\,M$_{\nu}$}&\multirow{2}{*}{$H_0/[\frac{\mathrm{km}}{\mathrm{s\,Mpc}}]$}&0.101&0.358&0.122&0.200\\
&&0.084&0.278&0.092&0.202\\
\cdashline{2-6}
\multirow{2}{*}{+\,w$_{0}$}&\multirow{2}{*}{$M_\nu/[\mathrm{meV}]$}&27&40&31&30\\
&&24&30&26&31\\
\cdashline{2-6}
\multirow{2}{*}{+\,w$_{\mathrm{a}}$}&\multirow{2}{*}{$w_{0}$}&0.0045&0.0285&0.0057&0.0114\\
&&0.0032&0.0214&0.0023&0.0115\\
\cdashline{2-6}
&\multirow{2}{*}{$w_{\mathrm{a}}$}&0.027&0.099&0.032&0.049\\
&&0.020&0.071&0.017&0.050\\
\cdashline{2-6}
\hline
\end{tabular}
\endgroup
}%

%% file: Conclusions.tex
\section{Discussion and conclusion} 

The production of robust and accurate forecasts on the sensitivity to cosmological parameters of future surveys like Euclid and the Square Kilometre Array is an important task in the context of the preparation phase of these surveys.   A major difficulty comes from the theoretical uncertainties arising on mildly non-linear scale, induced by the complexity of physical processes at play, such as non-linear clustering and baryonic feedback. In this paper, we take these uncertainties into account with a method that has lots of similarities with previous attempts in Refs.~\cite{Audren:2012vy,Baldauf:2016sjb}. We discuss a compromise between the fully uncorrelated theoretical error of Ref.~\cite{Audren:2012vy}, which is arguably too conservative, and the approach of Ref.~\cite{Baldauf:2016sjb} based on an exponentially-decaying error correlation, which is too expensive numerically for MCMC forecasts. Our method is at the same time realistic, firmly rooted on physical results from various astrophysical studies, and computationally tractable. It relies on an ansatz for the error amplitude and correlation length on non-linear scales up to $k_{\rm max} = 10 \ h {\rm Mpc}^{-1}$.

For the first time, we present forecasts based on Markov--Chain--Monte--Carlo simulations for 4 cosmological scenarios, 2 different modelings of the non-linear theoretical error (one conservative and one realistic) and 14 experimental configurations and combinations: overall more than 140 MCMC simulations, which required of the order of one hundred thousand CPU hours.
This work is therefore the most exhaustive analysis of this kind released so far, both for Euclid and the Square Kilometre Array, and the combination of them.
It also paves the way towards a realistic implementation of the non-linear theoretical errors, going beyond the usual cut-off scale method that is i) too pessimistic on non-linear scales because it cannot exploit all the power of the survey, ii) at the same time too optimistic because it does not account for the theoretical uncertainties on the power spectrum below the mildly non-linear cut-off scale.
Three observational probes have been considered:  galaxy clustering and weak lensing (for Euclid, SKA1 and SKA2), and HI intensity mapping (for SKA1) at low redshift (probing the already reionized universe).   Compared to the usual Fisher approach, a bayesian MCMC method allows to probe non-gaussian posteriors and is immune from the sometimes critical numerical stability issues that are linked to the choice of the step size for numerical derivatives.  

Our main findings arise from the impact of the non-linear theoretical uncertainty on the power spectrum, and are summarized thereafter:
\begin{enumerate}
\item Despite the fact that the theoretical error removes a lot of information from large wavenumbers, we find that for galaxy clustering, when considering scales up to $k_{\rm max} = 10 \ h {\rm Mpc}^{-1}$ with the theoretical error, we increase the lever arm to constraint $n_s$ and improve the sensitivity to this parameter by about a factor two.  
\item Because of the degeneracy breaking, the former point also leads to an improvement for $H_0$ (by about 50\% for Planck+SKA1 and 25\% for Planck+Euclid) and for the total neutrino mass $M_\nu$ (by about 40\% for Planck+SKA1 and 25\% for Planck+Euclid). We also observe a factor two improvement on $\omega_{\rm cdm}$ for SKA1-GC. 
\item  The sensitivity to the derived parameters $\sigma_8$, $\Omega_{\rm m}$ and $\Omega_{\Lambda}$ significantly improves for nearly all the experimental configurations.
\item Concerning the extended cosmological models, there is a remarkable improvement for the constraints on the dark energy equation of state parameters $w_0$ and $w_a$.
\item With Planck+SKA2, the neutrino mass could be constrained to $\sigma(M_\nu)=0.012 \, {\rm eV}$ with the baseline model and assuming a realistic theoretical error, leading to a $5\sigma$-detection.
\item The degeneracy of several parameters (including $M_\nu$) with $N_{\rm eff}$, $w_0$ and $w_{\rm a}$ is not as severe as previously thought. CS+GC data with the realistic theoretical error assumption can break the degeneracies. The final sensitivity to e.g.\,the total neutrino mass is stable at least against these simple extensions of the minimal cosmological model.
\end{enumerate}
It is worth noticing that any source of error on the power spectrum can be incorporated in our non-linear treatment, eventually at the price of modifying the overall shape of the theoretical error and the correlation length, if this source is dominant. The assumed error on the spectrum has strong impact on parameter sensitivities. Therefore, we recommend to use an efficient implementation of the theoretical error, as the one illustrated here, in the analysis of forthcoming surveys.

We have studied the constraining power of each individual probes and the advantage of combining them. In particular, we considered the combination of the Euclid survey with the SKA HI intensity mapping survey. The sensitivity to various parameters is reported in Table~\ref{tab:mnu} and~\ref{tab:ext}.  Our main conclusions are:
\begin{enumerate}
\item Overall, SKA1-IM is more constraining than SKA1 (CS + GC), while the performance of Euclid lies between SKA2 and SKA1-IM.
\item The forecasts on $w_0$ and $w_a$ improve by up to a factor five when including non-linear scales with a realistic modelling of uncertainties and when combining Euclid with SKA1-IM. In the most constraining scenario we obtain $\sigma(w_0) =0.0023 $ and $\sigma(w_a) = 0.017$, i.e.\,a factor forty of improvement compared to Planck alone. 
\end{enumerate}
These results emphasize the importance of combining Euclid with a survey extending the information down to very low redshift, such as SKA1 intensity mapping. This combination, together with the modelling of non-linear uncertainties, could make the difference between a $1.5\sigma$ hint and a stronger $2.3\sigma$ hint of the total neutrino mass when the DE is modelled with two free parameters. It is even more crucial for constraining the DE parameters themselves.

Our method to deal with the non-linear uncertainties is only a first proxy that could be made more accurate and updated with the results of future N-body simulations, with a better understanding of the baryonic feedback, or with analytical progress on any other source of error.  The suggested implementation can nevertheless be considered as a realistic target, given that it is based on the current understanding of those processes and on conservative assumptions about the expected precision of future N-body simulations.  For intensity mapping, our analysis could be refined by using a more precise foreground modelling.  

Realistic forecasts could be produced for other cosmological scenarios, e.g., specific dark-energy/modified-gravity parameterizations, or for assessing the accuracy of bayesian selection of inflationary models, see Refs.~\cite{Martin:2013nzq,Martin:2014rqa}. 

Obviously, extending the parameter space leads to more pronounced parameter degeneracies and call for additional data. Extra constraining power might come from independent probes (e.g., the 21cm intensity mapping from reionization provided by SKA), and also from the cross-correlations between different probes (e.g., between galaxy shear and clustering), left for further study.

%% file: AppGC.tex
\section{Galaxy clustering likelihood}
\label{Apendix:Galaxy clustering likelihood}
The galaxy power spectrum is defined as a function of a continuous density field, which represents
the probability density of finding a galaxy at some position $\bm{r}$. The galaxy density perturbation $\delta_g$ is then a perturbation of this probability density $p_g$:
\begin{equation}
p_g(\bm{r}) = \bar{n}(\bm{r})(1+\delta_g(\bm{r})) \ .
\end{equation}
$\bar{n}(\bm{r})$ is the expected number density of galaxies on a homogeneous background; it is calculated as the mean density over a sufficiently large volume. In our case this will be the volume corresponding to one redshift bin.

Starting from this idea, Ref.~\cite{Feldman:1993ky} derived a method to estimate the galaxy power spectrum $P_g$, with an error that will be used in the following to build up the likelihood. However, it is worth noting that this is a simplified approach ignoring the possible effect of galaxy properties such as luminosity (see e.g.~\cite{Smith:2015yya}). The result is a gaussian error:
\begin{equation}
\sigma_P^2 = \frac{(2\pi)^3}{V_k V_r} P_\mathrm{obs}^2(\bm{k})= \frac{(2\pi)^3}{V_k V_r}\left(P_g(\bm{k})+P_N\right)^{2} \ .
\end{equation}
The quantity in parentheses is the observable power spectrum of Eq.~(\ref{eq:pobs}) splitted into the part proportional to the matter power spectrum and the shot noise $P_N=1/\bar{n}=V_r/N$, where $N$ is the total number of detected galaxies inside the observed volume $V_r$. The volume $V_k$ of the shell in $k$-space over which the estimator is averaged has to be big enough to cancel the effects of performing a Fourier transform on a finite volume. It can be chosen to be as small as $V_k = \frac{(2\pi)^3}{V_r}$, where $V_r$ is the volume of the single redshift bin. Inside this volume Fourier modes are not independent. The likelihood is thus the product of as many gaussians as the independent Fourier modes (one per $V_k$):
\begin{equation}
{\cal L} = {\cal N} \exp \left[ -\frac{1}{2} \sum_{\text{independent }\bm{k}} 
\frac{V_k V_r}{(2\pi)^3}
\frac{ \left( \hat{P}_\mathrm{obs}(\bm{k}) -P_\mathrm{obs}(\bm{k})\right)^2}{P_\mathrm{obs}^2(\bm{k})}\right] \ .
\end{equation}
In the case of forecasts $\hat{P}_\mathrm{obs}$ does not represent actual measurements, but it is mock data corresponding to some fiducial cosmology, computed in the same way as the theoretical $P_\mathrm{obs}$. Hence, $\hat{P}_\mathrm{obs}$ and $P_\mathrm{obs}$ do not suffer from finite volume effects and are thus smooth, i.e. approximately constant, inside $V_k$. To replace the sum by an integral over the whole $k$-space an additional factor of 1/2 has to be introduced to account for the fact that the power spectrum is the Fourier transform of a real quantity, $P_\mathrm{obs}(\bm{k})=P_\mathrm{obs}(-\bm{k})$.
Reformulated in terms of $\chi^2=-2\ln{\cal L}$, the result is
\begin{equation}
\label{chi2-result}
\chi^2 = \sum_{\bar{z}}\int \dd^3k \frac{V_r(\bar{z})}{2(2\pi)^3}\frac{\left(\hat{P}_\mathrm{obs}(\bm{k},\bar{z})-P_\mathrm{obs}(\bm{k},\bar{z})\right)^2}{P_\mathrm{obs}^2(\bm{k},\bar{z})} \ .
\end{equation}
The dependencies of all quantities are shown here for clarity. Inside a redshift bin, all quantities are evaluated at the mean redshift $\bar{z}$ of that bin. In other words, anything is evaluated at the same time. This approximation has to be made to get information on the equal-time three-dimensional power spectrum $P_\mathrm{obs}(\bm{k},\bar{z})$.

The change of coordinates and Fourier modes depending on the choice of the cosmological model has already been discussed in Sec.~\ref{sec:GC}. Here, in order to estimate the $\chi^2$ we have to deal with three different cosmologies: The fiducial one used to compute $\hat{P}_\mathrm{obs}$, the one used to compute $P_\mathrm{obs}$ which we want to compare to the mock data and, finally, the one used to interpret the observations, in which all quantities of Eq.~\ref{chi2-result} are defined. The choice of the last one is arbitrary since all conversion factors in Eq.~\ref{chi2-result} cancel. Therefore, we can assume it to be equal to the fiducial one. To clarify what was done in the computation of forecasts, \cref{chi2-result} can be rewritten such that every quantity depending on the underlying cosmology is labelled with either $f$ (fiducial cosmology) or $s$ (sample cosmology):
\begin{equation}
\chi^2 = \sum_{\bar{z}}\int (k^f)^2\dd k^f \int_{-1}^{1} \dd \mu^f \frac{V_r^f}{2(2\pi)^2} \left[\frac{\dfrac{H^f }{(D_A^f)^2}P^f_\mathrm{obs}(k^f,\mu^f)-\dfrac{H^s }{(D_A^s)^2} P^s_\mathrm{obs}(k^s,\mu^s)}{\dfrac{H^s }{(D_A^s)^2}P^s_\mathrm{obs}(k^s,\mu^s)}\right]^2 \ .
\end{equation}
We can now replace the observed power spectrum as a function of the galaxy power spectrum and of the shot noise, and make use of the relation between the volumes in the two different spaces:
\begin{equation}
\dfrac{H^s }{(D_A^s)^2} \dfrac{V_r^s}{N}=\dfrac{H^f }{(D_A^f)^2}\dfrac{V_r^f}{N}~.
\end{equation}
Then the shot noise exactly cancels from the numerator and we are left with
\begin{equation}
\chi^2 = \sum_{\bar{z}}\int (k^f)^2\dd k^f \int_{-1}^{1} \dd \mu^f \frac{V_r^f}{2(2\pi)^2} \left[\frac{\dfrac{H^f }{(D_A^f)^2}P^f_g(k^f,\mu^f)-\dfrac{H^s }{(D_A^s)^2} P^s_g(k^s,\mu^s)}{\dfrac{H^s }{(D_A^s)^2}P^s_g(k^s,\mu^s)+\dfrac{H^f }{(D_A^f)^2}\dfrac{V_r^f}{N}}\right]^2 \ .
\end{equation}
The prefactor $f_{\text{AP}}$ of Eq.~\ref{P_g-def} has been written explicitely so every power spectrum can be evaluated in the same cosmology which was used to produce it.
To compute the integral, $k^s$ and $\mu^s$ must be expressed in terms of $k^f$ and $\mu^f$ as described in Eqs.~(\ref{k^s}, \ref{mu^s}).

Note that in this formula we correct some small inaccuracies present in the previous work by Ref.~\cite{Audren:2012vy}. In that reference, the pre-factor $V_r^f$ was incorrectly replaced by $V_r^s(1+\bar{z})^{-3}$, the second argument of $P_g^s$ was approximated as $\mu^f$ instead of $\mu^s$, and the volume in the last term of the denominator was $V_r^s$ instead of $V_r^f$. We checked explicitly that these inaccuracies led to slightly over-conservative error forecasts in Ref.~\cite{Audren:2012vy}.

%% file: AppCS.tex
\section{Cosmic shear likelihood}
\label{Appendix:lensing_likelihood}

The likelihood for lensing surveys is taken from Ref.~\cite{Audren:2012vy}:
\begin{equation}
\label{LensingLikelihood}
-2\ln{\cal L} \equiv \sum_l(2l+1)f_{\text{sky}}\left(\frac{d_l^{\text{mix}}}{d_l^{\text{th}}}+\ln\frac{d_l^{\text{th}}}{d_l^{\text{obs}}}-N\right) \ .
\end{equation}
$N$ is the number of considered redshift bins which is equal to the dimension of the $C_l$-matrices whose determinants are denoted with $d$. The determinant of these $N \times N$ symmetric matrices can be defined as follows:
\begin{align}
d_l^{\rm th} &= \det \left( C_l^{{\rm{th}}\,ij} + N_l^{ij} \right) \ , \\
d_l^{\rm obs} &= \det \left( C_l^{{\rm{fiducial}}\,ij} + N_l^{ij} \right) \ , \\
d_l^{\rm mix} &= \sum_k \det\left( N_l^{ij} + \begin{cases} C_l^{{\rm{th}}\,ij} &, j\neq k\\[10pt] C_l^{{\rm{fiducial}}\,ij} &, j=k \end{cases} \right)
\end{align}
Writing the theoretical angular power spectrum as $C_l$ and the observational one as $\hat{C}_l$, both including noise, \cref{LensingLikelihood} can be followed from a multivariate gaussian:
\begin{equation}
{\cal L} = {\cal N} \prod_{l,m} \left\lbrace \frac{1}{\sqrt{\det C_l}}\exp\left[ -\frac{1}{2} \sum_{i,j} a^{i \ast}_{lm} (C_l^{-1})^{ij} a^{j}_{lm} \right] \right\rbrace \ .
\end{equation}
The observed angular power spectrum is defined as
\begin{equation}
\hat{C}_l^{ij} = \frac{1}{2l+1}\sum_{m=-l}^l a^{i \ast}_{lm}a^{j}_{lm} \ .
\end{equation}
The inverse of a matrix $A$ can be replaced by $A^{-1}=\text{adj}(A)/\det (A)$ where $\text{adj}(A)$ is the adjugate which is the transpose of the cofactor matrix of $A$. The likelihood can thus be rewritten by executing the sum over $m$:
\begin{equation}
{\cal L} = {\cal N} \prod_{l} \left\lbrace \left(\frac{1}{\sqrt{\det C_l}}\right)^{2l+1}\exp\left[ -\frac{1}{2} \sum_{i,j}(2l+1)\hat{C}_l^{ij}\frac{(\text{adj}\,C_l)^{ij}}{\det C_l} \right] \right\rbrace \ .
\end{equation}
The remaining sum over $i$ and $j$ yields exactly $d_l^{\rm mix}$. This can best be explained graphically:
\begin{equation}
\sum_{i,j}\hat{C}_l^{ij}(\text{adj}\,C_l)^{ij} = \sum_{i,j} \hat{C}_l^{ij}
\begin{vmatrix}
... & 0 & ... \\[2pt]
0 & \hat{C}_l^{ij} & 0 \\[2pt]
... & 0 & ...
\end{vmatrix}
= \sum_{j}
\begin{vmatrix}
... & \hat{C}_l^{i_1 j} & ... \\[2pt]
... & \hat{C}_l^{i_2 j} & ... \\[2pt]
... & ... & ...
\end{vmatrix}
= d_l^{\rm mix} \ .
\end{equation}
We can now compute $\chi^2$:
\begin{equation}
\chi^2 = -2\ln{\cal L} = -2\ln{\cal N} + \sum_l (2l+1)\left( \ln d_l^{\rm th} + \frac{d_l^{\rm mix}}{d_l^{\rm th}} \right) \ .
\end{equation}
Our best fit model where theoretical results are the same as observational results,
\begin{equation}
\chi_0^2 = -2\ln{\cal L} = -2\ln{\cal N} + \sum_l (2l+1)\left( \ln d_l^{\rm obs} + \frac{N \times d_l^{\rm obs}}{d_l^{\rm obs}} \right) \ ,
\end{equation}
shall yield an effective $\chi^2$ of zero, so it has to be subtracted. After introducing an approximative correction for incomplete sky coverage\cite{Audren:2012vy}, Eq.~\ref{LensingLikelihood} is obtained.

%% file: AppIM.tex
\section{21cm intensity mapping likelihood}
\label{Appendix:21cm_likelihood}
The total brightness temperature at redshift $z$ is given by the background radiation field's temperature, with some fraction of it that is absorbed and re-emitted due to 21cm hyperfine transitions in neutral hydrogen atoms. The properties of HI in absorption and emission are described by the spin temperature $T_S$ and the optical depth $\tau$:
\begin{equation}
T_{\rm b} = T_S(1-e^{-\tau})+T_{\gamma}e^{-\tau} \ .
\end{equation}
Due to the low probability of a 21cm transition, the optical depth is typically small. The differential brightness temperature can thus be written linear in $\tau$:
\begin{equation}
\label{DTb_tau}
\Delta T_{\rm b} = \frac{T_S-T_{\gamma}}{1+z}\left(1-e^{-\tau}\right) \approx \frac{T_S-T_{\gamma}}{1+z}\tau \ .
\end{equation}
In order to compute $\tau$, the absorption coefficient $\alpha$ has to be determined through the equation of radiative transfer:
\begin{equation}
\label{radtransfer}
\frac{\dd I}{\dd s} = -\alpha I + j \ ,
\end{equation}
where $s$ is the radial distance (in physical units) and $I$ is the specific intensity, which is the energy flux per frequency and solid angle. Its radial derivative is given by
\begin{equation}
\label{dIds_dn0dt}
\frac{\dd I}{\dd s} = E_{10}\frac{\phi(\nu)}{4\pi}\frac{\dd n_0}{\dd t} \ .
\end{equation}
Each atom falling from the exited state 1 into the ground state 0 emits a photon of energy $E_{10}$. The radial derivative of the energy flux is hence proportional to the time derivative of the number of atoms in the ground state per unit of physical volume, i.e. the number density $n_0$. Under the assumption of isotropy, the derivative with respect to solid angle becomes a factor of $1/4\pi$. Due to line broadening, a single measured frequency corresponds to a small band of emitted frequencies described by the line profile $\phi(\nu)$ which is normalized to $\int \phi(\nu) \dd \nu = 1$.

In terms of Einstein coefficients the time derivative of the number density can be written as $\dd n_0/\dd t = -n_0 B_{01} I + n_1 B_{10} I + n_1 A_{10}$. In the next steps we use natural units where $\hslash = c = k_B = 1$ as well as the general relations $A_{10} = 4\pi\nu_0^3 B_{10}$ and $g_0 B_{01} = g_1 B_{10}$.  For 21cm hyperfine transitions, the statistical weights are $g_0 = 1$ and $g_1 = 3$ and one gets $T_S \gg E_{10}$.  Therefore, one gets the simplifications 
\begin{equation}
\frac{n_1}{n_0} = \frac{g_1}{g_0} \exp(-\frac{E_{10}}{T_S}) \approx 3(1-\frac{E_{10}}{T_S})~,
\end{equation} 
and $n_{\text{HI}} \equiv n_0+n_1 \simeq 4n_0 \simeq \frac{4}{3} n_1$.  Put together, one gets
\begin{equation}
\frac{\dd n_0}{\dd t} = -\frac{A_{10}}{4\pi\nu_0^3} \frac{3}{4}n_{\text{HI}}\frac{E_{10}}{T_S} I + \frac{3}{4} n_{\text{HI}} A_{10} \ ,
\end{equation}
which gives an expression for $\alpha$, 
\begin{equation}
\alpha = \frac{3A_{10}}{16T_S}\frac{\phi(\nu)}{\nu_0}n_{\text{HI}} \ .
\end{equation}
The line profile will be described by the simple model of a constant distribution over some range $\delta \nu$, corresponding to a small Doppler shift caused by constant velocity dispersion $\frac{dv}{ds}$ over a region of HI of radial extend $\delta s$:
\begin{equation}
\phi(\nu) = \frac{1}{\delta \nu} = \frac{1}{\frac{\dd v}{\dd s}\delta s \cdot\nu_0} \ .
\end{equation}
Averaged over big volumes, the approximation of a constant Hubble flow $\frac{dv}{ds} = H(z)$ can be used.  The optical depth is then given by
\begin{equation}
\tau \equiv \int_{\delta s} \alpha \dd s = \frac{3A_{10}}{16\nu_0^2 T_S}\frac{1}{H(z)}n_{\text{HI}} \ .
\end{equation}
The number density of neutral hydrogen can be written as its background value plus a perturbation in the HI density field,
\begin{equation}
n_{\text{HI}} = \frac{(1+z)^3}{m_{\text{H}}}\frac{3 H_0^2}{8\pi G}\Omega_{\text{HI}}(z)(1+\delta_{\text{HI}}) \ .
\end{equation}
With all constants written explicitly, the differential brightness temperature is given by
\begin{equation}
\Delta T_b = \frac{3A_{10}\cdot 3H_0}{16\nu_0^2\cdot 8\pi G h m_{\text{H}}} \frac{\hslash c^3}{k_B}\cdot \left(\frac{H_0 (1+z)^2}{H(z)}\right)\Omega_{\text{HI}}(z)(1+\delta_{\text{HI}}) h \left(1-\frac{T_{\gamma}}{T_S}\right) \ .
\end{equation}
The last term can be neglected because $T_S \gg T_{\gamma}$ inside galaxies, so that Eq.~\ref{DelTb189} is obtained. This modeling of the differential brightness temperature was e.g. used by Refs.~\cite{Battye:2012tg,Hall:2012wd,Bull:2014rha}.

When computing the power spectrum of fluctuations in the differential brightness temperature, it is convenient to neglect the local fluctuations of $H(z)$.
As a consequence, the power spectrum $P_{21}$ is proportional to the power spectrum $P_{\text{HI}}$ of HI density fluctuations, as described in \cref{subsection:21cm power spectrum}.

By considering $\delta(\bm{k})$ as a set of independent gaussian random realizations, the variance is simply given by the power spectrum squared. In terms of independent modes of a finite volume survey, it is corrected by the volume of a single independent mode $(2\pi)^3/V_r$ and by the averaging volume $V_k$ which determines the grid of sampled modes:
\begin{equation}
\sigma_P^2(\bm{k}) = \frac{(2\pi)^3}{V_k V_r}P^2(\bm{k}) \ .
\end{equation}
So the same formalism as described in \cref{Apendix:Galaxy clustering likelihood} can be used.  Note that the noise power may take mathematically the role of the shot noise of galaxy power spectra but is part of the power spectrum itself while shot noise is only an artefact caused by the discrete nature of the signal, in contrast to the theoretical distribution used to describe it. In fact, there is also a shot noise in the case of intensity mapping since the signal still originates from discretely spaced galaxies. Nevertheless, it is negligible (see e.g. \cite{Olivari:2017bfv}) because of the huge number of observed galaxies, when no selection process is reducing their number.